\documentclass{aa}  
\usepackage{mathtools}
\usepackage{lscape}
\usepackage{longtable,threeparttable}
\usepackage{rotating} 
\usepackage{graphicx}
\usepackage{multicol}
\usepackage{wasysym}
\usepackage{tablefootnote}
\newcommand{\Msun}{\mbox{$\mathrm{M}_{\odot}$}}
\usepackage{txfonts,booktabs}
\usepackage{hyperref}
\hypersetup{
	unicode=false,
	pdftoolbar=true,
	pdfmenubar=true,
	pdffitwindow=false,
	pdfstartview={FitH},
	pdftitle={My title},
	pdfauthor={Author},
	pdfsubject={Subject},
	pdfcreator={Creator},
	pdfproducer={Producer},
	pdfkeywords={keyword1} {key2} {key3},
	pdfnewwindow=true,
	colorlinks=true,
	linkcolor=red,
	citecolor=blue,
	filecolor=magenta,
	urlcolor=cyan
}

\usepackage{txfonts}
\begin{document}

	\title{The EREBOS project -- Investigating the effect of substellar and low-mass stellar companions on late stellar evolution}
	\subtitle{Survey, target selection and atmospheric parameters}
	\titlerunning{The EREBOS project}
	\authorrunning{V. Schaffenroth et al.}
	\author{V. Schaffenroth \inst{1} \and B.N. Barlow\inst{2} \and S. Geier\inst{1} \and M. Vu{\v c}kovi{\'c}\inst{3} \and D. Kilkenny\inst{4} \and M. Wolz\inst{5}   \and T. Kupfer\inst{6}  \and U. Heber\inst{5} \and H. Drechsel\inst{5}  \and S. Kimeswenger\inst{7,8}  \and T. Marsh\inst{9} \and M. Wolf\inst{10} \and I. Pelisoli\inst{1} \and J. Freudenthal\inst{11} \and S. Dreizler\inst{11} \and S. Kreuzer\inst{5} \and E. Ziegerer\inst{5} } 
	\institute{
		{Institut für Physik und Astronomie, Universität Potsdam, Haus 28, Karl-Liebknecht-Str. 24/25, 14476, Potsdam-Golm, Germany
			\email{schaffenroth@astro.physik.uni-potsdam.de}}
		\and
		Department of Physics, High Point University, One University Parkway, High Point, NC 27268, USA
		\and  Instituto de F\'{i}sica y Astronom\'{i}a, Facultad de Ciencias, Universidad de Valpara\'{i}so, Gran Breta\~{n}a 1111, Playa Ancha, Valpara\'{i}so 2360102, Chile
		\and Department of Physics \& Astronomy, University of the Western Cape, Private Bag X17, Bellville 7535, South Africa
		\and
		Dr.\,Remeis-Observatory \& ECAP, Astronomical Institute, Friedrich-Alexander University Erlangen-N\"urnberg, Sternwartstr.~7, 96049
		Bamberg, Germany
		\and        Kavli Institute for Theoretical Physics, University of California, Santa Barbara, CA 93106, USA     
		\and
		Institut f{\"u}r Astro- und Teilchenphysik, Universit{\"a}t Innsbruck, Technikerstrasse 25/8, 6020 Innsbruck, Austria
		\and Instituto de Astronom{\'i}a, Universidad Cat{\'o}lica del Norte, Av. Angamos 0610, Antofagasta, Chile
		\and
		Department of Physics, University of Warwick, Coventry CV4 7AL, UK
		\and Astronomical Institute, Faculty of Mathematics and Physics, Charles University, V Holešovičkách 2, 180 00, Praha 8, Czech Republic  
		\and
		Institut f\"ur Astrophysik, Georg-August Universit\"at G\"ottingen, Friedrich-Hund-Platz 1, 37077 G\"ottingen, Germany
	}
	\date{Received 4 June 2019 / Accepted 19 July 2019}
	
	
	
	
	\abstract{
		Eclipsing post-common envelope binaries are highly important for resolving the poorly understood, very short-lived common envelope phase of stellar evolution.
		Most hot subdwarfs (sdO/Bs) are the bare helium-burning cores of red giants which have lost almost all of their hydrogen envelopes. This mass loss is often triggered by common envelope
		interactions with close stellar or even sub-stellar companions. 
		Cool companions to hot subdwarf stars such as late-type stars and brown dwarfs are
		detectable from characteristic light curve variations -- reflection effects and often eclipses. 
		In the recently published catalog of eclipsing binaries in the Galactic Bulge and in the ATLAS (Asteroid Terrestrial-impact Last Alert System) survey, we discovered 125 new eclipsing systems showing a reflection effect by visual inspection of the light curves and using a machine-learning algorithm, in addition to the 36 systems discovered before by the OGLE (Optical Gravitational Lesing Experiment) team. The EREBOS (Eclipsing Reflection Effect Binaries from Optical Surveys) project
		aims at analyzing all newly discovered eclipsing binaries of the HW Vir type (hot subdwarf + close, cool companion) based on a spectroscopic and photometric follow up to derive the mass distribution of the companions, constrain the fraction of sub-stellar companions and determine the minimum mass needed to strip off the red-giant envelope.
		To constrain the nature of the primary we derived the absolute magnitude and the reduced proper motion of all our targets with the help of the parallaxes and proper motions measured by the Gaia mission and compared those to the Gaia white dwarf candidate catalogue. For a sub-set of our targets with observed spectra the nature could be derived by measuring the atmospheric parameter of the primary confirming that less than 10\% of our systems are not sdO/Bs with cool companions but white dwarfs or central stars of planetary nebula. This large sample of eclipsing hot subdwarfs with cool companions allowed us to derive a significant period distribution for hot subdwarfs with cool companions for the first time showing that the period distribution is much broader than previously thought and ideally suited to find the lowest mass companions to hot subdwarf stars. The comparison with related binary populations shows that the period distribution of HW Vir systems is very similar to WD+dM systems and central stars of planetary nebula with cool companions. In the future several new photometric surveys will be carried out, which will increase the sample of this project even more giving the potential to test many aspects of common envelope theory and binary evolution.
	}
	
	\maketitle
	%
	
	\section{Introduction}
	
	Most subluminous B stars (sdBs) are core helium-burning stars with very thin hydrogen
	envelopes and masses around $0.5$ \Msun\, \citep{heber:2009,heber:2016}. To form such an object, the hydrogen
	envelope of the red-giant progenitor must be stripped off almost entirely. Since a high fraction of sdB
	stars are members of short-period binaries \citep{maxted:2001}, common envelope ejection
	triggered by a close stellar companion is generally regarded as the most probable formation channel for many of the sdB stars.
	
	There is, however, increasing evidence that sub-stellar companions might also have a significant influence on sdB star formation (which is still poorly understood) and it has been proposed that planets and brown dwarfs could be responsible for the loss of envelope mass in the red-giant phase of sdB progenitors \citep{soker}. As soon as the primary star evolves to become a red giant, close sub-stellar companions must enter a common envelope. Whether those objects are able to eject the envelope and survive, evaporate, or merge with the stellar core depends mostly on their masses. While planets below $10\,{\rm M}_{\rm Jup}$ might not survive the interactions, companions exceeding this mass might be able to eject the envelope and survive as close companions \citep{soker}.
	
	The best evidence for interactions with sub-stellar companions is provided by the discovery of three close, eclipsing sdB binaries
	with brown dwarf companions. Photometric and spectroscopic follow-up observations of the sdB binary J162256+473051 revealed that the system is eclipsing with a period of 0.069 d and the companion is probably a brown dwarf with a mass of 0.064 ${\rm M}_{\rm \odot}$ \citep{vs:2014_I}. The short period system J082053+000843 (0.096 d) is also eclipsing, and the companion has a mass in the range 0.045 - 0.067 ${\rm M}_{\rm \odot}$ \citep{geier}. \citet{vs:2015a}  discovered  another, specially interesting eclipsing hot subdwarf, showing pulsations, with a brown dwarf companion with a mass of 0.069 ${\rm M}_{\rm \odot}$ (V2008-1753, 0.065 d). Additionally, two sdB systems with candidate brown dwarf companions have been detected \citep[periods $ \sim 0.3\rm\,d$,][]{vs:2014_II}, but since they do not eclipse, only minimum companion masses --- both below the hydrogen burning limit --- can be derived (0.048 and 0.027 \Msun).
	
	The most successful way to detect eclipsing binaries with cool stellar or sub-stellar companions (HW Vir systems) is by inspecting their light curves, which show besides the eclipses a characteristic quasi-sinusoidal variation caused by the so--called reflection effect \citep[see][and references therein]{muchfuss_photo}. This effect is observed in any close binary systems consisting of a hot primary and cool companion. As the secondaries in these systems are supposed to orbit synchronously, the hemisphere of the cool companion facing the hot primary is constantly irradiated, which leads to an increased flux over the orbital phase as the heated side of the secondary comes in the view \citep[see][for a detailed discussion of this effect]{wilson:1990,budaj}.
	
	The amplitude of the reflection effect scales with the temperature ratio and the radii of the primary and secondary stars, as well as the inverse orbital separation \citep{wilson:1990,budaj}.
	Hence, the reflection effect is strongest when both components of a close binary system have a very small separation, similar radii, and a high temperature difference. In systems consisting of a low-mass main sequence star or brown dwarf and a hot, compact star like a hot subdwarf, these conditions are fulfilled. 
	Very hot white dwarf binaries with cool companions, such as NN Serpentis \citep[]{NN_ser} or post-AGB stars with cool, main sequence companions, including central stars of planetary nebula \citep[e.g., UU Sge,][]{pn} show the same features.
	
	Due to the short periods and similar radii of the components, sdB binaries with low-mass companions also have a high probability of eclipsing. Such systems are of great value because they allow the determination of the masses and radii of both components, as well as their separation, with an accuracy up to a few percent using  combined photometric and spectroscopic analyses. 
	The known sample of eclipsing and non-eclipsing systems is quite inhomogeneous, with most having been found in photometric surveys by their characteristic light curve variations \citep[e.g.,][]{vs} or from light curves only covering a few hours aimed at searching for pulsations \citep[e.g.,][]{jeffrey:2014}. 
	These published systems have orbital periods between 0.069 d and 0.26 d \citep[][and references therein]{muchfuss_photo}.
	
	A photometric follow-up of spectroscopically selected targets \citep{muchfuss_photo} also allowed us to determine the fraction of sub-stellar companions to sdBs. We derived a value of > 8\% for sub-stellar companions in close orbits around sdBs. Moreover, they seem to be at least as frequent as low-mass stellar companions, as two of the four reflection effect binaries we discovered are sdBs with sub-stellar companions.
	
	This shows that close sub-stellar companions are able to eject a red-giant envelope and that they are much more frequent than predicted by standard binary evolution theory. Due to their high fraction of close sub-stellar companions, hot subdwarfs are well suited to study the interactions between stars and brown dwarfs or giant planets.  To understand both the common envelope phase under extreme conditions and the role of close-in planets and brown dwarfs for late stellar evolution, we need to study a large and homogeneously selected sample of eclipsing sdB binaries.
	\section{Project overview}
	\label{erebos}
	The increasing number of photometric surveys provides us with a huge source to find more eclipsing hot subdwarf stars.
	Thirty-six new HW\,Vir candidates have been discovered by the Optical Gravitational Lensing Experiment (OGLE) project \citep{ogle,ogle_II}, almost tripling the number of such objects known and providing the first large sample of eclipsing sdBs. These systems have been identified by their blue colors, short orbital periods, and characteristic light curves in the I-band.
	
	To investigate this unique sample of HW Vir candidates, we conduct the EREBOS (\textbf{E}clipsing \textbf{R}eflection \textbf{E}ffect \textbf{B}inaries from \textbf{O}ptical \textbf{S}urveys) project, which aims to measure orbital, atmospheric, and fundamental parameters.
	
	Key questions we want to answer over the course of the EREBOS project include the following: What is the minimum mass of the companion necessary to eject the common envelope? Is there a well defined minimum mass or a continuum ranging from the most massive brown dwarfs down to hot Jupiter planets? What is the fraction of close sub-stellar companions to sdBs and how does it compare to the possible progenitor systems such as main sequence stars with brown dwarf or hot Jupiter companions? To address these issues and understand both the common envelope phase under extreme conditions and the role of close-in planets and brown dwarfs for late stellar evolution, we need to know the parameters of post-common envelope systems over as much of the period distribution as possible.

	In the following sections we will discuss our target selection, try to constrain the nature of the primary star, and give first results based on the orbital parameters from the light curves and atmospheric parameters from the spectra we have taken so far.
	\section{Target selection}
	\label{selection}
	As already described, hot subdwarf binaries with cool, low mass companions exhibit a unique light curve shape, with their strong eclipses and reflection effect. By visual inspection of light curves, the OGLE team identified 36 new HW Vir candidates in their sample \citep{ogle,ogle_II}. In the course of the OGLE survey, approximately 450\,000 new eclipsing binaries have been published to date \citep{ogle_III}. More light curve surveys monitoring billions of stars are available. 
	As we are searching for systems with a defined set of characteristics, we developed a set of criteria
	to select a limited number of potential targets for our own visual inspection.
	
	\subsection{Light curve surveys}
	\label{photo_obs}
	We used two different surveys
	to search for light curves with the typical properties that we require.
	\subsubsection{OGLE project}
	OGLE is a long-term, large scale photometric sky survey focused on variability. Its original purpose was the detection of micro-lensing events and it is therefore observing fields with high stellar densities.
	A detailed description of the fourth phase of the project can be found in \citet{ogle_proj}. OGLE-IV is conducted at the Las Campanas Observatory in Chile with a 1.3m telescope dedicated to the project.
	
	More than a billion sources are regularly monitored in different fields in the Galactic Bulge, the Small and Large Magellanic Clouds, and the Galactic disc.
	The OGLE-IV camera is equipped with V- and I-band interference filter sets. The OGLE I-band filter resembles very closely the standard Johnson I-band filter; the OGLE-IV V-band filter is similar to the standard Johnson filter but extends slightly less far into the red. Most of the observations are performed in the I filter and the resulting light curves have from several hundred to more than a thousand data points with an integration time of 100 s. The OGLE-IV photometry covers the range of $12 < I < 20.5$ mag \citep{udalski}. The light curves are published in different catalogs together with an ephemeris for each star.
	
	\subsubsection{ATLAS project}
	ATLAS is a high cadence all-sky survey system designed to find dangerous near-Earth asteroids. ATLAS achieved first light in June 2015 and now consists of two independent units, one on Haleakala (HKO), and one on Mauna Loa (MLO) in the Hawai‘ian islands. Details of the project can be found in \citet{atlas}. Each telescope is a 0.65m Schmidt
	observing in a cyan filter (c, covering 420-650 nm) and an orange filter (o, 560-820 nm). The fisheye camera takes 32 second exposures on a 40 second cadence and ATLAS covers the entire accessible sky with a cadence of 2 days over a magnitude range 0 < m < 20. As of the end of January 2018, ATLAS had taken about 600\,000 exposures resulting in 240 million light curves with more than 100 epochs.
	
	\subsection{Color selection and period constraints}
\begin{figure}
		\includegraphics[angle=-90,width=\linewidth]{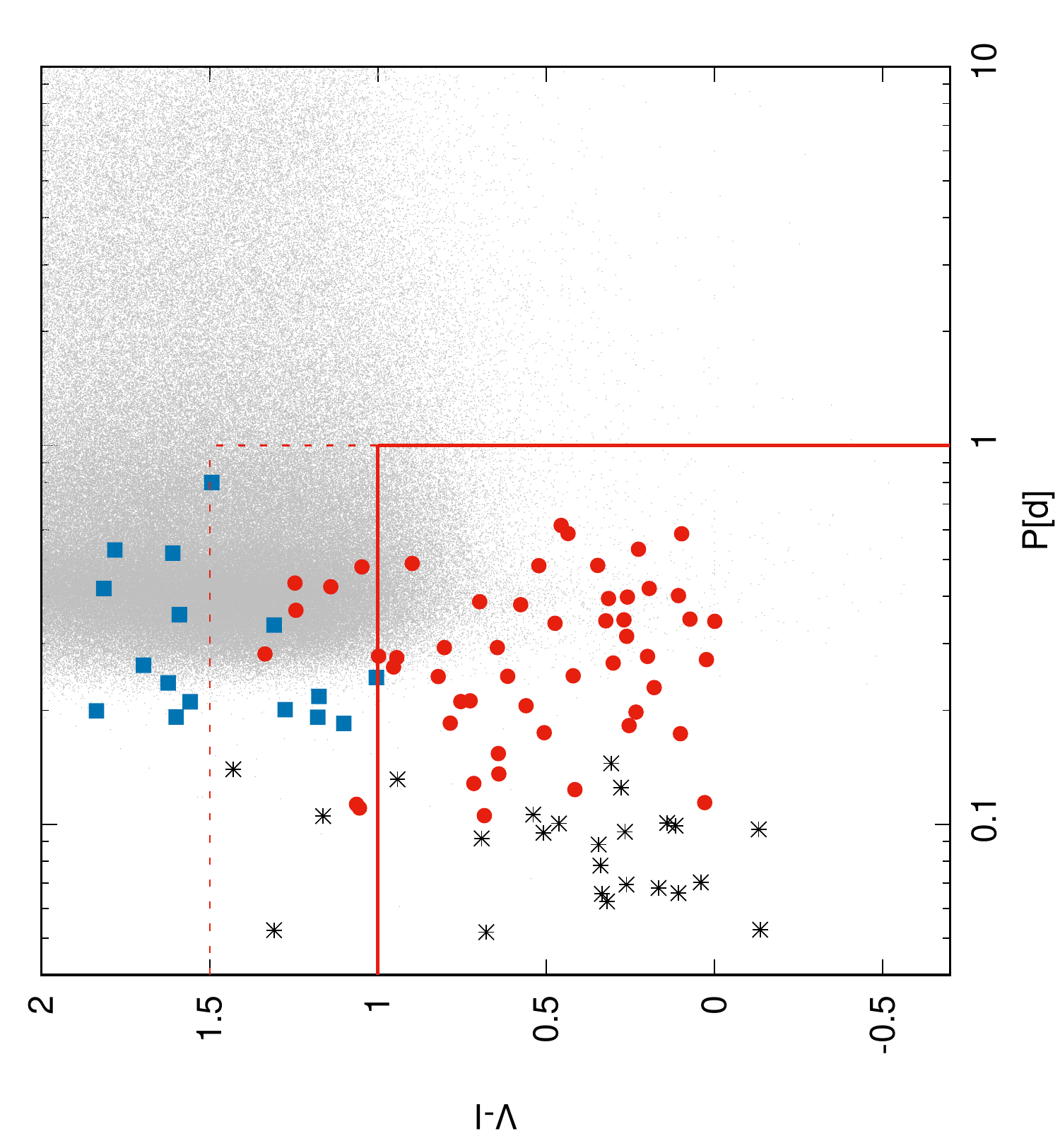}
		\caption{Period and color selections we applied to the eclipsing systems published by the OGLE team \citep{ogle_III}. HW Vir systems found by the OGLE team are shown as black stars; red circles represent those found by visual inspection and blue squares those found by machine learning. It can be seen that we select short-period binaries with the bluest colors.}
		\label{v-i} 
	\end{figure}

	Because sdBs are hot and hence blue, we limited our search to the bluest targets in OGLE. As OGLE observes not only in the I-band but also in the V-band for most targets (more details in Sect. \ref{photo_obs}), we could use the color index $V-I$ for the selection. The Galactic Bulge is a very dense region and has very patchy and substantial reddening. Because of this, we cannot simply apply color cuts characteristic for hot subdwarfs. Instead, we investigated the colors of the HW Vir candidates identified by \citet{ogle} and \citet{ogle_II} and decided to limit our search to targets with $V-I<1$ (see Fig. \ref{v-i}), as only three systems were found outside these limits. The longest period HW Vir system previously known is AA Dor with 0.26 d. The longest period reflection effect system has a period of 0.8 d \citep{jeffrey:2014}. Consequently, we focused our search on systems with orbital periods less than one day. Later, we also started to extended our search up to $V-I<1.5$.
	
	Figure \ref{v-i} shows the color and period selection we applied and the systems we found. These criteria left us with 2200 systems, each of which was phased with the ephemeris provided by the OGLE team. We inspected the light curves of all 2200 systems visually and found 51 new HW Vir candidates with periods between 0.1 d and 0.6 d. Most of the other light curves were consistent with contact systems or $\beta$ Lyrae type binaries. It is unsurprising that for the shortest periods almost all blue objects are HW Vir system candidates, as blue main sequence stars are much larger and have longer periods.
	\subsection{Light curve selection with machine learning}
	To identify additional systems not covered by the initial color selection we used
	the light curves of the 87 identified systems as a training set for machine learning. 
	
	As this set already includes systems with a large period range, variety of inclinations, and different S/N, it is ideally suited for training. For this we used the support vector machine (SVM) provided by the \textsc{python} package \textsc{sklearn}. We performed a C-Support Vector Classification (SVC) with the default squared exponential (rbf) kernel with a penalty parameter, $C$, of $10^5$ and a Kernel coefficient, $\gamma$, of $10^{-2}$. All 450\,000 eclipsing binaries were phased with the OGLE ephemerides. The SVM was then trained with the previously identified systems and applied to all OGLE light curves.
	
	2613 light curves were selected in this way. Those were inspected again visually. In this way 20 new systems were found, which have redder colors than our initial color selection. Two of those systems have no observations in the V band. Only six out of 87 systems, which have been found by our colour criteria before, were not detected by the SVM. All of those systems have very poor S/N. This means the number of false positives is still quite high, but we are dealing with a small enough number of systems to investigate them all by eye. In the future we would like to improve this process by creating a sample of synthetic light curves with different inclinations, S/N, and orbital periods.
	\subsection{Cross-matching the Gaia catalog of hot subdwarf stars with ATLAS}
	\citet{gaia_catalog} published a catalog of candidate hot subdwarf stars of the complete sky. This catalog was used to search for more HW Vir systems by cross-matching it with ATLAS.
	With the cross-match, 1600 objects from the Gaia hot subdwarf catalog were found in ATLAS. All of them were again phased after identifying the period with a Lomb-Scargle algorithm \citep{lomb,scargle} and then inspected visually. 50 additional new HW Vir candidates were found in the ATLAS data, as well as several known systems. The light curves of all systems in our sample can be found in Figs. \ref{lc1} and \ref{lc2}.
	\begin{figure*}
		\centering
		\includegraphics[width=0.8\linewidth]{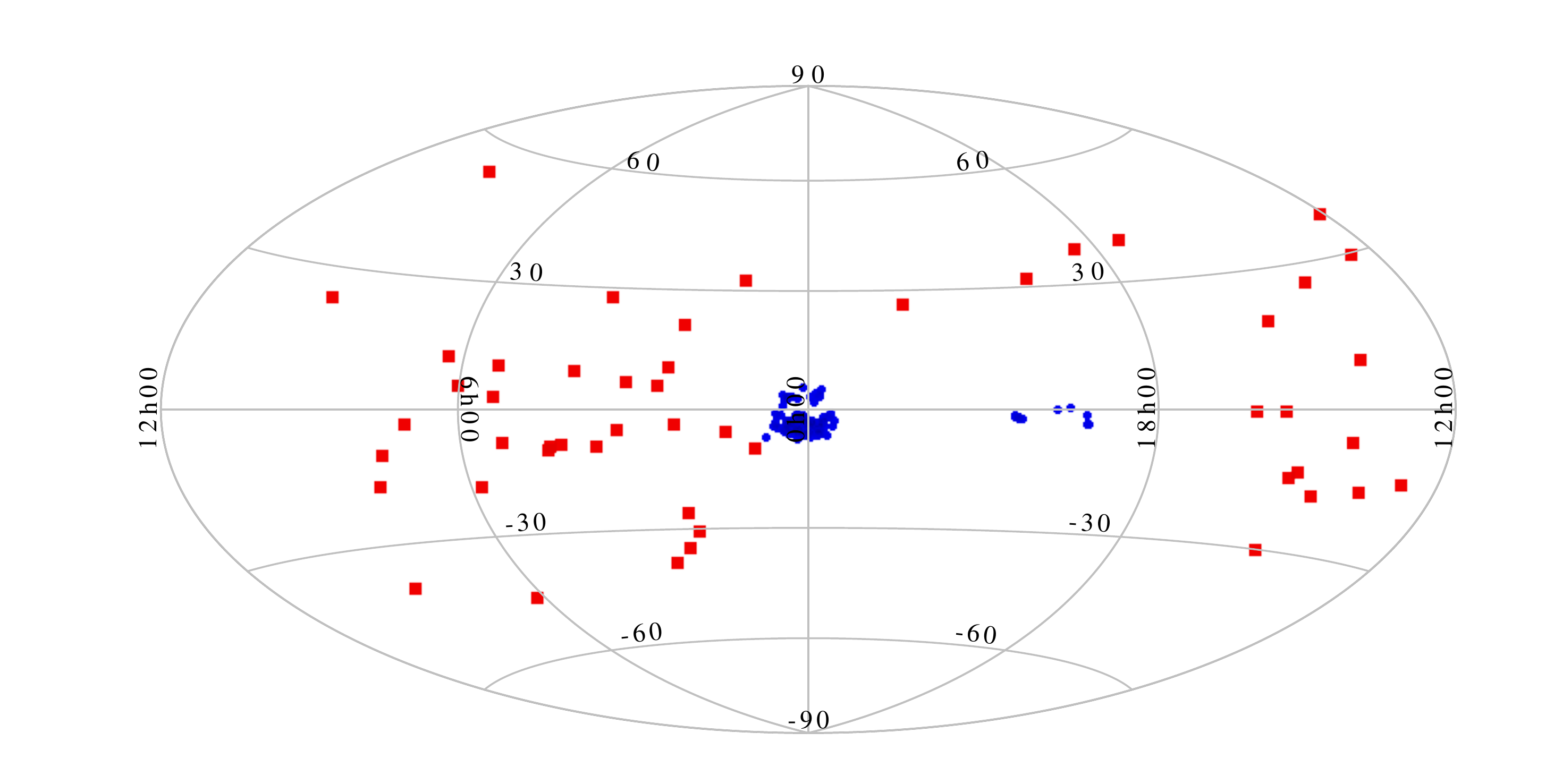}
		\caption{Distribution of the OGLE targets (blue circles) and the ATLAS targets (red squares) on the sky. The OGLE fields in the Galactic disc and the Galactic Bulge are clearly visible. The ATLAS targets are distributed over the whole sky accessible from the Northern hemisphere.}
		\label{distribution}
	\end{figure*}
	
	Figure \ref{distribution} shows the Galactic distribution of the new HW Vir candidates. The OGLE and Gaia magnitudes of all our targets, along with those of the published HW Vir binaries, can be found in Table \ref{orbit} and \ref{orbit_wd}. The OGLE targets, shown with blue circles, are found in the Bulge and the Galactic disc fields of OGLE. The ATLAS targets, shown with red squares, are distributed over the complete sky accessible from the Northern hemisphere.
	
\begin{figure}
		\includegraphics[width=\linewidth]{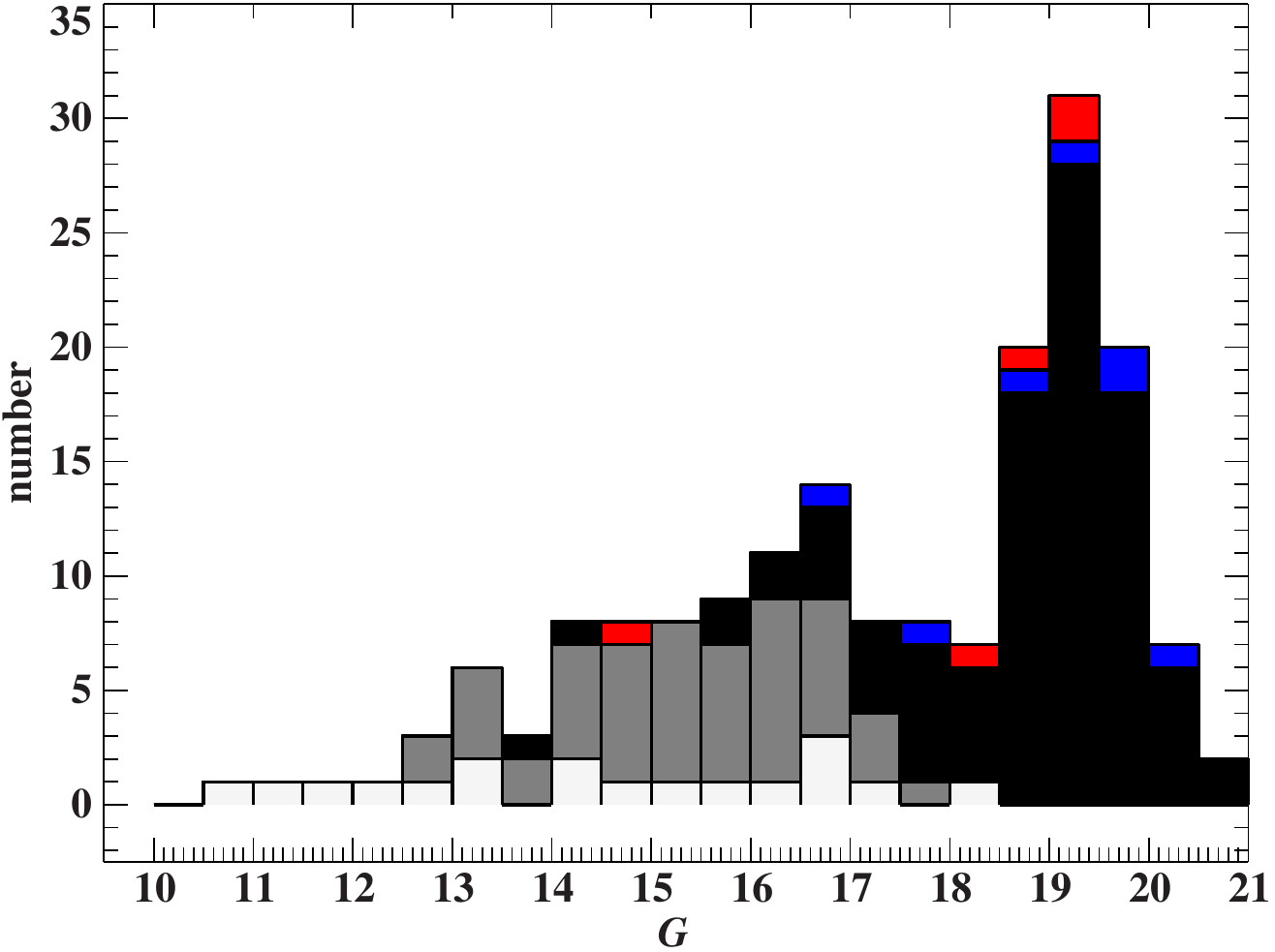}
		\caption{The magnitude distribution of the published HW Vir systems is shown in white, the ATLAS HW Vir candidates in grey, the OGLE HW Vir candidates in black, the eclipsing systems with candidate white dwarf primaries in blue and the central stars of planetary nebula are in red (more details are given in Sect. \ref{nature} and Tables \ref{orbit}, \ref{orbit_wd}, \ref{orbit_pn}).}
		\label{mag}
	\end{figure}

	The magnitude distribution of our targets is illustrated in Fig. \ref{mag}. The OGLE targets peak at a brightness of 19 mag, which is much fainter than any previously known systems and makes the follow-up more difficult. However, we found several brighter systems in the ATLAS survey.
	
	\section{Spectral parameters of the EREBOS targets}
\begin{figure*}
		\includegraphics[width=1.05\linewidth]{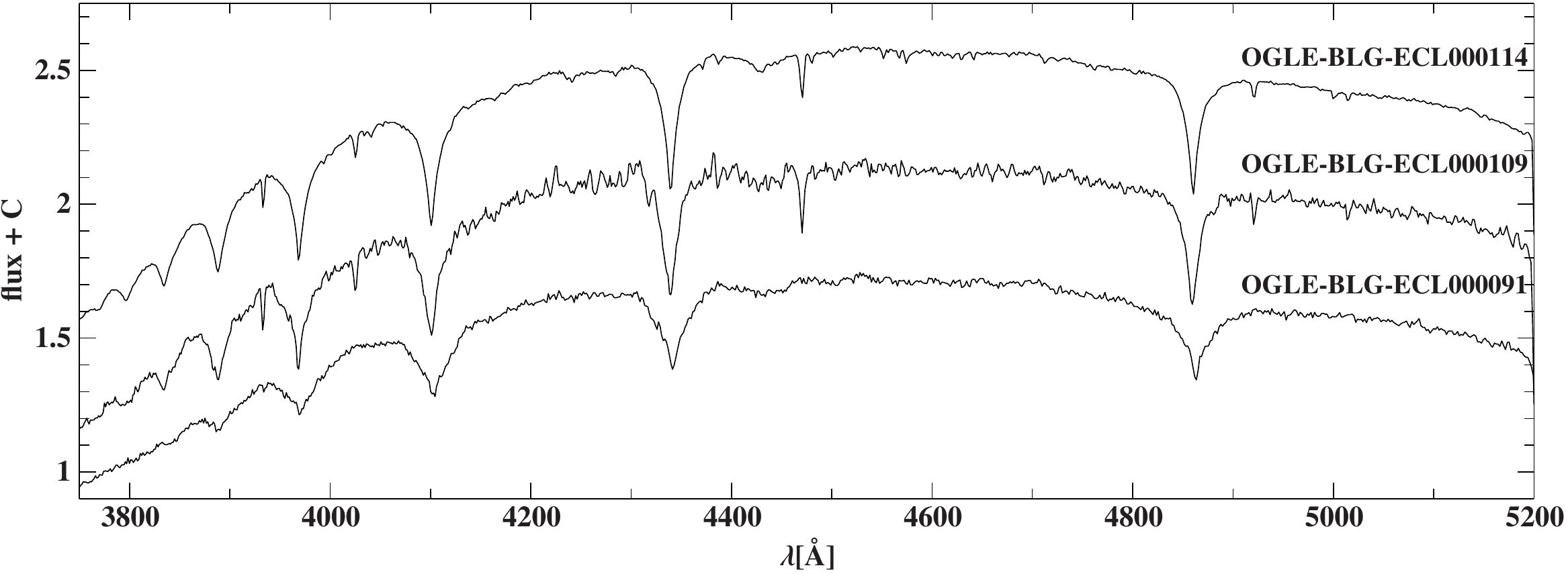}
		\caption{Three example spectra showing a typical sdB (OGLE-BLG-ECL-000114), a DA white dwarf (OGLE-BLG-ECL-000091), and a pre-He white dwarf (OGLE-BLG-ECL-000109). SdBs and pre-He white dwarfs show very similar spectra and can only be distingushed by deriving the atmospheric parameters.}
		\label{spectra}
	\end{figure*}

	From the light curves we can obtain very accurate orbital parameters (the orbital period, the inclination of the system, and the relative radii) but the light curve analysis suffers from degeneracies due to several coupled parameters, so it is essential to constrain as many of these parameters as possible using time-resolved spectroscopy.
	For 27 of our targets, we already have spectroscopic follow-up to confirm the nature of the primary star. All our observing runs can be found in Table \ref{obs}. 
	In the following we will give more details of the spectroscopic observations and show the first atmospheric parameters derived. Figure \ref{spectra} shows three examples of our co-added spectra from the ESO-VLT/FORS2 spectrograph.
	\begin{table*}
		\centering
		\caption{Spectroscopic follow-up observations}
		\begin{tabular}{llllc}
			\toprule\toprule
			Date & nights& Run &Telescope \& Instrument &Observers\\\toprule
			09/10 May 2015 & 2 &095.D-0167(A)&ESO-VLT/FORS2 & S. Kimeswenger\\
			Oct 2015-June 2017&12.5&196.D-0214(A-D) & ESO-VLT/FORS2 & Service\\
			04/05 Aug 2016 & 2&NOAO 2016B-0283 & SOAR/Goodman & B. Barlow\\
			30/31 Mar 2016 & 2&NOAO 2016A-0259 & SOAR/Goodman & B. Barlow\\
			07/08 Jun 2016 & 2&NOAO 2016A-0259 & SOAR/Goodman & B. Barlow\\
			01-05 Feb 2014 & 4&092.D-0040(A) &ESO-NTT/EFOSC2 & S. Geier \\
			01-04 Jul 2017 & 3&099.D-0217(A) &ESO-NTT/EFOSC2 & E. Ziegerer \\
			30 Jul - 02 Aug 2018 &3& 0101.D-0791(A)&ESO-NTT/EFOSC2 & S. Kreuzer\\
			26 Feb 2019 & 1&SO2018B-002 & SOAR/Goodman & I. Pelisoli\\

			\bottomrule
		\end{tabular}
		\label{obs}
	\end{table*}
	
	\subsection{Spectroscopic observations}
	
	\subsubsection{ESO-NTT/EFOSC2}
	For the brighter systems we used the EFOSC2 spectrograph mounted at the 3.58m ESO/NTT telescope. Nine of our systems have been observed in several runs (092.D-0040(A), 099.D-0217(A), 0101.D-0791(A)). We always took several spectra per star in grism Gr\#19 (4441-5114 $\AA$) with a 1'' slit to achieve a resolution of $R\sim 3000$ to derive a radial velocity curve covering the whole orbit. This grism includes $\rm H\beta$ as well as HeI 4472 and 4922. Usually, several objects were observed in a repeating sequence but always with exposure times less than 5-10\% of the orbital period to avoid orbital smearing. To derive the atmospheric parameters, moreover, one spectrum per star was taken with grism  Gr\#07 (3270-5240 $\AA$) with resolution $R\sim 500-700$. This grism covers the Balmer jump.
	\subsubsection{ESO-VLT/FORS2}
	Most of the OGLE target periods are quite short and, as we are interested especially in the shortest periods, this limits the exposure times greatly; EFOSC2 is not adequate for observation of those targets. 
	We applied successfully for an ESO large program with ESO-VLT/FORS (196.D-0214(A-D)) for fainter, short period systems, after a feasibility study in visitor mode (095.D-0167(A)). In our first run in 10 May 2015 we observed one 20-mag object (OGLE-GD-ECL-10384) for a half a night taking 24 spectra using Grism GRIS\_600B+22 (330 - 621 $\AA$, 1'' slit, R$\sim$780) with an exposure time of 600 s each and covering two full orbits. The ESO large program was executed in service mode over the course of two years. We divided all observations into 1 hour observing blocks (OBs), in which time-resolved spectroscopy of one target was performed per block. As the periods of many of our targets are very short, a significant part of the orbit can be covered in one hour. For these observations we used Grism GRIS\_1200B+97 (3660-5110, 1'' slit, R$\sim$1420). We limited the exposure time to $\sim$5\% of the orbital period to prevent orbital smearing. About 6-8 spectra were taken per OB. For each target several OBs were taken to cover the whole orbit distributed over one semester. 
	
	The observation of the Bulge targets and the data analysis turned out to be less straightforward than expected, as we have problems with the enormous crowding in the Bulge field, which complicates photometry as well as spectroscopy. Already, the identification of the targets was a huge challenge at the beginning, because of the lack of good finding charts in the visual, which lead to miss-identification of the target in some cases.
		Additionally, the ESO-VLT/FORS2 spectrograph was found not to be as stable as expected, because it was not designed for the determination of radial velocities. This complicates the determination of accurate radial velocity curves. Hence, we concentrate only on the determination of the atmospheric parameters for this paper, as more work is needed to derive the radial velocity curves from the FORS2 spectra. The radial velocities of all of our targets will be presented in future papers.
	
	Individual spectra were then co-added after being corrected for radial velocity so that we could use the co-added spectra for atmospheric analyses. Radial velocities were determined with the \textsc{iraf}\footnote{\citet{IRAF}} task \textsc{fxcorr} for cross-correlation against model spectra and co-added spectra to measure the radial velocity shifts in the Balmer and Helium lines.
	
	\subsubsection{SOAR/Goodman}
	We have collected time--resolved spectroscopy of many of the brighter, short-period systems with the Goodman Spectrograph on the 4.1m SOuthern Astrophysical Research (SOAR) telescope \citep{clemens:2004}. Four of our targets have been observed to date with time allocated through the National Optical Astronomy Observatory (NOAO Proposal IDs 2016B-0283, 2016A-0259,SO2018B-002). Our standard observing configuration uses a 930 mm$^{-1}$ VPH grating from Syzygy Optics, LLC, in conjunction with a 1.03\arcsec\ longslit to achieve a spectral resolution of 2.9 \AA\ ($R\sim 1500$) over the wavelength range 3600--5300 \AA. We observed each target using a series of consecutive spectra covering one full orbital period. In order to maximize our duty cycle, the spectral images were binned 2x2, resulting in a pixel scale of 0.3\arcsec\ per binned pixel in the spatial direction and 0.84 \AA\ per binned pixel in the dispersion direction. We again kept the integration times to less than 5\% of the orbital period to avoid phase smearing. 
	We use Gaussian fits to the Balmer lines to determine radial velocities and created a co--added spectrum for atmospheric modeling after individually correcting their velocities.

	\subsection{Atmospheric parameters}
	\label{atmo}
	\begin{center}
		\begin{figure*}
			\centering
			\includegraphics[width=0.4\linewidth]{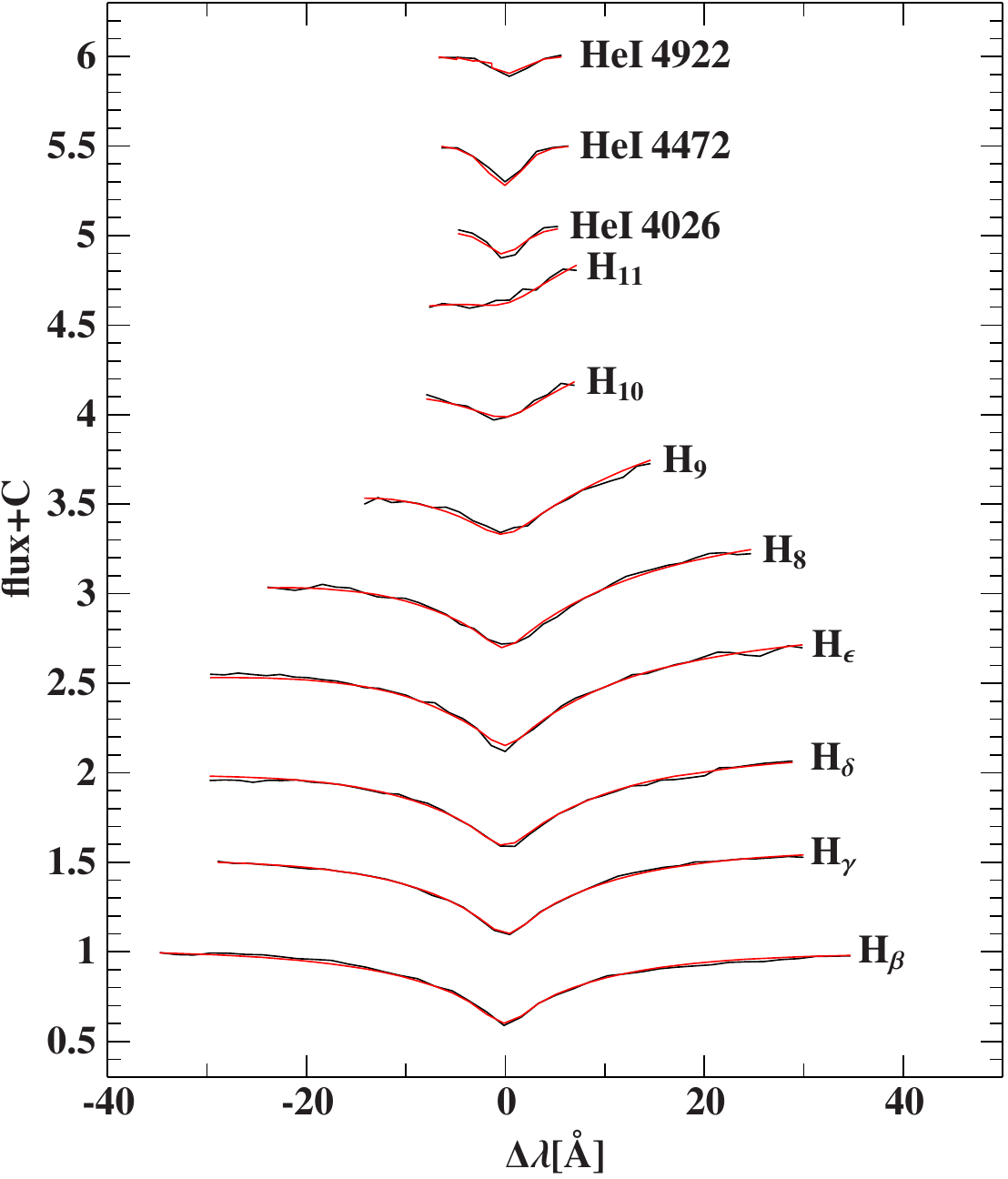}\hspace{1cm}
			\includegraphics[width=0.4\linewidth]{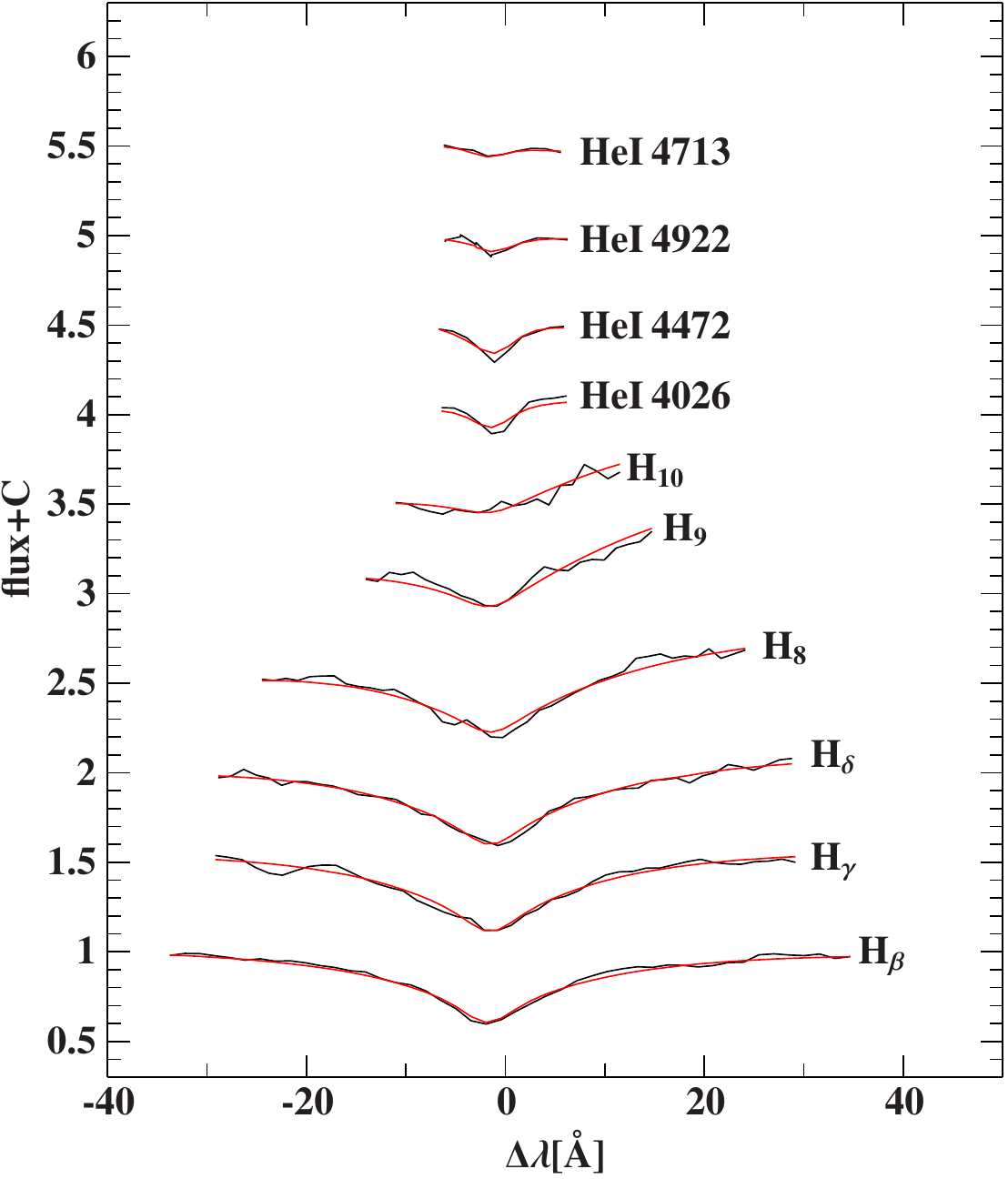}
			\caption{Example fits of hydrogen and helium lines with model spectra for a typical sdB (OGLE-BLG-ECL-000103. left panel), and a pre-He WD star (OGLE-BLF-ECL-000109. right
				panel). The atmospheric parameters of these stars are given in Table \ref{param}.}
			\label{lineplot}
		\end{figure*}
	\end{center}
	\begin{table}
		\setlength{\tabcolsep}{4pt}
		\caption{Atmospheric parameters of our observed targets from the EREBOS project spectroscopy with statistical errors}
		\label{param}
		\small
		\begin{tabular}{c|ccc}\hline\hline
			target name & $T_{\mathrm{eff}}$ & $\log{g}$ & $\log{y}$\\
			& [$10^3$K] & [cgs] & \\
			\toprule
			OGLE-BLG-ECL-000114$^{a,c}$ & 29.2 $\pm$ 0.5 & 5.55 $\pm$ 0.07 &-2.28 $\pm$ 0.10\\
			OGLE-BLG-ECL-000139$^c$ & 29.6 $\pm$ 0.5 & 5.50 $\pm$ 0.06 &-2.97 $\pm$ 0.13\\
			OGLE-BLG-ECL-000103$^a$ & 29.5 $\pm$ 0.4 & 5.70 $\pm$ 0.05 &-1.75 $\pm$ 0.11\\
			OGLE-BLG-ECL-000163$^c$ & 28.0 $\pm$ 0.3 & 5.37 $\pm$ 0.04 &-1.96 $\pm$ 0.10\\
			OGLE-BLG-ECL-000124$^a$ & 26.3 $\pm$ 0.7 & 5.46 $\pm$ 0.06 &-2.02 $\pm$ 0.16\\
			OGLE-BLG-ECL-000010$^a$ & 27.9 $\pm$ 0.7 & 5.32 $\pm$ 0.06 &-2.59 $\pm$ 0.18\\
			OGLE-BLG-ECL-000202$^a$ & 36.3 $\pm$ 0.5 & 5.25 $\pm$ 0.09 &-3.17 $\pm$ 0.12\\
			OGLE-BLG-ECL-000110$^a$ & 23.6 $\pm$ 0.5 & 5.35 $\pm$ 0.05 &-2.30 $\pm$ 0.13\\
			OGLE-BLG-ECL-000109$^a$ & 29.3 $\pm$ 0.3 & 6.05 $\pm$ 0.05 &-1.96 $\pm$ 0.10\\
			OGLE-BLG-ECL-000212$^a$ & 30.1 $\pm$ 0.5 & 5.38 $\pm$ 0.07 &-2.64 $\pm$ 0.11\\
			OGLE-BLG-ECL-000207$^a$ & 24.4 $\pm$ 0.4 & 5.54 $\pm$ 0.06 &-2.15 $\pm$ 0.14\\
			OGLE-BLG-ECL-173411$^b$ & 26.0 $\pm$ 0.1 & 5.28 $\pm$ 0.03 &-2.48 $\pm$ 0.11\\
			OGLE-BLG-ECL-361688$^b$ & 27.2 $\pm$ 1.3 & 5.20 $\pm$ 0.08 &-2.39 $\pm$ 0.02\\
			OGLE-BLG-ECL-416194$^b$ & 35.6 $\pm$ 0.9 & 5.34 $\pm$ 0.20 &-2.21 $\pm$ 0.35\\
			OGLE-BLG-ECL-017842$^b$ & 29.7 $\pm$ 0.8 & 5.81 $\pm$ 0.17 &-2.65 $\pm$ 0.28\\
			OGLE-BLG-ECL-280838$^b$ & 28.3 $\pm$ 0.6 & 5.55 $\pm$ 0.08 &-2.78 $\pm$ 0.25\\
			OGLE-BLG-ECL-412658$^b$ & 36.7 $\pm$ 0.8 & 5.48 $\pm$ 0.08 &-2.49 $\pm$ 0.33\\
			OGLE-GD-ECL-08577$^b$ & 28.4 $\pm$ 1.0 & 5.43 $\pm$ 0.15 &-2.01 $\pm$ 0.27\\
			OGLE-GD-ECL-10834$^a$ & 27.6 $\pm$ 0.8 & 5.64 $\pm$ 0.16 &-2.54 $\pm$ 0.18\\
			OGLE-GD-ECL-11388$^b$ & 29.0 $\pm$ 0.3 & 5.56 $\pm$ 0.04 &-2.77 $\pm$ 0.05\\
			OGLE-GD-ECL-11471$^a$ & 28.4 $\pm$ 0.5 & 5.71 $\pm$ 0.10 &-2.17 $\pm$ 0.08\\	
			\hline
			J282.4644-13.6762$^b$ & 27.5 $\pm$ 0.6 & 5.54 $\pm$ 0.07 &-2.25 $\pm$ 0.21\\
			J351.7186+12.5060$^b$ & 29.0 $\pm$ 0.4 & 5.68 $\pm$ 0.08 &-1.86 $\pm$ 0.14\\
			J315.0724-14.190$^b$ & 30.2 $\pm$ 2.0 & 5.96 $\pm$ 0.38 &-2.00 $\pm$ 0.16\\
			J079.5290-23.1458$^b$ & 30.9 $\pm$ 0.7 & 5.75 $\pm$ 0.09 &-1.96 $\pm$ 0.14\\
			J129.0542-08.0399$^c$ & 31.1 $\pm$ 0.3 & 5.49 $\pm$ 0.06 & -2.85\\
			\toprule
		\end{tabular}
		\tablefoot{\\$^a$ ESO-VLT/FORS2\\$^b$ ESO-NTT/EFOSC2\\$^c$ SOAR/Goodman}
	\end{table}	
	
	\begin{figure}
		\includegraphics[width=\linewidth]{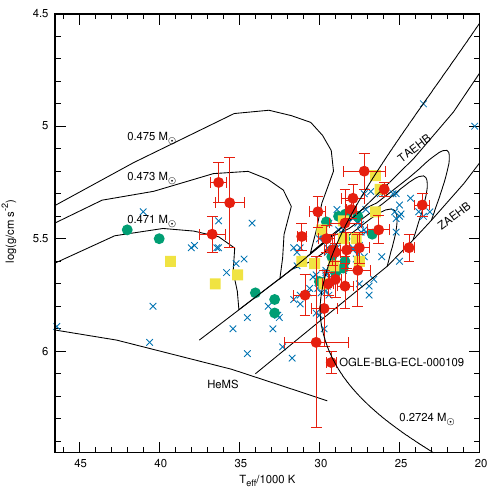}
		\caption{$\rm T_{eff}-\log{g}$ diagram of the HW Virginis systems. The zero-age EHB,
			ZAEHB, and the terminal-age EHB, TAEHB are superimposed by
			evolutionary tracks by \citet{Dorman:1993} for sdB masses of 0.471, 0.473, and 0.475 $\rm M_{\odot}$ with one track for an extremely low mass white dwarf of a mass of 0.2724 $\rm M_{\odot}$ by \citet{althaus}. The newly found systems are shown as red circles with error bars. The published HW Vir systems are shown as green circles, the reflection effect systems without eclipses as yellow squares. The blue crosses represent other sdB binaries \citep{kupfer:2015}, either with white dwarf or unknown companions.}
		\label{teff-logg}
	\end{figure}

	Atmospheric parameters were determined by calculating synthetic spectra using LTE model atmospheres with solar metallicity and metal line blanketing \citep{heber:2000} and fitting these to the Balmer and helium lines using SPAS \citep{hirsch}. In this way, we determined the atmospheric parameters (effective temperature $T_{\rm eff}$, surface gravity $\log{g}$ and ihelium abundance $\log{y}$) for 25 systems, which can be found in Table \ref{param}. Some example line fits are showed in Fig. \ref{lineplot}. The errors given are only statistical errors. As the contribution of the companion to the visible flux varies over the orbital phase due to the reflection effect, apparent variations of the atmospheric parameters are found in some eclipsing sdB binaries with cool companions with high S/N spectra on the order of 1500 K and 0.1 dex over the whole orbital phase \citep{vs:2014_I}. This means that in case of systems with very low statistical errors we have to adopt an uncertainty of 750 K in temperature and 0.05 dex in $\log(g)$.
	
	The parameters show that the primaries are mostly typical for HW Vir systems \citep[see][and references therein]{muchfuss_photo}. OGLE-BLG-ECL-000091 and J186.9106-30.7203 are the only exceptions, as both clearly look like DA white dwarfs with very broad Balmer lines (see Fig. \ref{spectra} for an example).
	
	However, not only the tracks of He-core burning objects cross the extreme horizontal branch, but also post-RGB objects with masses too low to burn He in the core (see Fig.~\ref{teff-logg}). Those evolve directly into He-WD and are, hence, also called pre-He WDs. The lifetime of pre-He WDs crossing the EHB is only 1/100 of the lifetime of a core-He burning sdB on the EHB. Therefore, we expect them to be much rarer and that most objects on the EHB are hot subdwarfs. This depends, however, also on the birth rate of these systems, which is unknown. This means in most cases it is not possible to distinguish between sdBs and pre-He WDs from the atmospheric parameters alone.
	
	To compare the atmospheric parameters of our targets to other sdB binaries we have plotted the $\rm T_{eff}-\log{g}$ diagram of the reflection effect systems and other sdB binaries from \citet{kupfer:2015} in Fig. \ref{teff-logg}. The published HW Vir and reflection effect systems seem to cluster in a relatively small region of the $\rm T_{eff}-\log{g}$ diagram. Very few are found on the EHB at higher temperatures. 
	Our new systems almost double the number of atmospheric parameter determinations for sdBs with cool, low mass companions. It appears now that the EHB is well populated with such systems. Only near the He-main sequence (HeMS) we still do not find any HW Vir systems. Moreover, we find three systems which have already evolved away from the EHB. 
	
	An interesting system is OGLE-BLG-ECL-000109, which lies clearly below the He-main sequence. This means it cannot be a He-core burning object and is most probably a post-RGB object with a mass too low to burn He in the core. 
	The position of OGLE-BLG-ECL-000109 agrees best with  a track for a very low-mass white dwarf with a mass near 0.27 $\rm M_{\rm \odot}$ \citep{althaus}.
	Some HW Vir systems have been found where the analysis could not unambiguously distinguish between a helium-core burning object on the extreme horizontal branch and a pre-He WD \citep[e.g., J082053+000843 and HS 2231+2441,][]{geier,almeida}. 
	This is the first HW Vir system lying significantly below the EHB and can only be explained by being a pre-He white dwarf.

	\section{Constraining the nature of the primary star}
	\label{nature}
	We selected our candidates purely based on the light curve shapes.
	To learn about the population of our targets, we need to constrain the nature of the primary star. The easiest way to do this is with spectroscopy  and so far we have spectroscopic confirmation for 25 systems with a most likely hot subdwarf primary. Only two of our targets have a white dwarf primary (see Sect. \ref{atmo}).	
	
	For the rest of our sample, we have not yet obtained  spectroscopy. As described in Sect. \ref{erebos}, white dwarfs or post-AGB objects can be possible contaminants in our sample. 
	Of course, such systems are no less interesting, but it is important to have a homogeneous sample from which to draw conclusions.
	
	The ESA Gaia Data Release 2 \citep{gaia_dr2} provided us for the first time with tools to constrain the nature of the primary star. All results and Gaia parameters for our systems can be found in Tables \ref{orbit}, \ref{orbit_wd}, \ref{gaia} and \ref{gaia_wd}. 75 of the 107 OGLE targets and all ATLAS targets have parallaxes and proper motions.
	
	\subsection{Absolute magnitudes and distances}
	As seen in Fig~\ref{teff-logg} it is not possible to uniquely distinguish sdBs from pre-He WDs from the atmospheric parameters alone, however, the masses and radii of pre-He WDs are expected to be lower, and therefore, the absolute magnitudes are fainter.
	Hot subdwarf stars have radii around 0.15-0.2 $\rm R_{\odot}$, comparable to M dwarf stars. White dwarfs have radii that are much smaller -- comparable to Earth ($\sim$0.01 $\rm R_{\odot}$) -- and usually have much fainter absolute magnitudes. 
	
	\begin{figure}
		\includegraphics[width=\linewidth]{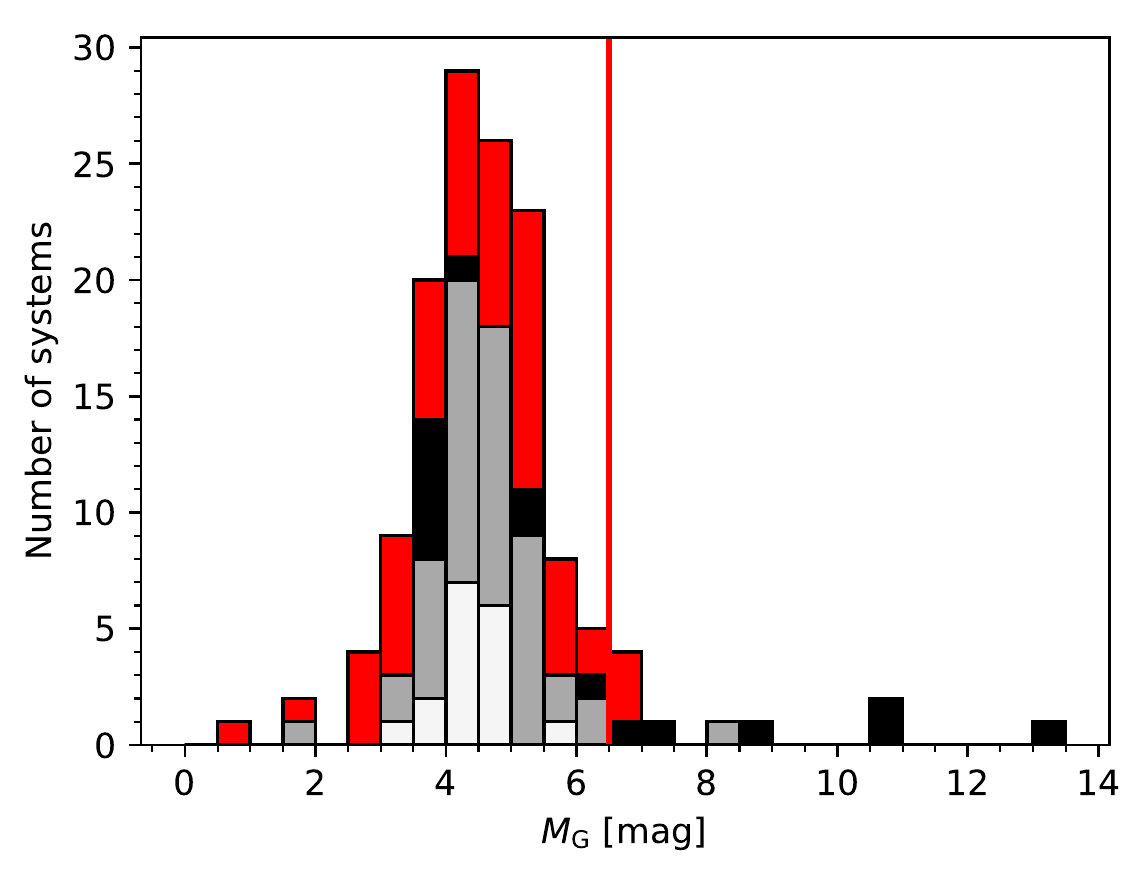}
		\caption{Distribution of the absolute G magnitude ($M_{\rm G}$) of our targets. The published HW Vir stars are shown in white; the ATLAS targets are displayed in gray; and the OGLE targets with parallaxes with errors less than 25\% are displayed in black. For the rest of the OGLE targets we used the distances by \citet{distances}, shown in red. The red line marks $M_{\rm G}=6.5$ mag, above which the primary stars are faint enough that they are more likely to be white dwarfs.}
		\label{gabs}
	\end{figure}
	In Fig. \ref{gabs} the distribution of the absolute $G$ magnitude ($\rm M_{\rm G}$)  is shown. 
	Absolute magnitudes were calculated via the distance modulus $G-M_{\rm G}= 5\log_{10}d-5+(A_{\rm G})$.  We used the distances published by \citet{distances}. Distances with large parallax errors are mostly based on the length scale model they use, which is unclear for sdBs, so they have to be taken with caution. As all our systems show photometric variability the uncertainty in $G$ will also be larger than given in the DR2 Gaia catalogue. On the other hand outliers are neglected for the determination of $G$, hence the mean Gaia magnitude should give a good estimate of the correct sdB magnitude \citep{gaia_process_photometry}.

\begin{figure}
		\includegraphics[width=\linewidth]{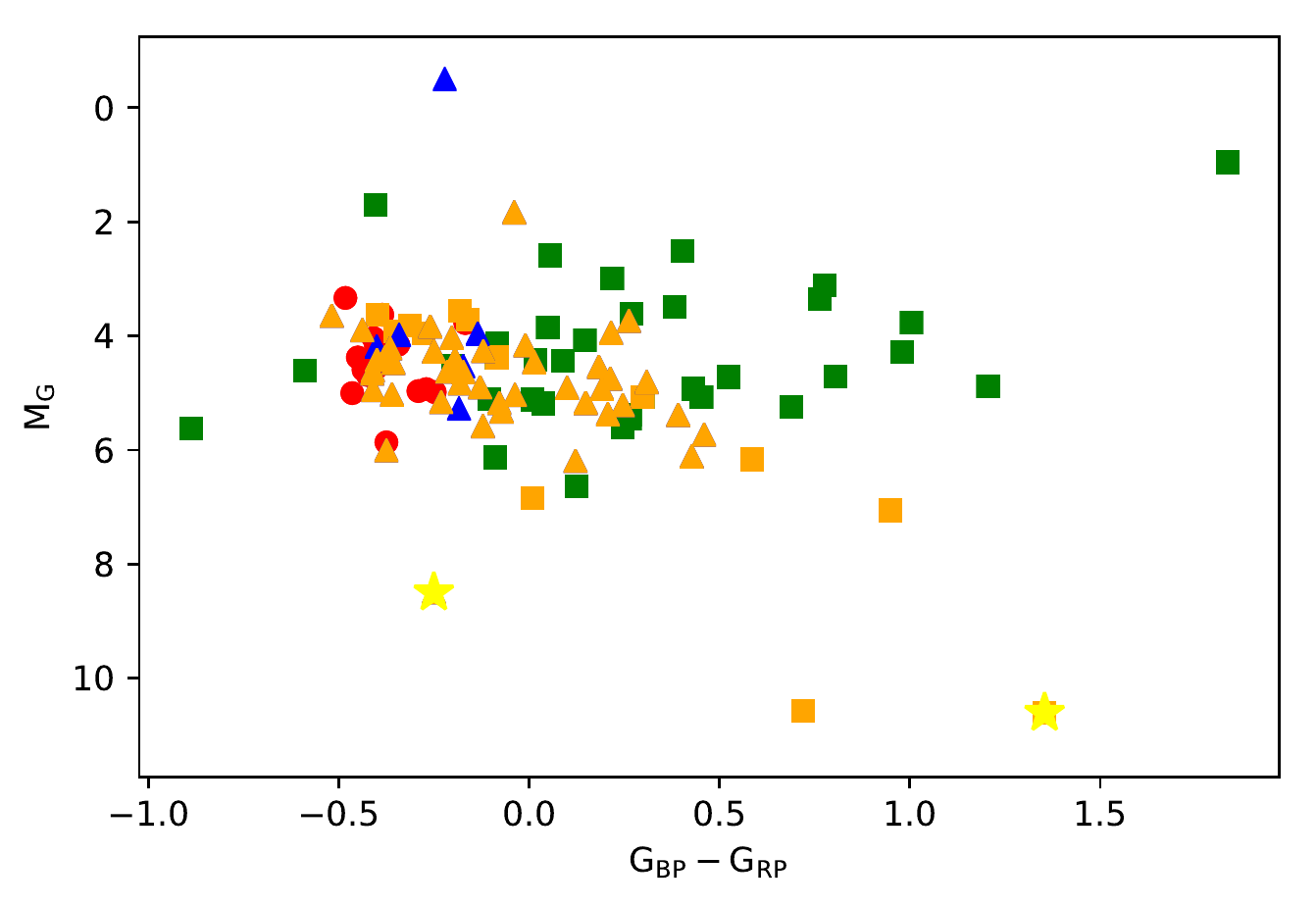}
		\caption{Color-magnitude diagram of all our targets in Gaia colors. Red circles represent the published HW Vir systems for comparison. Squares indicate the OGLE targets and triangles the ATLAS targets. Targets with parallaxes better than 25\% are marked in orange. For the green OGLE targets and the blue ATLAS targets we used the distances by \citet{distances}. The yellow stars mark our confirmed system with a white dwarf primary.}\label{cmd}
	\end{figure}

	For targets in the Bulge, the reddening cannot be neglected, and we  constrained the reddening $E(B-V)$ by using Stilism\footnote{\url{https://stilism.obspm.fr/}}\citep{stilism}, which gives the reddening at certain coordinates depending on the distance. 
	In most cases, we derive only a lower limit for the reddening because the dust maps do not extend far enough ($<1-2 \rm \,kpc$). To obtain the Gaia G-band extinction coefficient $A_{\rm G}$ we used eq. 1 from the \citet{extinction}, which uses $G$, $BP$, $RP$ and $E(B-V)$ as input parameters. For some of the OGLE targets, no $BP$ or $RP$ was given in the Gaia data, so we assumed similar colors to those found for the other OGLE targets (see Fig. \ref{cmd}). 
	
	When looking at the distribution of absolute G magnitudes of all targets,
	it is obvious that the different subsamples show the same distribution peaking at $M_{\rm G} = 4.5$ mag -- as expected for sdBs. We checked the absolute magnitudes of the white dwarf candidates from the Gaia white dwarf catalog \citep{gaia_wd} and the Gaia hot subdwarf catalog \citep{gaia_catalog}. Most of the objects with $M_{\rm G}$ $>$ 6.5 are classified as white dwarf stars, and therefore a white dwarf primary is more likely. Nine of our objects have absolute magnitudes fainter than 6.5 mag and two of them have been spectroscopically confirmed as a white dwarf. On the other hand, three of them have been confirmed as sdBs. As mentioned before, the reddening we applied is only a lower limit, and therefore the absolute magnitude of the OGLE targets are upper limits. The temperatures of those three targets is rather low for sdBs with $24000-26000$ K. It is also possible that they are pre-He WDs instead of sdBs.
	
	Another criterion is the color of the targets, which we can combine with absolute magnitude in a color magnitude diagram, shown in Fig. \ref{cmd}. The published HW Vir stars (see Table \ref{orbit} and \ref{gaia}) are concentrated at $BP-RP<0$. The ATLAS targets show a larger spread, but for those we neglected reddening, which might be important for more distant targets. Due to the high reddening, which can only be poorly constrained in the bulge, the OGLE targets have an even wider distribution in color.
	
\begin{figure}
		\includegraphics[width=1\linewidth]{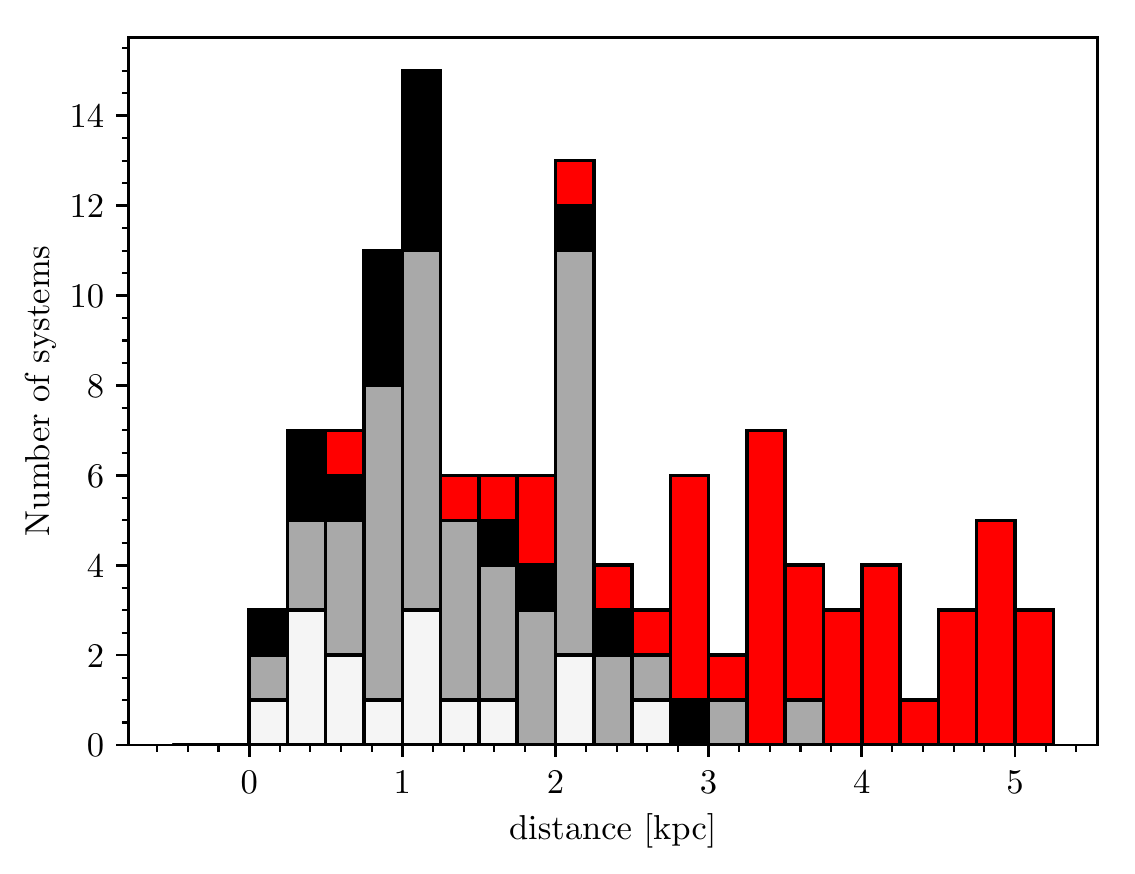}
		\caption{Distance distribution of our targets. The published HW Vir stars are shown in white; the ATLAS targets are displayed in gray and the OGLE targets with parallaxes having errors less than 25\% are displayed in black. For the rest of the OGLE targets, shown in red, we used the distances by \citet{distances}}
		\label{distances}
	\end{figure}

	From the Gaia parallaxes it is also possible to determine the distances of the objects; this is shown in Fig. \ref{distances}. The distance distribution agrees nicely with the distribution of the Gaia hot subdwarf catalog \citep{gaia_catalog}. The only difference is that it looks like there are too few objects observed at distances between 1.25~kpc and 2~kpc. This bi-modality is mainly seen in the ATLAS targets. As expected from their fainter magnitudes, the OGLE targets are mostly further away (80\% have distances $>$ 2.5 kpc). Almot all targets but two with distances larger than 3~kpc have large parallax errors and hence uncertain distances.
	\subsection{Reduced proper motions}
	As the reddening cannot be constrained perfectly for the more distant objects, and because they have parallaxes with large errors, the absolute magnitudes of these targets are not reliable. However, to distinguish hot white dwarfs from hot subdwarf stars, reduced proper motions can be used. The reduced proper motion, defined as $H_{\rm G} = G+5(\log \mu+1)$, can be used as a proxy for the distance of an object, since more distant objects exhibit less movement, on average. \citet{sdss_wd} showed that this selection method is well suited to separate hot subdwarfs from white dwarf candidates.
	
\begin{figure}
		\includegraphics[width=1.1\linewidth]{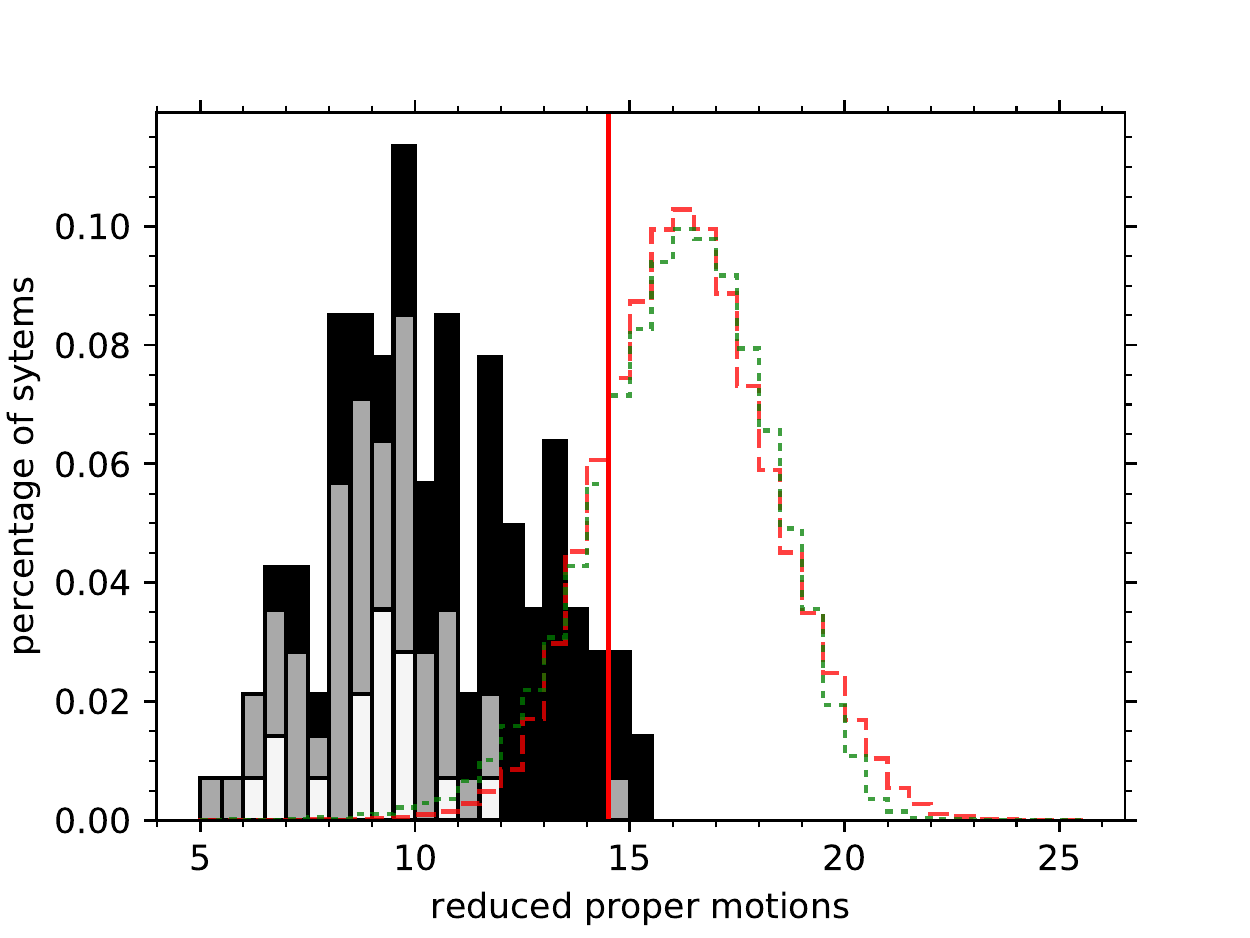}
		\caption{Normalized distribution of the reduced proper motions of our targets (white are the published HW Vir systems, gray the ATLAS targets and black the OGLE targets). Superimposed are the reduced proper motions of the objects with a white dwarf probability $>$ 90\% from the Gaia white dwarf catalog \citep{gaia_wd}, shown with the red dashed line. The green dotted line shows the subsample of the targets with spectral classification from SDSS spectra. The red line marks our reduced proper motion cut of $H_G<14.5$; at higher values, targets are less likely to be hot subdwarfs.}
		\label{red_pm}
	\end{figure}

	Figure \ref{red_pm} shows the distribution of the reduced proper motions of our targets superimposed with the distribution for white dwarfs taken from the Gaia white dwarf catalog \citep{gaia_wd}. It is clear that the subdwarfs are found at smaller reduced proper motions. However, there is a region where both distributions overlap and the nature of the primary cannot be determined unambiguously. Combing the selection from the absolute magnitude with the reduced proper motion and the spectroscopically confirmed targets, we decided to define the cut at $H_G<14.5$. This leaves us with seven objects, which have higher probabilities to be white dwarf binaries with cool, low mass companions out of 123 objects with Gaia parallaxes and proper motions. All white dwarf binary candidates can be found in Table \ref{orbit_wd} and \ref{gaia_wd}. This means that the white dwarf binaries are only a very small part of our sample and most targets are indeed hot subdwarf binaries.
	
\begin{figure}
		\includegraphics[width=1.1\linewidth]{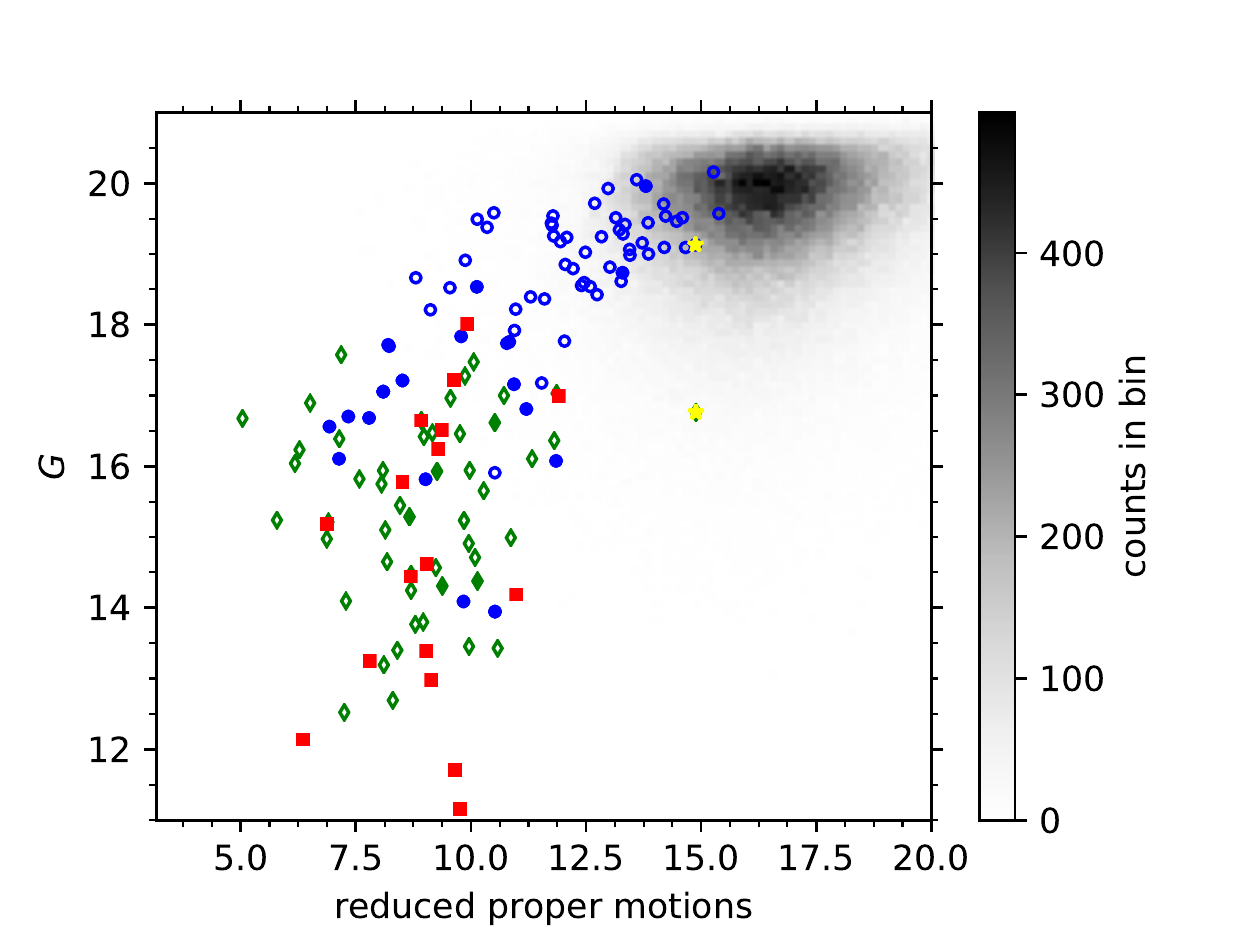}
		\caption{Relation between reduced proper motion and apparent $G$ magnitude. Filled symbols mark objects with spectral confirmation. Red squares represent the published HW Vir systems,  green diamonds the ATLAS targets, and blue circles the OGLE targets. The yellow star marks the confirmed white dwarf binary. Superimposed is the number distribution of objects with a white dwarf probability $>$ 90\% from the Gaia white dwarf catalog \citep{gaia_wd}.}
		\label{gmag_red_pm}
	\end{figure}

	This becomes even more clear when we look at the relation between reduced proper motion and apparent $G$ magnitude (Fig. \ref{gmag_red_pm}). The bulk of white dwarfs is seen at larger reduced proper motions than the sdBs and also at fainter apparent magnitudes. It is not surprising that most white dwarfs are fainter as they have much smaller radii. Only very few white dwarfs are found in the overlapping region. However, they are much more frequent, so an uncertainty for some OGLE targets remains.
	
	\subsection{Binary central stars of planetary nebula}
	Another class of object that can have light curves similar to HW Vir systems are binary central stars of planetary nebula (bCSPN) with cool, low-mass companions. For those objects it is under debate whether the the nebulae are ejected AGB or RGB envelopes or ejected common envelopes
	\citep[e.g.,][]{pn_rgb}. With our selection criteria we cannot really distinguish them from sdBs. However, they have quite short lifetimes and are hence much rarer \citep{post-agb}. In our absolute magnitude distribution, we have a few targets which are much brighter than the rest. For them the probability of being a post-AGB binary with an extremely hot primary is higher. 
	OGLE-BLG-ECL-412658 with an absolute magnitude of $1.69_{-0.98}^{+0.92}$ was found to be an sdB star with a temperature of 35600 K evolving away from the EHB. Two of our targets (J265.3145+29.5881 and OGLE-BLG-ECL-149869) seem to have an absolute magnitude brighter than 1 mag, which is much brighter than one would expect for a hot subdwarf. However, both have a very small and uncertain parallax. J171.4930-20.1447 on the other hand has an absolute magnitude of $1.66\pm0.22$ with a small parallax error of only 10\% and is hence the best candidate to be a post-AGB object.
	
	\citet{cat_pn} performed a survey to find planetary nebulae in the direction of the Galactic Bulge. Four objects from our target list were confirmed as planetary nebulae by them. Those can be found in Table \ref{orbit_pn} and \ref{gaia_pn}. Another known planetary nebula was found in the cross-match of the Gaia hot subdwarf catalog with ATLAS. Currently only 11 eclipsing central stars of planetary nebula showing a reflection effect are known (\url{http://www.drdjones.net/bCSPN/}).
	
	\subsection{Nature of the primary star}
	By combining all criteria, it is safe to say that we have only a small level of contamination by white dwarfs or post-AGB binaries, probably less than 10\%. For most of our target sample, an sdB primary star is most likely. This is also supported by the fact that only 2 out of 28 targets were confirmed not to have an sdB primary (see Sect. \ref{atmo}).
	
	\section{Results}
	
	\subsection{The period distribution of eclipsing hot subdwarfs with cool companion}
\begin{figure}
		\includegraphics[width=\linewidth]{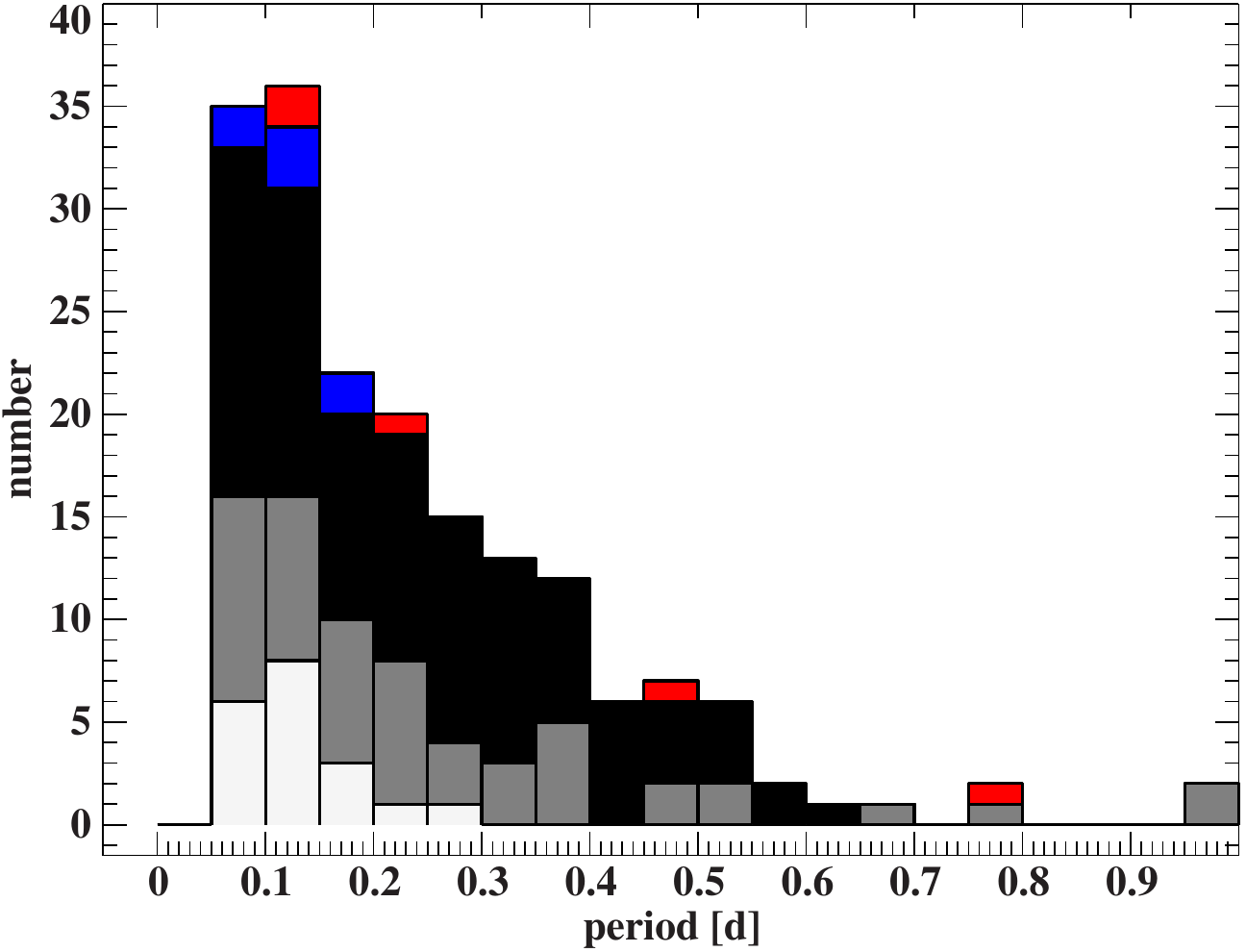}
		\caption{Period distribution for all our targets. The currently published HW Vir systems are shown in white, with the ATLAS targets in gray, and the OGLE targets in black. In blue we marked those systems which have a higher probability to be white dwarf binaries, in red the central stars of planetary nebula.}
		\label{period}
	\end{figure}

	With all the newly discovered candidate HW Vir systems presented here, the number of known systems has increased from 20 to 170 systems.
	The most straight-forward parameter to derive from light curves of eclipsing binaries is the orbital period. It can be found in Table \ref{orbit}. All light curves are displayed in Fig. \ref{lc1} and \ref{lc2}. The period distribution is shown in Fig. \ref{period}. The previously known period distribution of the HW Vir systems covered a range of 0.07 to 0.26 d with a sharp peak around 0.1 d; the distribution of the new systems is much broader, now spanning periods from just 0.05 d to more than one day. 
	
	As the eclipsing probability strongly correlates with the period, it is not surprising that most systems are found at periods around 0.1 d -- as seen in the smaller sample. The number of systems at shorter periods has increased, but we also find a significant number at longer periods -- up to 0.5 d. This population was completely unknown before, but most of the previously known systems were found while looking for short-period pulsations with light curves of perhaps only an hour or two, so long-period systems would not necessarily have been detected. Apart from the smaller chance of detecting eclipses, reflection effects will tend to become much smaller as the component separation and period increase.
	
\begin{figure}
		\includegraphics[width=\linewidth]{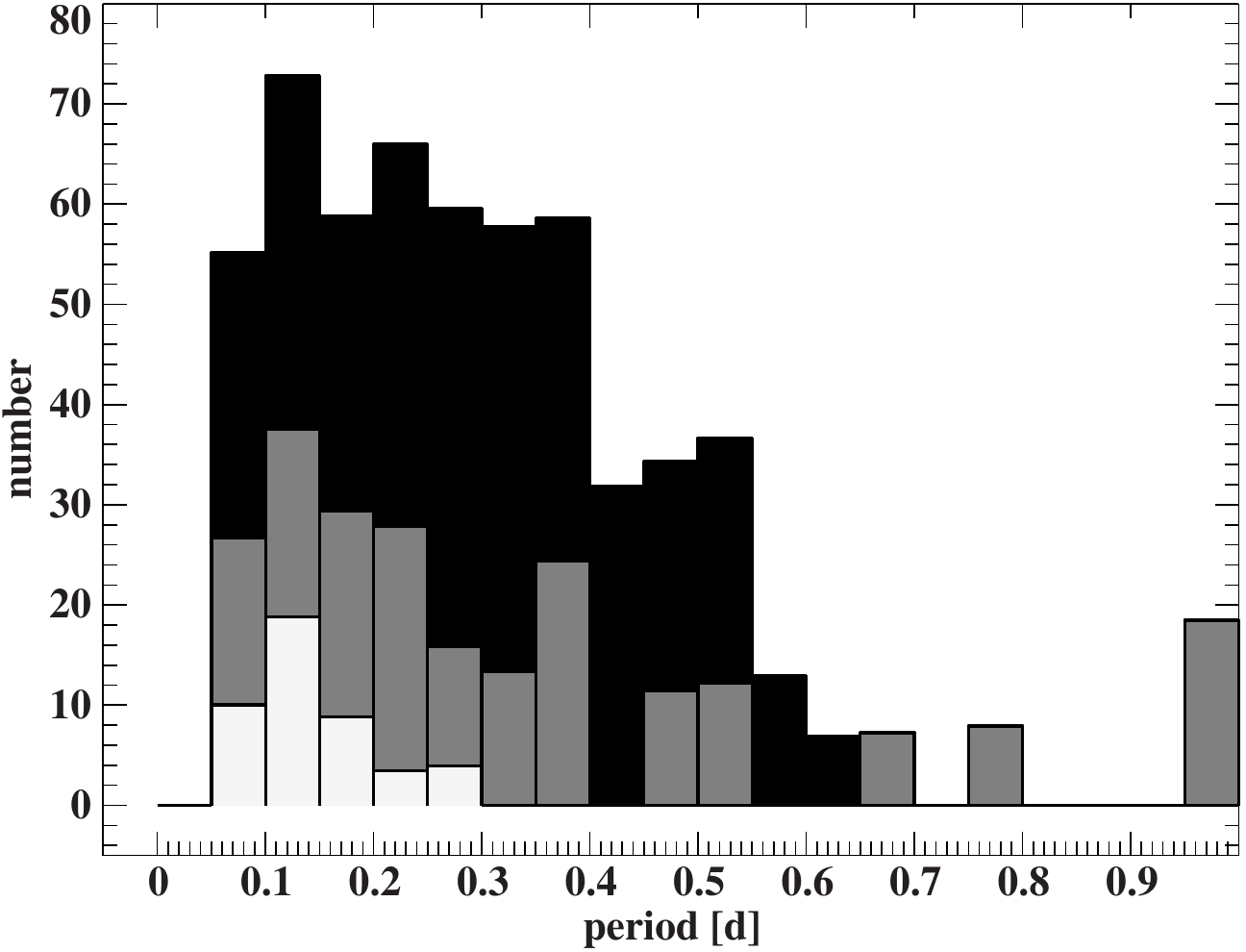}
		\caption{Period distribution for all our targets corrected for the eclipsing probability. The currently published HW Vir systems are shown in white, the ATLAS targets in gray and the OGLE targets in black.
		}
		\label{period_corrected}
	\end{figure}

	As we limit ourselves to eclipsing systems,
	we also tried to constrain the true period distribution. For this we had to correct for the number of systems which are not selected by EREBOS because they do not show eclipses. The probability of eclipses occurring is dependent on the relative radius of both stars and therefore correlates with the orbital separation $a$, ($p_{\rm ecl}=\frac{r_1+r_2}{a}$). The separation $a$ can be calculated from the masses and the period ($a=(G\frac{m_1+m_2}{4\pi^2})^{1/3}\cdot P^{2/3}$). We assumed for the eclipsing probability the median masses and radii of the published HW Vir systems\citep[see][]{muchfuss_photo}. The corrected period distribution can be found in Fig. \ref{period_corrected}. Up to a period of 0.35 d it appears to be a fairly flat distribution. For longer periods -- up to one day -- the number of systems drops significantly. Above 0.55 d we found only 8 systems. This small number of systems does not allow to draw significant conclusions on the number of systems for periods longer than 13 hours.
		
	\subsection{Selection effects} For interpreting the period distribution it is important to understand the selection effects that limit the detection of HW Vir systems at certain periods. To detect the shortest-period systems, a short cadence of observation of the light curve is essential. The light curve surveys we used have quite random cadence. The shortest-period eclipsing systems released by the OGLE team has 0.052 d period and is part of our target sample. The shortest-period system we found in the ATLAS survey has 0.062 d. In OGLE three systems with significantly shorter periods have been found. We could find shorter period systems around 0.05~d when phasing the ATLAS light curves of the sdB candidates from the hot subdwarf Gaia catalogue \citep{gaia_catalog}. The question is, why did we not find such short-period systems in ATLAS? One explanation is that they might be very rare. From the OGLE catalogue it is hard to say what the minimum period of a HW Vir system is, as there were no shorter-period systems released. To solve the question about the minimum period, we have to wait for photometric surveys observing a large number of systems with a better cadence and more epochs.  
	
	The eclipsing probability decreases substantially as the period increases, so it is not surprising that we do not find systems with periods longer than about a day. The eclipsing probability for those systems is less than 10\%.  The reflection effect also gets weaker with increasing period. We expect to find only systems with an extremely high reflection effect for periods larger than one day assuming the quality of the light curves from the OGLE and ATLAS survey. Such systems are much rarer than the typical systems (see the light curves of our targets shown in Fig.~\ref{lc1} and \ref{lc2}). To find longer-period sdB binaries with cool companions we need a larger sample and have to include the systems showing only the reflection effect, as well as observations with very accurate light curves.
	
	\section{Discussion}
	For population synthesis, the criteria for the ejection of the common envelope are crucially determined by the orbital period distribution of post-CE binaries \citep{han:2002}. Hence, a period distribution as complete as possible is important for the understanding of those parameters.
	\subsection{The companion masses and periods of the known reflection effect binaries}
	\begin{figure}
		\includegraphics[width=\linewidth]{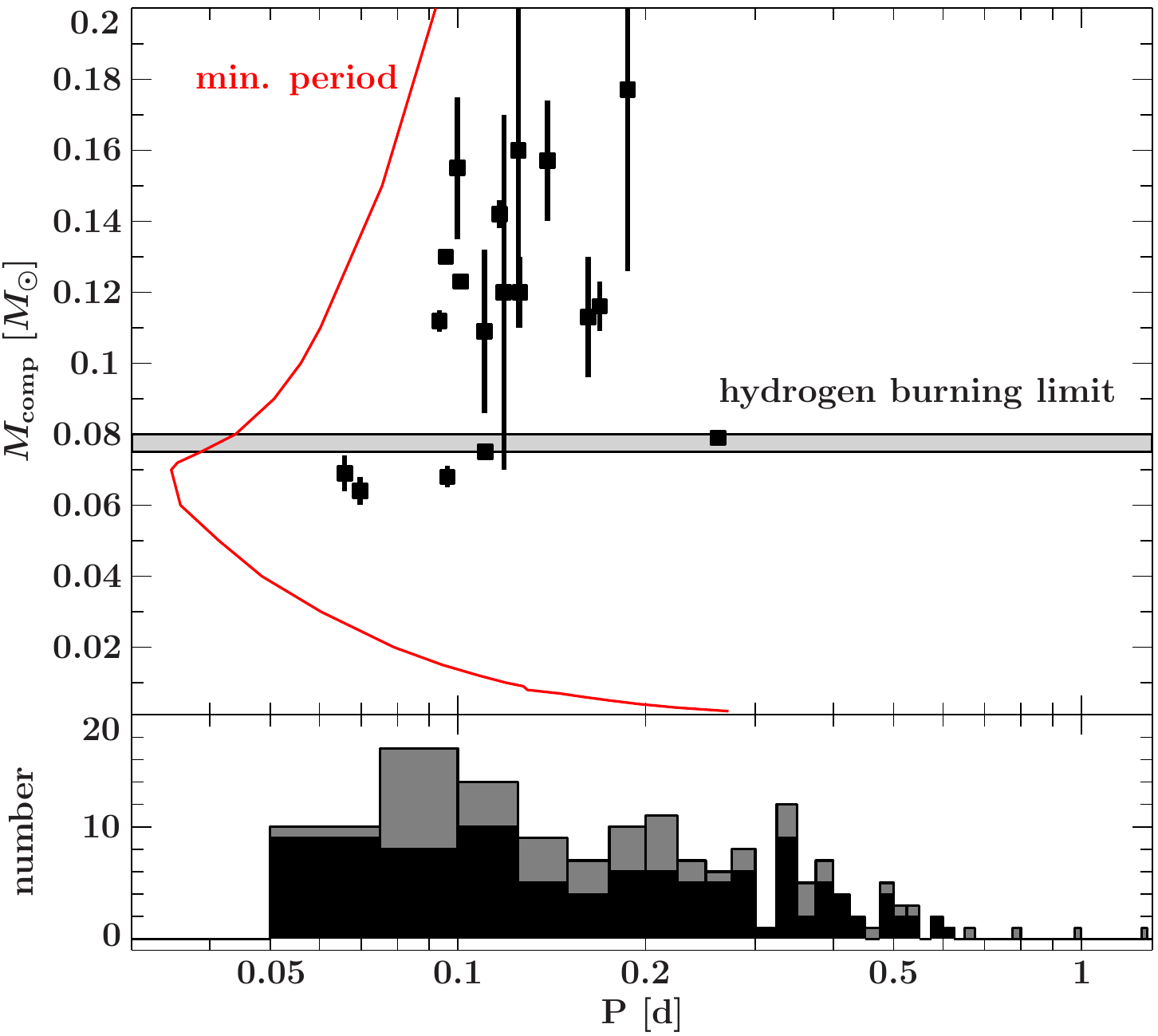}
		\caption{Period/companion mass diagram of the published HW Vir systems. A summary of the parameters can be found in \citet{muchfuss_photo} and references therein. The gray area marks the hydrogen-burning limit. Companions with smaller masses are sub-stellar. The solid red line denotes the minimum period at which a companion of a certain mass could exist in orbit around a canonical--mass sdB, assuming the companion cannot exceed its Roche radius. In the lower panel the period distribution of the newly discovered HW Vir systems presented here is shown for comparison (OGLE targets in black and ATLAS targets in grey).}
		\label{logp-mcomp}
	\end{figure}
	
	Figure \ref{logp-mcomp} shows the companion masses of the published HW Vir systems \citep[][and references therein for a summary of the orbital parameters]{muchfuss_photo} plotted against their orbital
	periods -- there is no obvious correlation between companion mass and period. However, the confirmed sub-stellar companions seem to be found preferentially in the shortest-period systems. The minimum period possible for a system consisting of an sdB and a companion of a certain mass can be calculated by assuming that the radius of the companion cannot exceed its Roche radius. To derive the Roche radius we used the formula by \citet{eggleton}, which depends on the mass ratio and separation of the binary. For the radius of the companion we used the mass-radius relation by \citet{baraffe:98}, and for the sdB mass we adopt the canonical value ($\sim 0.47$ \Msun). 
	
	The minimum orbital period is reached for an sdB with a cool, low-mass companion in the brown dwarf mass range.
	When we look at the period distribution of the newly discovered systems, which is shown in Fig. \ref{logp-mcomp} as comparison, we can see that the number of short-period systems below 0.1 d has increased substantially. Especially in the period range below 0.1 d, where all the currently confirmed brown dwarf companions have been found, we discovered 30 new systems, down to almost the minimal possible period at 0.04 d \citep[see also][]{min_period}. 
	
	To find possible Jupiter mass planets we have to search at longer periods, because they get destroyed during the common-envelope phase if they are too close to the star. We estimate that they can only survive at periods longer than 0.2 to 0.25~d (see Fig. \ref{logp-mcomp}). The only eclipsing sdB binary with a cool, low mass companion known to have such a period was AA Dor. The newly discovered systems increase the range of periods up to more than one day, allowing a search for close Jupiter mass objects around hot subdwarf stars, which could be responsible for the formation of the sdB. This newly discovered sample gives us a unique opportunity to study the parameters of post-common envelope systems over a large period range.

	\begin{figure}
		\includegraphics[width=\linewidth]{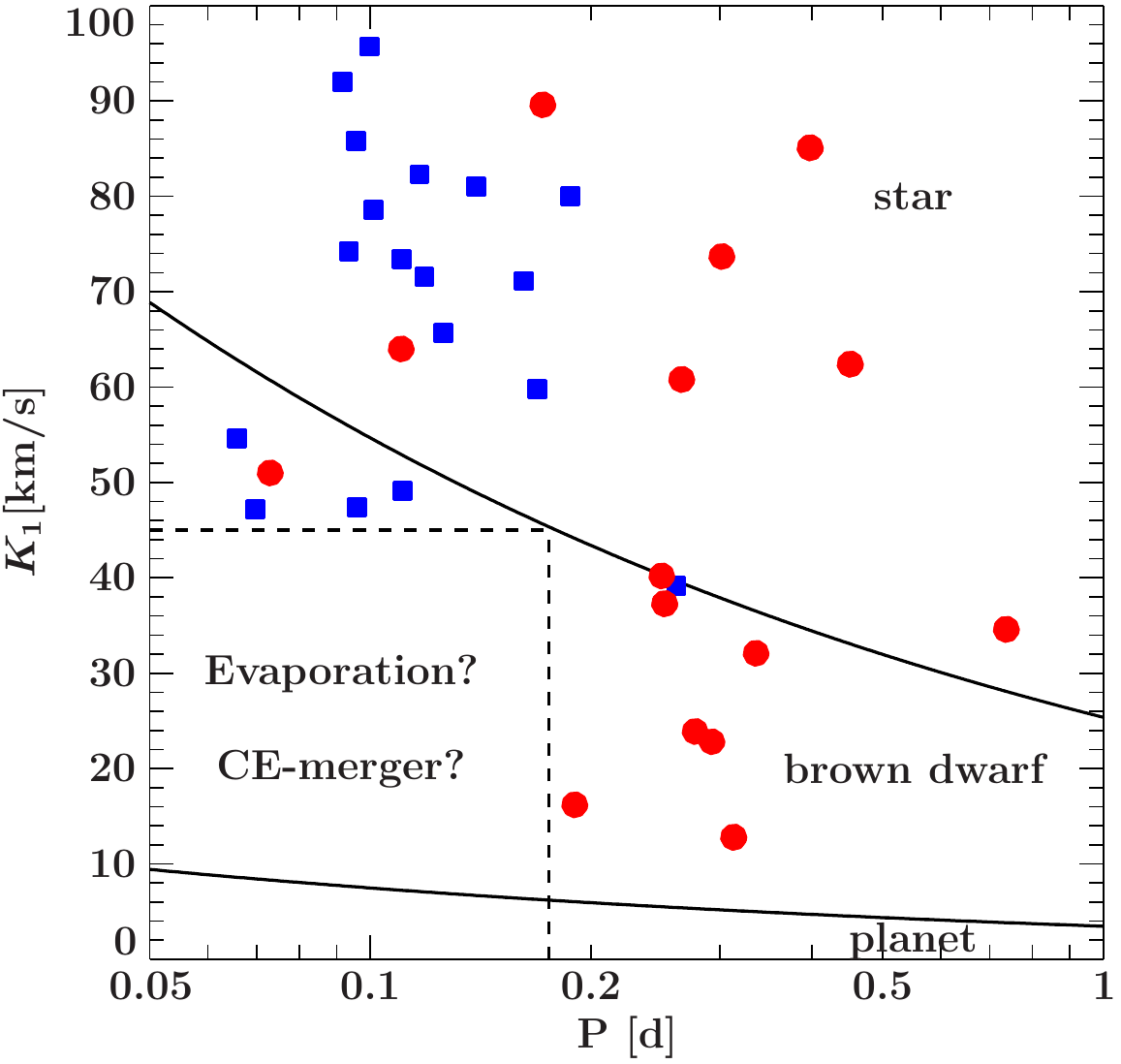}
		\caption{This figure is an updated version of Fig. 4 from \citet{vs:2014_II} adding the HW Vir systems published since then \citep[see][for a table of the parameters of all published HW Vir systems]{muchfuss_photo}. It shows the RV semi-amplitudes of all known sdB binaries with reflection
			effects and spectroscopic solutions plotted against their orbital periods. Blue squares mark eclipsing sdB binaries
			of HW Vir type where the companion mass is well constrained; red circles
			systems without eclipses, where only lower limits can be derived for
			the companion masses. The dashed lines mark the regions to the right
			where the minimum companion masses derived from the binary mass
			function (assuming 0.47\Msun\,for the sdBs) exceed 0.01 \Msun (lower curve)
			and 0.08 \Msun (upper curve).}
		\label{k-m2min}
	\end{figure}
	
	For non-eclipsing systems, the absolute mass of the companion cannot be determined. Some first tests show, however, that the inclination and radii can be constrained with space-based quality light curves (Schaffenroth et al. in prep), which will allow us to constrain the companion masses.  Assuming the canonical sdB mass, a minimum mass for all companions can be derived. An overview of the 33 known sdB binaries
	with reflection effects and known orbital parameters is shown in Fig. \ref{k-m2min}. Although only minimum masses can
	be derived for most of the companions, we can use this sample to do some statistics. While most companions
	are late M-dwarfs with masses close to $\sim$0.1 \Msun, there is no sharp drop below the hydrogen-burning limit. The
	fraction of close sub-stellar companions is substantial. An obvious feature in Fig. \ref{k-m2min} is the lack of binaries with
	periods shorter than $\sim 0.18$ d and $K < 47\,\rm kms^{-1}$ corresponding to companion masses of less than $\sim 0.06\, {\rm M}_{\rm \odot}$.
	This feature is not caused by selection effects: the comparable radii of giant planets and stars close to the hydrogen-burning limit means that their eclipse depths would be similar, and their shorter orbital periods would mean the reflection effect should be as strong or stronger than typical HW Vir binaries. An explanation for the lack of objects at short-periods could be that
	the companion triggered the ejection of the envelope, but was destroyed during the common-envelope phase. This could also explain the formation of some of the single sdBs. Our target sample is ideal to study this region in the diagram more thoroughly, as it increases the number of known systems with periods smaller than 0.15~d significantly.

	\subsection{Comparison with related (eclipsing) binary populations}
	\begin{figure*}
		\includegraphics[width=0.5\linewidth]{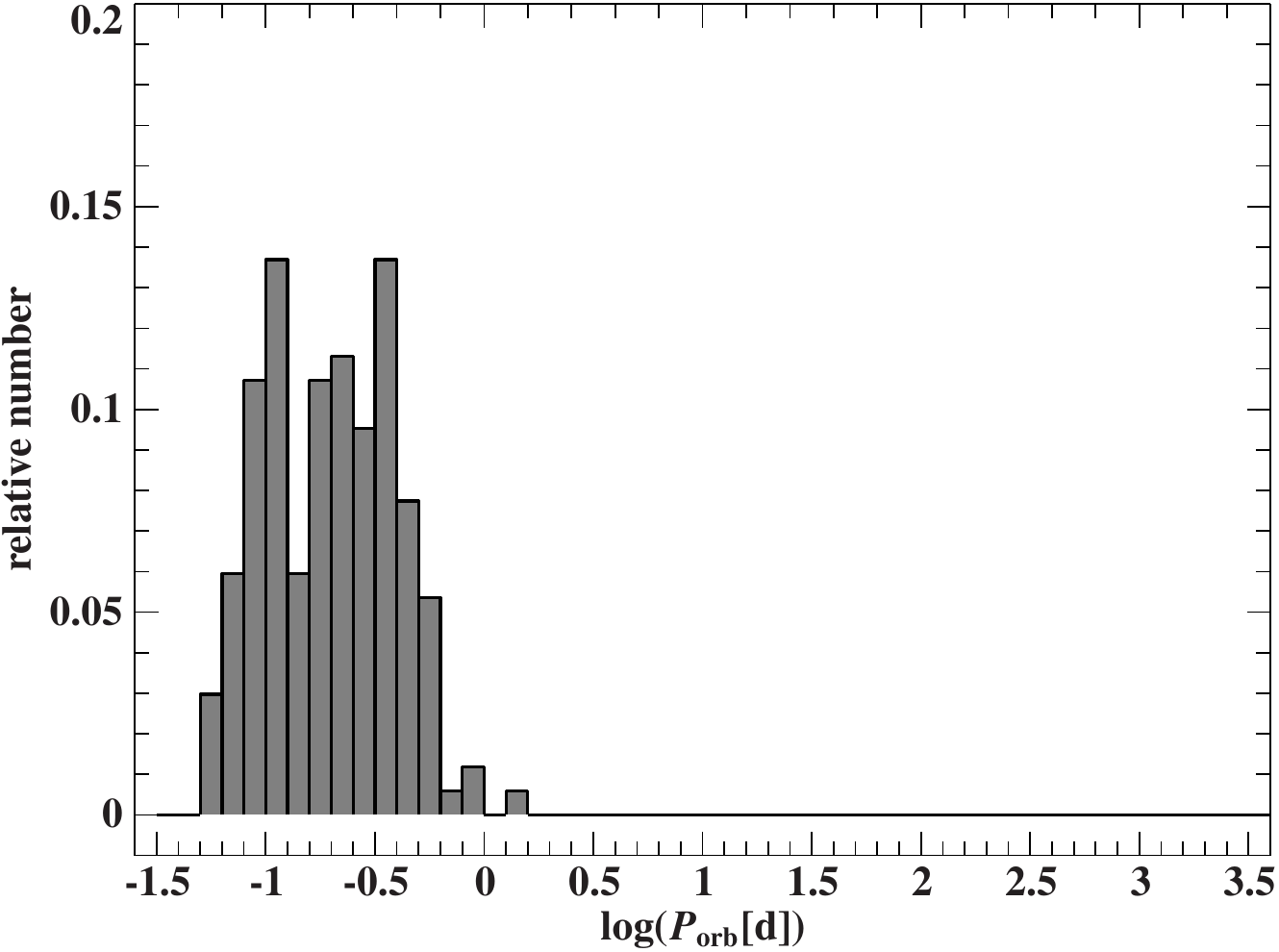}
		\includegraphics[width=0.5\linewidth]{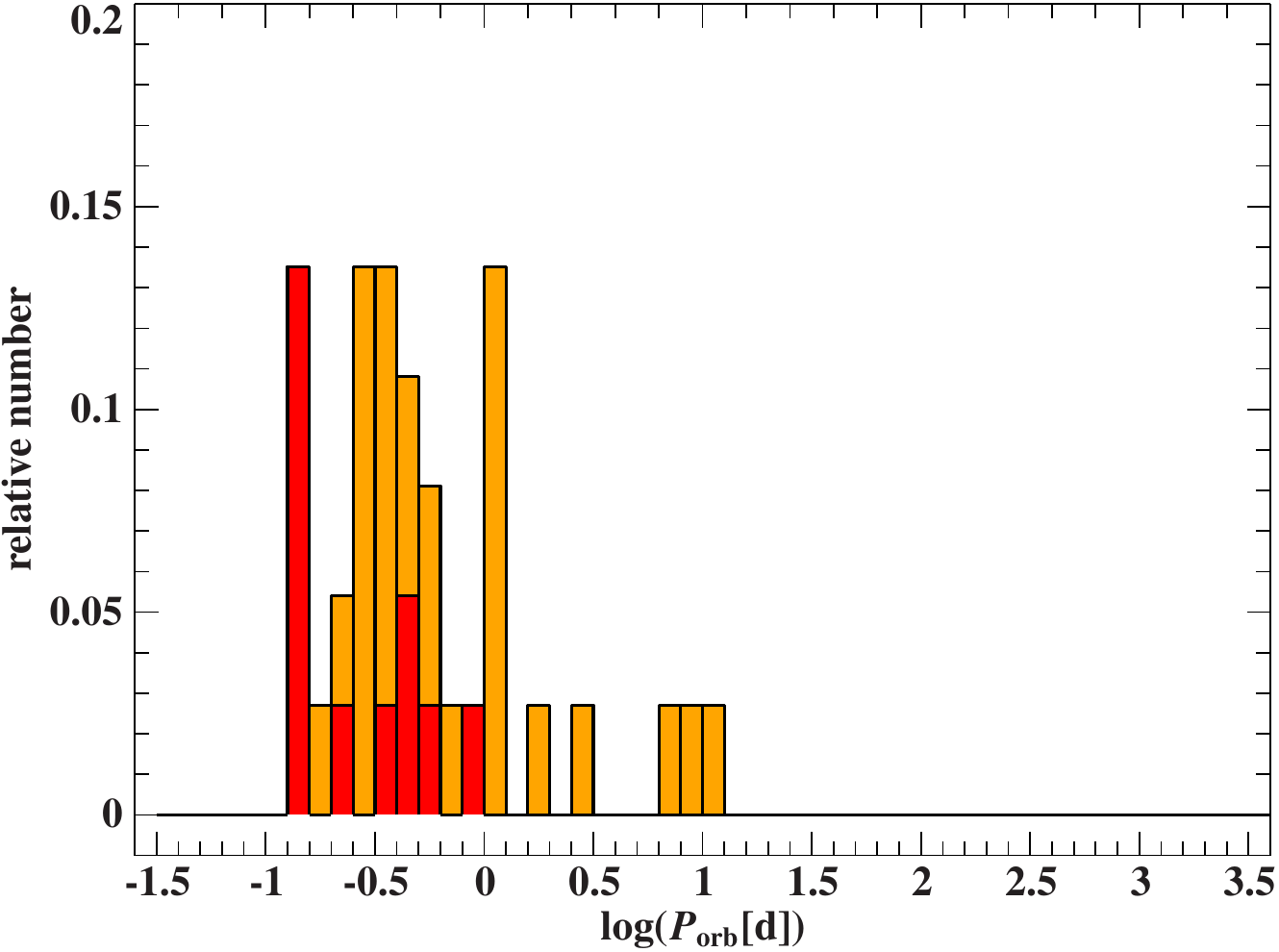}
		\includegraphics[width=0.5\linewidth]{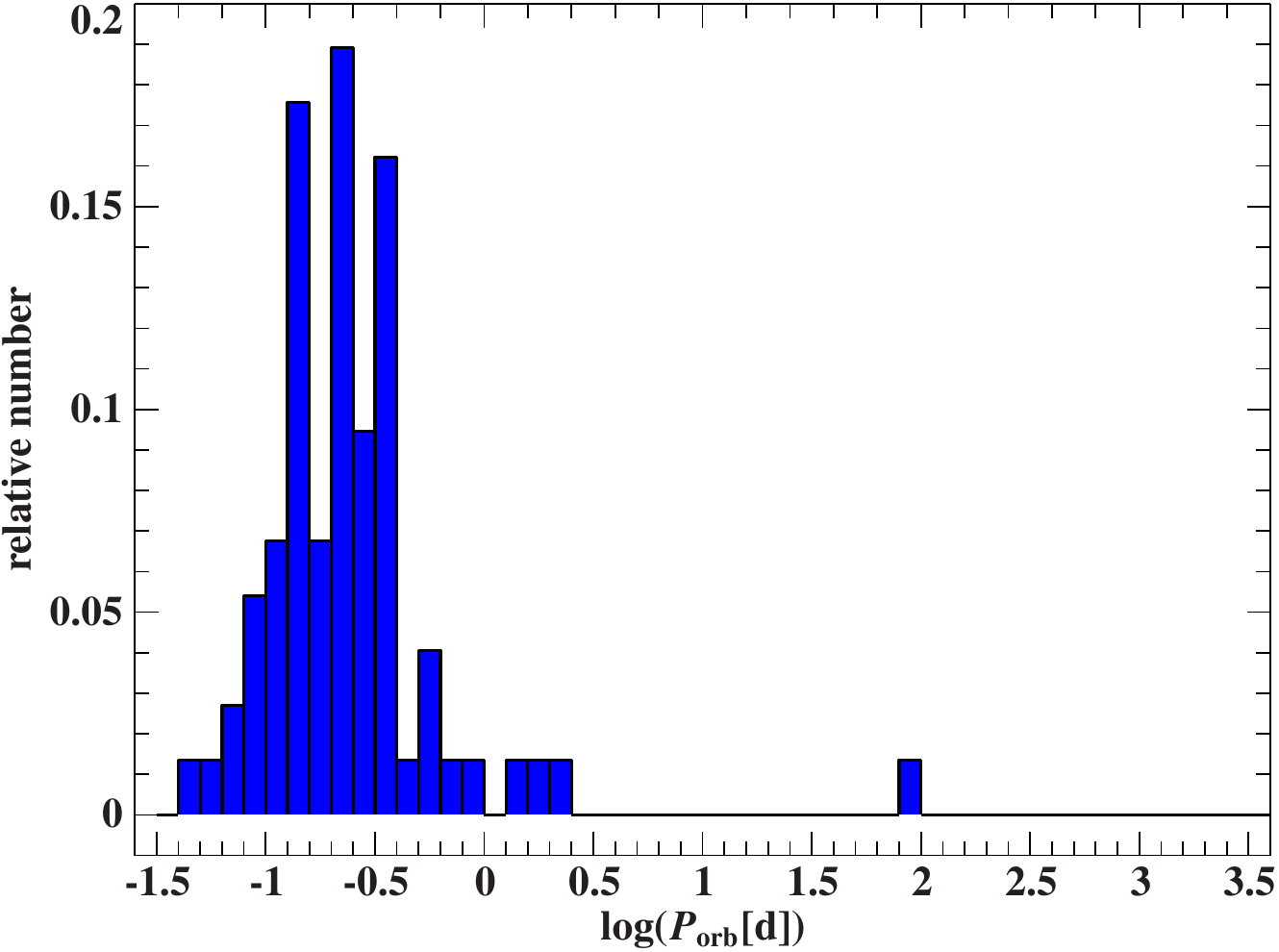}
		\includegraphics[width=0.5\linewidth]{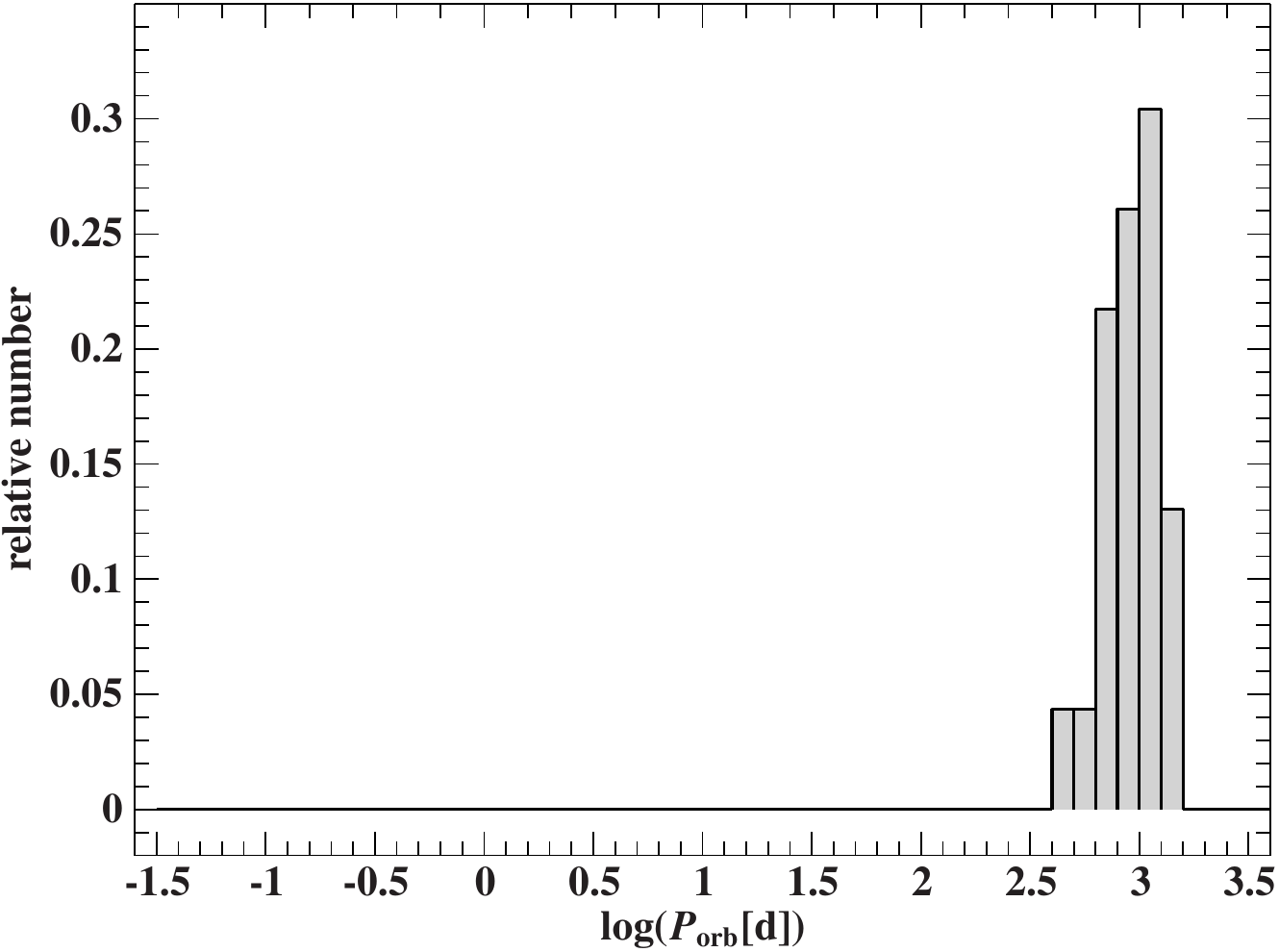}
		\includegraphics[width=0.5\linewidth]{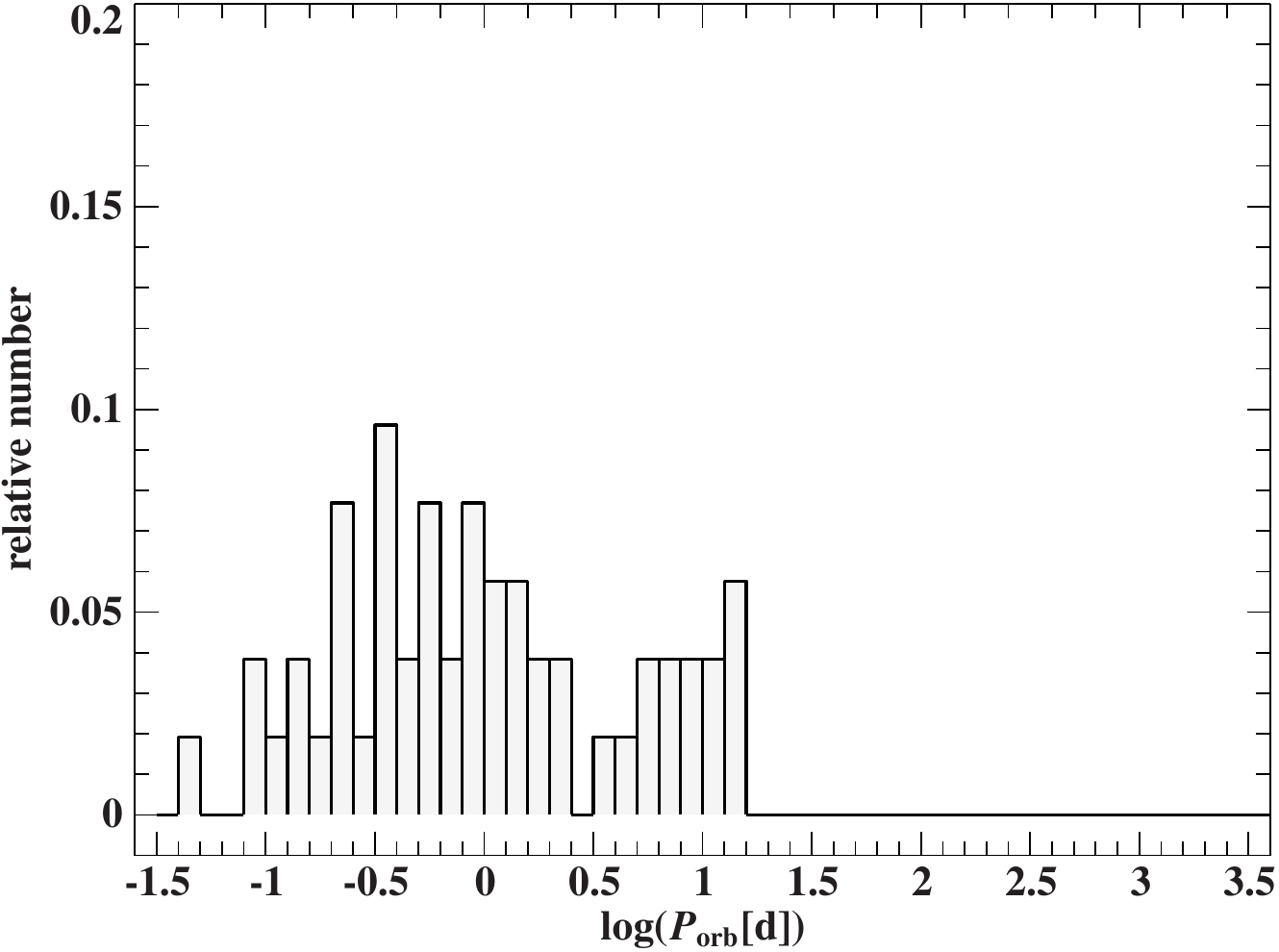}
		\includegraphics[width=0.5\linewidth]{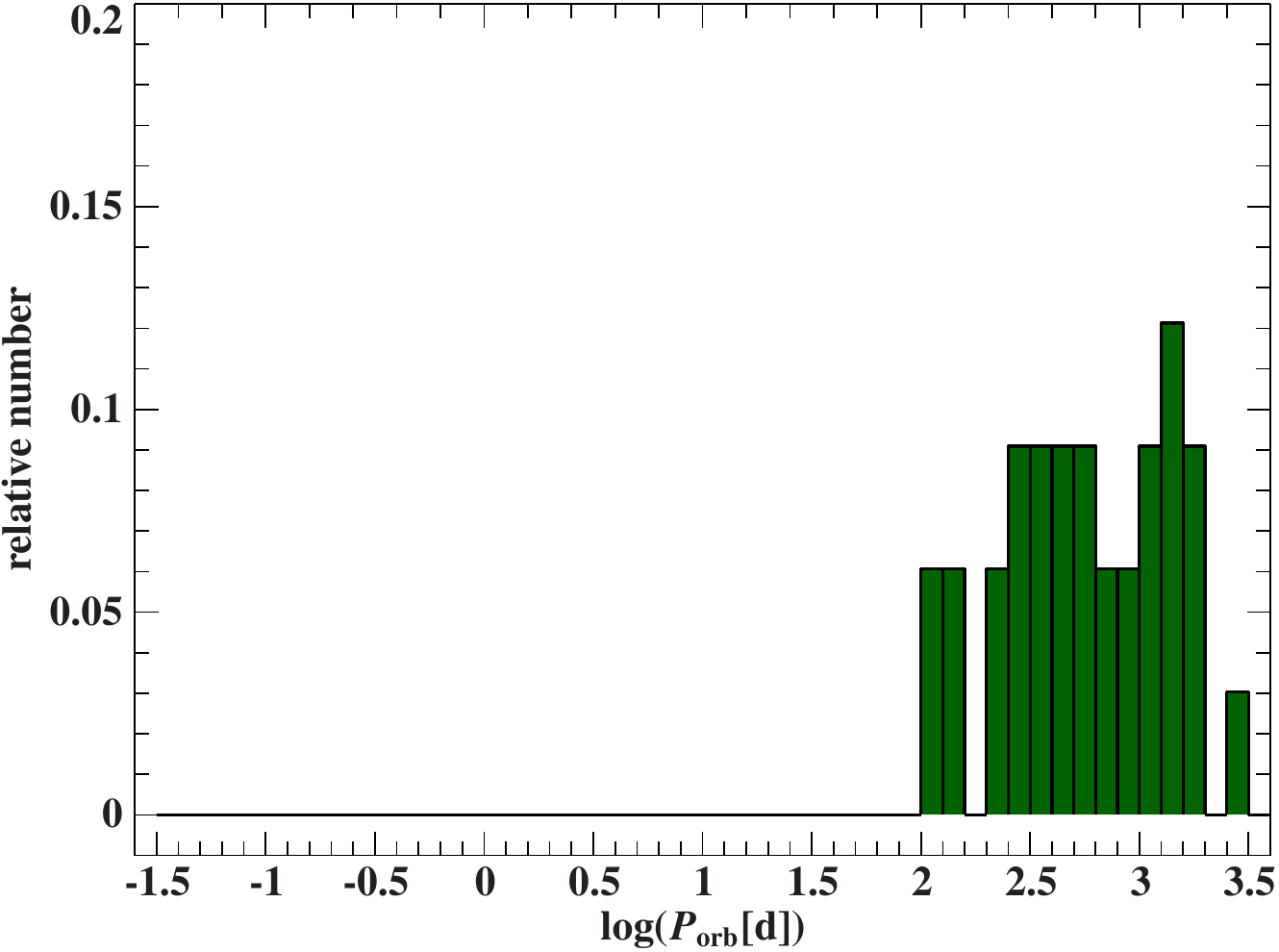}
		\caption{Period distribution of different kinds of post-common envelope systems. \textbf{Left:} \textit{top}: in grey all HW Vir systems including the HW Vir candidate systems from this paper; \textit{middle}:  in blue the known eclipsing WD+dM/BD systems from \citet{wd_ecl}; \textit{bottom:} in white the known sdB+WD systems from \citet{kupfer:2015}. \textbf{Right:} \textit{top:} in red the known eclipsing binary central stars of planetary nebula showing a reflection effect and in orange the non-eclipsing binary central stars of planetary nebula showing a reflection effect (\url{http://www.drdjones.net/bCSPN/}); \textit{middle:} the known sdB + main sequence companions from the Roche lobe overflow channel \citep{long_sdb} in light-gray.; \textit{bottom:} the post-AGB binaries from \citet{post-agb-binaries} in darkgreen.}
		\label{period_ecl}
	\end{figure*}
	It is also interesting to compare our period distribution with the distribution of other types of post-common envelope systems. This is shown in Fig. \ref{period_ecl}. The eclipsing WD+dM/BD systems show a very similar distribution. However, the longest period systems have periods of 2.3 d. All the systems with periods $>0.5$~d have WD primaries with masses more than 0.55 $\rm M_{\odot}$ and have to be post-AGB stars. The longest period system KOI-3278 with a period of 88 d is a post-common envelope system of a CO-WD with a G type companion \citep{parsons:2015}. It is the longest period eclipsing post-common envelope system known. As the primary is a CO white dwarf it has to be a post-AGB binary. The white dwarf is much smaller than the subdwarf, which means that the eclipsing probability is much smaller for white dwarf binaries. However, they are much more common than hot subdwarf stars.
	
	As already discussed in \citet{kupfer:2015} the period distribution of the known sdB+WD systems resulting from the second common envelope channel is much broader than the period distribution of the reflection effect systems with a secondary peak at periods of several days. 
	
	Another type of post-common envelope systems are the binary central stars of planetary nebula . As only 11 eclipsing systems are known, we also added systems which only show the reflection effect. The period distribution of the eclipsing bCSPNs with cool companions looks very similar to the distribution of the HW Vir systems. However, systems showing only the reflection effect and no eclipses are found up to periods of several days, almost as long as the sdB+WD systems. The primary can be a very hot WD or sdO and so the amplitude of the reflection effect is quite large in some but not all cases. The bCSPNs can either be post-RGB or post-AGB systems. 
	
	The longest period post-common envelope system with an sdB primary has a period of 27.8 d, but the nature of the companion has not yet been determined. 
	For comparison we also added the group of long-period sdB systems with FGK companions showing composite spectra. They are found at periods of a few 100 d which means they were formed through the Roche lobe overflow channel. The shortest period for such sdB binaries is 479 d \citep{long_sdb}. No systems with periods in the range $\sim 28-480$ d have been found yet. 
	
	\citet{post-agb-binaries} published a sample of 33 post-AGB binaries. They have periods similar to the long-period sdB binaries and show other characteristics also seen in the long-period sdB binaries (e.g. significant eccentricities in many systems); population synthesis, on the other hand, predicted periods of a few days for common-envelope systems \citep{nie}. 
	\citet{post-agb-binaries} claim that post-AGB stars with periods less than 100~d should fill their Roche lobes and therefore evolve into bCSPNs quickly or contain hotter post-AGB primaries. Those should be distinguishable from post-RGB stars quite easily, as they are expected to be much more luminous.
	
	\section{Summary and outlook}
	
	In the EREBOS project we study a large sample of homogeneously selected HW Vir systems. We investigated two photometric surveys to find more such systems and increased the number of known systems by a factor of almost 10. We plan a photometric and spectroscopic follow-up of as many targets as possible to determine the fundamental stellar ($M$, $R$), atmospheric ($T_{\rm eff}$, $\log{g}$) and binary parameters ($a$, $P$). At the moment we already have spectroscopic follow-up for 28 objects. For several of our systems we took photometric follow-up in several bands ($B,V,R$), which is essential for modelling the reflection effect. 
	
	For 25 of our targets, it has been spectroscopically confirmed that they are indeed systems consisting of a hot subdwarf primary and a cool, low mass companion; only two targets have a DA primary. Four systems in our sample were confirmed to be central stars of planetary nebula. This means that $\sim$90\% of our target sample are most likely to be eclipsing hot subdwarf binaries.
	
	The main goal of EREBOS is to investigate HW Vir systems over the whole range of the period distribution. We hope to improve the understanding of the common envelope phase by investigating a large number of post-common envelope binaries. Moreover, we are especially interested in the influence of the lowest mass companions -- brown dwarfs or massive planets -- on stellar evolution.  A future goal is an improved physical model of the reflection effect, which we hope to achieve with this huge sample of reflection effect binaries.
	
	A byproduct that will emerge from our sample is a mass-radius relation for cool, low-mass objects which are highly irradiated by a hot companion. This should shed light on the question of how much such objects are inflated.
	
	The TESS misson, which is observing at the moment, provides 27 d light curves of a each field and, for a few bright targets, light curves with a 2 min cadence are transmitted. The full-frame images are transferred every 30 min, allowing us to derive the light curves of all targets. This will give 27 d light curves with space-quality, allowing us to detect  reflection effects with periods of several days and to find the longest period reflection effect systems. It will also provide excellent light curves to expand our target sample further. Also several ground-based surveys are or will provide excellent light curve data in several photometric bands, which will find a large number of new HW Vir systems (e.g., ZTF and BlackGEM, where we became a member).
	
	\begin{acknowledgements}
		Based on observations collected at the European Organisation for Astronomical Research in the Southern Hemisphere under ESO programme(s) 092.D-0040(A), 095.D-0167(A), 196.D-0214(A-D), 099.D-0217(A), 0101.D-0791(A). Also based on observations obtained at the Southern Astrophysical Research (SOAR) telescope, which is a joint project of the Minist\'{e}rio da Ci\^{e}ncia, Tecnologia, Inova\c{c}\~{o}es e Comunica\c{c}\~{o}es (MCTIC) do Brasil, the U.S. National Optical Astronomy Observatory (NOAO), the University of North Carolina at Chapel Hill (UNC), and Michigan State University (MSU).
		We would like to thank the Chilean Time Allocation Committee for awarding us time for complementing the EREBOS Project with the proposal CL-2016B-018. Unfortunately, there was an error on our side and the finder chart was wrongly marked so that the data were not used in the analysis presented in this paper.
		V.S. is supported by the Deutsche Forschungsgemeinschaft (DFG) through grant GE 2506/9-1. BB is supported by the National Science Foundation grant AST-1812874. DK acknowledges financial support from the University of the Western Cape and the National Research Foundation of South Africa. IP acknowledges funding by the Deutsche Forschungsgemeinschaft under grant GE2506/12-1. V.S and S.G were supported by the DAAD PPP USA for this project. Thereby, we would also like to thank the Kavli Institute for Theoretical Physics, UC Santa Barbara for hosting V.S and S.G. during the stay funded by DAAD PPP USA, meanwhile a large part of the manuscript was written.
		This research has made use of ISIS functions (ISISscripts) provided by ECAP/Remeis observatory and MIT (http://www.sternwarte.uni-erlangen.de/isis/).  
		
	\end{acknowledgements}

	\bibliography{aabib}

\begin{thebibliography}{51}
\expandafter\ifx\csname natexlab\endcsname\relax\def\natexlab#1{#1}\fi

\bibitem[{{Af{\c{s}}ar} \& {Ibano{\v{g}}lu}(2008)}]{pn}
{Af{\c{s}}ar}, M. \& {Ibano{\v{g}}lu}, C. 2008, \mnras, 391, 802

\bibitem[{{Almeida} {et~al.}(2017){Almeida}, {Damineli}, {Rodrigues},
  {Pereira}, \& {Jablonski}}]{almeida}
{Almeida}, L.~A., {Damineli}, A., {Rodrigues}, C.~V., {Pereira}, M.~G., \&
  {Jablonski}, F. 2017, \mnras, 472, 3093

\bibitem[{{Althaus} {et~al.}(2013){Althaus}, {Miller Bertolami}, \&
  {C{\'o}rsico}}]{althaus}
{Althaus}, L.~G., {Miller Bertolami}, M.~M., \& {C{\'o}rsico}, A.~H. 2013,
  \aap, 557, A19

\bibitem[{{Bailer-Jones} {et~al.}(2018){Bailer-Jones}, {Rybizki}, {Fouesneau},
  {Mantelet}, \& {Andrae}}]{distances}
{Bailer-Jones}, C.~A.~L., {Rybizki}, J., {Fouesneau}, M., {Mantelet}, G., \&
  {Andrae}, R. 2018, VizieR Online Data Catalog, 1347

\bibitem[{{Baraffe} {et~al.}(1998){Baraffe}, {Chabrier}, {Allard}, \&
  {Hauschildt}}]{baraffe:98}
{Baraffe}, I., {Chabrier}, G., {Allard}, F., \& {Hauschildt}, P.~H. 1998, \aap,
  337, 403

\bibitem[{{Budaj}(2011)}]{budaj}
{Budaj}, J. 2011, \aj, 141, 59

\bibitem[{{Clemens} {et~al.}(2004){Clemens}, {Crain}, \&
  {Anderson}}]{clemens:2004}
{Clemens}, J.~C., {Crain}, J.~A., \& {Anderson}, R. 2004, in \procspie, Vol.
  5492, Ground-based Instrumentation for Astronomy, ed. A.~F.~M. {Moorwood} \&
  M.~{Iye}, 331--340

\bibitem[{{Dorman} {et~al.}(1993){Dorman}, {Rood}, \&
  {O'Connell}}]{Dorman:1993}
{Dorman}, B., {Rood}, R.~T., \& {O'Connell}, R.~W. 1993, APJ, 419, 596

\bibitem[{{Eggleton}(1983)}]{eggleton}
{Eggleton}, P.~P. 1983, \apj, 268, 368

\bibitem[{{Gaia Collaboration} {et~al.}(2018{\natexlab{a}}){Gaia
  Collaboration}, {Babusiaux}, {van Leeuwen}, {Barstow}, {Jordi}, {Vallenari},
  {Bossini}, {Bressan}, {Cantat-Gaudin}, {van Leeuwen}, \& et~al.}]{extinction}
{Gaia Collaboration}, {Babusiaux}, C., {van Leeuwen}, F., {et~al.}
  2018{\natexlab{a}}, \aap, 616, A10

\bibitem[{{Gaia Collaboration} {et~al.}(2018{\natexlab{b}}){Gaia
  Collaboration}, {Brown}, {Vallenari}, {Prusti}, {de Bruijne}, {Babusiaux},
  {Bailer-Jones}, {Biermann}, {Evans}, {Eyer}, {Jansen}, {Jordi}, {Klioner},
  {Lammers}, {Lindegren}, {Luri}, {Mignard}, {Panem}, {Pourbaix}, {Randich},
  {Sartoretti}, {Siddiqui}, {Soubiran}, {van Leeuwen}, {Walton}, {Arenou},
  {Bastian}, {Cropper}, {Drimmel}, {Katz}, {Lattanzi}, {Bakker}, {Cacciari},
  {Casta{\~n}eda}, {Chaoul}, {Cheek}, {De Angeli}, {Fabricius}, {Guerra},
  {Holl}, {Masana}, {Messineo}, {Mowlavi}, {Nienartowicz}, {Panuzzo},
  {Portell}, {Riello}, {Seabroke}, {Tanga}, {Th{\'e}venin}, {Gracia-Abril},
  {Comoretto}, {Garcia-Reinaldos}, {Teyssier}, {Altmann}, {Andrae}, {Audard},
  {Bellas-Velidis}, {Benson}, {Berthier}, {Blomme}, {Burgess}, {Busso},
  {Carry}, {Cellino}, {Clementini}, {Clotet}, {Creevey}, {Davidson}, {De
  Ridder}, {Delchambre}, {Dell'Oro}, {Ducourant},
  {Fern{\'a}ndez-Hern{\'a}ndez}, {Fouesneau}, {Fr{\'e}mat}, {Galluccio},
  {Garc{\'\i}a-Torres}, {Gonz{\'a}lez-N{\'u}{\~n}ez}, {Gonz{\'a}lez-Vidal},
  {Gosset}, {Guy}, {Halbwachs}, {Hambly}, {Harrison}, {Hern{\'a}ndez},
  {Hestroffer}, {Hodgkin}, {Hutton}, {Jasniewicz}, {Jean-Antoine-Piccolo},
  {Jordan}, {Korn}, {Krone-Martins}, {Lanzafame}, {Lebzelter}, {L{\"o}ffler},
  {Manteiga}, {Marrese}, {Mart{\'\i}n-Fleitas}, {Moitinho}, {Mora}, {Muinonen},
  {Osinde}, {Pancino}, {Pauwels}, {Petit}, {Recio-Blanco}, {Richards},
  {Rimoldini}, {Robin}, {Sarro}, {Siopis}, {Smith}, {Sozzetti}, {S{\"u}veges},
  {Torra}, {van Reeven}, {Abbas}, {Abreu Aramburu}, {Accart}, {Aerts},
  {Altavilla}, {{\'A}lvarez}, {Alvarez}, {Alves}, {Anderson}, {Andrei},
  {Anglada Varela}, {Antiche}, {Antoja}, {Arcay}, {Astraatmadja}, {Bach},
  {Baker}, {Balaguer-N{\'u}{\~n}ez}, {Balm}, {Barache}, {Barata}, {Barbato},
  {Barblan}, {Barklem}, {Barrado}, {Barros}, {Barstow}, {Bartholom{\'e}
  Mu{\~n}oz}, {Bassilana}, {Becciani}, {Bellazzini}, {Berihuete}, {Bertone},
  {Bianchi}, {Bienaym{\'e}}, {Blanco-Cuaresma}, {Boch}, {Boeche}, {Bombrun},
  {Borrachero}, {Bossini}, {Bouquillon}, {Bourda}, {Bragaglia}, {Bramante},
  {Breddels}, {Bressan}, {Brouillet}, {Br{\"u}semeister}, {Brugaletta},
  {Bucciarelli}, {Burlacu}, {Busonero}, {Butkevich}, {Buzzi}, {Caffau},
  {Cancelliere}, {Cannizzaro}, {Cantat-Gaudin}, {Carballo}, {Carlucci},
  {Carrasco}, {Casamiquela}, {Castellani}, {Castro-Ginard}, {Charlot},
  {Chemin}, {Chiavassa}, {Cocozza}, {Costigan}, {Cowell}, {Crifo}, {Crosta},
  {Crowley}, {Cuypers}, {Dafonte}, {Damerdji}, {Dapergolas}, {David}, {David},
  {de Laverny}, {De Luise}, {De March}, {de Martino}, {de Souza}, {de Torres},
  {Debosscher}, {del Pozo}, {Delbo}, {Delgado}, {Delgado}, {Di Matteo},
  {Diakite}, {Diener}, {Distefano}, {Dolding}, {Drazinos}, {Dur{\'a}n},
  {Edvardsson}, {Enke}, {Eriksson}, {Esquej}, {Eynard Bontemps}, {Fabre},
  {Fabrizio}, {Faigler}, {Falc{\~a}o}, {Farr{\`a}s Casas}, {Federici},
  {Fedorets}, {Fernique}, {Figueras}, {Filippi}, {Findeisen}, {Fonti},
  {Fraile}, {Fraser}, {Fr{\'e}zouls}, {Gai}, {Galleti}, {Garabato},
  {Garc{\'\i}a-Sedano}, {Garofalo}, {Garralda}, {Gavel}, {Gavras}, {Gerssen},
  {Geyer}, {Giacobbe}, {Gilmore}, {Girona}, {Giuffrida}, {Glass}, {Gomes},
  {Granvik}, {Gueguen}, {Guerrier}, {Guiraud}, {Guti{\'e}rrez-S{\'a}nchez},
  {Haigron}, {Hatzidimitriou}, {Hauser}, {Haywood}, {Heiter}, {Helmi}, {Heu},
  {Hilger}, {Hobbs}, {Hofmann}, {Holland}, {Huckle}, {Hypki}, {Icardi},
  {Jan{\ss}en}, {Jevardat de Fombelle}, {Jonker}, {Juh{\'a}sz}, {Julbe},
  {Karampelas}, {Kewley}, {Klar}, {Kochoska}, {Kohley}, {Kolenberg},
  {Kontizas}, {Kontizas}, {Koposov}, {Kordopatis}, {Kostrzewa-Rutkowska},
  {Koubsky}, {Lambert}, {Lanza}, {Lasne}, {Lavigne}, {Le Fustec}, {Le
  Poncin-Lafitte}, {Lebreton}, {Leccia}, {Leclerc}, {Lecoeur-Taibi},
  {Lenhardt}, {Leroux}, {Liao}, {Licata}, {Lindstr{\o}m}, {Lister}, {Livanou},
  {Lobel}, {L{\'o}pez}, {Managau}, {Mann}, {Mantelet}, {Marchal}, {Marchant},
  {Marconi}, {Marinoni}, {Marschalk{\'o}}, {Marshall}, {Martino}, {Marton},
  {Mary}, {Massari}, {Matijevi{\v{c}}}, {Mazeh}, {McMillan}, {Messina},
  {Michalik}, {Millar}, {Molina}, {Molinaro}, {Moln{\'a}r}, {Montegriffo},
  {Mor}, {Morbidelli}, {Morel}, {Morris}, {Mulone}, {Muraveva}, {Musella},
  {Nelemans}, {Nicastro}, {Noval}, {O'Mullane}, {Ord{\'e}novic},
  {Ord{\'o}{\~n}ez-Blanco}, {Osborne}, {Pagani}, {Pagano}, {Pailler},
  {Palacin}, {Palaversa}, {Panahi}, {Pawlak}, {Piersimoni}, {Pineau}, {Plachy},
  {Plum}, {Poggio}, {Poujoulet}, {Pr{\v{s}}a}, {Pulone}, {Racero}, {Ragaini},
  {Rambaux}, {Ramos-Lerate}, {Regibo}, {Reyl{\'e}}, {Riclet}, {Ripepi}, {Riva},
  {Rivard}, {Rixon}, {Roegiers}, {Roelens}, {Romero-G{\'o}mez}, {Rowell},
  {Royer}, {Ruiz-Dern}, {Sadowski}, {Sagrist{\`a} Sell{\'e}s}, {Sahlmann},
  {Salgado}, {Salguero}, {Sanna}, {Santana-Ros}, {Sarasso}, {Savietto},
  {Schultheis}, {Sciacca}, {Segol}, {Segovia}, {S{\'e}gransan}, {Shih},
  {Siltala}, {Silva}, {Smart}, {Smith}, {Solano}, {Solitro}, {Sordo}, {Soria
  Nieto}, {Souchay}, {Spagna}, {Spoto}, {Stampa}, {Steele},
  {Steidelm{\"u}ller}, {Stephenson}, {Stoev}, {Suess}, {Surdej}, {Szabados},
  {Szegedi-Elek}, {Tapiador}, {Taris}, {Tauran}, {Taylor}, {Teixeira},
  {Terrett}, {Teyssand ier}, {Thuillot}, {Titarenko}, {Torra Clotet}, {Turon},
  {Ulla}, {Utrilla}, {Uzzi}, {Vaillant}, {Valentini}, {Valette}, {van Elteren},
  {Van Hemelryck}, {van Leeuwen}, {Vaschetto}, {Vecchiato}, {Veljanoski},
  {Viala}, {Vicente}, {Vogt}, {von Essen}, {Voss}, {Votruba}, {Voutsinas},
  {Walmsley}, {Weiler}, {Wertz}, {Wevers}, {Wyrzykowski}, {Yoldas},
  {{\v{Z}}erjal}, {Ziaeepour}, {Zorec}, {Zschocke}, {Zucker}, {Zurbach}, \&
  {Zwitter}}]{gaia_dr2}
{Gaia Collaboration}, {Brown}, A.~G.~A., {Vallenari}, A., {et~al.}
  2018{\natexlab{b}}, \aap, 616, A1

\bibitem[{{Geier} {et~al.}(2019){Geier}, {Raddi}, {Gentile Fusillo}, \&
  {Marsh}}]{gaia_catalog}
{Geier}, S., {Raddi}, R., {Gentile Fusillo}, N.~P., \& {Marsh}, T.~R. 2019,
  \aap, 621, A38

\bibitem[{{Geier} {et~al.}(2011){Geier}, {Schaffenroth}, {Drechsel}, {Heber},
  {Kupfer}, {Tillich}, {{\O}stensen}, {Smolders}, {Degroote}, {Maxted},
  {Barlow}, {G{\"a}nsicke}, {Marsh}, \& {Napiwotzki}}]{geier}
{Geier}, S., {Schaffenroth}, V., {Drechsel}, H., {et~al.} 2011, APJ, 731, L22+

\bibitem[{{Gentile Fusillo} {et~al.}(2015){Gentile Fusillo}, {G{\"a}nsicke}, \&
  {Greiss}}]{sdss_wd}
{Gentile Fusillo}, N.~P., {G{\"a}nsicke}, B.~T., \& {Greiss}, S. 2015, \mnras,
  448, 2260

\bibitem[{{Gentile Fusillo} {et~al.}(2019){Gentile Fusillo}, {Tremblay},
  {G{\"a}nsicke}, {Manser}, {Cunningham}, {Cukanovaite}, {Hollands}, {Marsh},
  {Raddi}, {Jordan}, {Toonen}, {Geier}, {Barstow}, \& {Cummings}}]{gaia_wd}
{Gentile Fusillo}, N.~P., {Tremblay}, P.-E., {G{\"a}nsicke}, B.~T., {et~al.}
  2019, \mnras, 482, 4570

\bibitem[{{Han} {et~al.}(2002){Han}, {Podsiadlowski}, {Maxted}, {Marsh}, \&
  {Ivanova}}]{han:2002}
{Han}, Z., {Podsiadlowski}, P., {Maxted}, P.~F.~L., {Marsh}, T.~R., \&
  {Ivanova}, N. 2002, MNRAS, 336, 449

\bibitem[{{Heber}(2009)}]{heber:2009}
{Heber}, U. 2009, ARA{\rm \&}A, 47, 211

\bibitem[{{Heber}(2016)}]{heber:2016}
{Heber}, U. 2016, ArXiv e-prints

\bibitem[{{Heber} {et~al.}(2000){Heber}, {Reid}, \& {Werner}}]{heber:2000}
{Heber}, U., {Reid}, I.~N., \& {Werner}, K. 2000, \aap, 363, 198

\bibitem[{{Hillwig} {et~al.}(2017){Hillwig}, {Frew}, {Reindl}, {Rotter},
  {Webb}, \& {Margheim}}]{pn_rgb}
{Hillwig}, T.~C., {Frew}, D.~J., {Reindl}, N., {et~al.} 2017, \aj, 153, 24

\bibitem[{Hirsch(2009)}]{hirsch}
Hirsch, H. 2009, Phd thesis, Friedrich Alexander Universit\"at Erlangen
  N\"urnberg

\bibitem[{{Jeffery} \& {Ramsay}(2014)}]{jeffrey:2014}
{Jeffery}, C.~S. \& {Ramsay}, G. 2014, \mnras, 442, L61

\bibitem[{{Kupfer} {et~al.}(2015){Kupfer}, {Geier}, {Heber}, {{\O}stensen},
  {Barlow}, {Maxted}, {Heuser}, {Schaffenroth}, \&
  {G{\"a}nsicke}}]{kupfer:2015}
{Kupfer}, T., {Geier}, S., {Heber}, U., {et~al.} 2015, \aap, 576, A44

\bibitem[{{Lallement} {et~al.}(2014){Lallement}, {Vergely}, {Valette},
  {Puspitarini}, {Eyer}, \& {Casagrande}}]{stilism}
{Lallement}, R., {Vergely}, J.-L., {Valette}, B., {et~al.} 2014, \aap, 561, A91

\bibitem[{{Lomb}(1976)}]{lomb}
{Lomb}, N.~R. 1976, \apss, 39, 447

\bibitem[{{Maxted} {et~al.}(2001){Maxted}, {Heber}, {Marsh}, \&
  {North}}]{maxted:2001}
{Maxted}, P.~f.~L., {Heber}, U., {Marsh}, T.~R., \& {North}, R.~C. 2001, MNRAS,
  326, 1391

\bibitem[{{Miller Bertolami}(2016)}]{post-agb}
{Miller Bertolami}, M.~M. 2016, \aap, 588, A25

\bibitem[{{Miszalski} {et~al.}(2009){Miszalski}, {Acker}, {Moffat}, {Parker},
  \& {Udalski}}]{cat_pn}
{Miszalski}, B., {Acker}, A., {Moffat}, A.~F.~J., {Parker}, Q.~A., \&
  {Udalski}, A. 2009, VizieR Online Data Catalog, 349

\bibitem[{{Nelson} {et~al.}(2018){Nelson}, {Schwab}, {Ristic}, \&
  {Rappaport}}]{min_period}
{Nelson}, L., {Schwab}, J., {Ristic}, M., \& {Rappaport}, S. 2018, \apj, 866,
  88

\bibitem[{{Nie} {et~al.}(2012){Nie}, {Wood}, \& {Nicholls}}]{nie}
{Nie}, J.~D., {Wood}, P.~R., \& {Nicholls}, C.~P. 2012, \mnras, 423, 2764

\bibitem[{{Oomen} {et~al.}(2018){Oomen}, {Van Winckel}, {Pols}, {Nelemans},
  {Escorza}, {Manick}, {Kamath}, \& {Waelkens}}]{post-agb-binaries}
{Oomen}, G.-M., {Van Winckel}, H., {Pols}, O., {et~al.} 2018, \aap, 620, A85

\bibitem[{{Parsons} {et~al.}(2015{\natexlab{a}}){Parsons}, {Agurto-Gangas},
  {G{\"a}nsicke}, {Rebassa-Mansergas}, {Schreiber}, {Marsh}, {Dhillon},
  {Littlefair}, {Drake}, {Bours}, {Breedt}, {Copperwheat}, {Hardy}, {Buisset},
  {Prasit}, \& {Ren}}]{wd_ecl}
{Parsons}, S.~G., {Agurto-Gangas}, C., {G{\"a}nsicke}, B.~T., {et~al.}
  2015{\natexlab{a}}, \mnras, 449, 2194

\bibitem[{{Parsons} {et~al.}(2015{\natexlab{b}}){Parsons}, {Agurto-Gangas},
  {G{\"a}nsicke}, {Rebassa-Mansergas}, {Schreiber}, {Marsh}, {Dhillon},
  {Littlefair}, {Drake}, {Bours}, {Breedt}, {Copperwheat}, {Hardy}, {Buisset},
  {Prasit}, \& {Ren}}]{parsons:2015}
{Parsons}, S.~G., {Agurto-Gangas}, C., {G{\"a}nsicke}, B.~T., {et~al.}
  2015{\natexlab{b}}, \mnras, 449, 2194

\bibitem[{{Parsons} {et~al.}(2010){Parsons}, {Marsh}, {Copperwheat}, {Dhillon},
  {Littlefair}, {G{\"a}nsicke}, \& {Hickman}}]{NN_ser}
{Parsons}, S.~G., {Marsh}, T.~R., {Copperwheat}, C.~M., {et~al.} 2010, \mnras,
  402, 2591

\bibitem[{{Pietrukowicz} {et~al.}(2013){Pietrukowicz}, {Mr{\'o}z},
  {Soszy{\'n}ski}, {Udalski}, {Poleski}, {Szyma{\'n}ski}, {Kubiak},
  {Pietrzy{\'n}ski}, {Wyrzykowski}, {Ulaczyk}, {Koz{\l}owski}, \&
  {Skowron}}]{ogle}
{Pietrukowicz}, P., {Mr{\'o}z}, P., {Soszy{\'n}ski}, I., {et~al.} 2013, \actaa,
  63, 115

\bibitem[{{Riello} {et~al.}(2018){Riello}, {De Angeli}, {Evans}, {Busso},
  {Hambly}, {Davidson}, {Burgess}, {Montegriffo}, {Osborne}, {Kewley},
  {Carrasco}, {Fabricius}, {Jordi}, {Cacciari}, {van Leeuwen}, \&
  {Holland}}]{gaia_process_photometry}
{Riello}, M., {De Angeli}, F., {Evans}, D.~W., {et~al.} 2018, \aap, 616, A3

\bibitem[{{Scargle}(1982)}]{scargle}
{Scargle}, J.~D. 1982, \apj, 263, 835

\bibitem[{{Schaffenroth} {et~al.}(2015){Schaffenroth}, {Barlow}, {Drechsel}, \&
  {Dunlap}}]{vs:2015a}
{Schaffenroth}, V., {Barlow}, B.~N., {Drechsel}, H., \& {Dunlap}, B.~H. 2015,
  \aap, 576, A123

\bibitem[{{Schaffenroth} {et~al.}(2014{\natexlab{a}}){Schaffenroth}, {Classen},
  {Nagel}, {Geier}, {Koen}, {Heber}, \& {Edelmann}}]{vs:2014_II}
{Schaffenroth}, V., {Classen}, L., {Nagel}, K., {et~al.} 2014{\natexlab{a}},
  \aap, 570, A70

\bibitem[{{Schaffenroth} {et~al.}(2013){Schaffenroth}, {Geier}, {Drechsel},
  {Heber}, {Wils}, {{\O}stensen}, {Maxted}, \& {di Scala}}]{vs}
{Schaffenroth}, V., {Geier}, S., {Drechsel}, H., {et~al.} 2013, \aap, 553, A18

\bibitem[{{Schaffenroth} {et~al.}(2018){Schaffenroth}, {Geier}, {Heber},
  {Gerber}, {Schneider}, {Ziegerer}, \& {Cordes}}]{muchfuss_photo}
{Schaffenroth}, V., {Geier}, S., {Heber}, U., {et~al.} 2018, \aap, 614, A77

\bibitem[{{Schaffenroth} {et~al.}(2014{\natexlab{b}}){Schaffenroth}, {Geier},
  {Heber}, {Kupfer}, {Ziegerer}, {Heuser}, {Classen}, \& {Cordes}}]{vs:2014_I}
{Schaffenroth}, V., {Geier}, S., {Heber}, U., {et~al.} 2014{\natexlab{b}},
  \aap, 564, A98

\bibitem[{{Soker}(1998)}]{soker}
{Soker}, N. 1998, \aj, 116, 1308

\bibitem[{{Soszy{\'n}ski} {et~al.}(2016){Soszy{\'n}ski}, {Pawlak},
  {Pietrukowicz}, {Udalski}, {Szyma{\'n}ski}, {Wyrzykowski}, {Ulaczyk},
  {Poleski}, {Koz{\l}owski}, {Skowron}, {Skowron}, {Mr{\'o}z}, \&
  {Hamanowicz}}]{ogle_III}
{Soszy{\'n}ski}, I., {Pawlak}, M., {Pietrukowicz}, P., {et~al.} 2016, \actaa,
  66, 405

\bibitem[{{Soszy{\'n}ski} {et~al.}(2015){Soszy{\'n}ski}, {St{\c e}pie{\'n}},
  {Pilecki}, {Mr{\'o}z}, {Udalski}, {Szyma{\'n}ski}, {Pietrzy{\'n}ski},
  {Wyrzykowski}, {Ulaczyk}, {Poleski}, {Koz{\l}owski}, {Pietrukowicz},
  {Skowron}, \& {Pawlak}}]{ogle_II}
{Soszy{\'n}ski}, I., {St{\c e}pie{\'n}}, K., {Pilecki}, B., {et~al.} 2015,
  \actaa, 65, 39

\bibitem[{{Tody}(1986)}]{IRAF}
{Tody}, D. 1986, in \procspie, Vol. 627, Instrumentation in astronomy VI, ed.
  D.~L. {Crawford}, 733

\bibitem[{{Tonry} {et~al.}(2018){Tonry}, {Denneau}, {Heinze}, {Stalder},
  {Smith}, {Smartt}, {Stubbs}, {Weiland}, \& {Rest}}]{atlas}
{Tonry}, J.~L., {Denneau}, L., {Heinze}, A.~N., {et~al.} 2018, \pasp, 130,
  064505

\bibitem[{{Udalski} {et~al.}(2008){Udalski}, {Pont}, {Naef}, {Melo}, {Bouchy},
  {Santos}, {Moutou}, {D{\'{\i}}az}, {Gieren}, {Gillon}, {Hoyer}, {Mayor},
  {Mazeh}, {Minniti}, {Pietrzy{\'n}ski}, {Queloz}, {Ramirez}, {Ruiz},
  {Shporer}, {Tamuz}, {Udry}, {Zoccali}, {Kubiak}, {Szyma{\'n}ski},
  {Soszy{\'n}ski}, {Szewczyk}, {Ulaczyk}, \& {Wyrzykowski}}]{udalski}
{Udalski}, A., {Pont}, F., {Naef}, D., {et~al.} 2008, \aap, 482, 299

\bibitem[{{Udalski} {et~al.}(2015){Udalski}, {Szyma{\'n}ski}, \&
  {Szyma{\'n}ski}}]{ogle_proj}
{Udalski}, A., {Szyma{\'n}ski}, M.~K., \& {Szyma{\'n}ski}, G. 2015, \actaa, 65,
  1

\bibitem[{{Vos} {et~al.}(2019){Vos}, {Vu{\v c}kovi{\'c}}, {Chen}, {Han},
  {Boudreaux}, {Barlow}, {{\O}stensen}, \& {N{\'e}meth}}]{long_sdb}
{Vos}, J., {Vu{\v c}kovi{\'c}}, M., {Chen}, X., {et~al.} 2019, \mnras, 482,
  4592

\bibitem[{{Wilson}(1990)}]{wilson:1990}
{Wilson}, R.~E. 1990, \apj, 356, 613

\end{thebibliography}
	\bibliographystyle{aa}
	
	\appendix
	\onecolumn
	\section{Light curves, orbital parameters, magnitudes, Gaia parallaxes and proper motions of all our targets}
	\begin{figure}
		\caption{Phased light curves of all our HW Vir candidates from the OGLE survey.}
		\label{lc1}
		\includegraphics[width=0.25\linewidth]{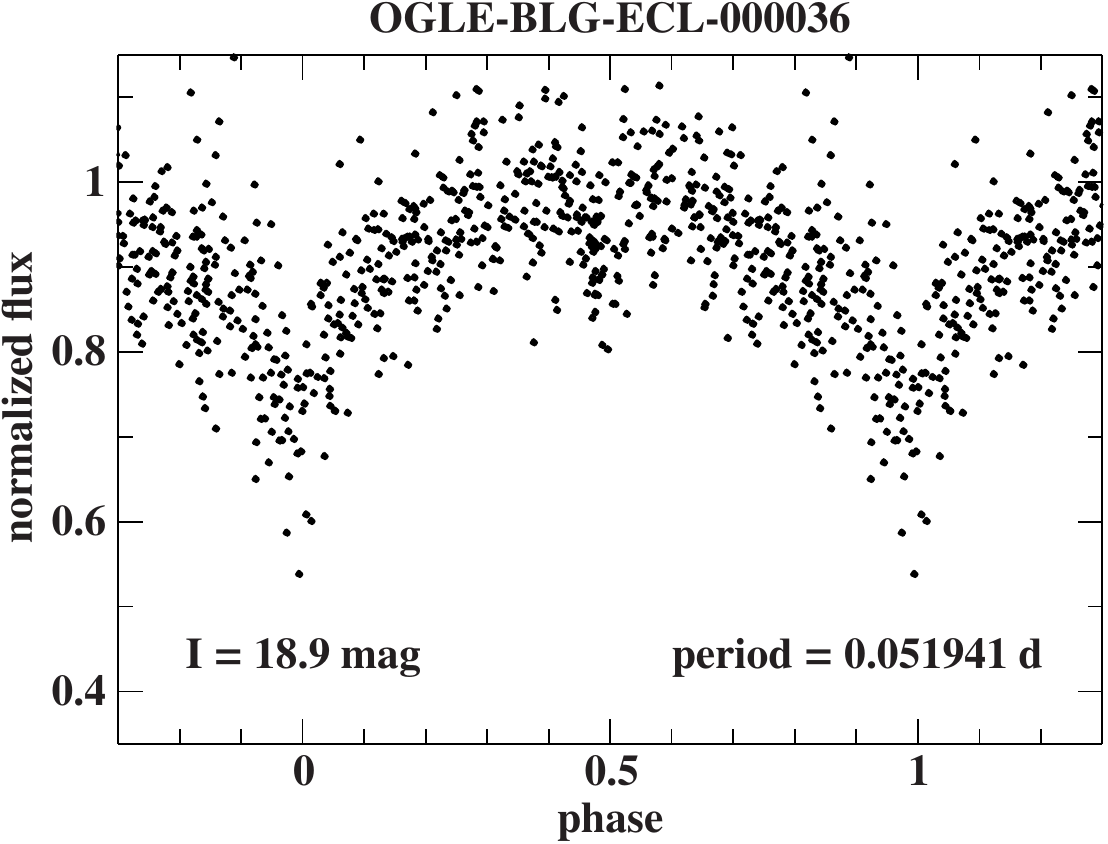}\hfill
		\includegraphics[width=0.25\linewidth]{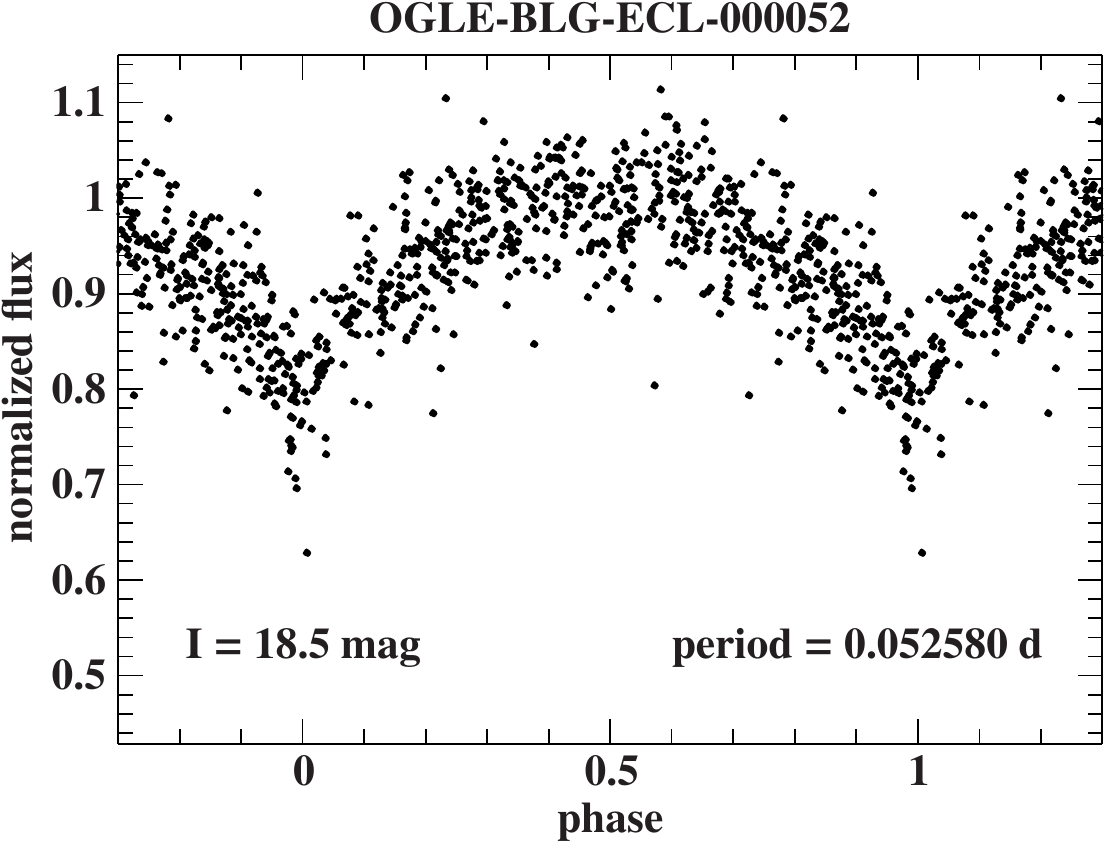}\hfill
		\includegraphics[width=0.25\linewidth]{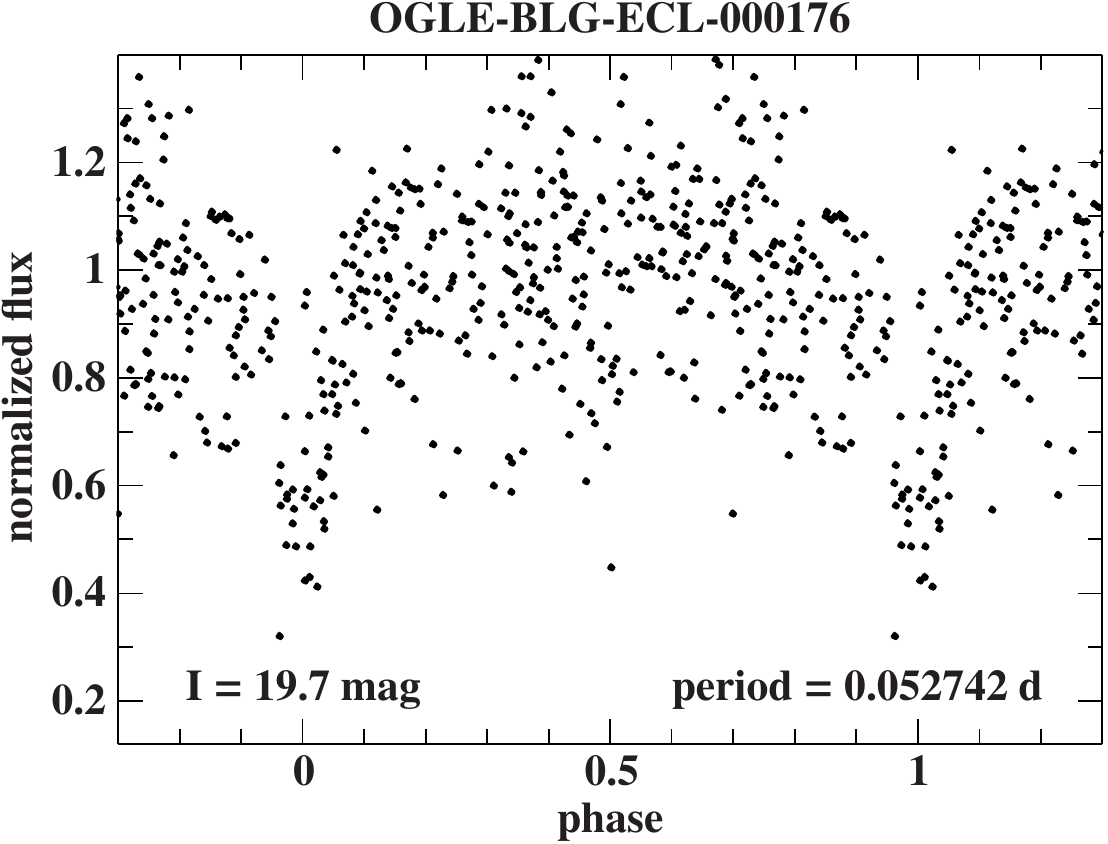}\hfill
		\includegraphics[width=0.25\linewidth]{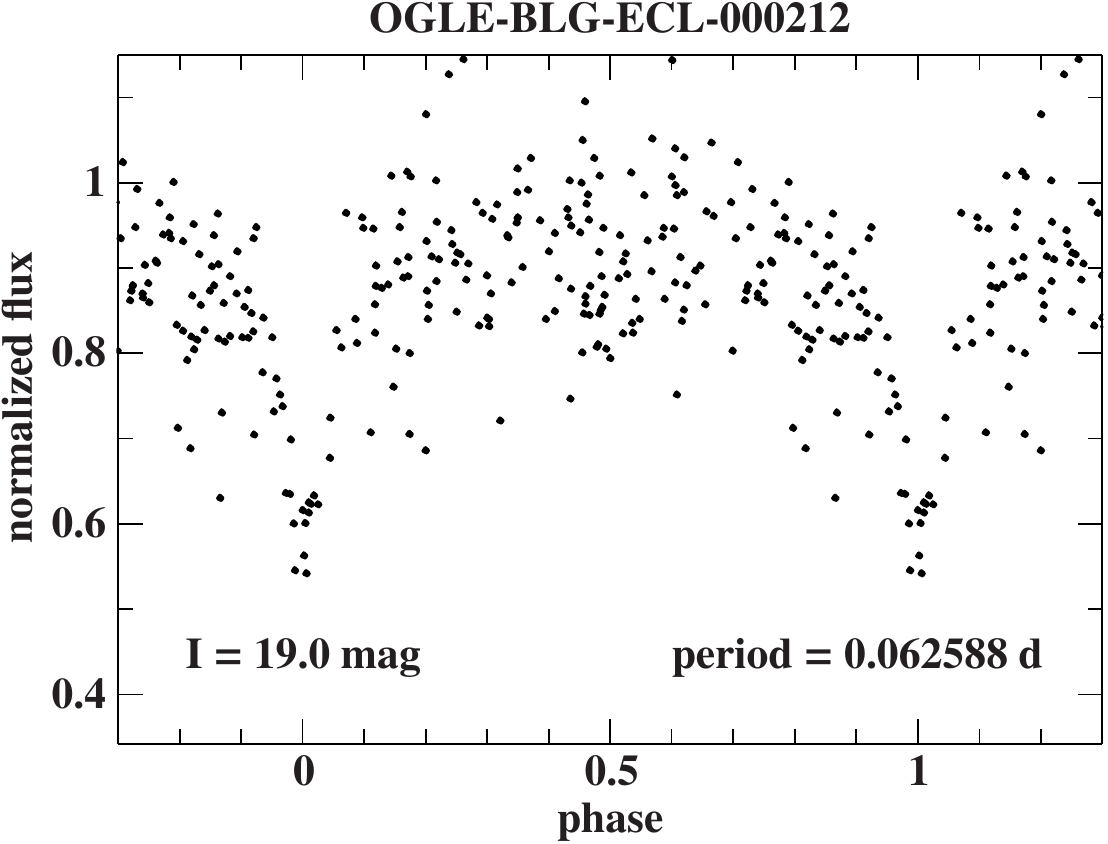}\hfill
		\includegraphics[width=0.25\linewidth]{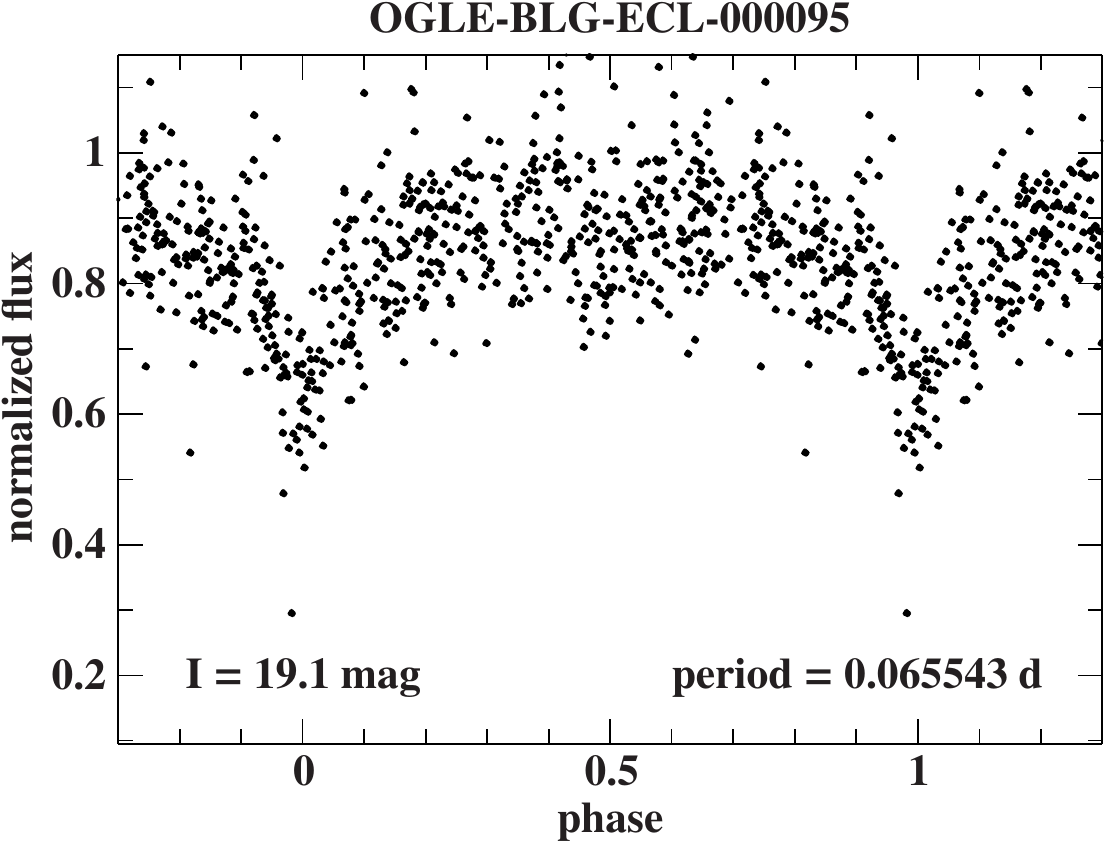}\hfill
		\includegraphics[width=0.25\linewidth]{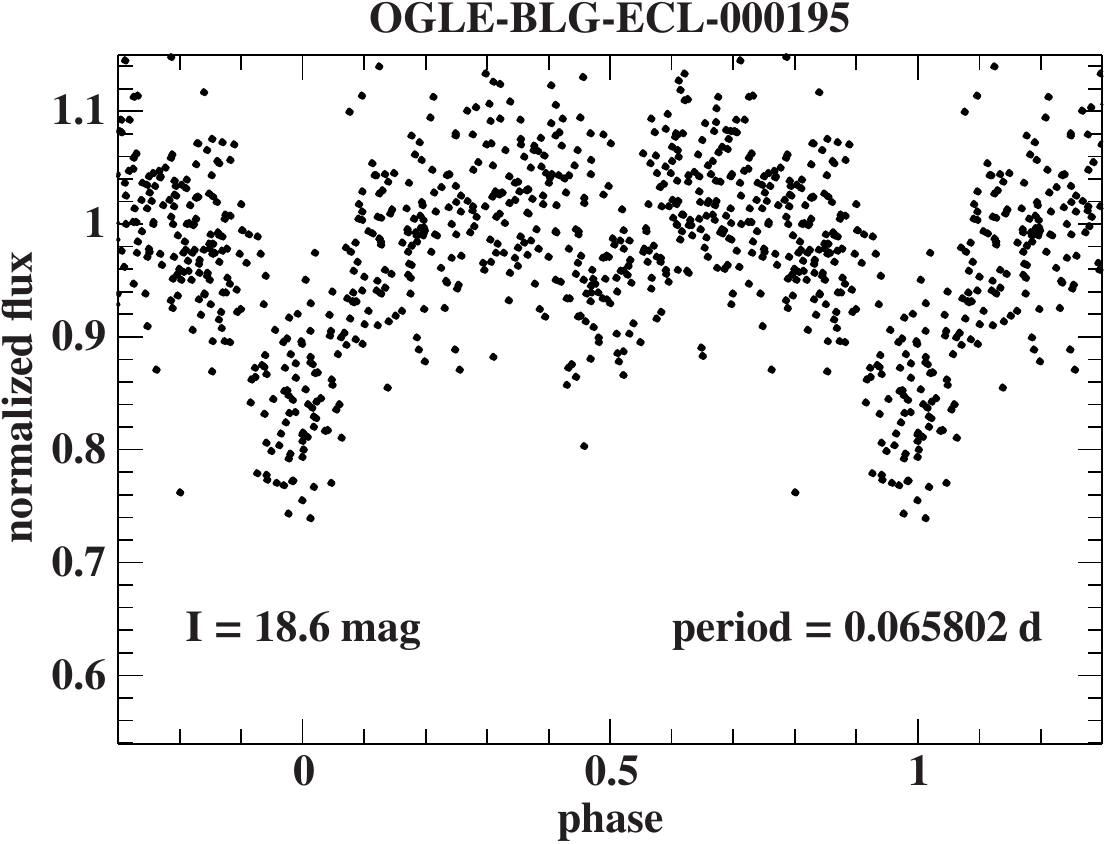}\hfill
		\includegraphics[width=0.25\linewidth]{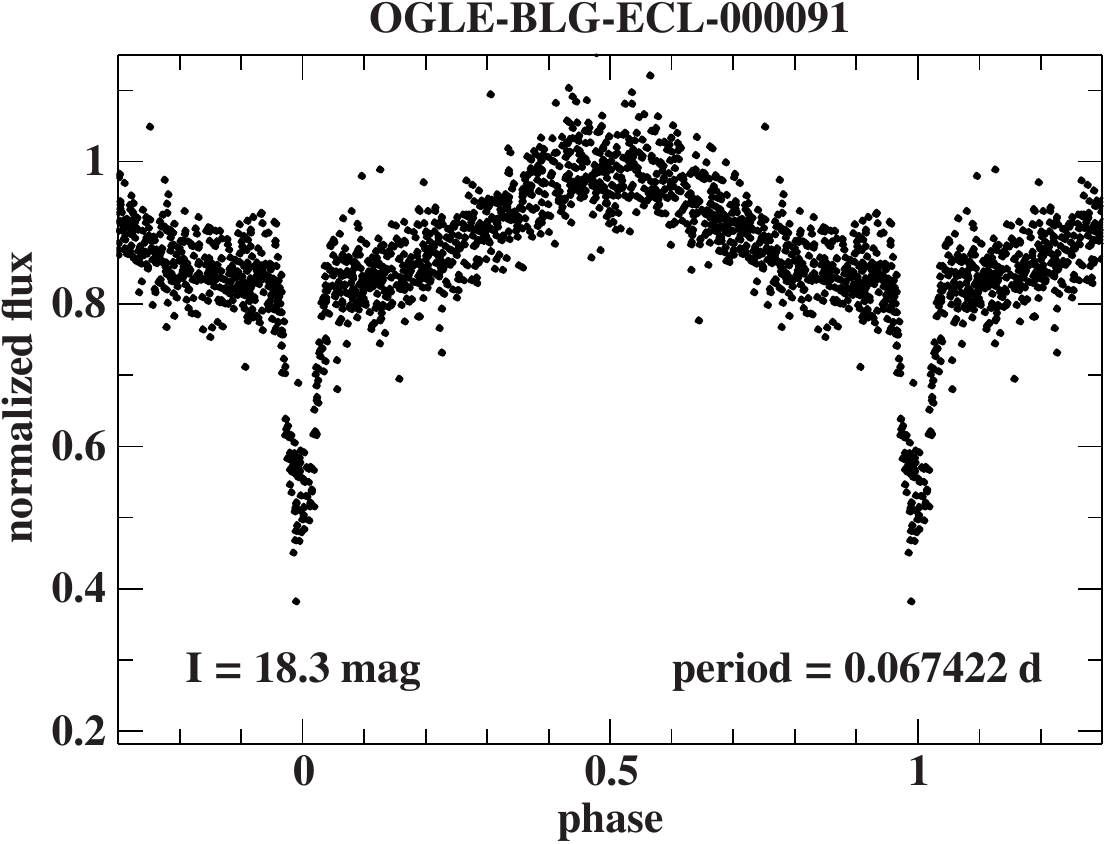}\hfill
		\includegraphics[width=0.25\linewidth]{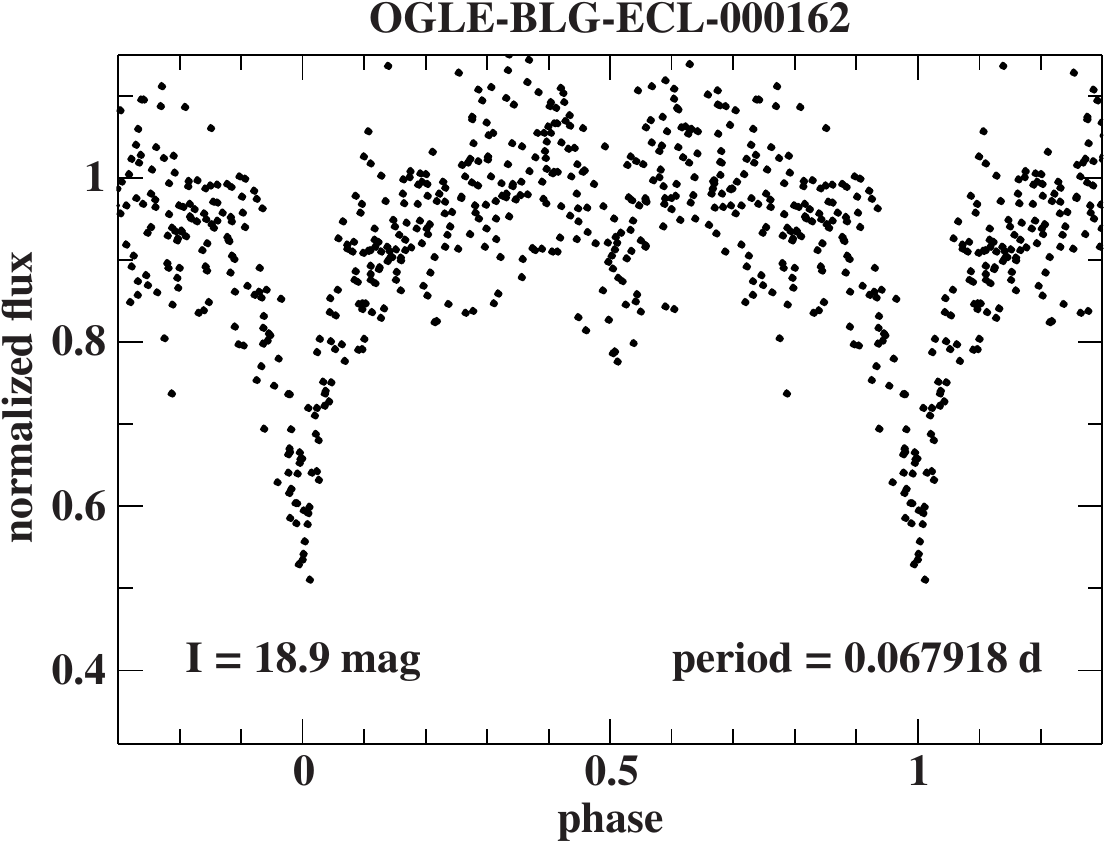}\hfill
		\includegraphics[width=0.25\linewidth]{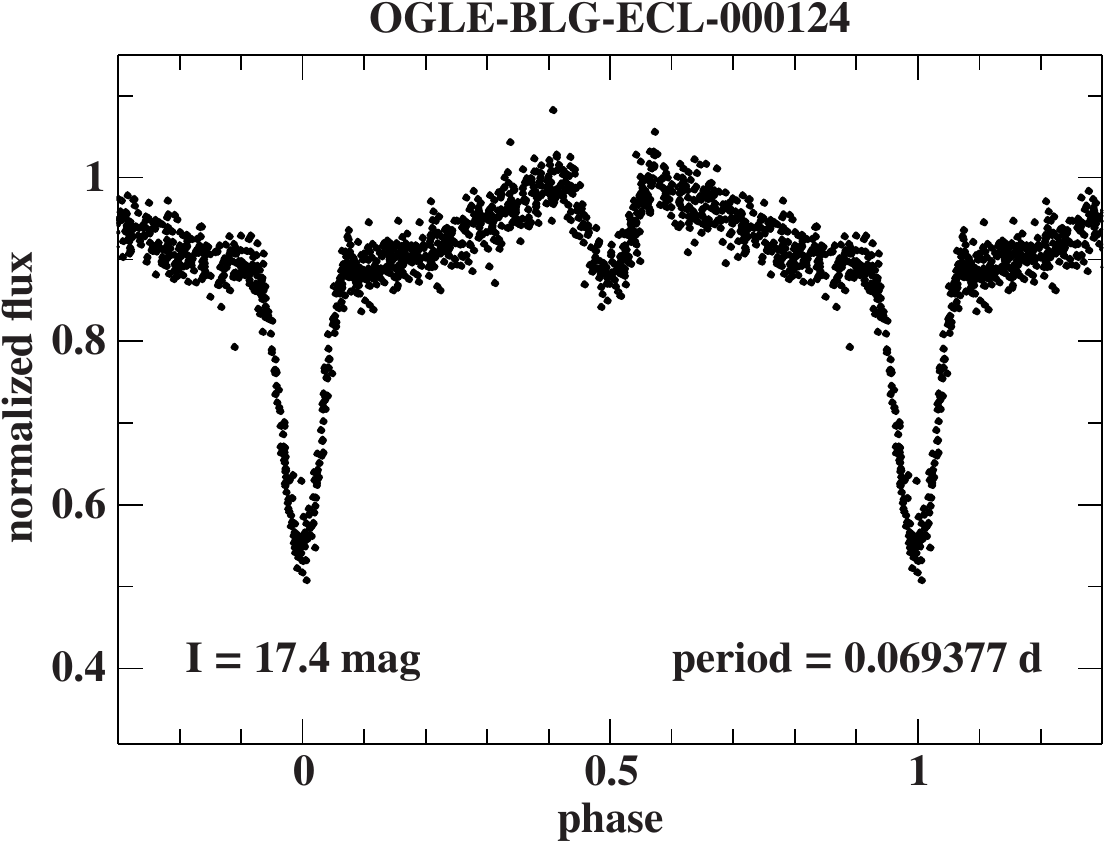}\hfill
		\includegraphics[width=0.25\linewidth]{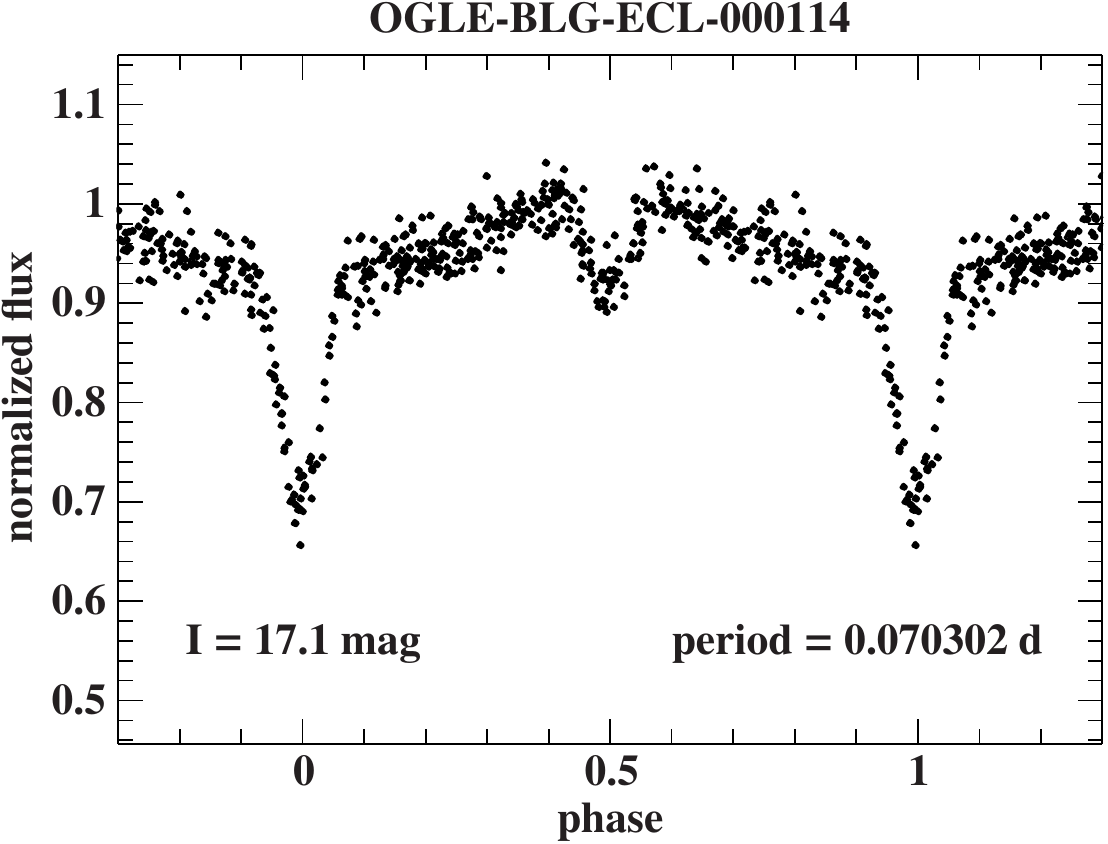}\hfill
		\includegraphics[width=0.25\linewidth]{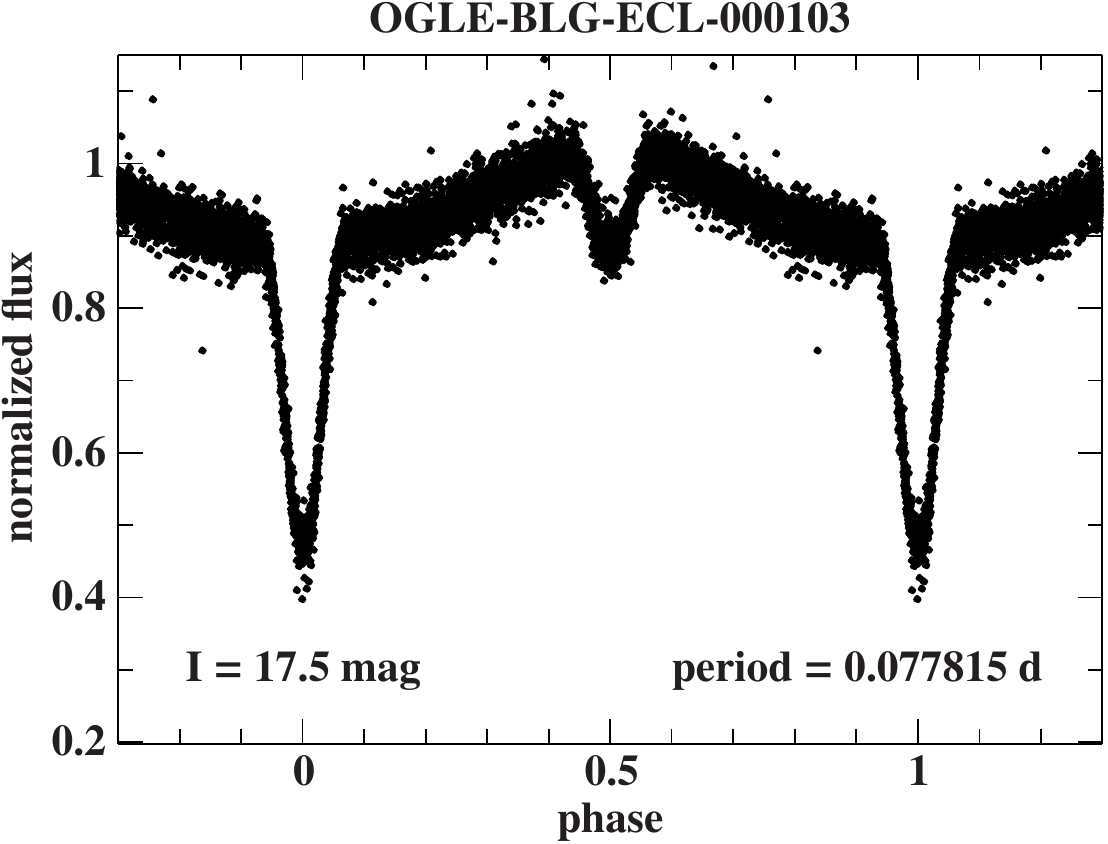}\hfill
		\includegraphics[width=0.25\linewidth]{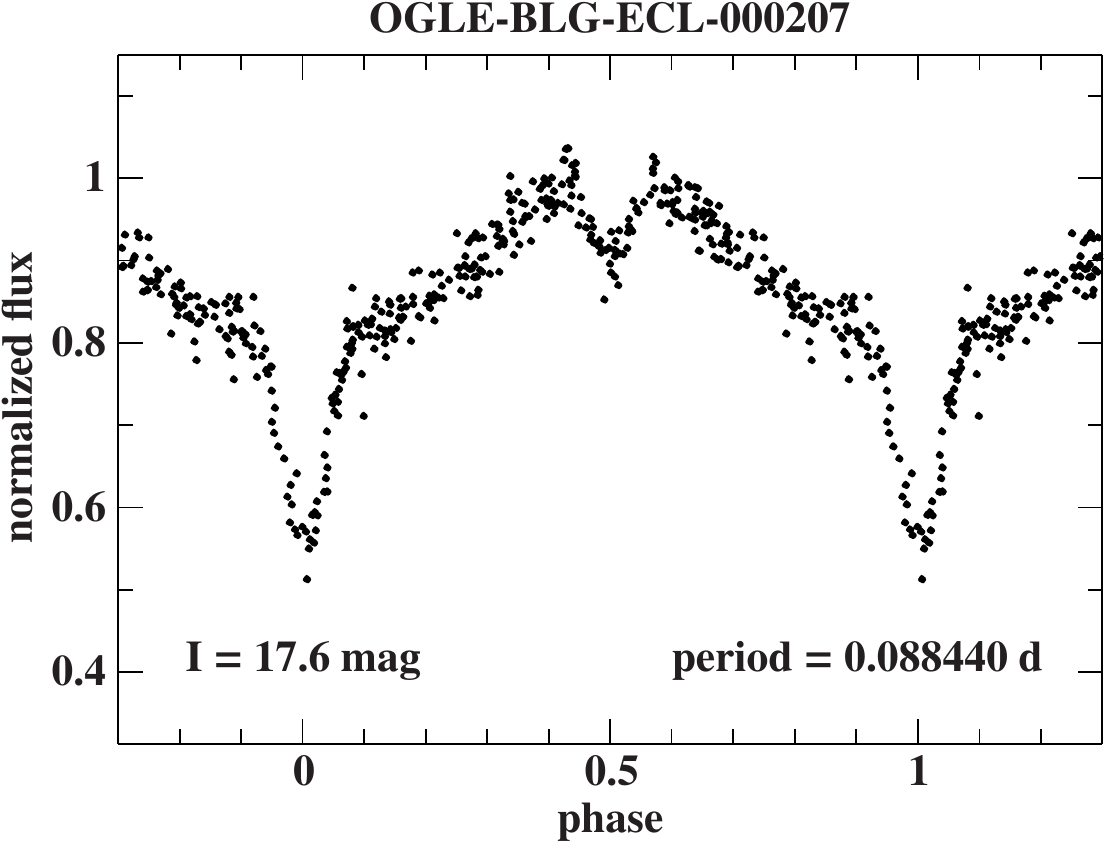}\hfill
		\includegraphics[width=0.25\linewidth]{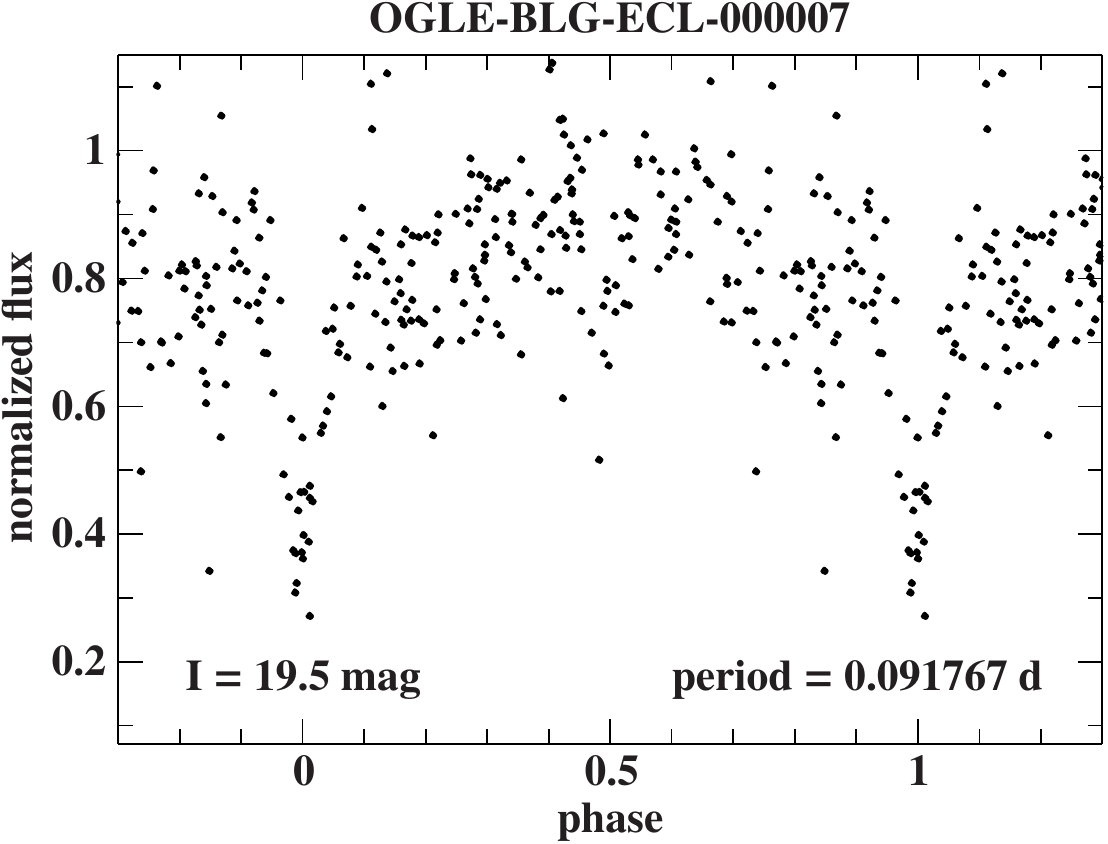}\hfill
		\includegraphics[width=0.25\linewidth]{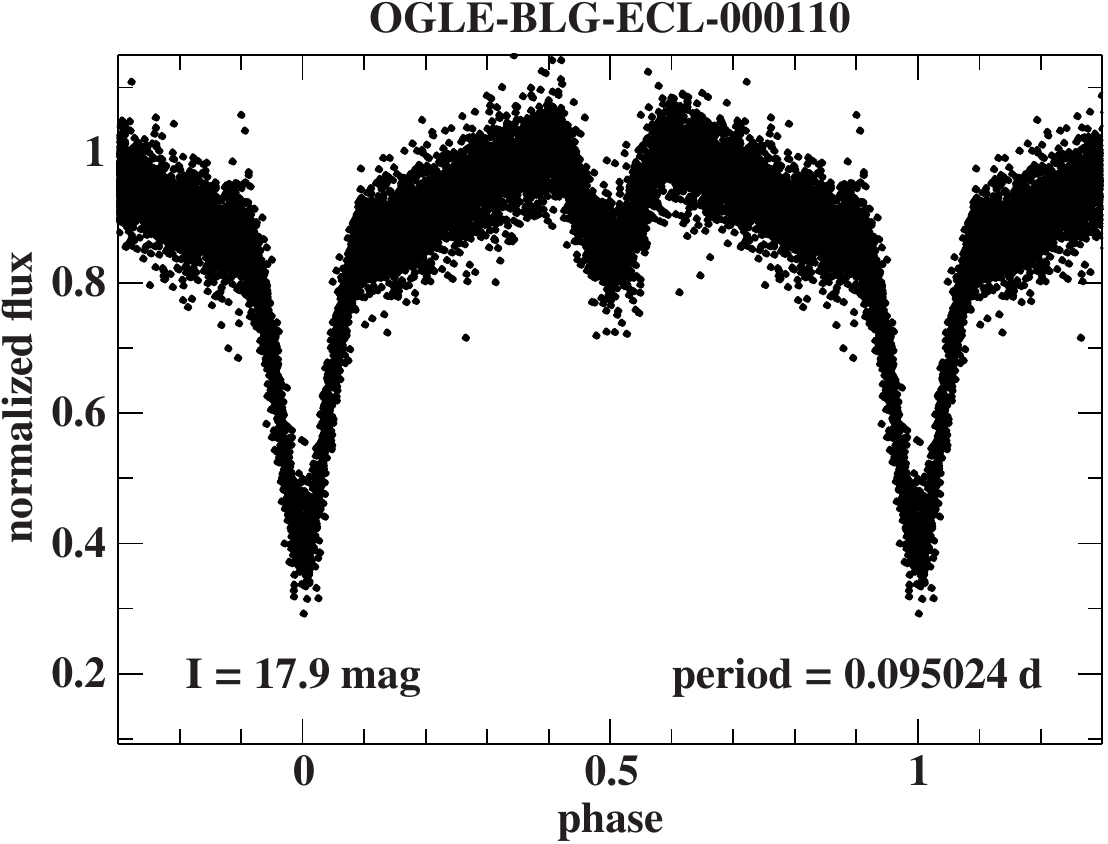}\hfill
		\includegraphics[width=0.25\linewidth]{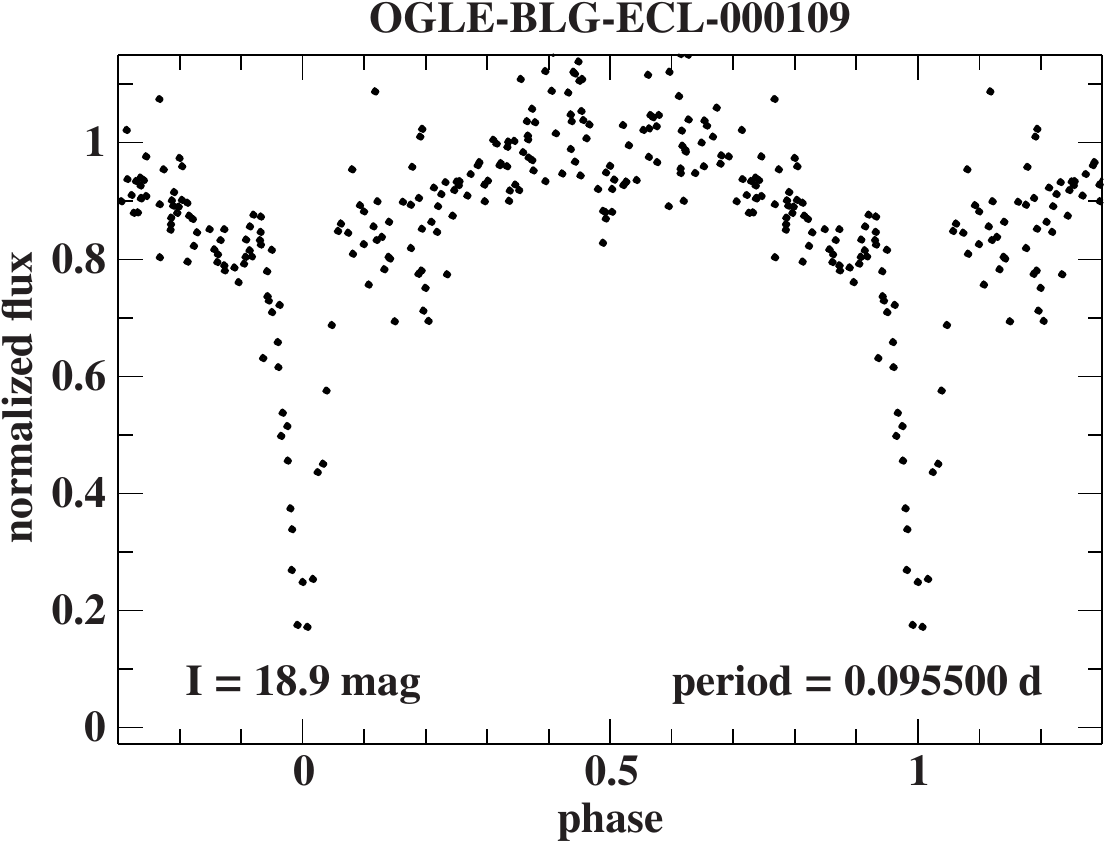}\hfill
		\includegraphics[width=0.25\linewidth]{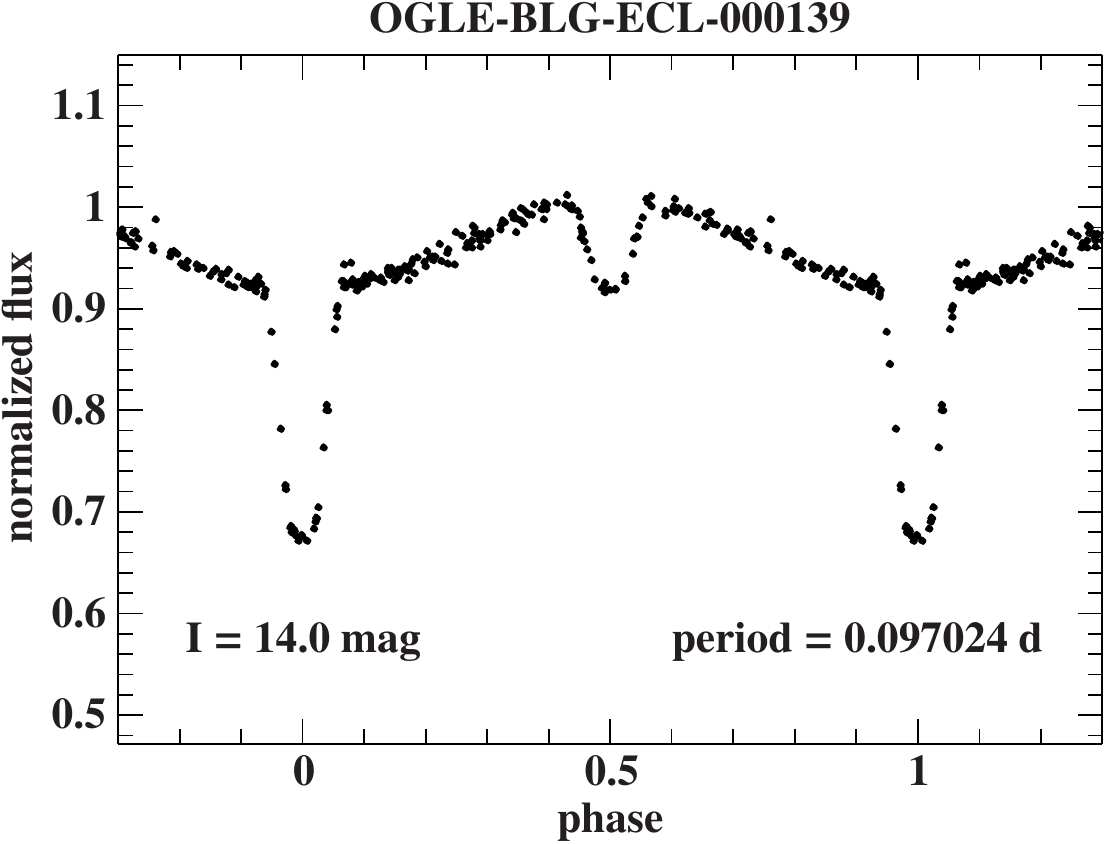}\hfill
		\includegraphics[width=0.25\linewidth]{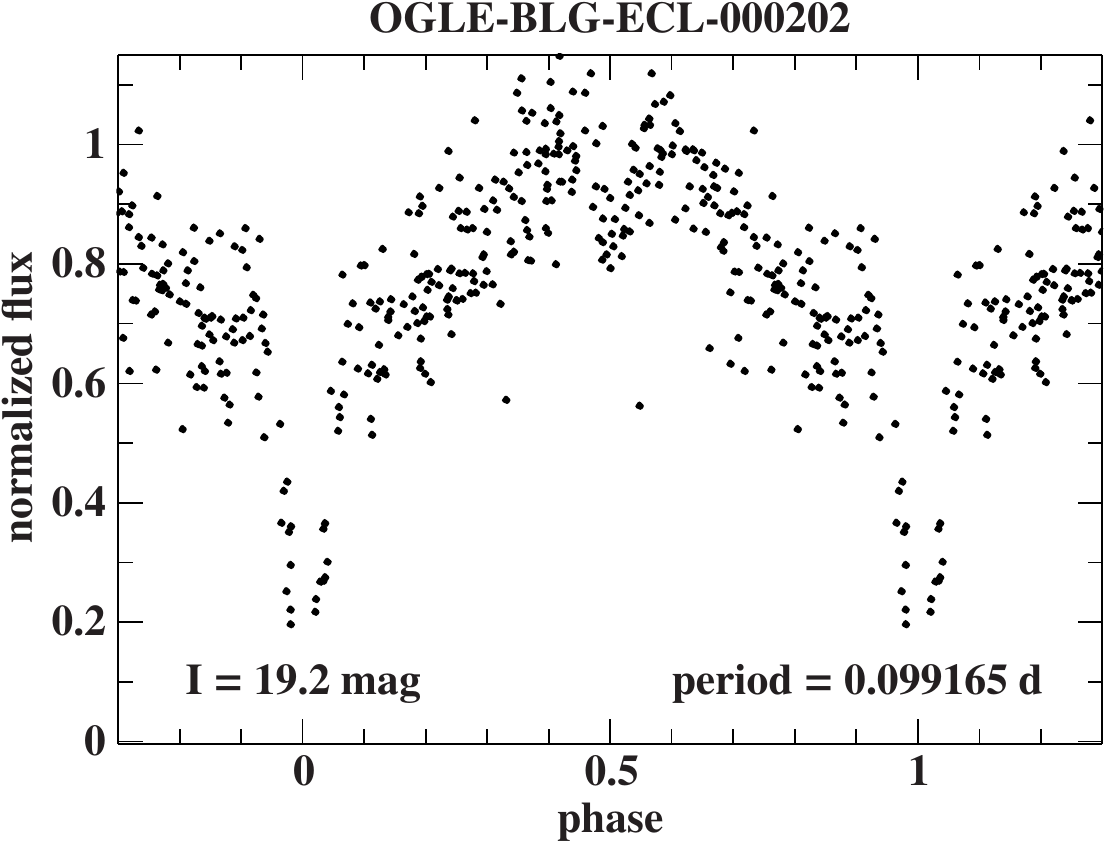}\hfill
		\includegraphics[width=0.25\linewidth]{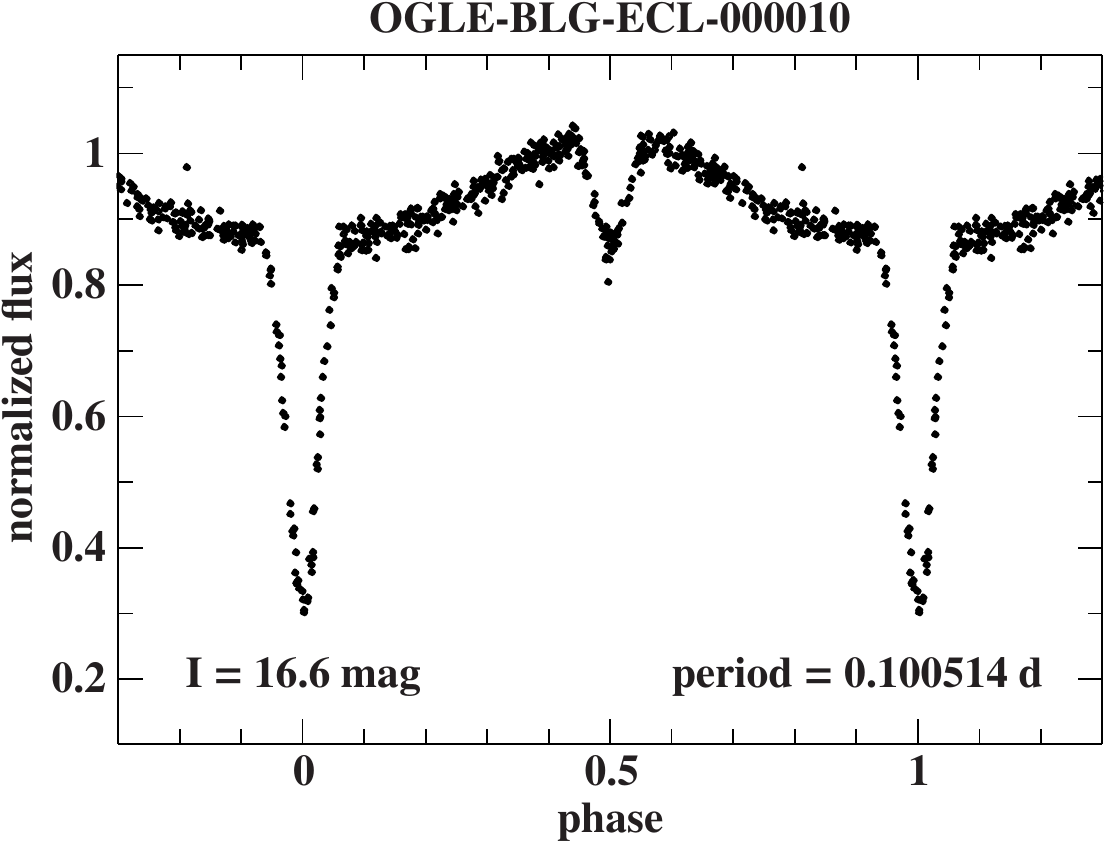}\hfill
		\includegraphics[width=0.25\linewidth]{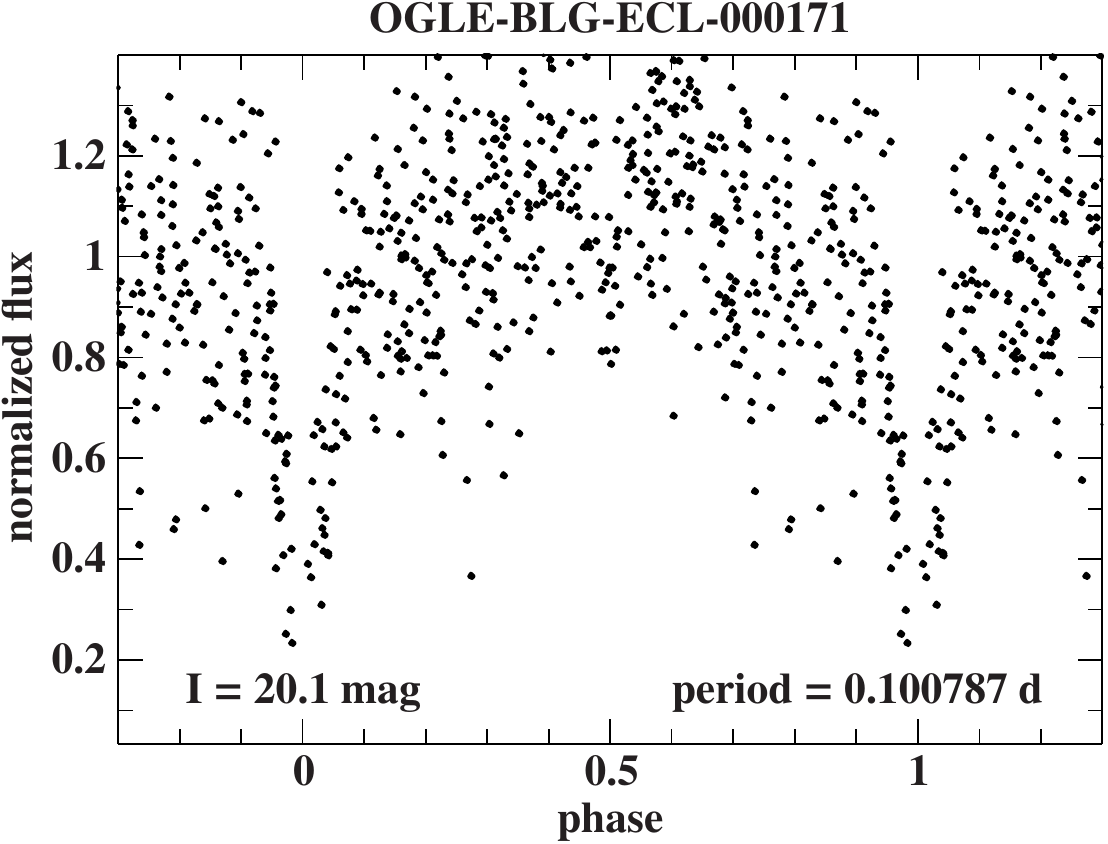}\hfill
		\includegraphics[width=0.25\linewidth]{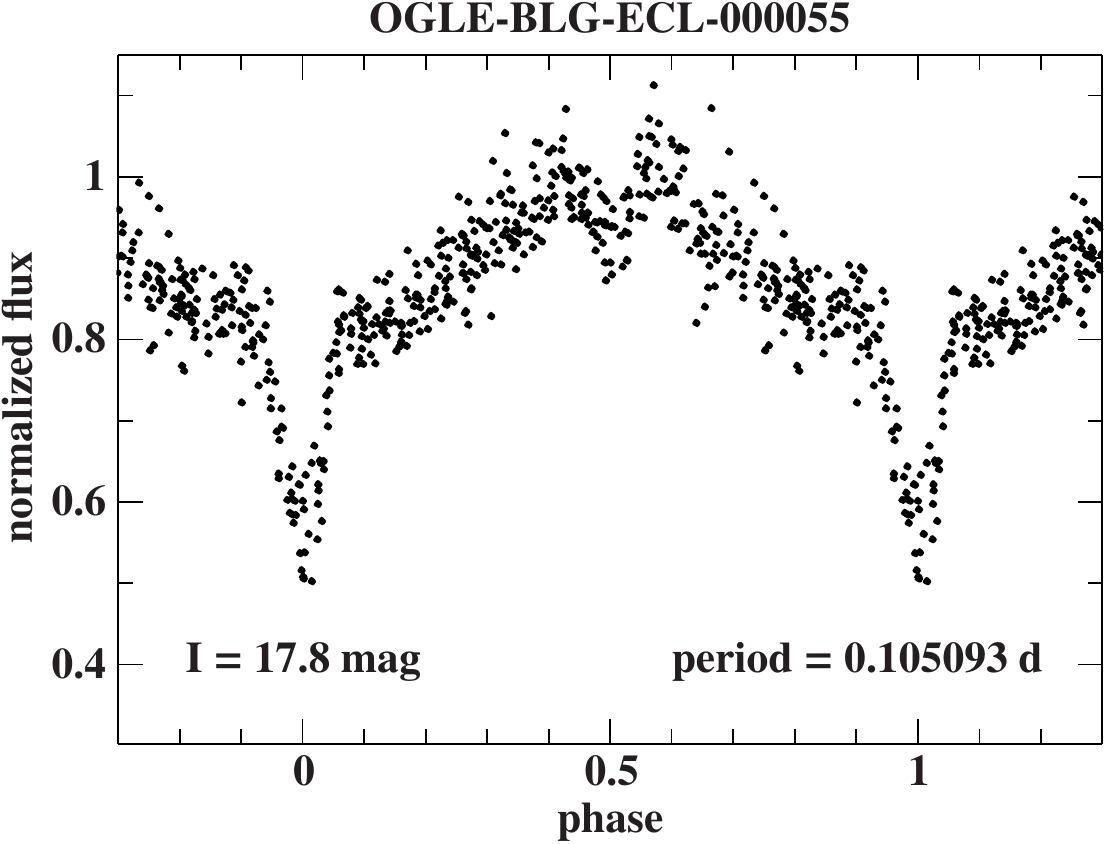}\hfill
		\includegraphics[width=0.25\linewidth]{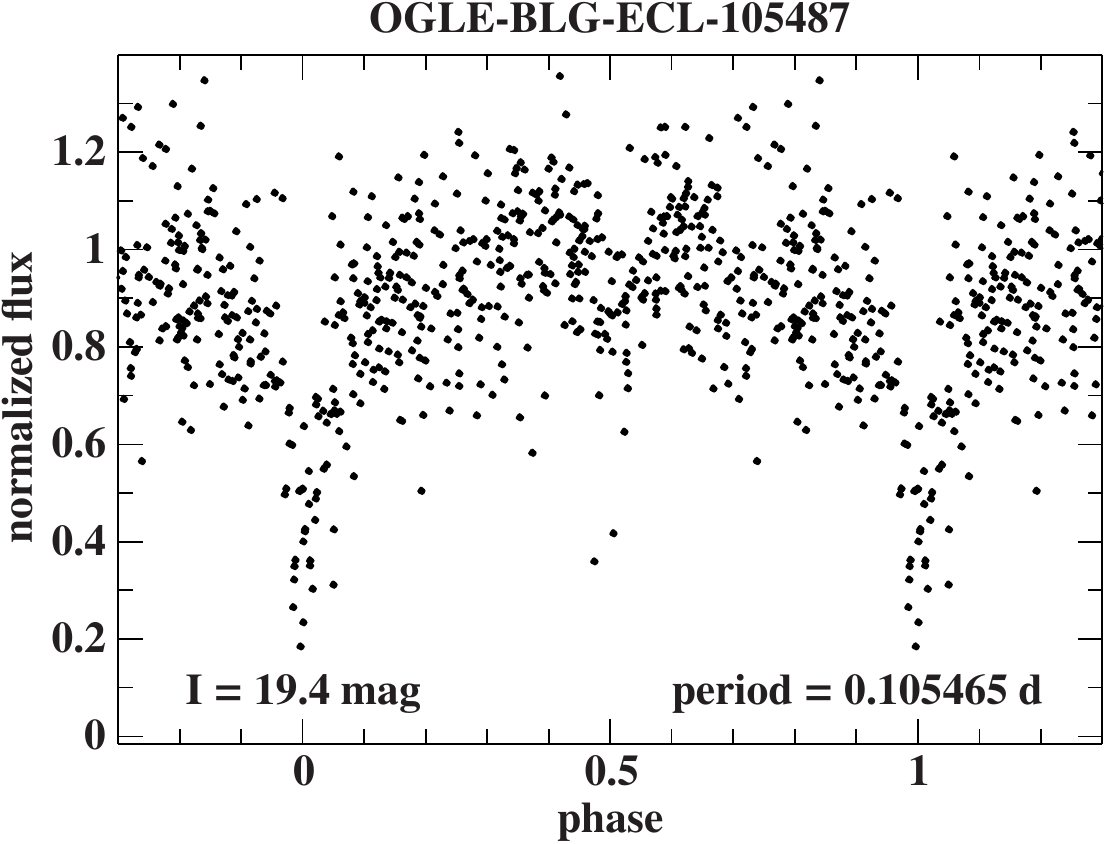}\hfill
		\includegraphics[width=0.25\linewidth]{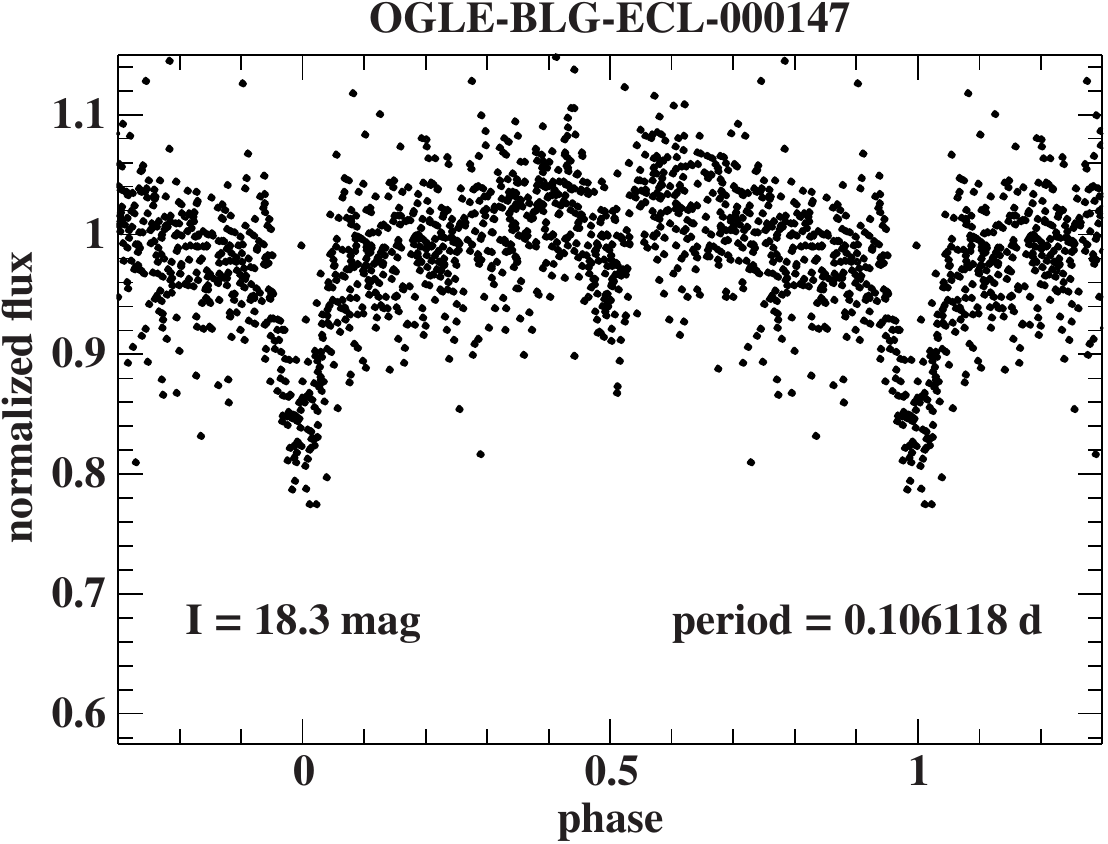}\hfill
		\includegraphics[width=0.25\linewidth]{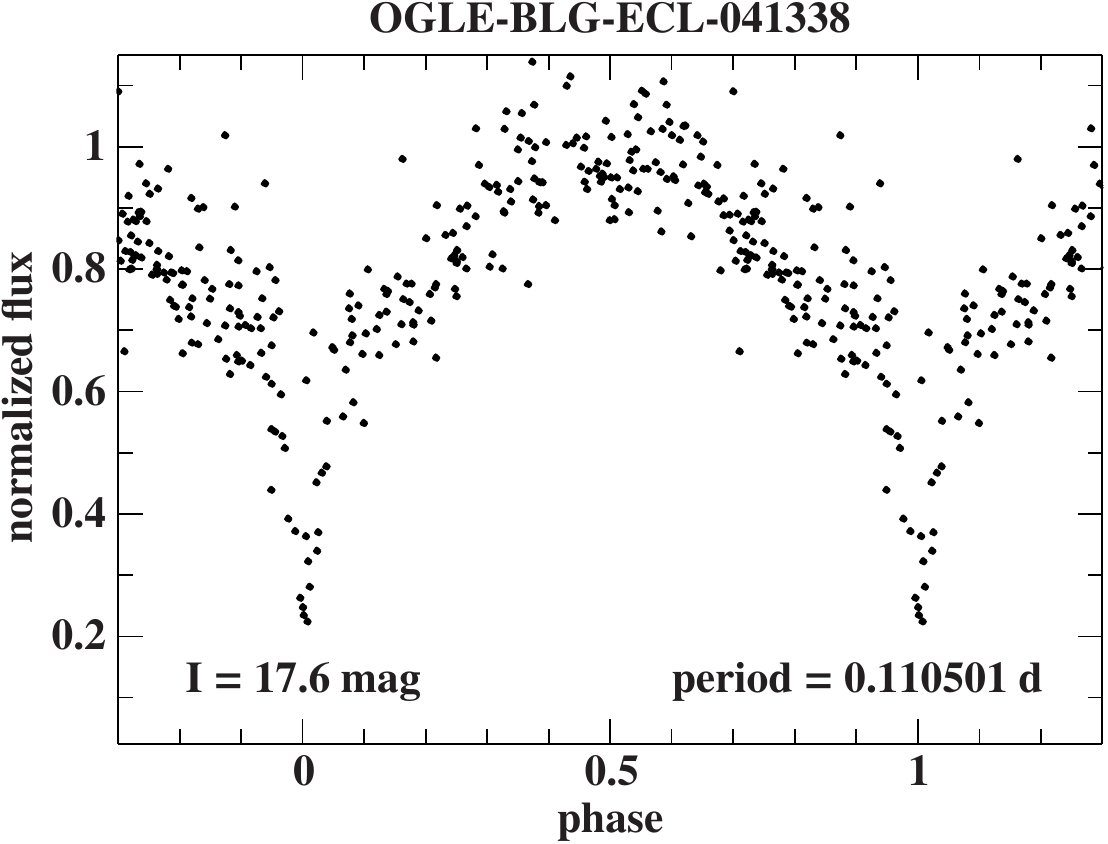}\hfill
		\includegraphics[width=0.25\linewidth]{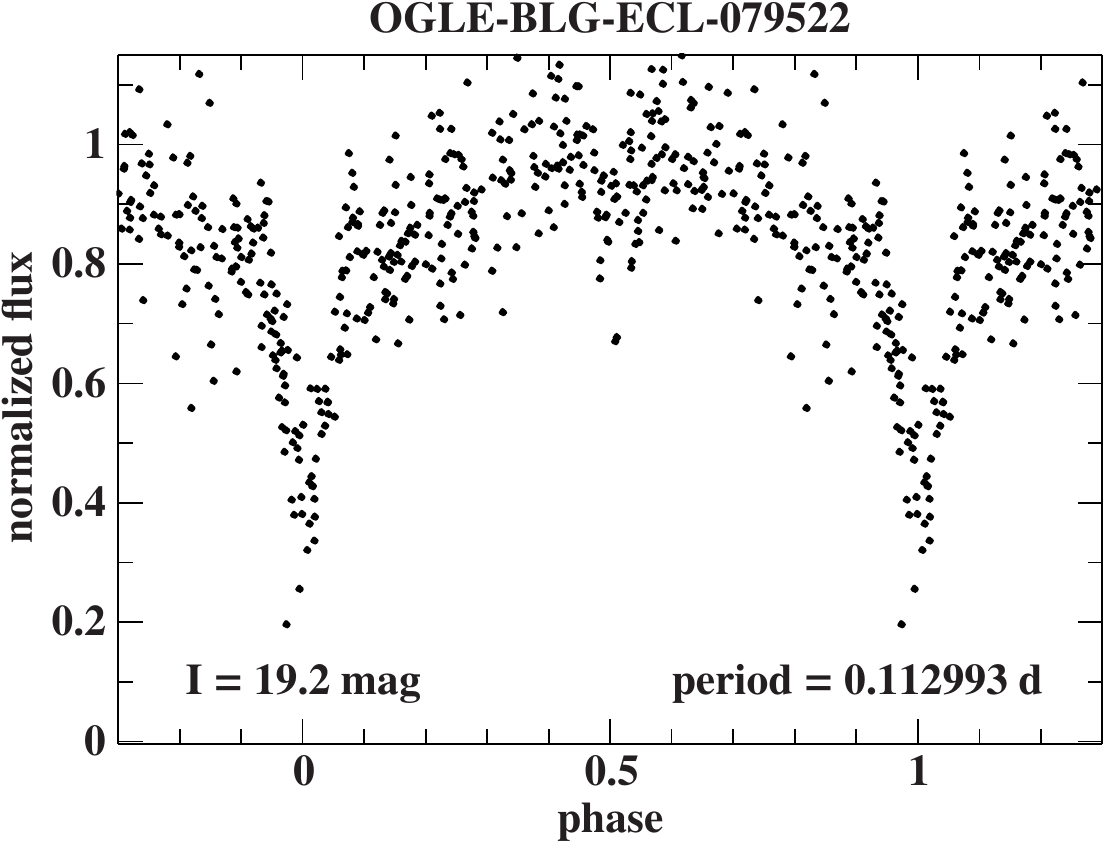}\hfill
		\includegraphics[width=0.25\linewidth]{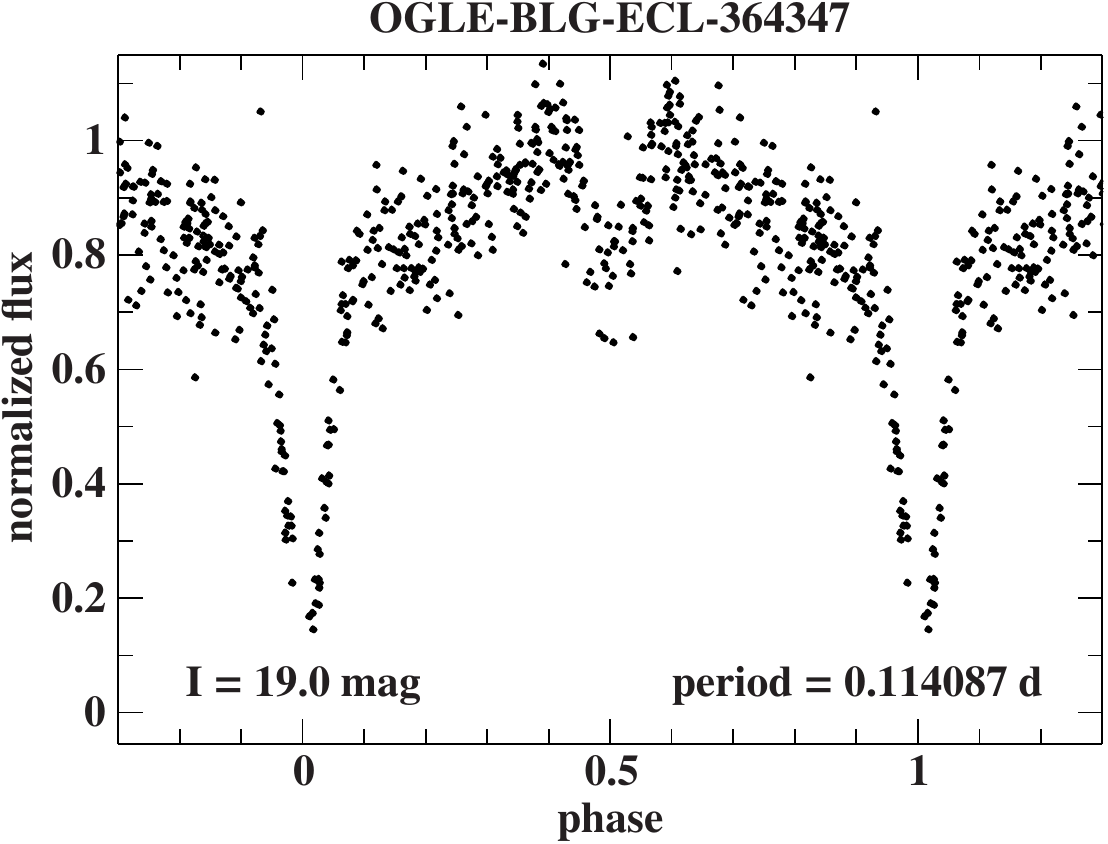}\hfill
		\includegraphics[width=0.25\linewidth]{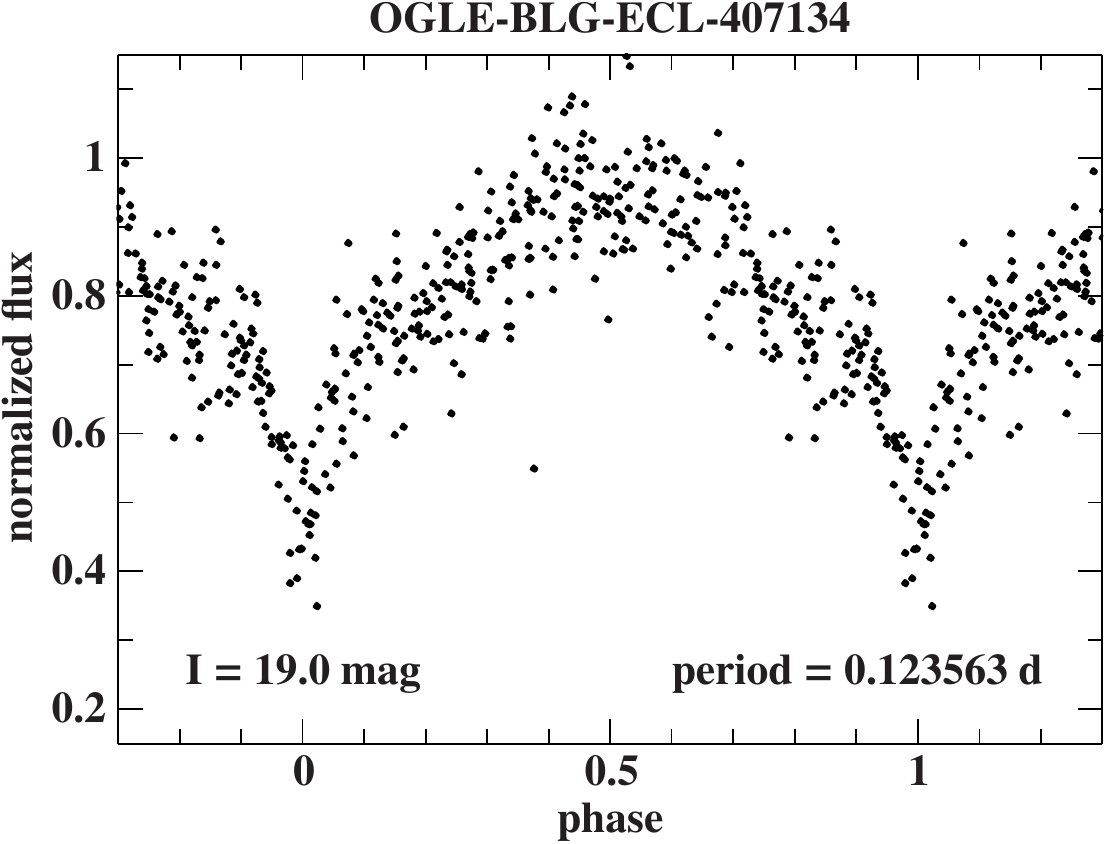}\hfill
		\includegraphics[width=0.25\linewidth]{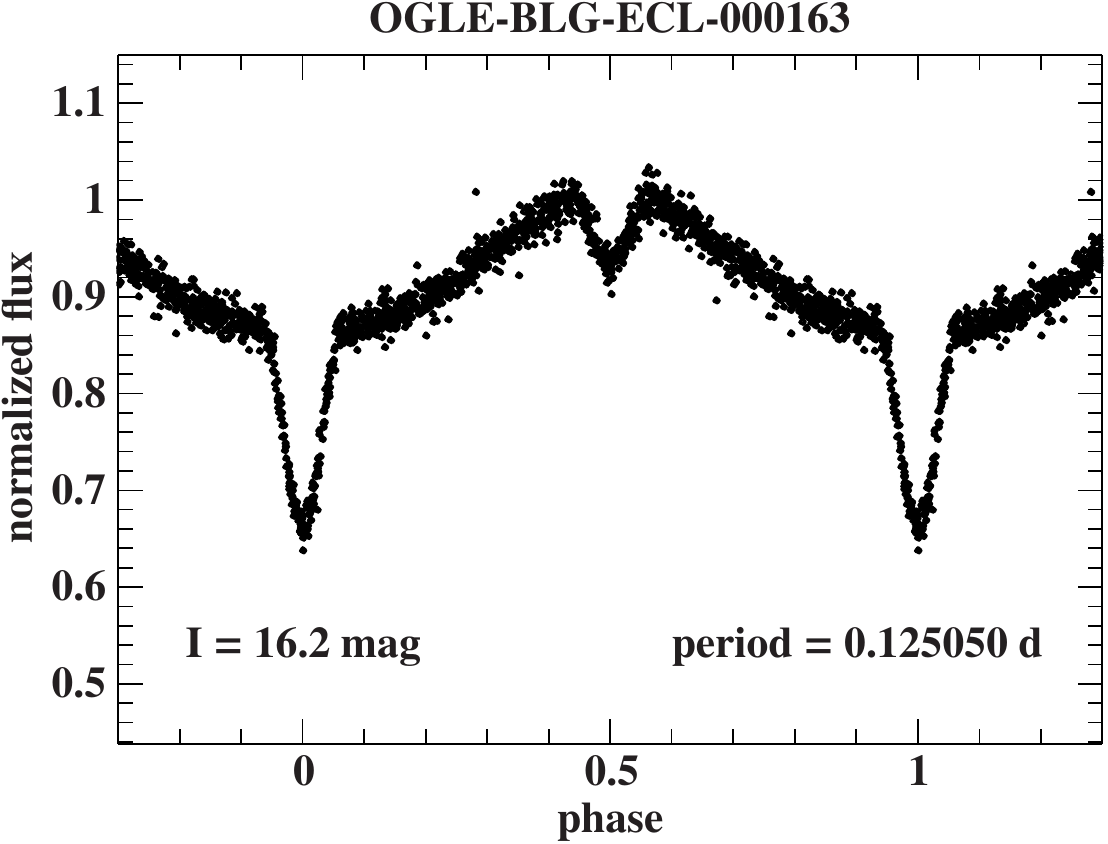}\hfill
		\includegraphics[width=0.25\linewidth]{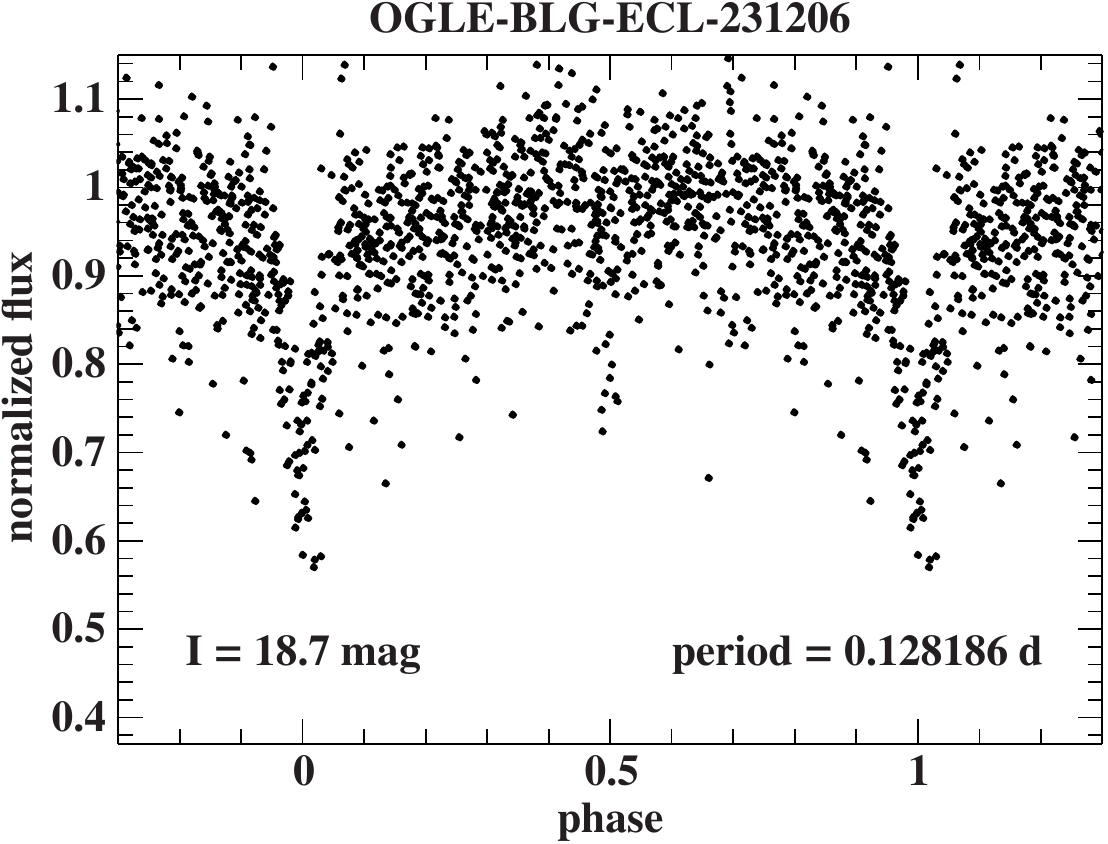}\hfill
	\end{figure}
	\begin{figure}
		\includegraphics[width=0.25\linewidth]{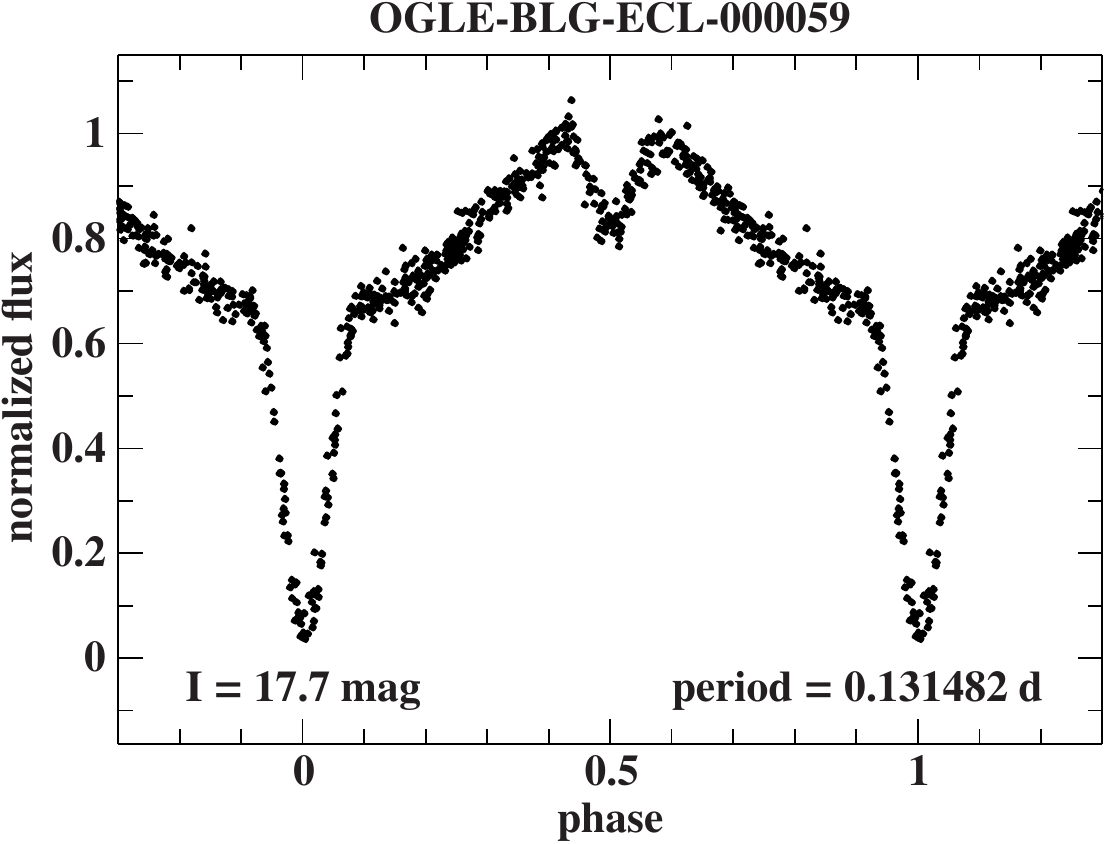}\hfill
		\includegraphics[width=0.25\linewidth]{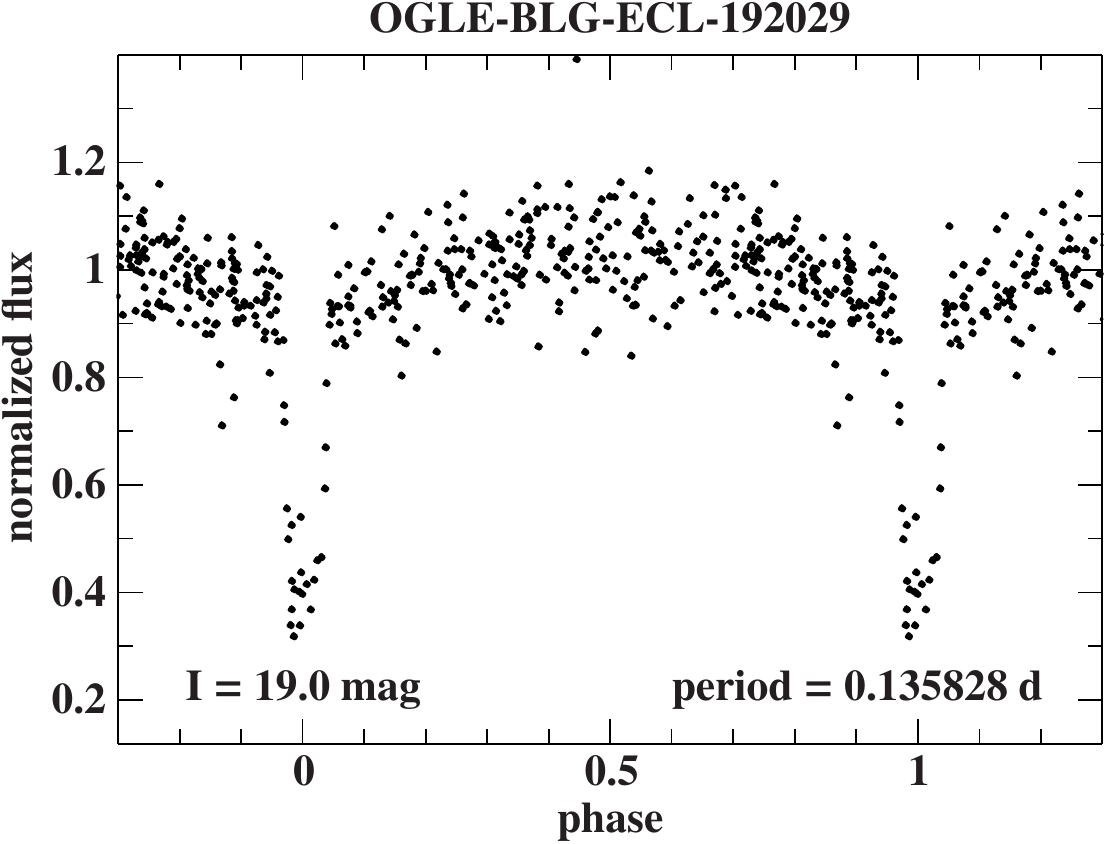}\hfill
		\includegraphics[width=0.25\linewidth]{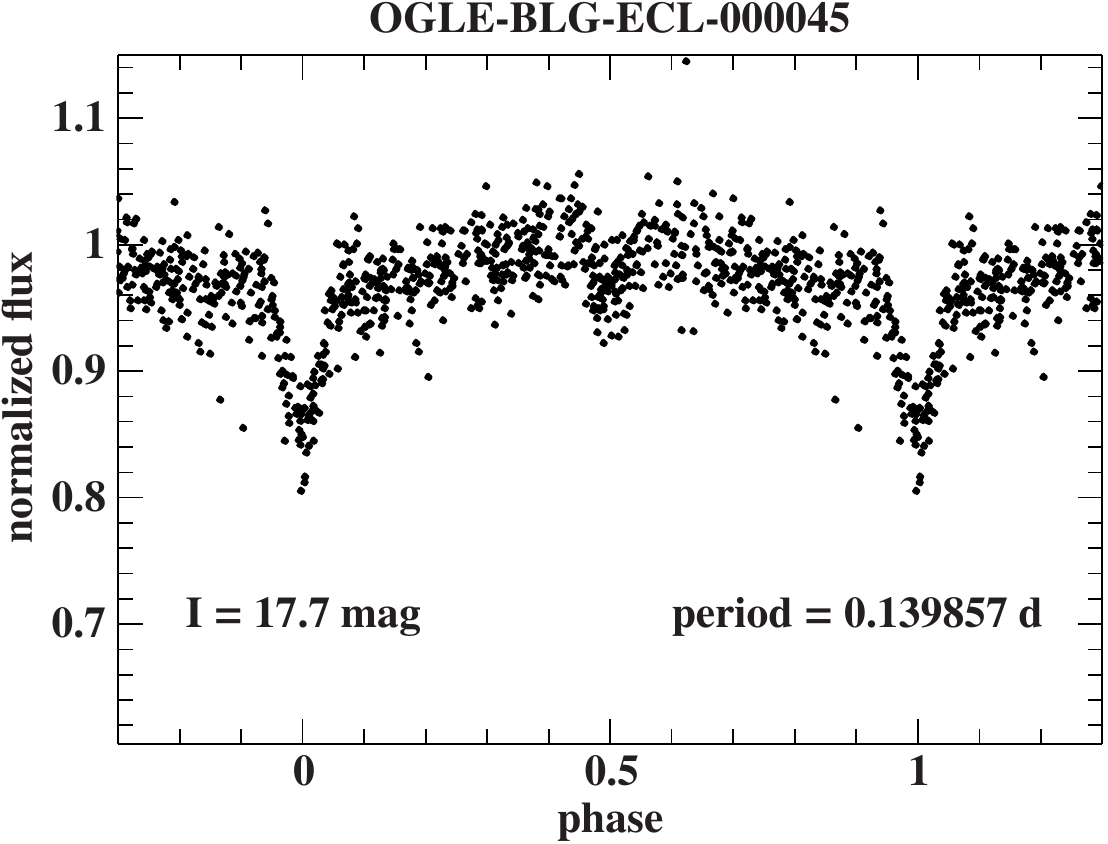}\hfill
		\includegraphics[width=0.25\linewidth]{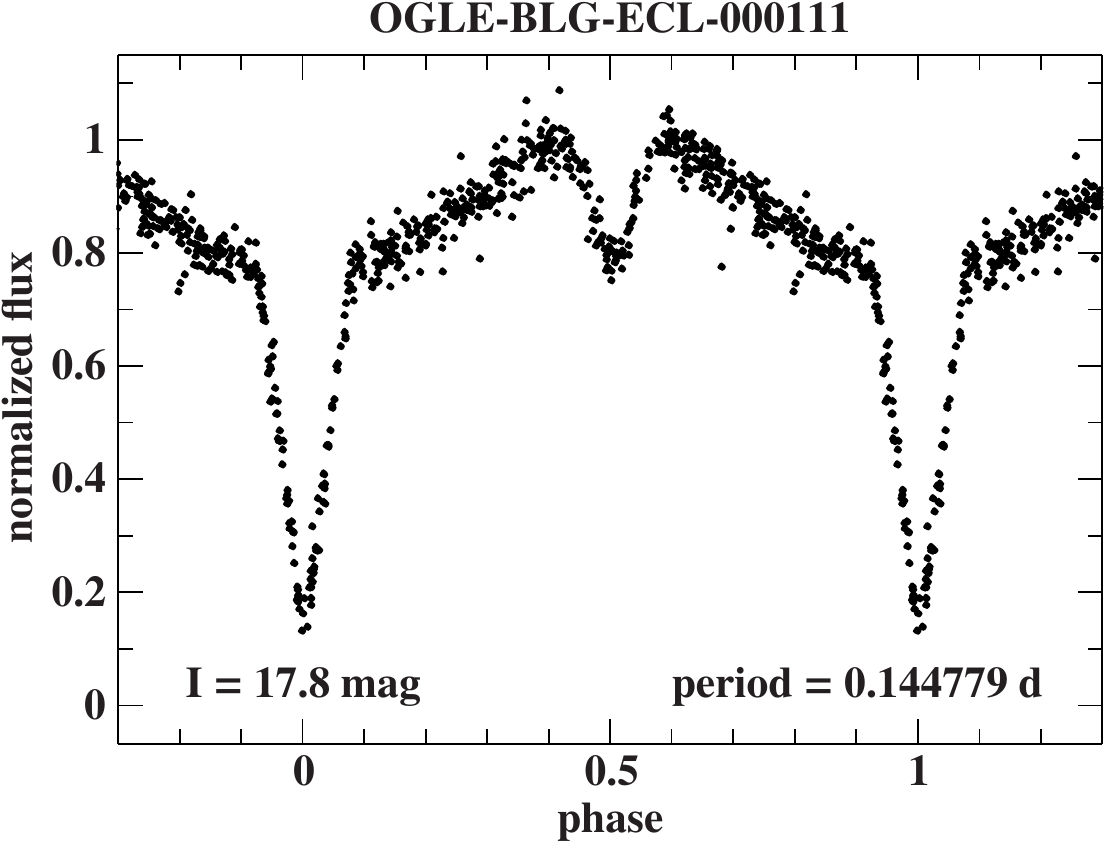}\hfill
		\includegraphics[width=0.25\linewidth]{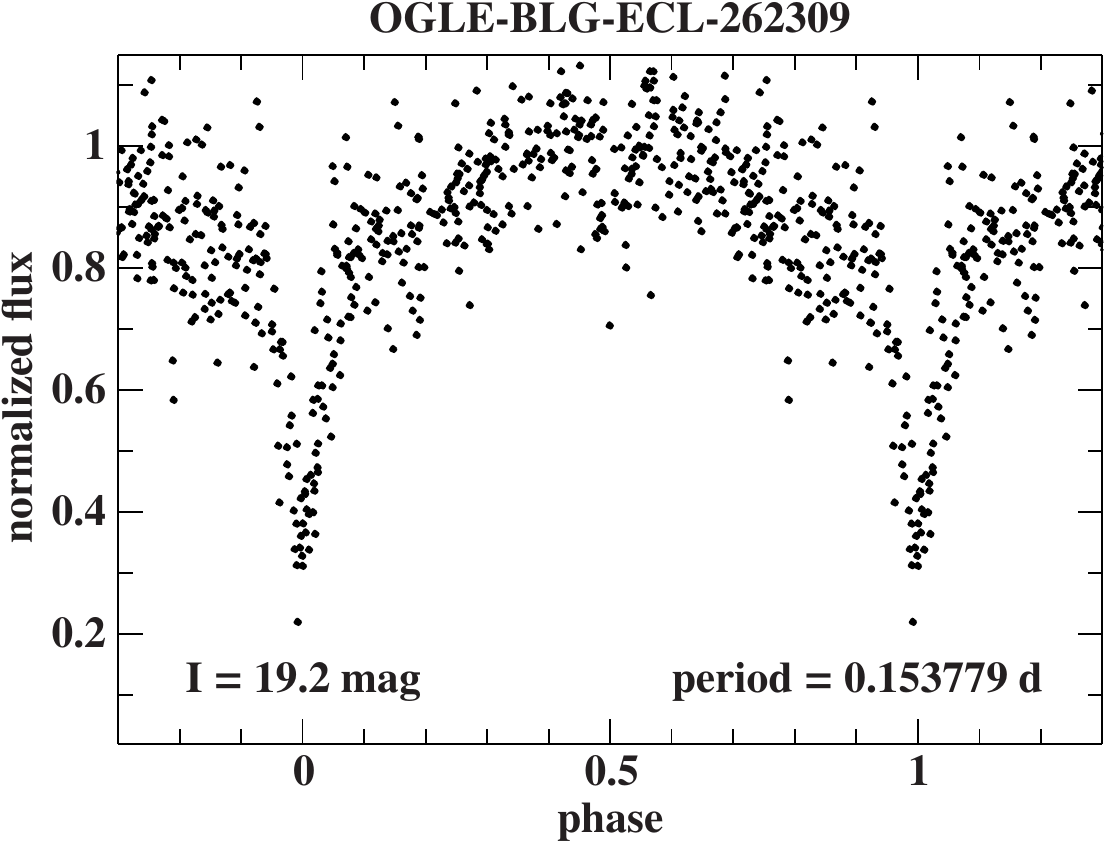}\hfill
		\includegraphics[width=0.25\linewidth]{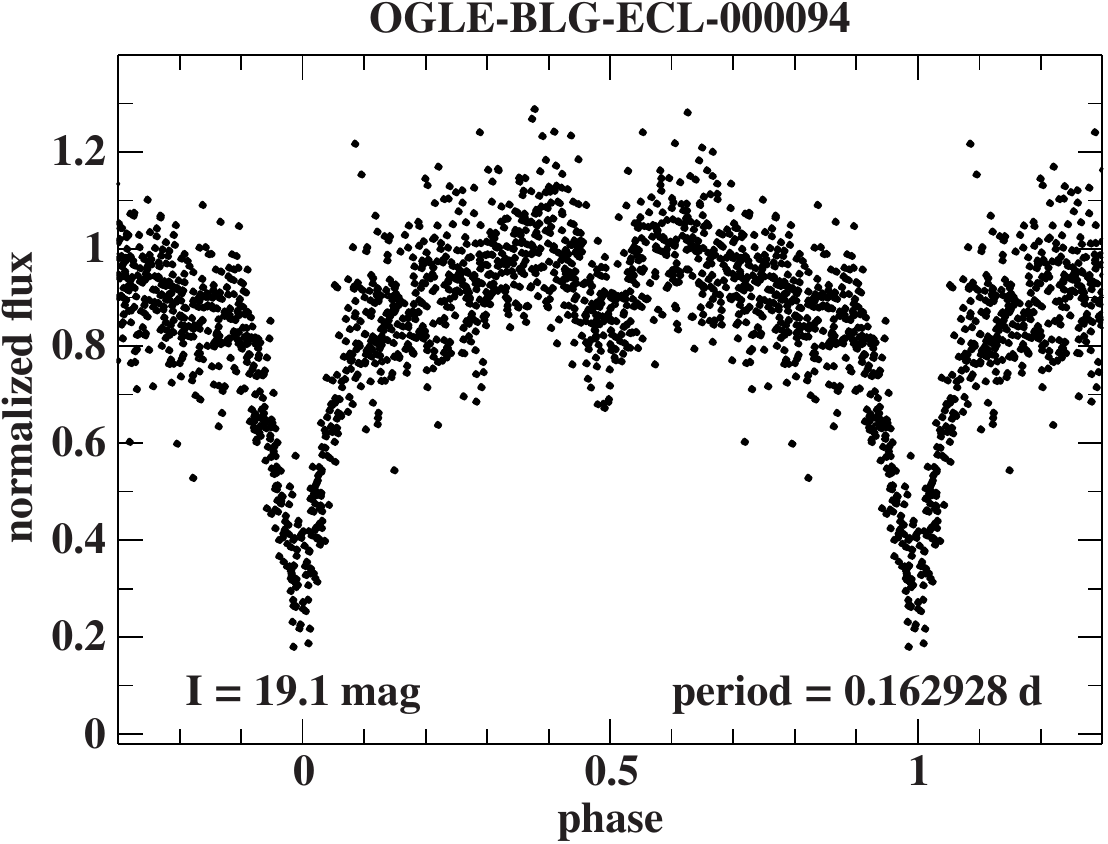}\hfill
		\includegraphics[width=0.25\linewidth]{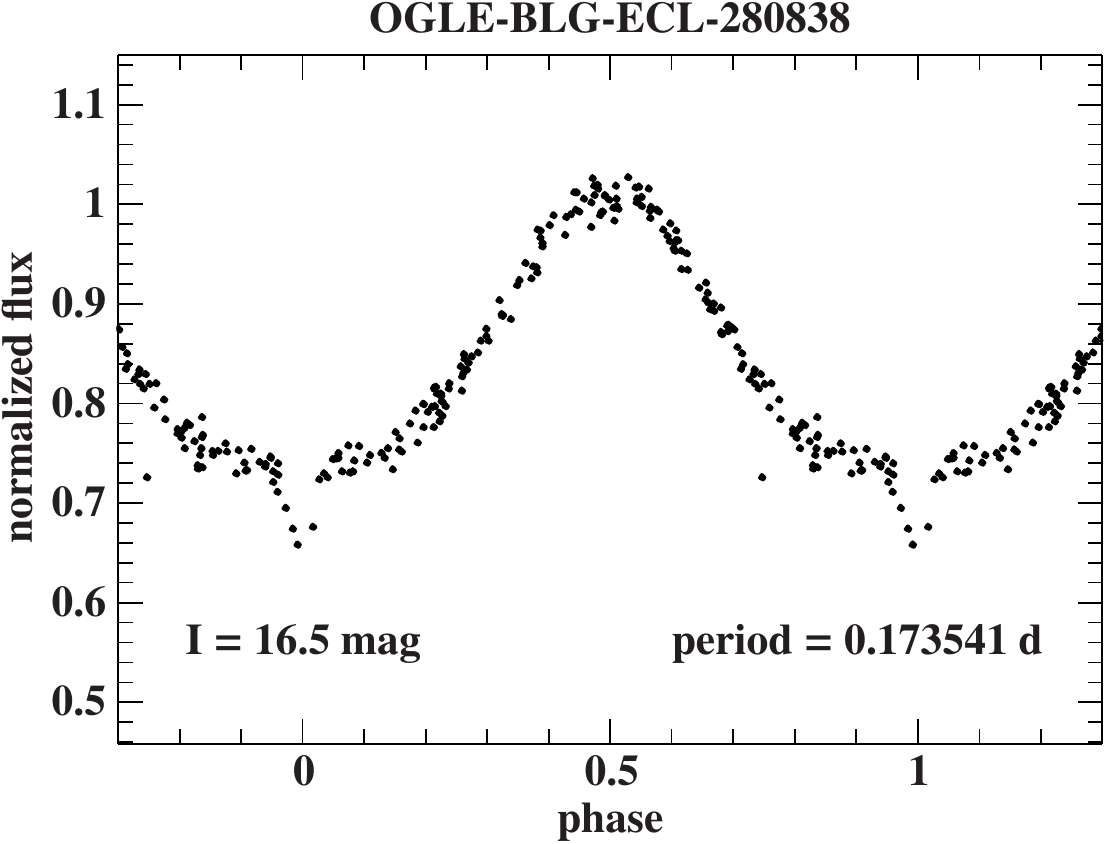}\hfill
		\includegraphics[width=0.25\linewidth]{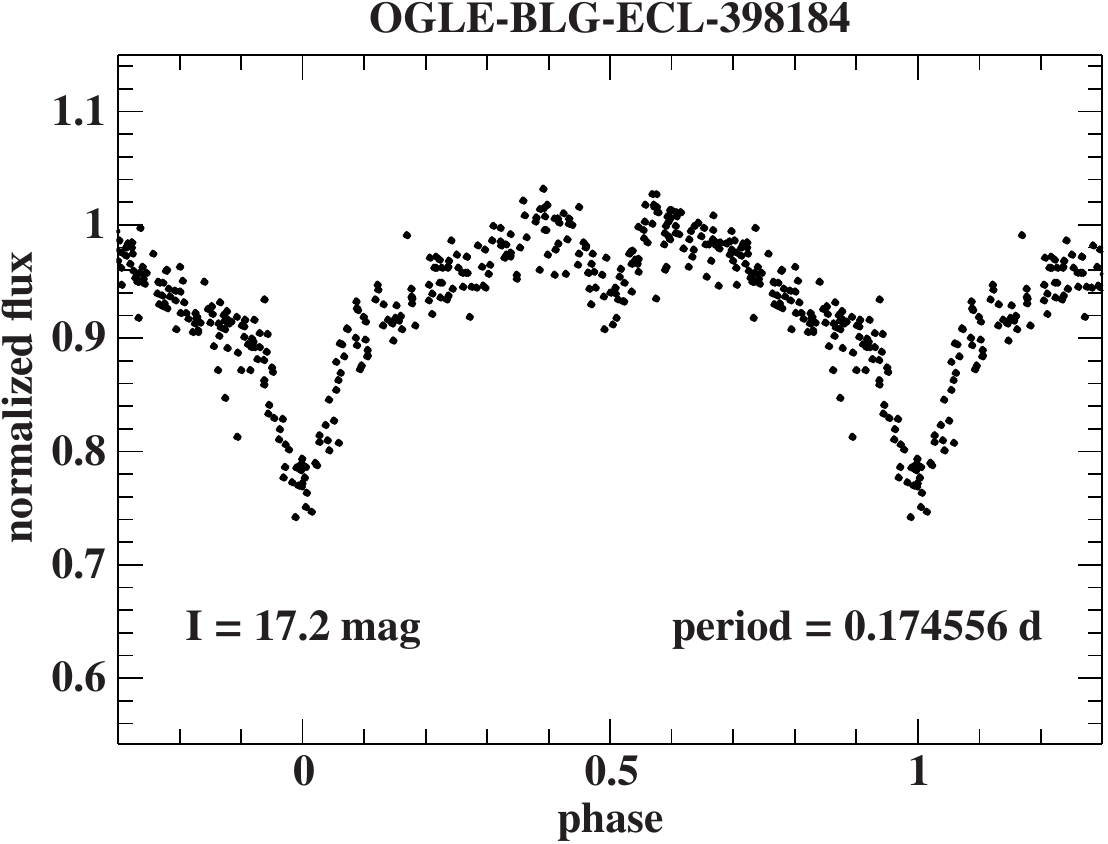}\hfill
		\includegraphics[width=0.25\linewidth]{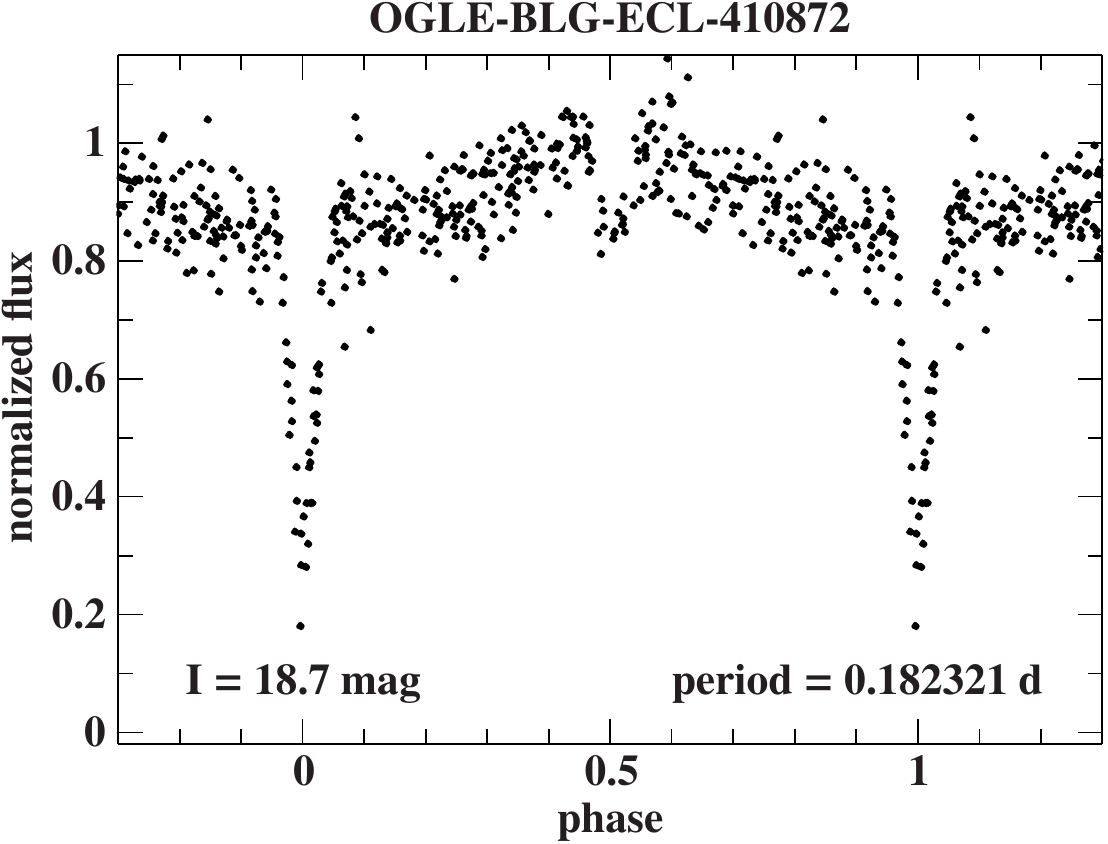}\hfill
		\includegraphics[width=0.25\linewidth]{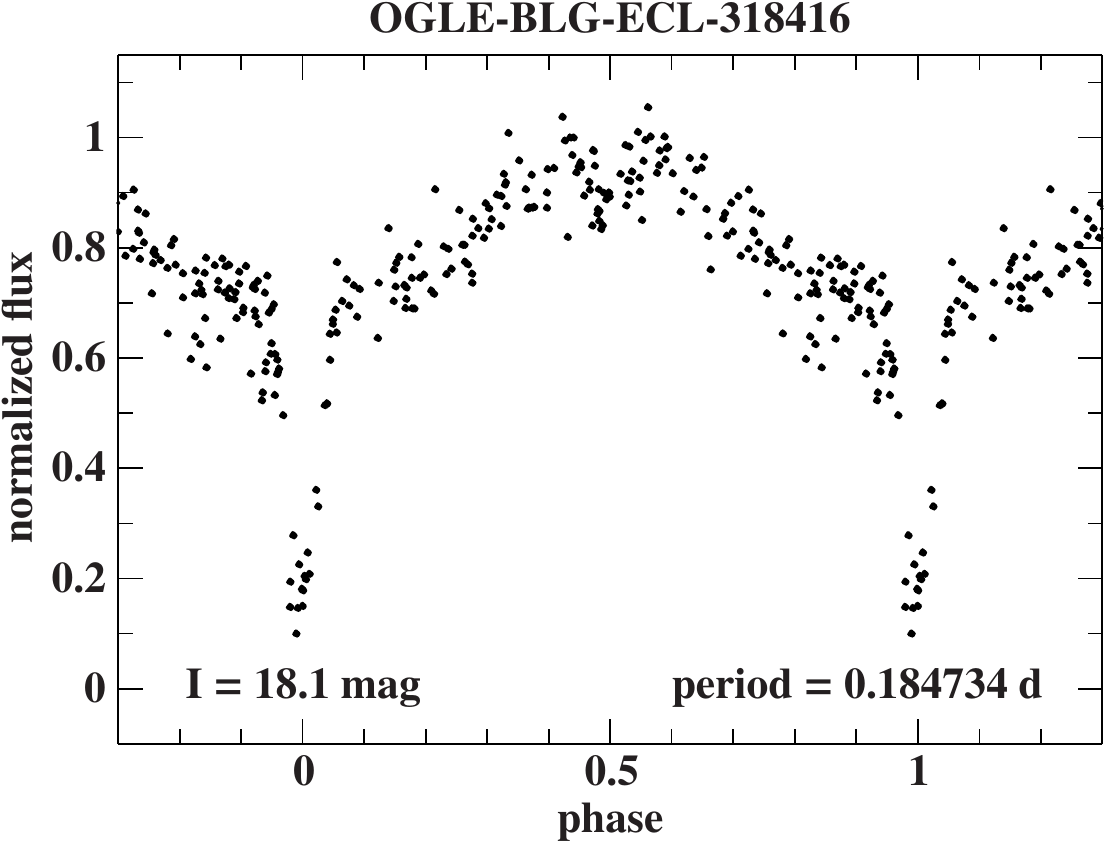}\hfill
		\includegraphics[width=0.25\linewidth]{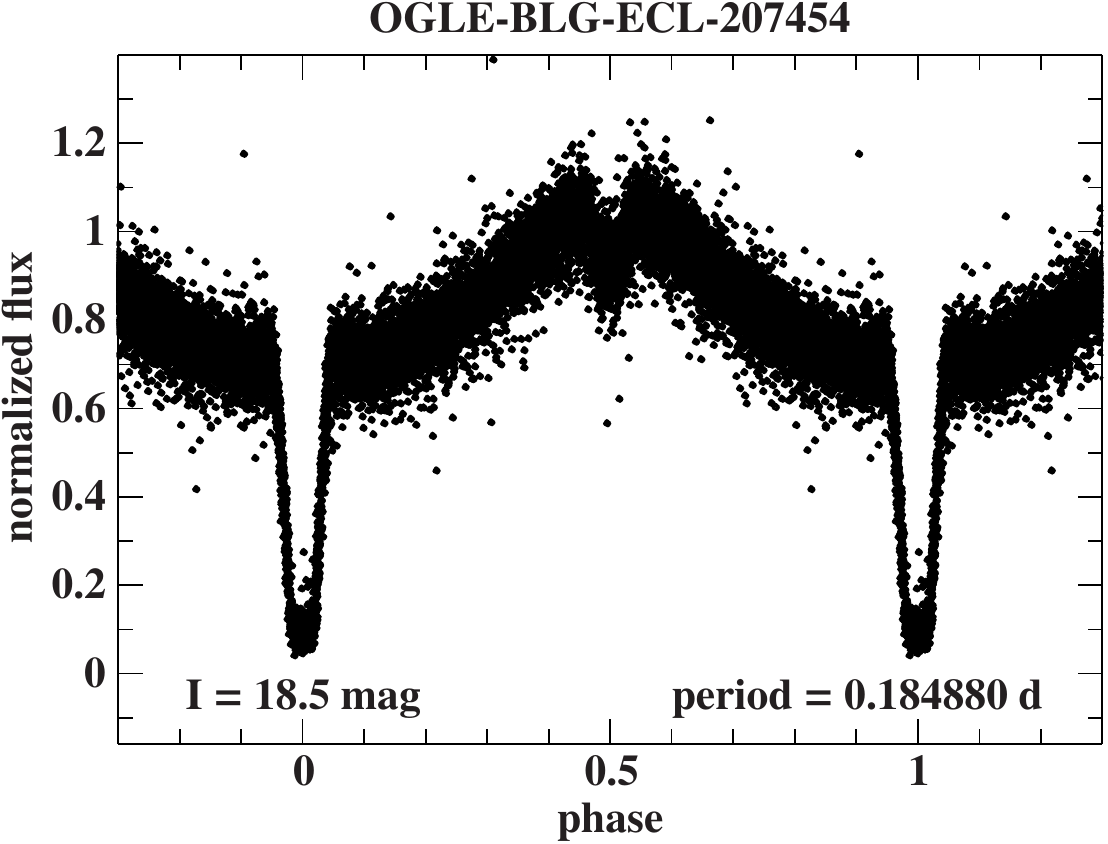}\hfill
		\includegraphics[width=0.25\linewidth]{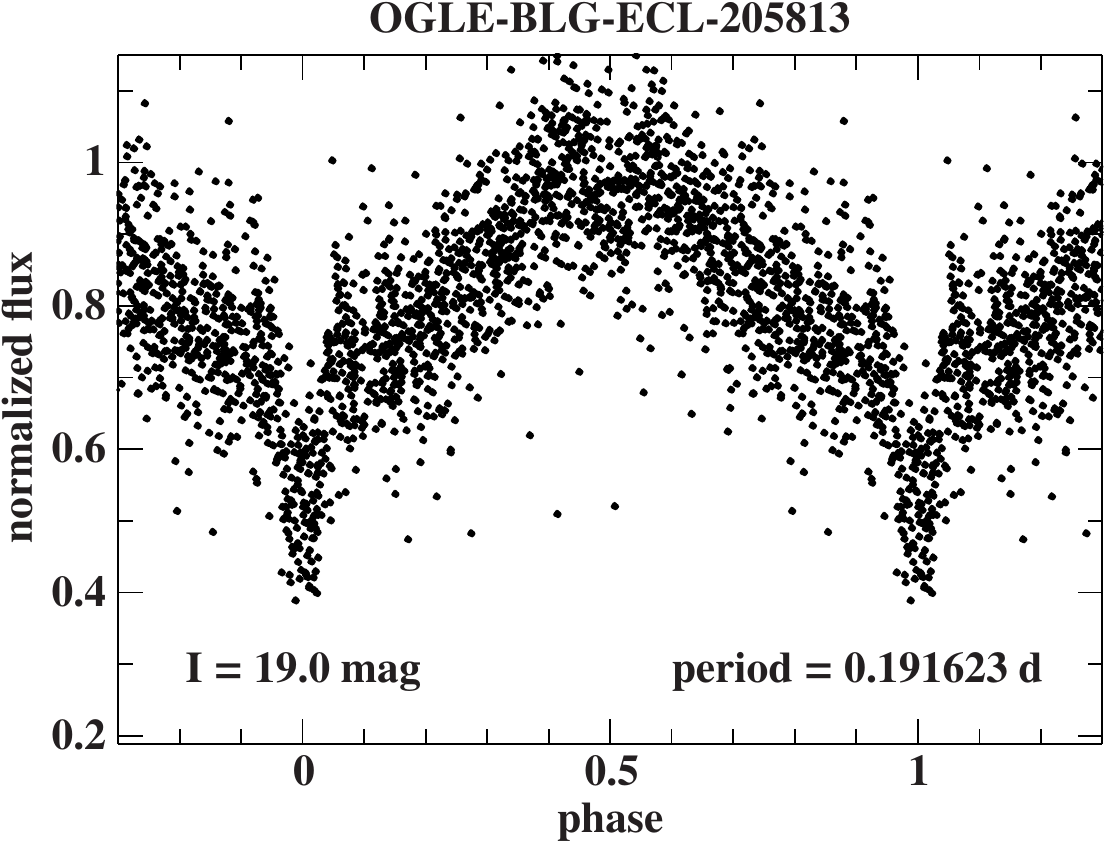}\hfill
		\includegraphics[width=0.25\linewidth]{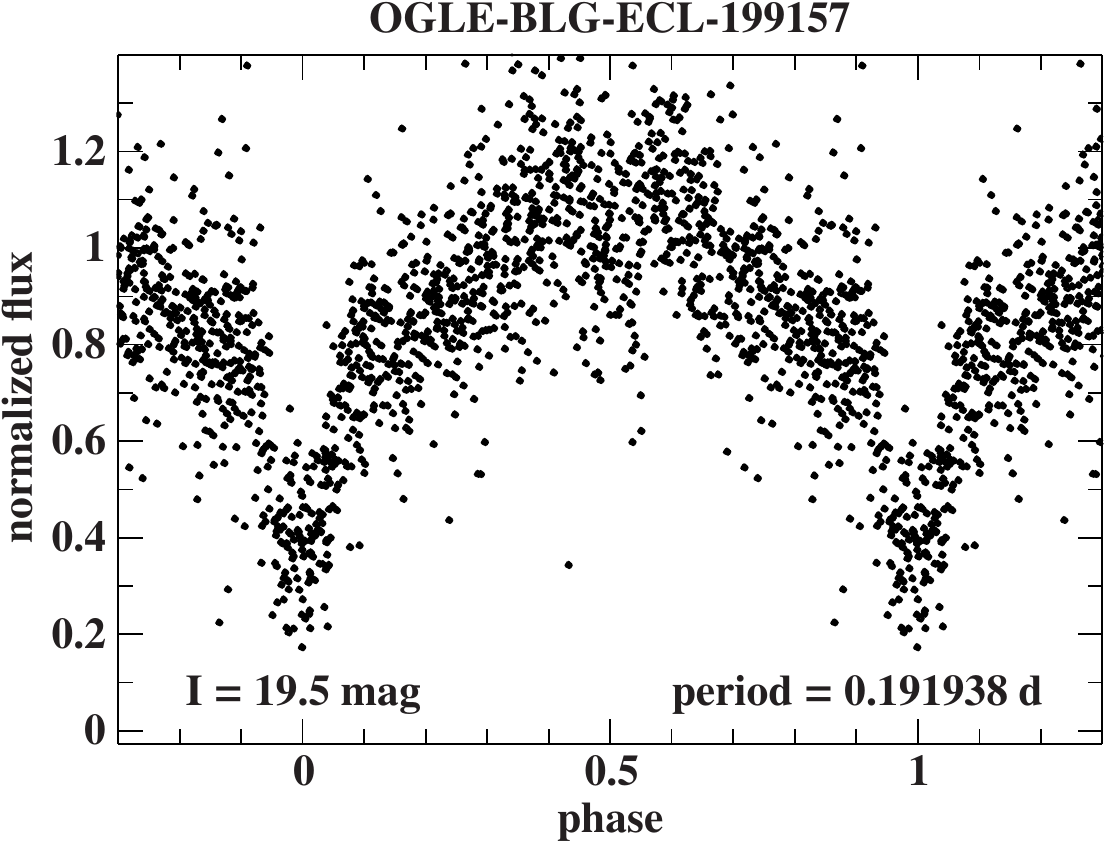}\hfill
		\includegraphics[width=0.25\linewidth]{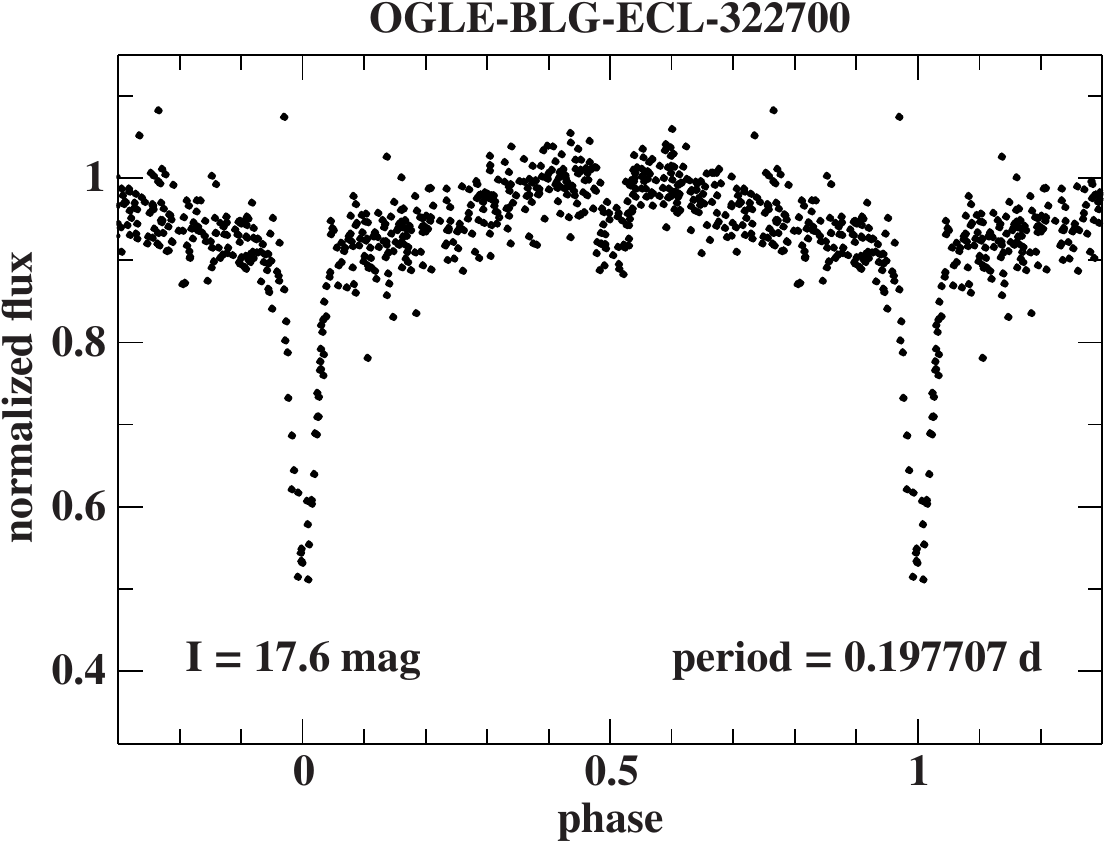}\hfill
		\includegraphics[width=0.25\linewidth]{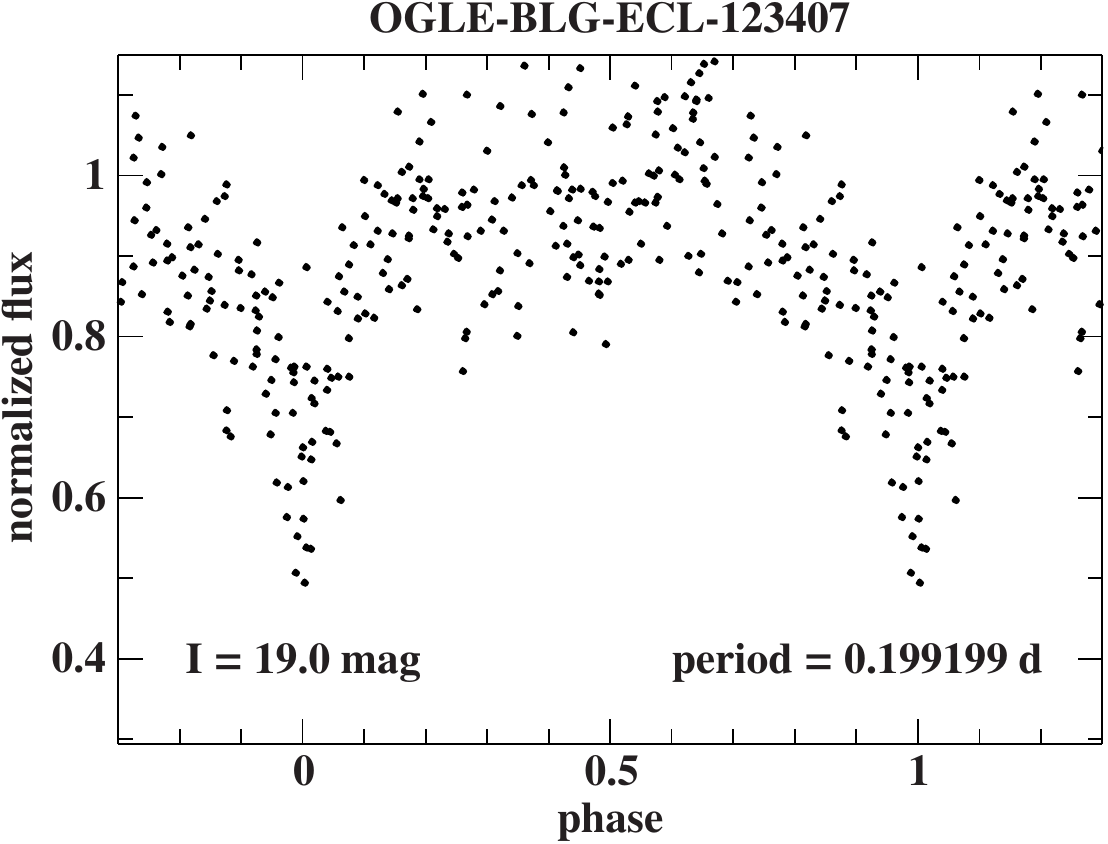}\hfill
		\includegraphics[width=0.25\linewidth]{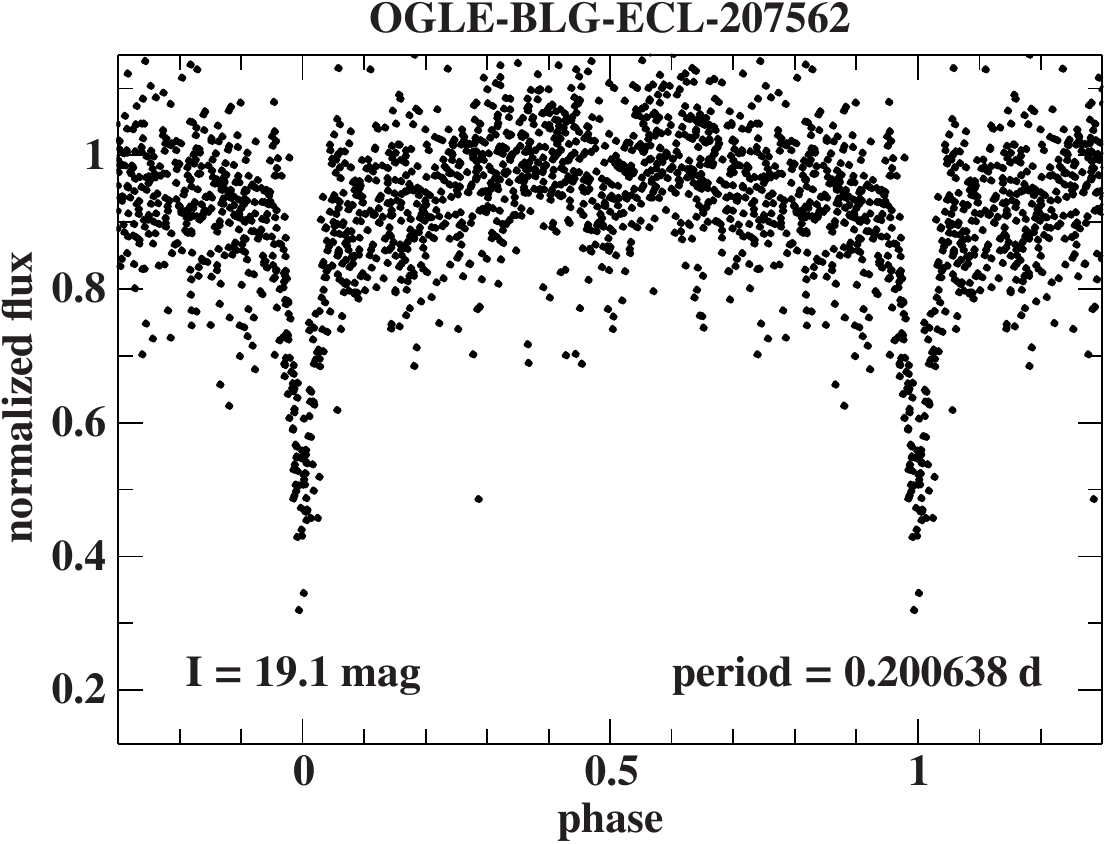}\hfill
		\includegraphics[width=0.25\linewidth]{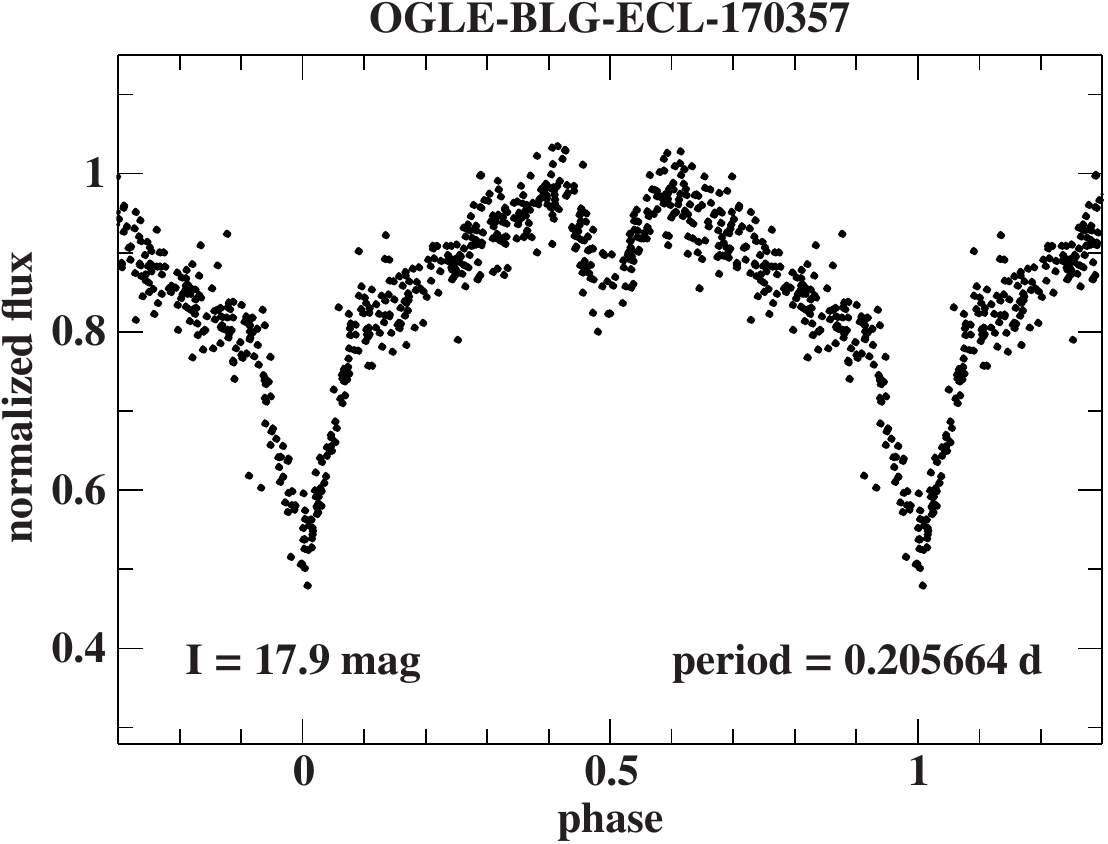}\hfill
		\includegraphics[width=0.25\linewidth]{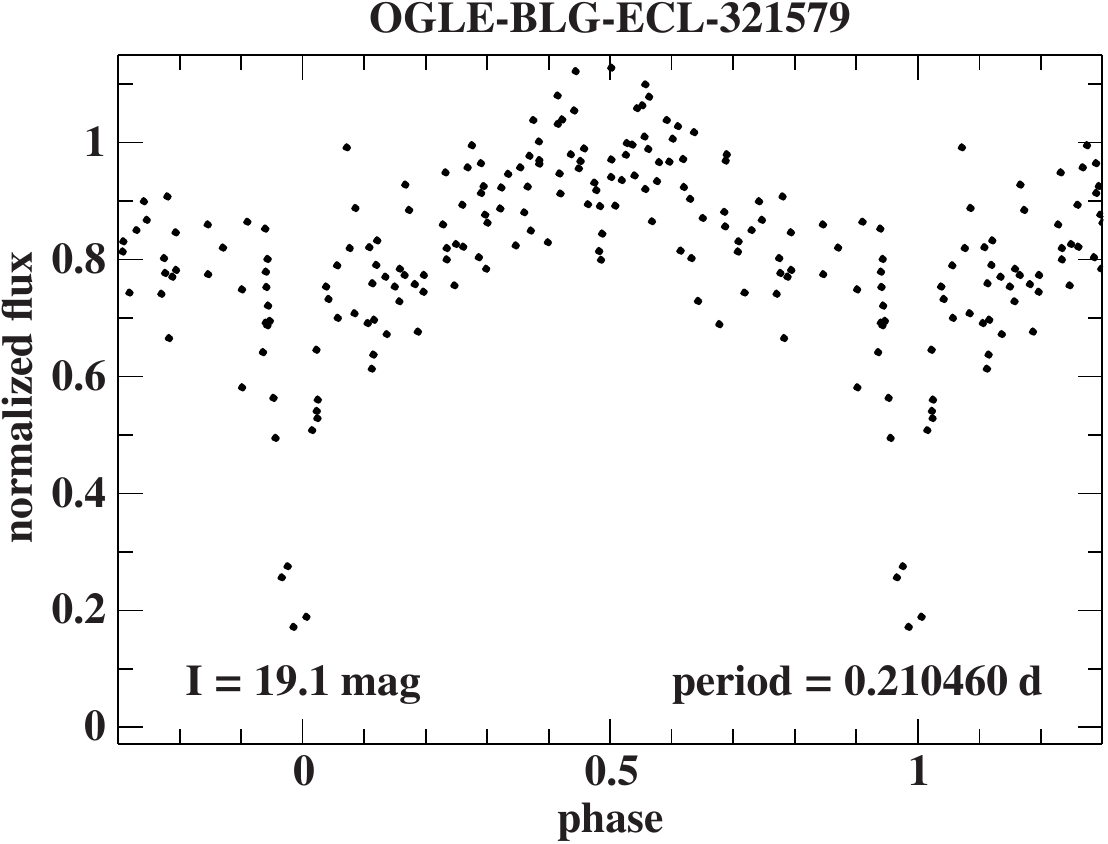}\hfill
		\includegraphics[width=0.25\linewidth]{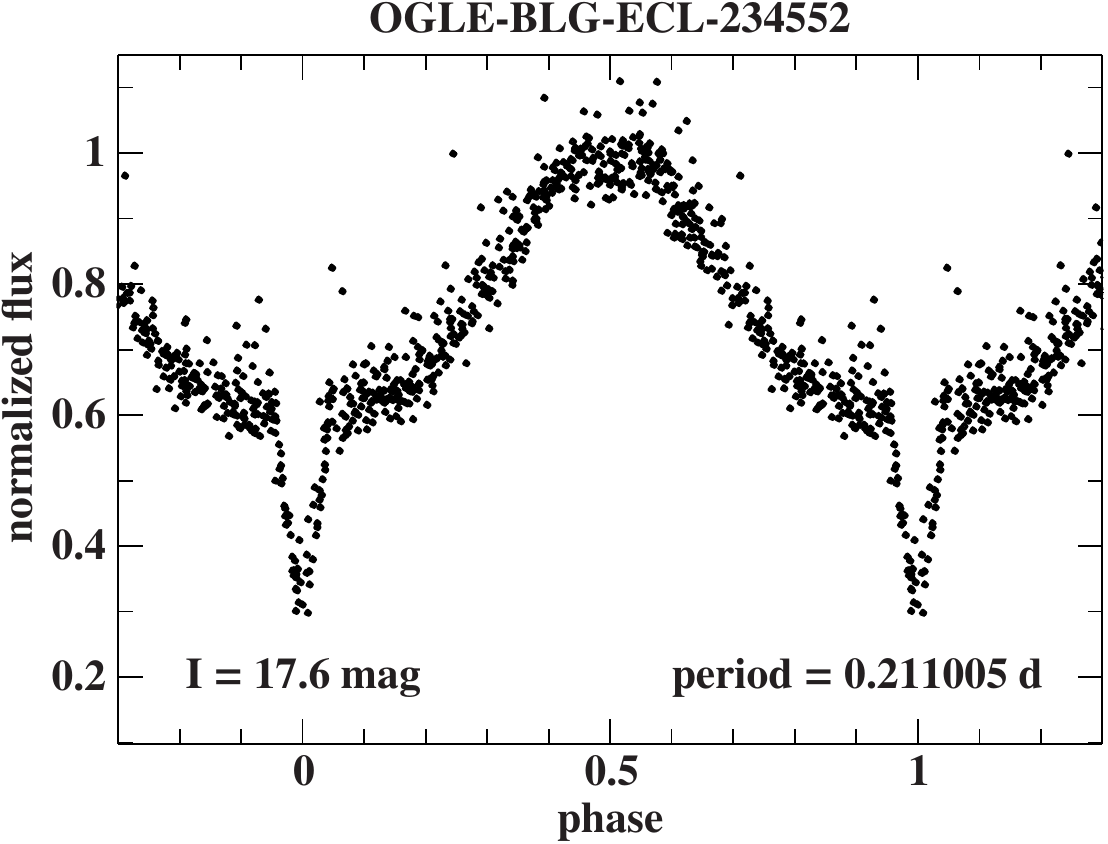}\hfill
		\includegraphics[width=0.25\linewidth]{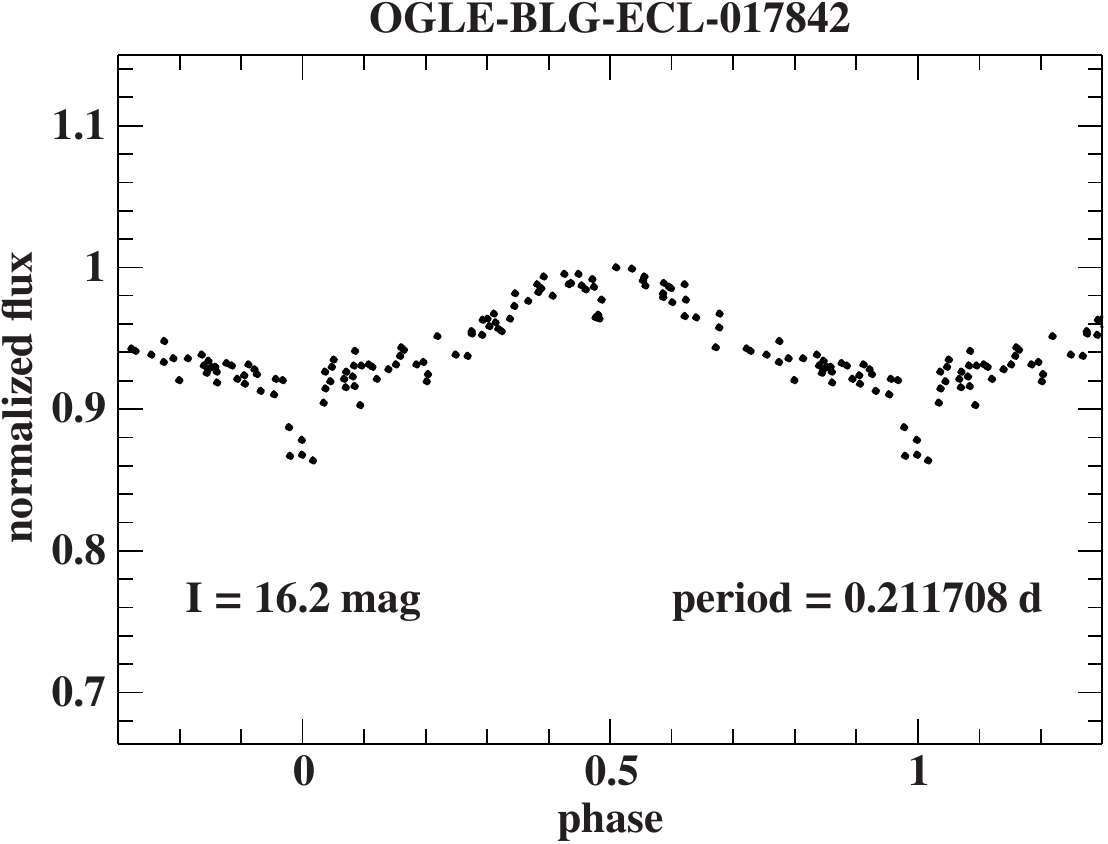}\hfill
		\includegraphics[width=0.25\linewidth]{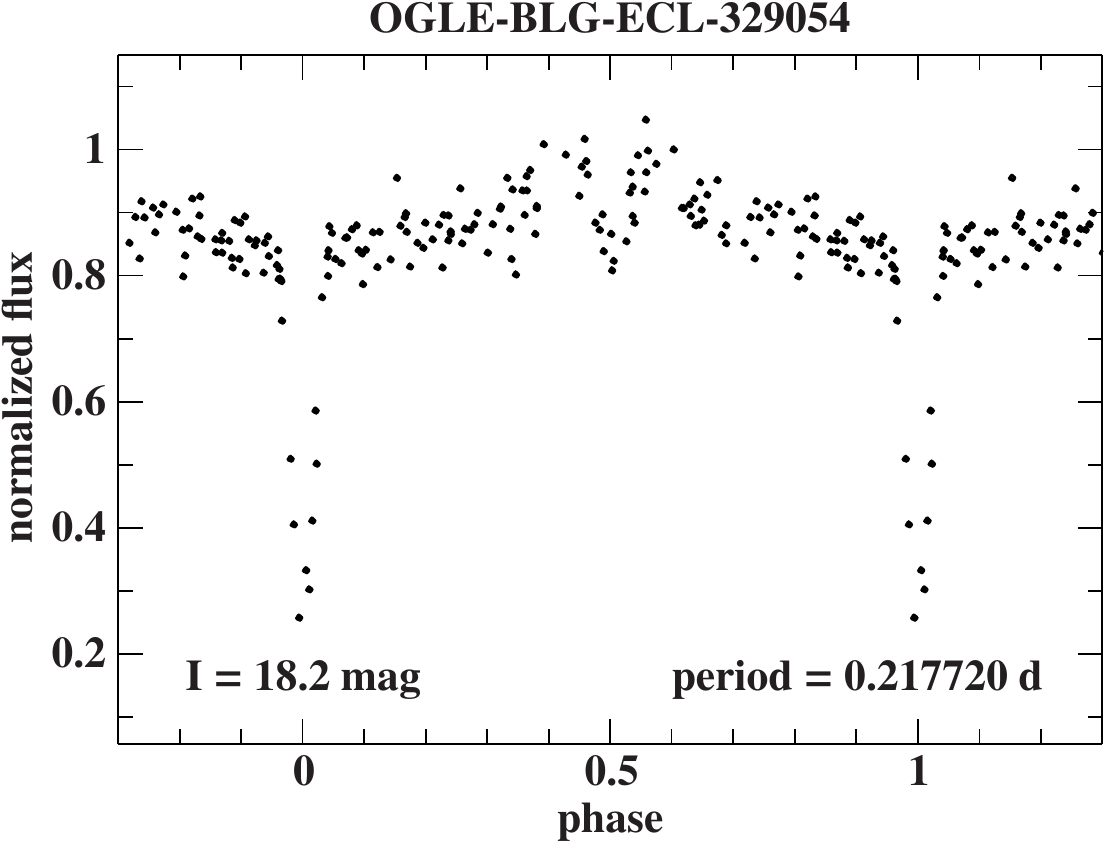}\hfill
		\includegraphics[width=0.25\linewidth]{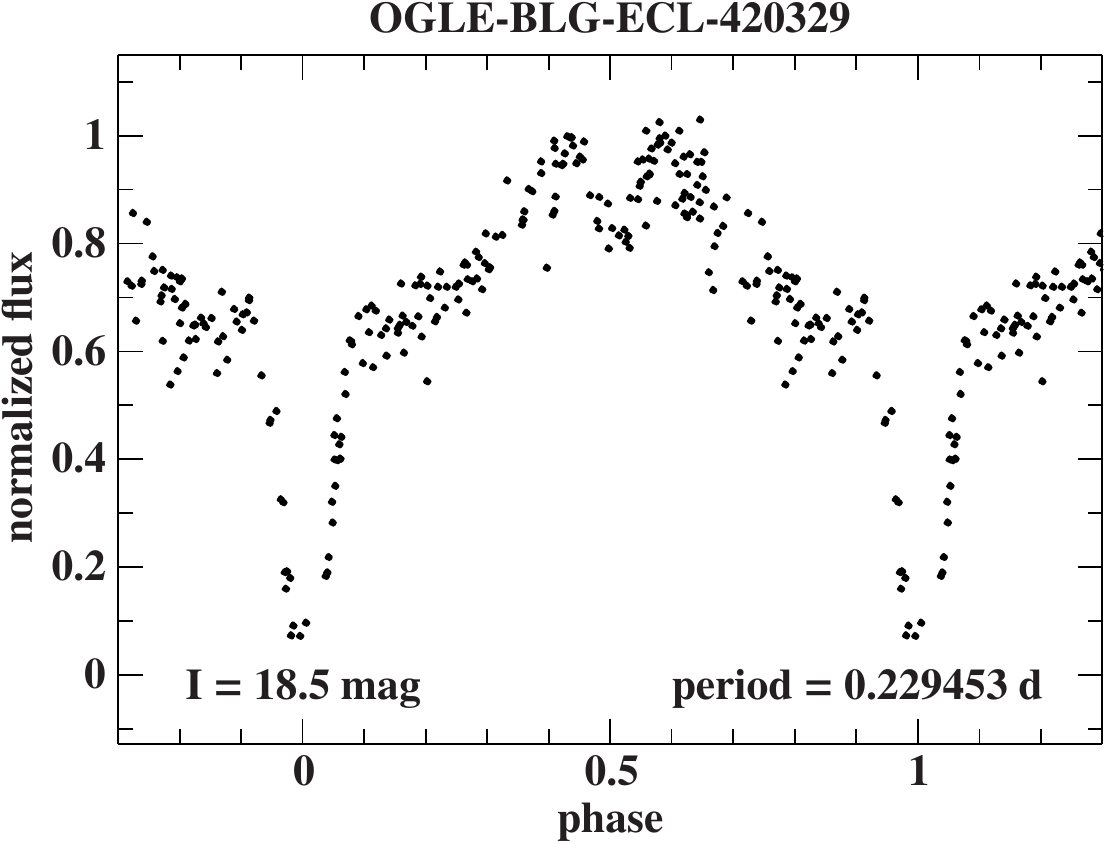}\hfill
		\includegraphics[width=0.25\linewidth]{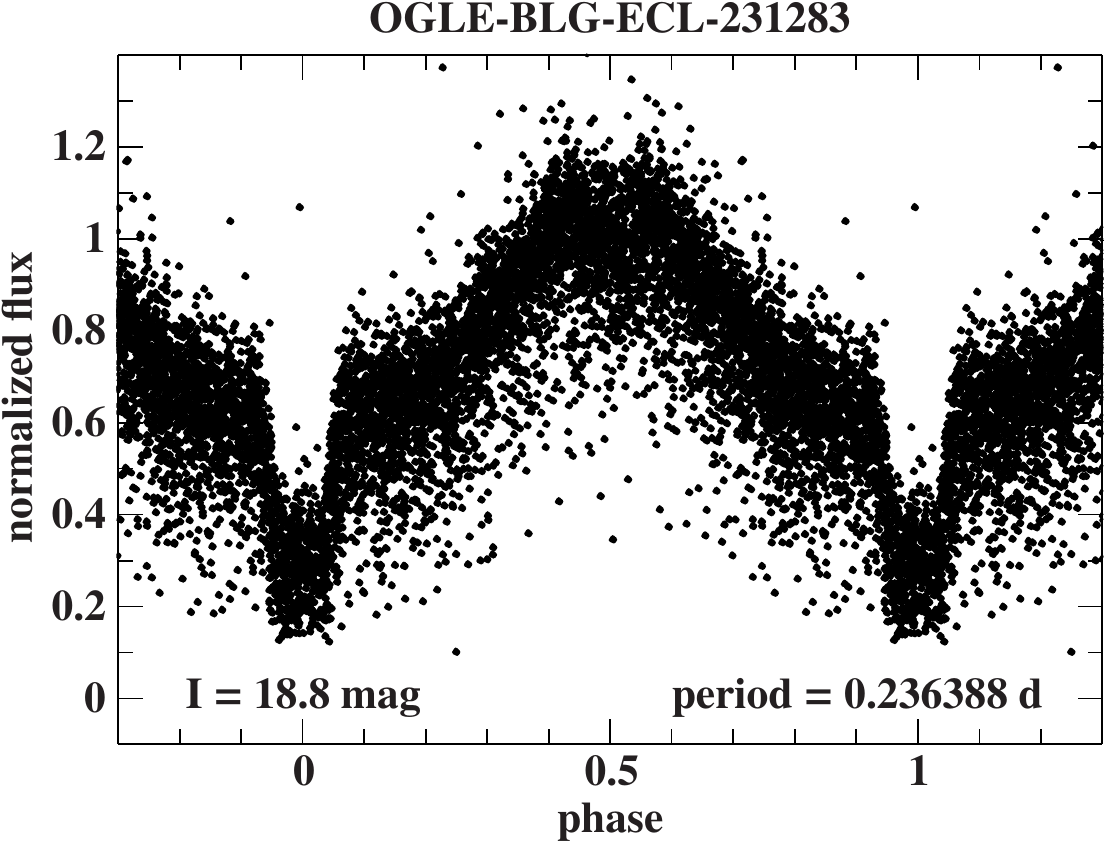}\hfill
		\includegraphics[width=0.25\linewidth]{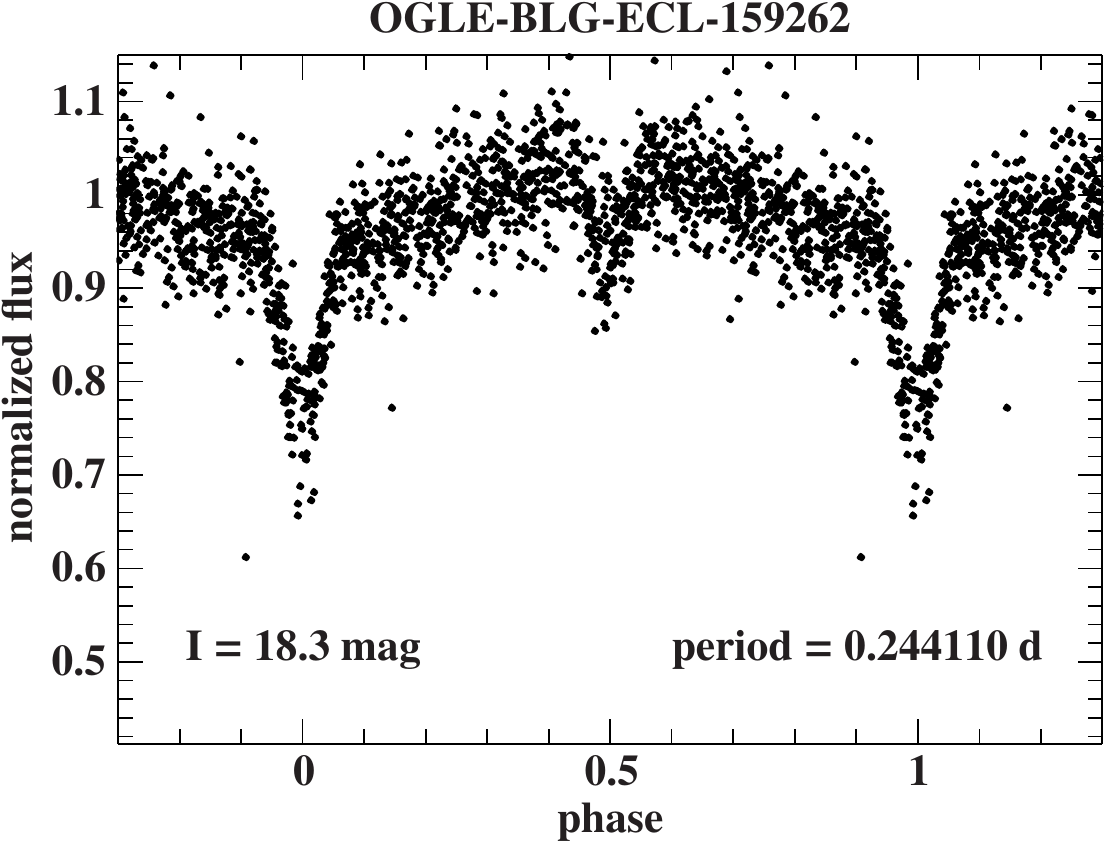}\hfill
		\includegraphics[width=0.25\linewidth]{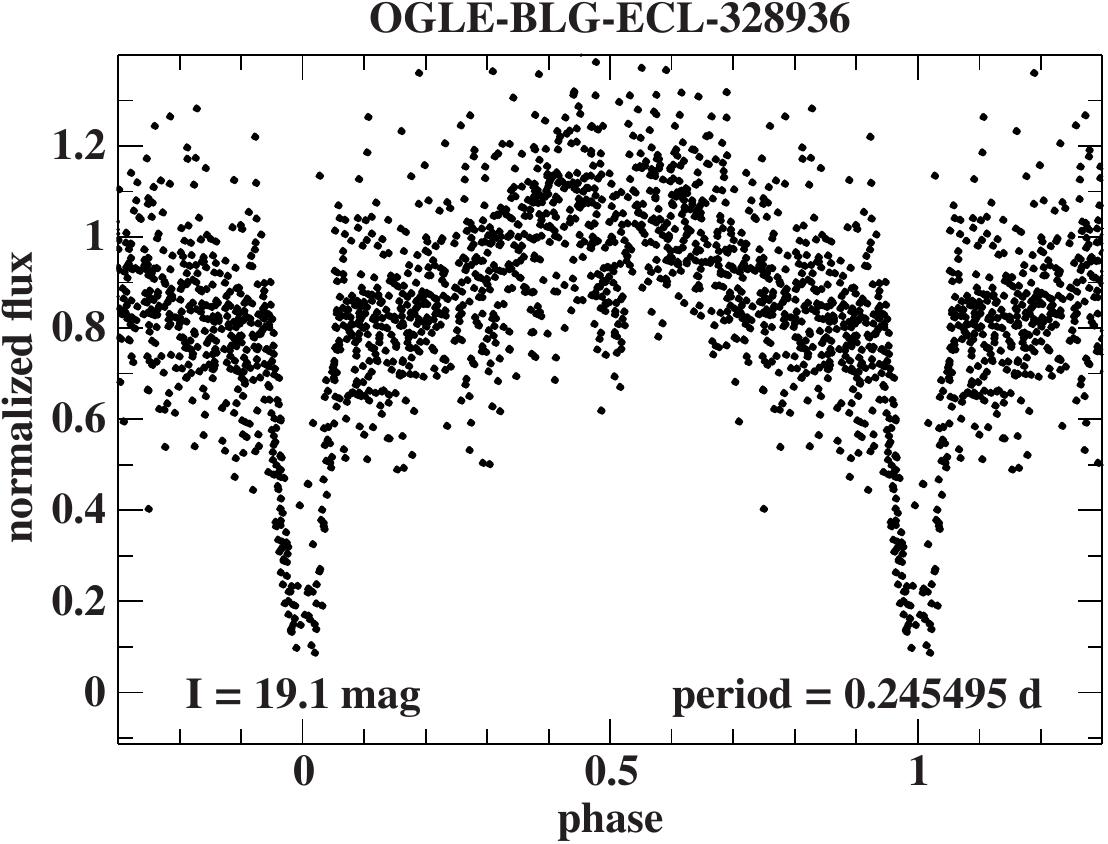}\hfill
		\includegraphics[width=0.25\linewidth]{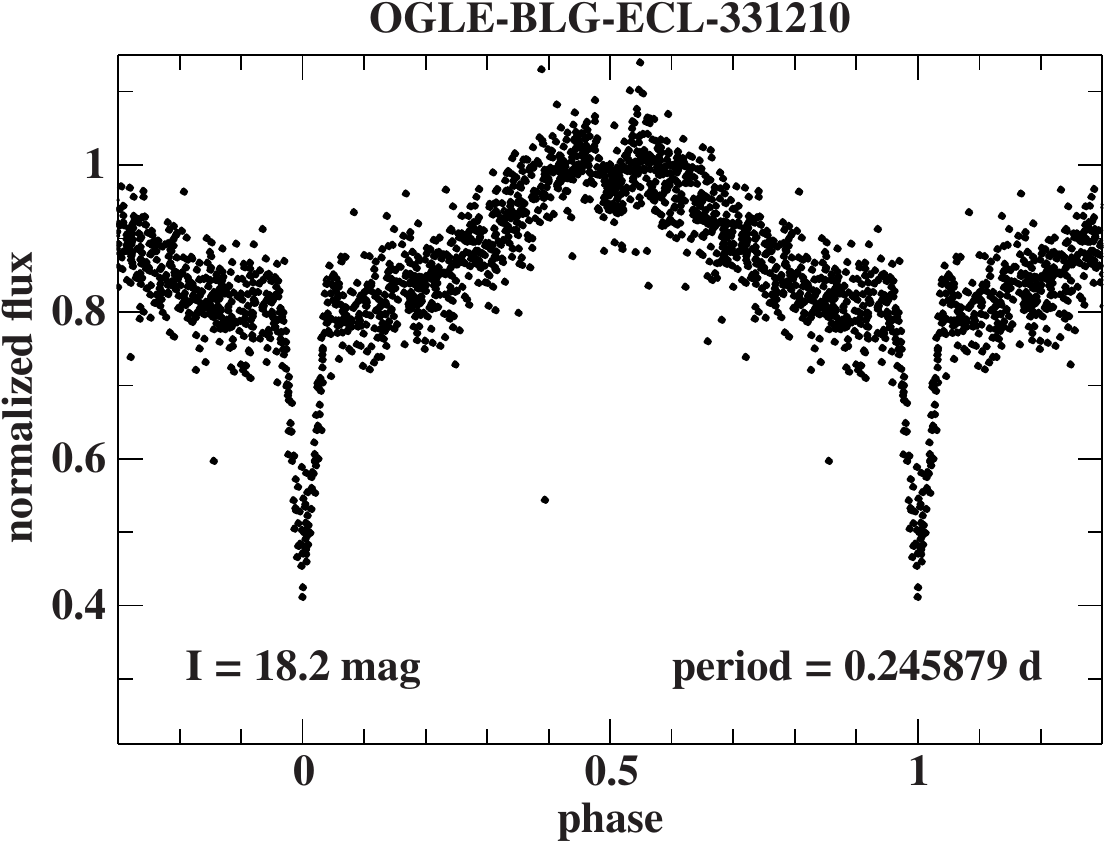}\hfill
		\includegraphics[width=0.25\linewidth]{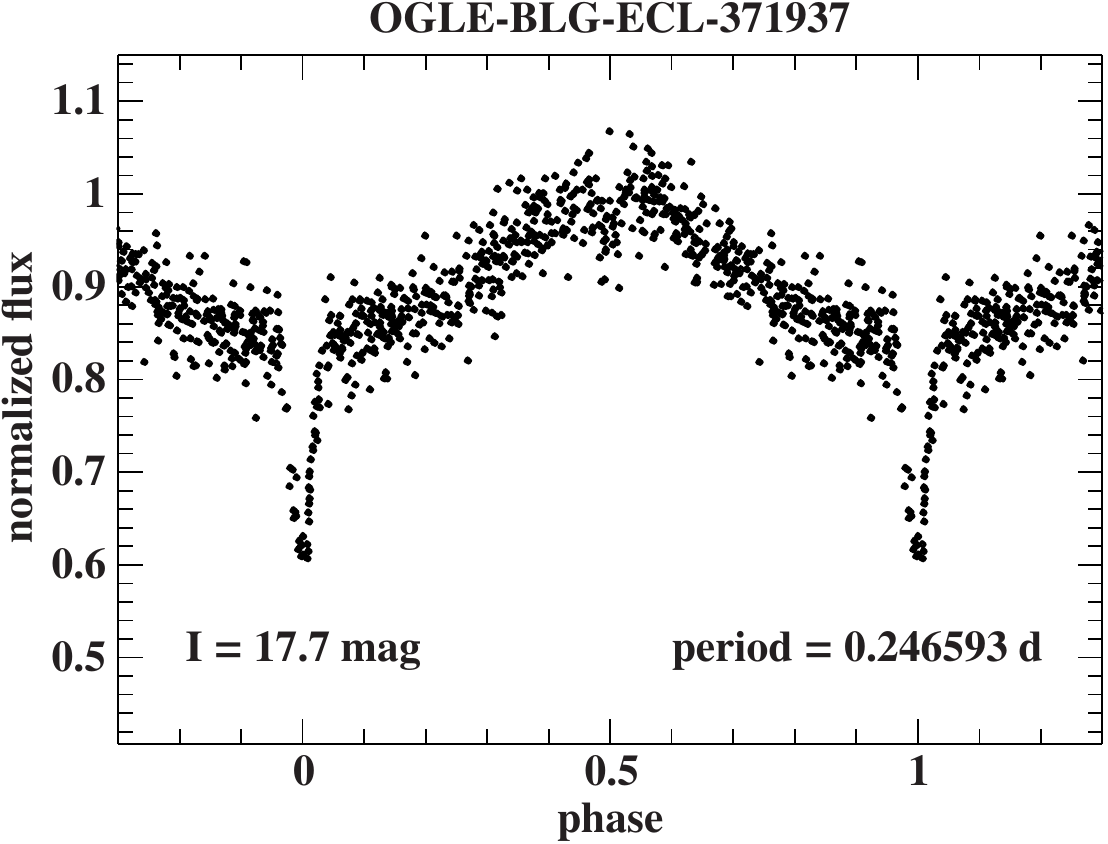}\hfill
		\includegraphics[width=0.25\linewidth]{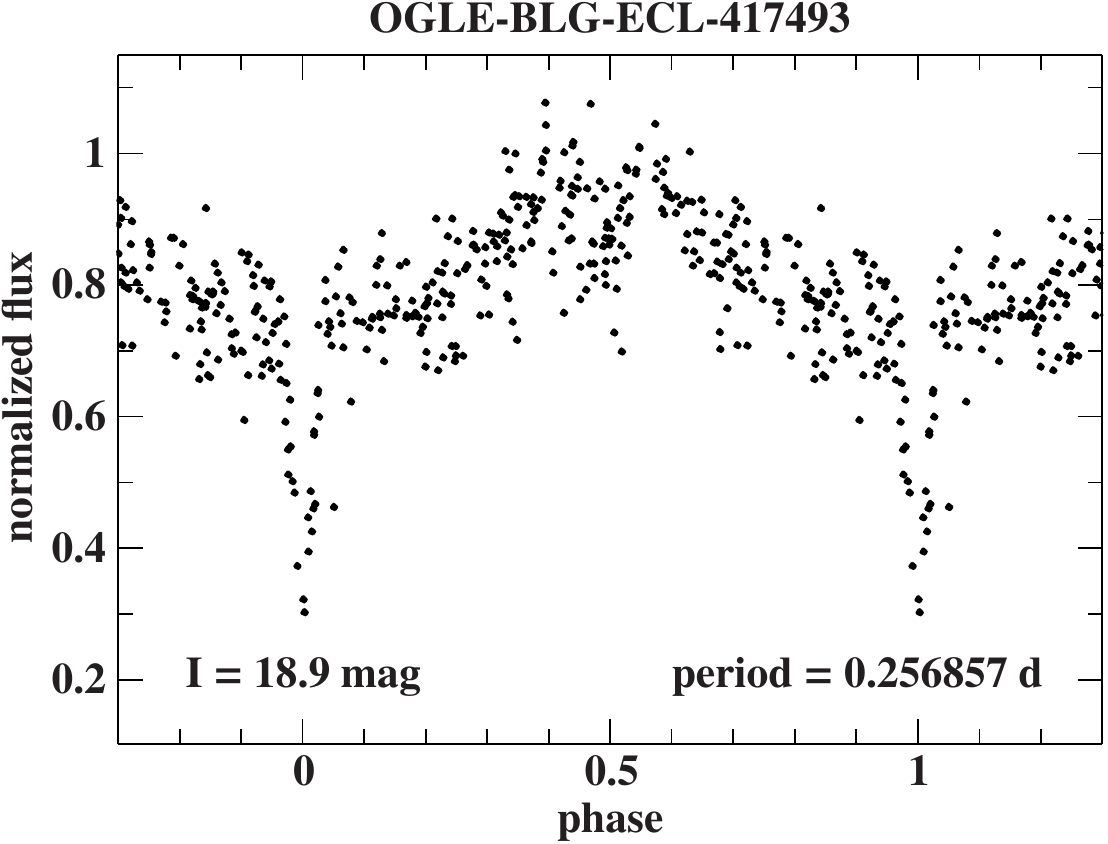}\hfill
	\end{figure}
	\begin{figure}
		\includegraphics[width=0.25\linewidth]{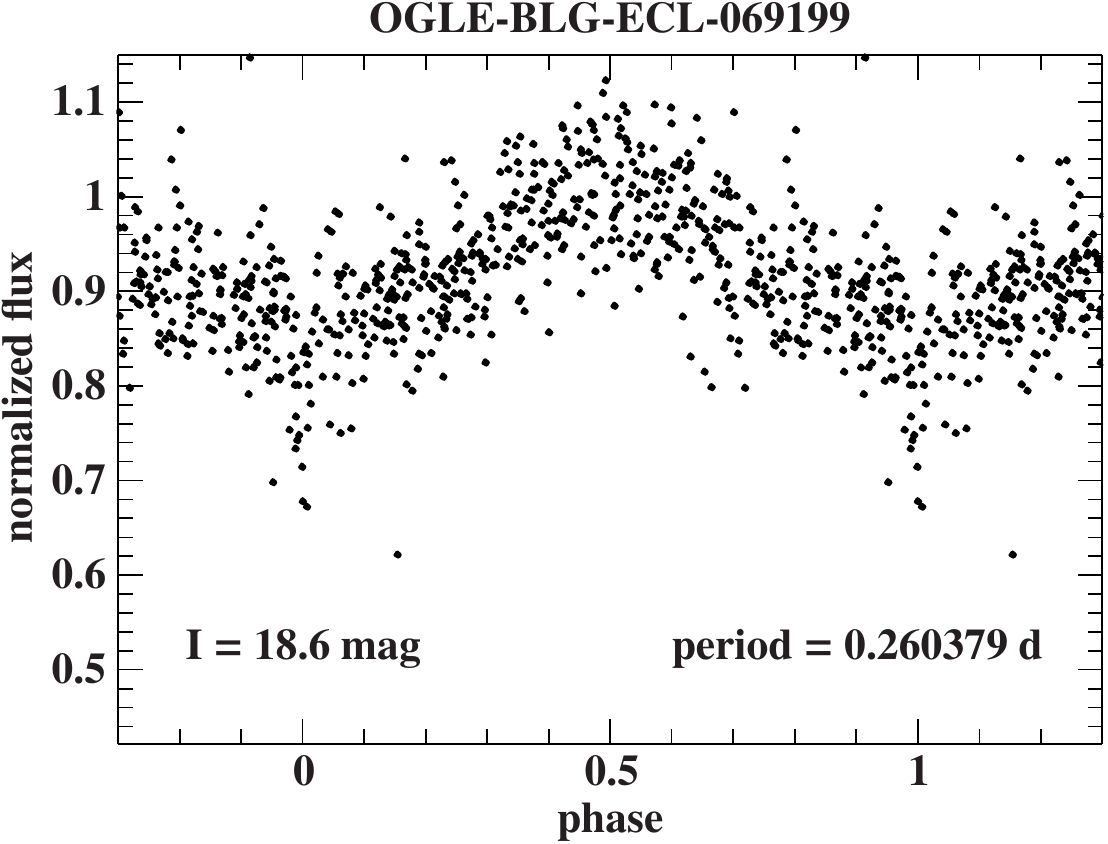}\hfill
		\includegraphics[width=0.25\linewidth]{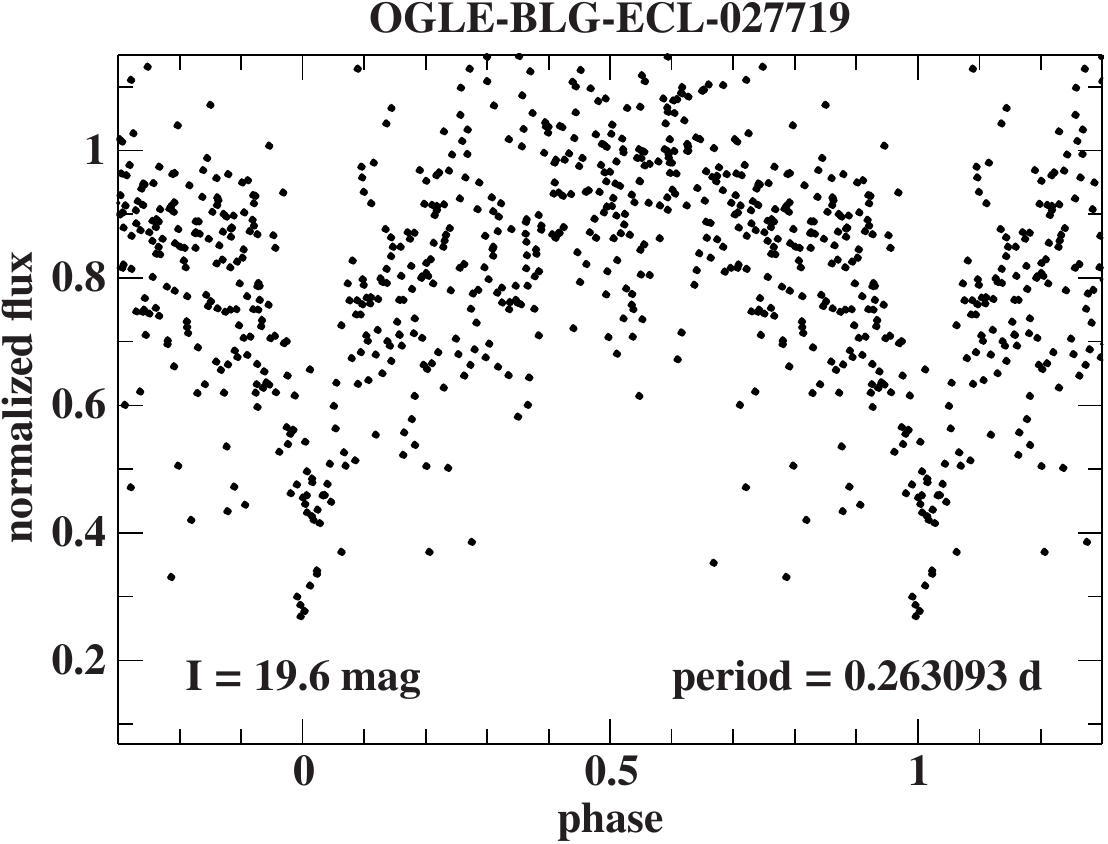}\hfill
		\includegraphics[width=0.25\linewidth]{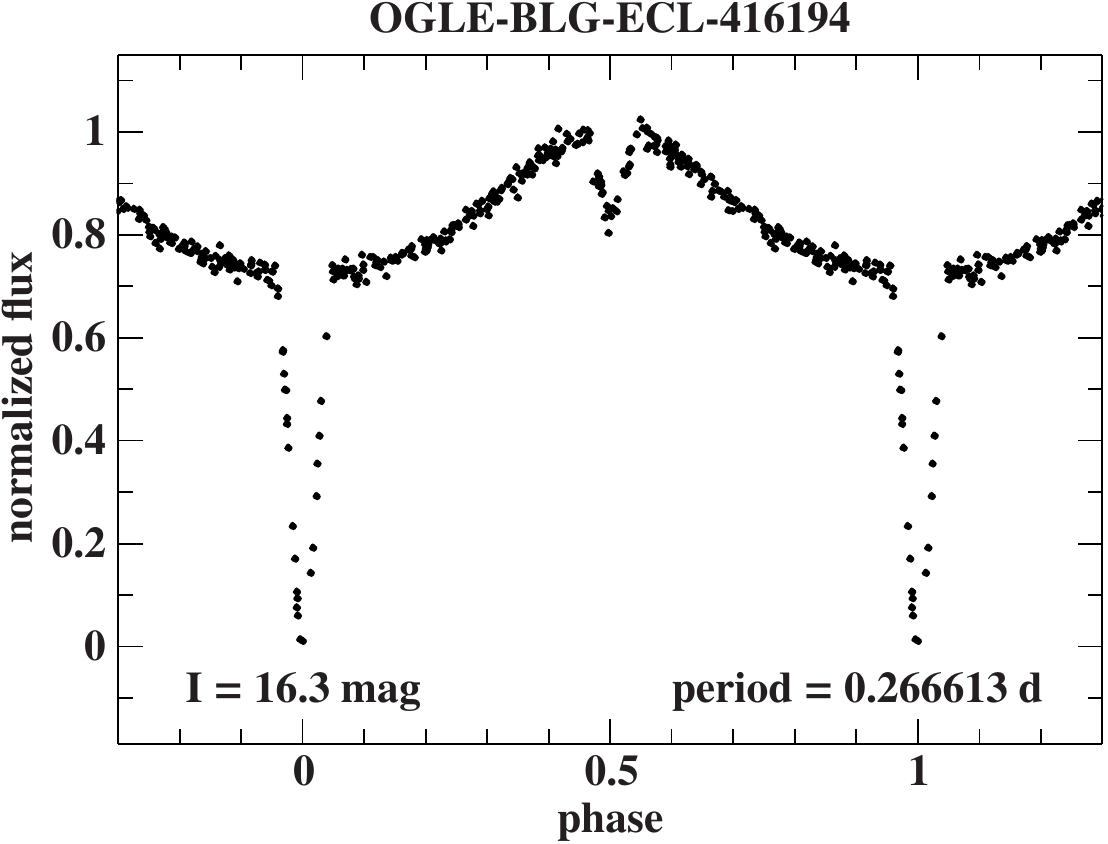}\hfill
		\includegraphics[width=0.25\linewidth]{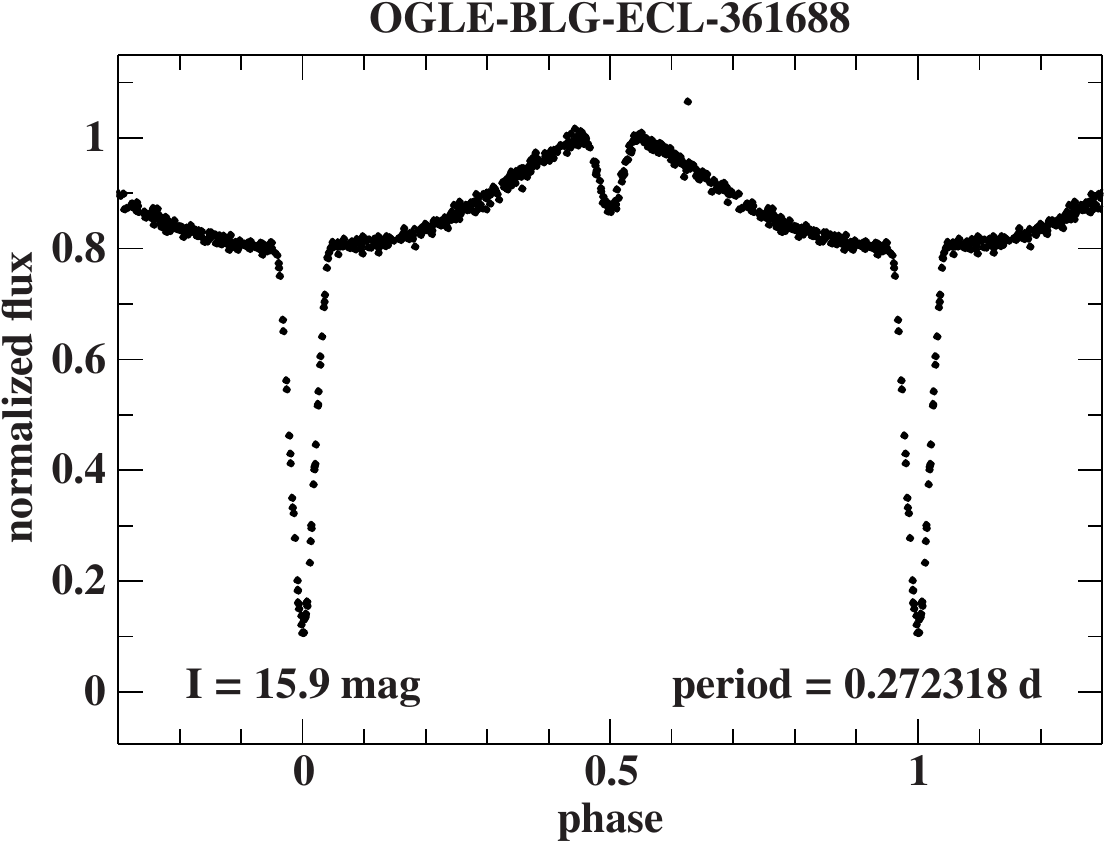}\hfill
		\includegraphics[width=0.25\linewidth]{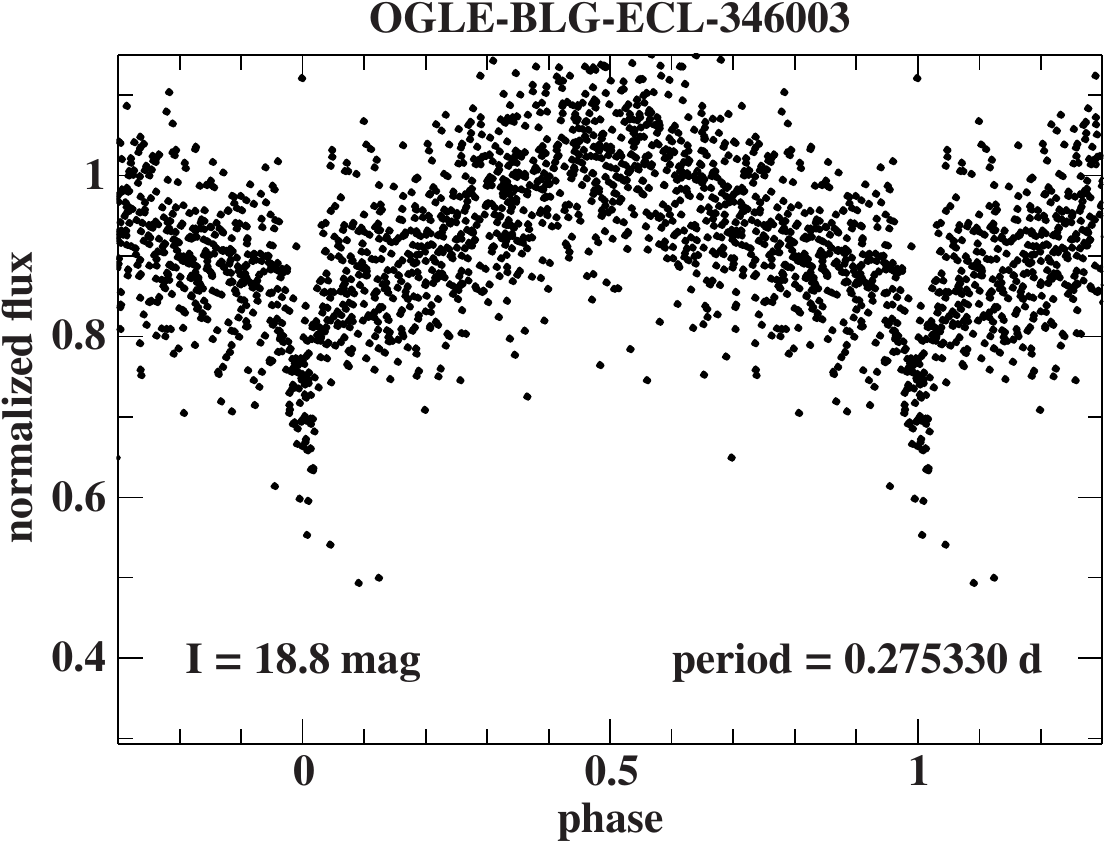}\hfill
		\includegraphics[width=0.25\linewidth]{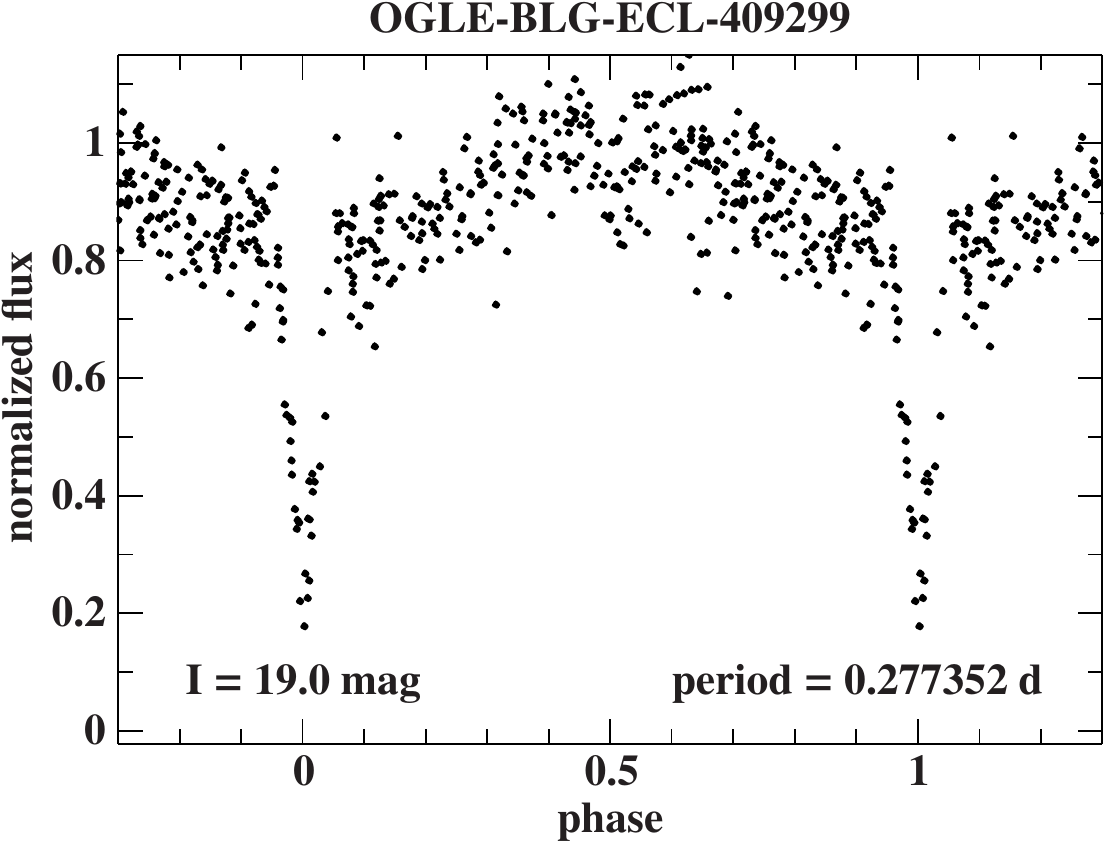}\hfill
		\includegraphics[width=0.25\linewidth]{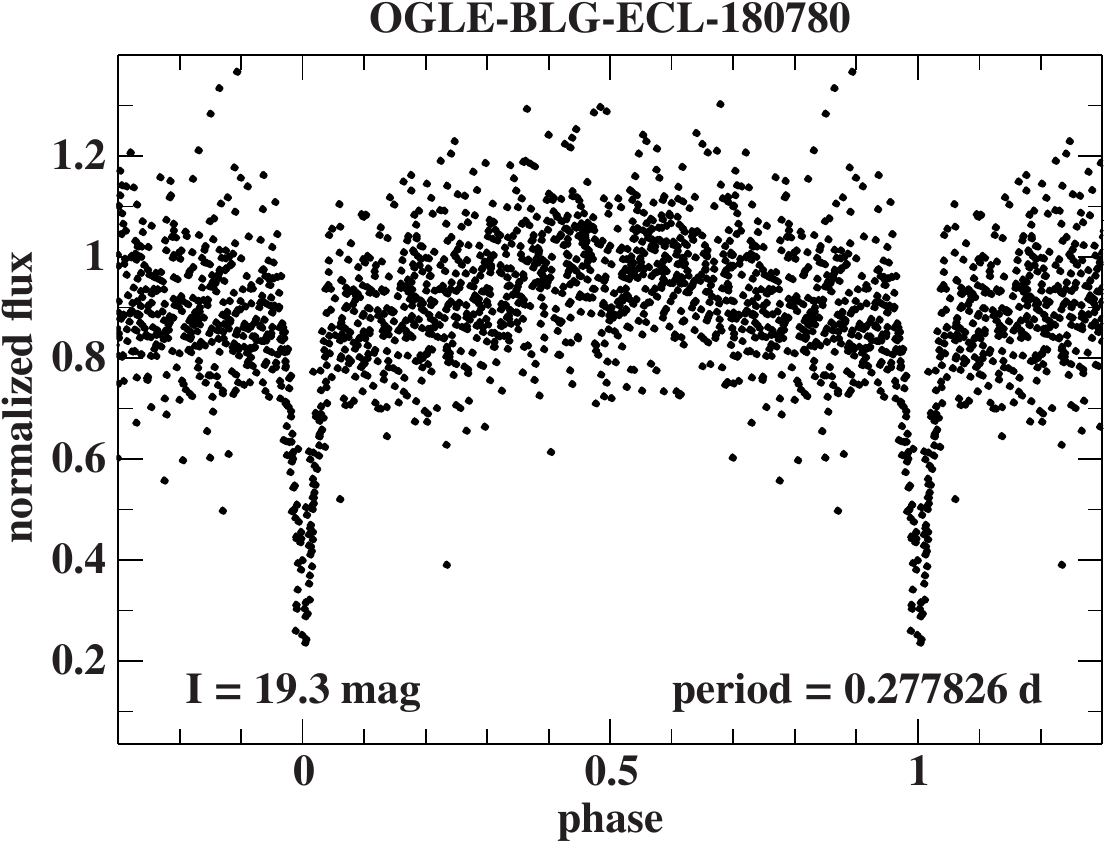}\hfill
		\includegraphics[width=0.25\linewidth]{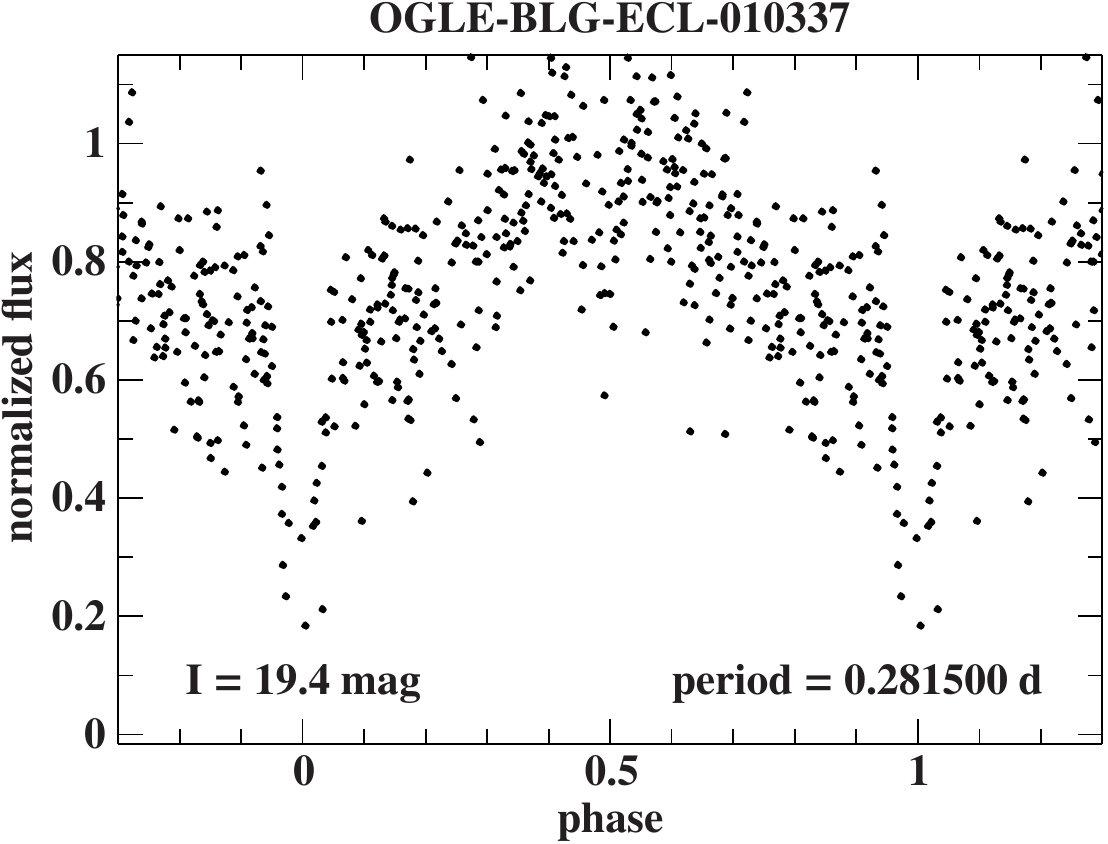}\hfill
		\includegraphics[width=0.25\linewidth]{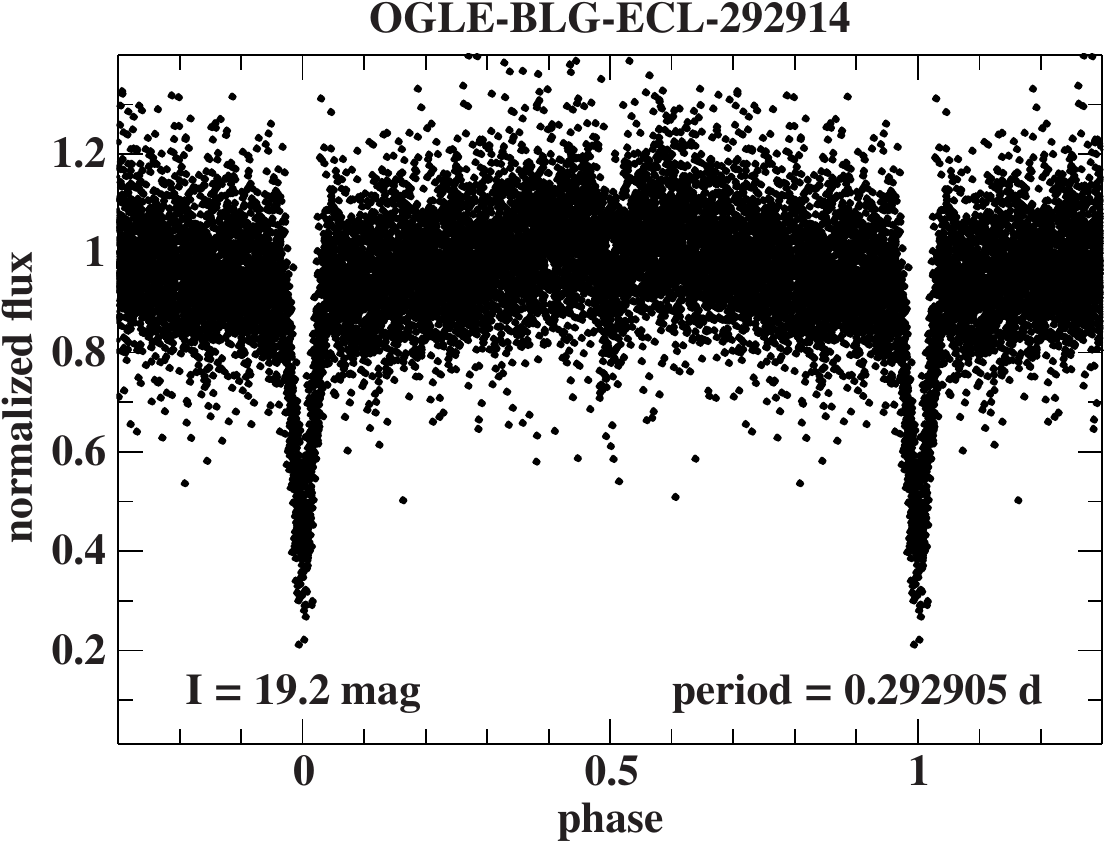}\hfill
		\includegraphics[width=0.25\linewidth]{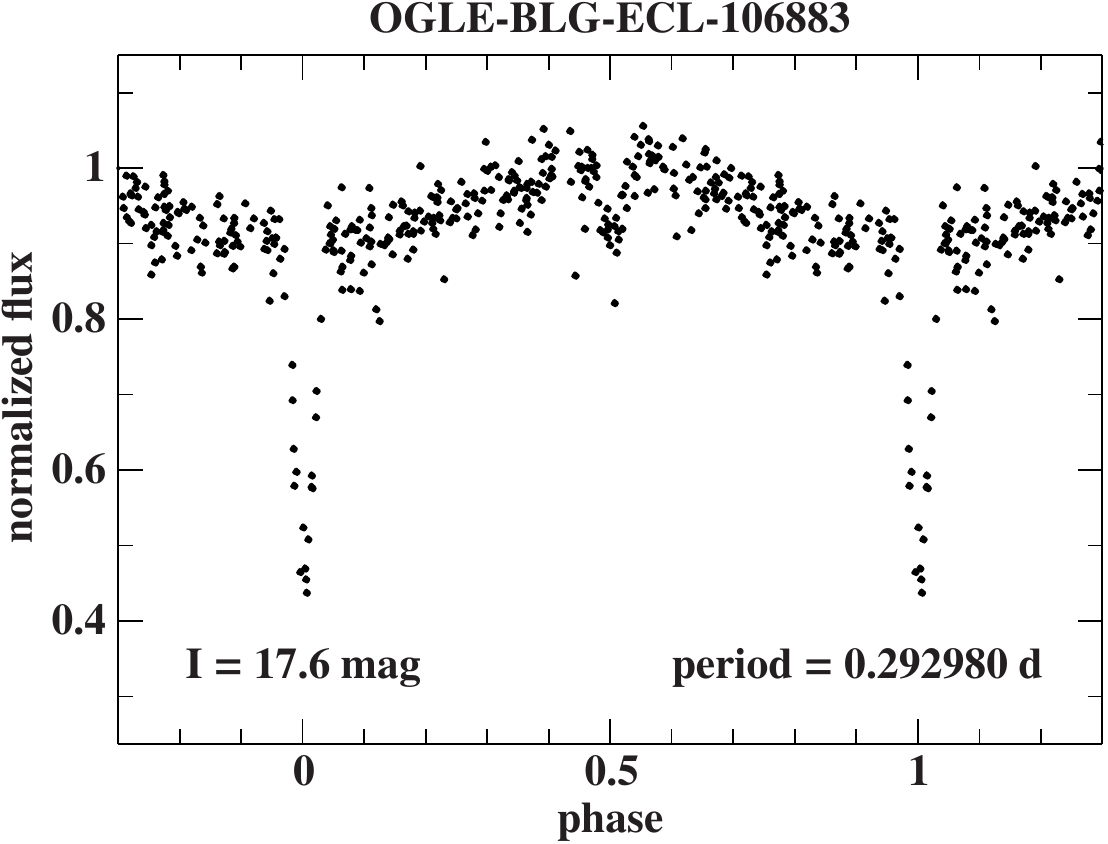}\hfill
		\includegraphics[width=0.25\linewidth]{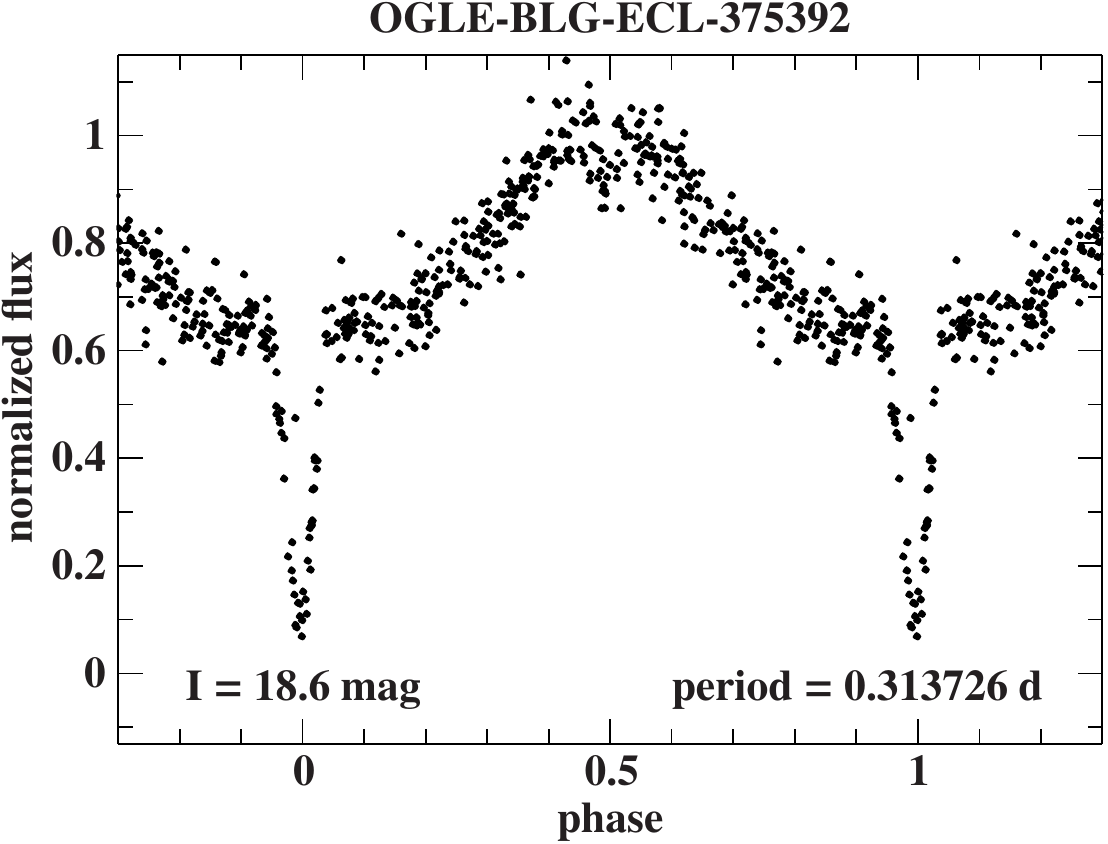}\hfill
		\includegraphics[width=0.25\linewidth]{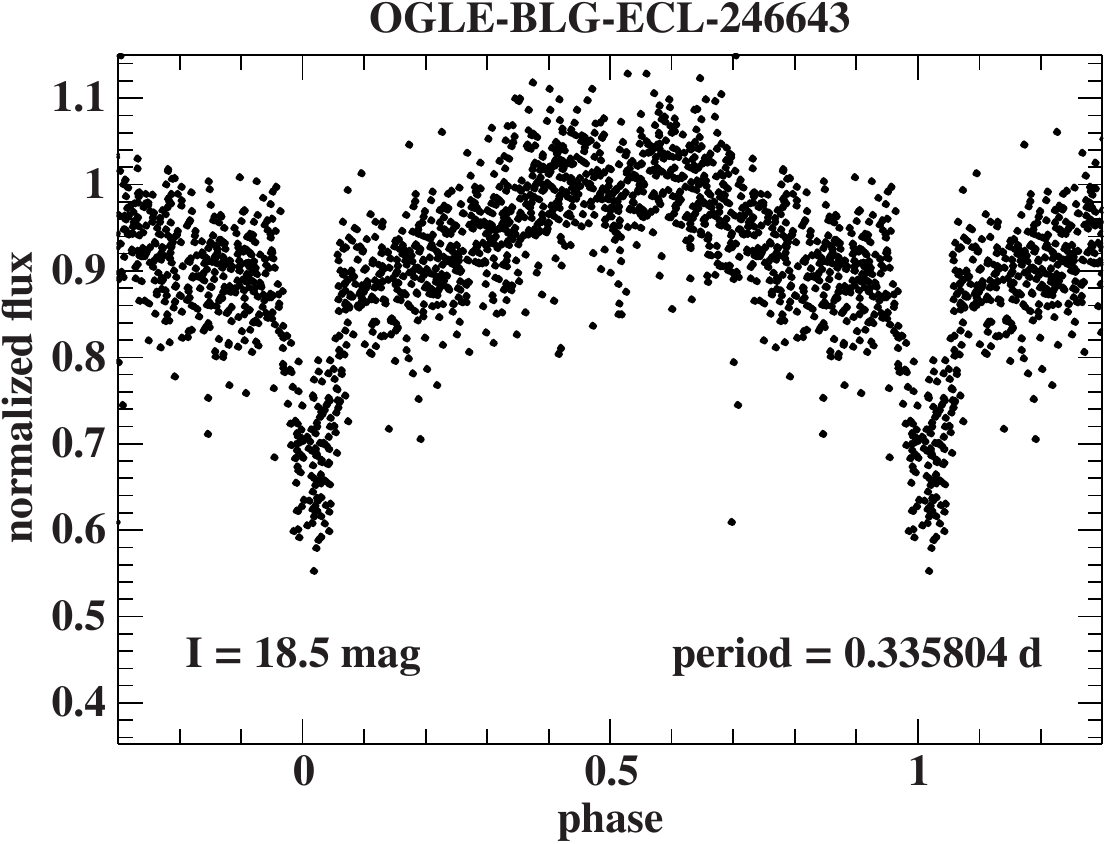}\hfill
		\includegraphics[width=0.25\linewidth]{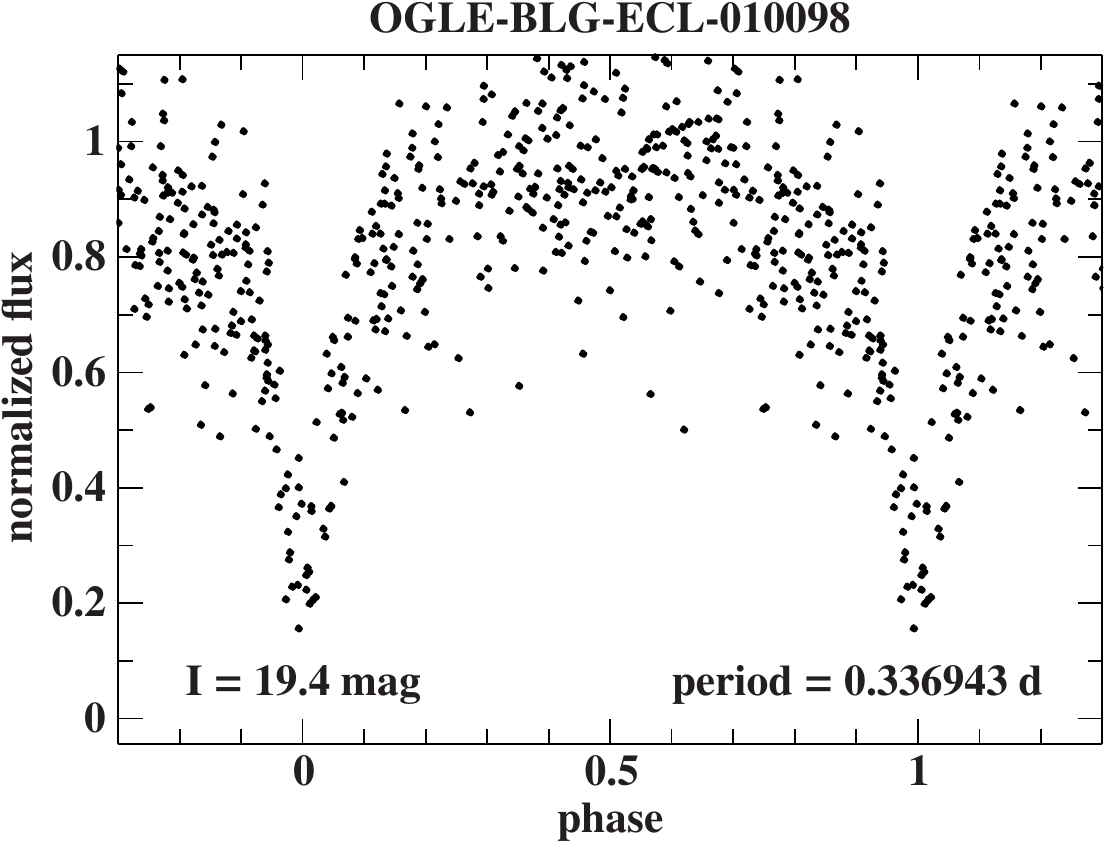}\hfill
		\includegraphics[width=0.25\linewidth]{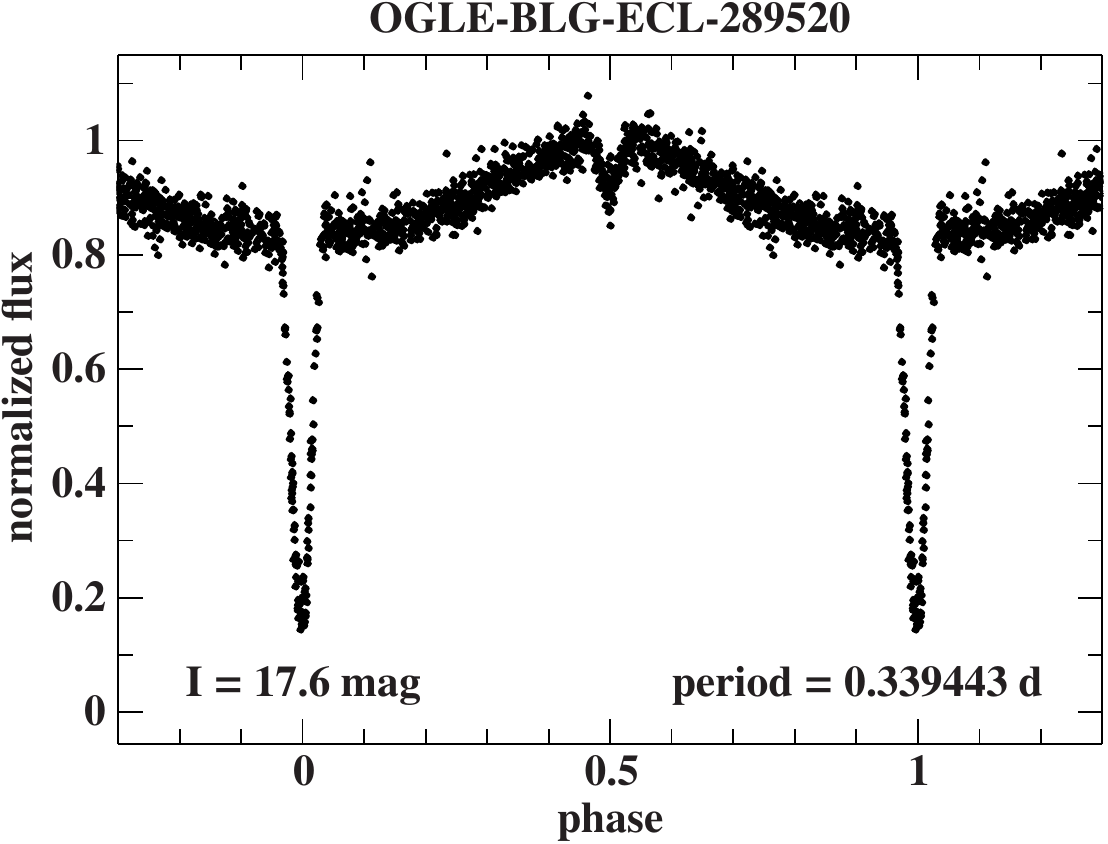}\hfill
		\includegraphics[width=0.25\linewidth]{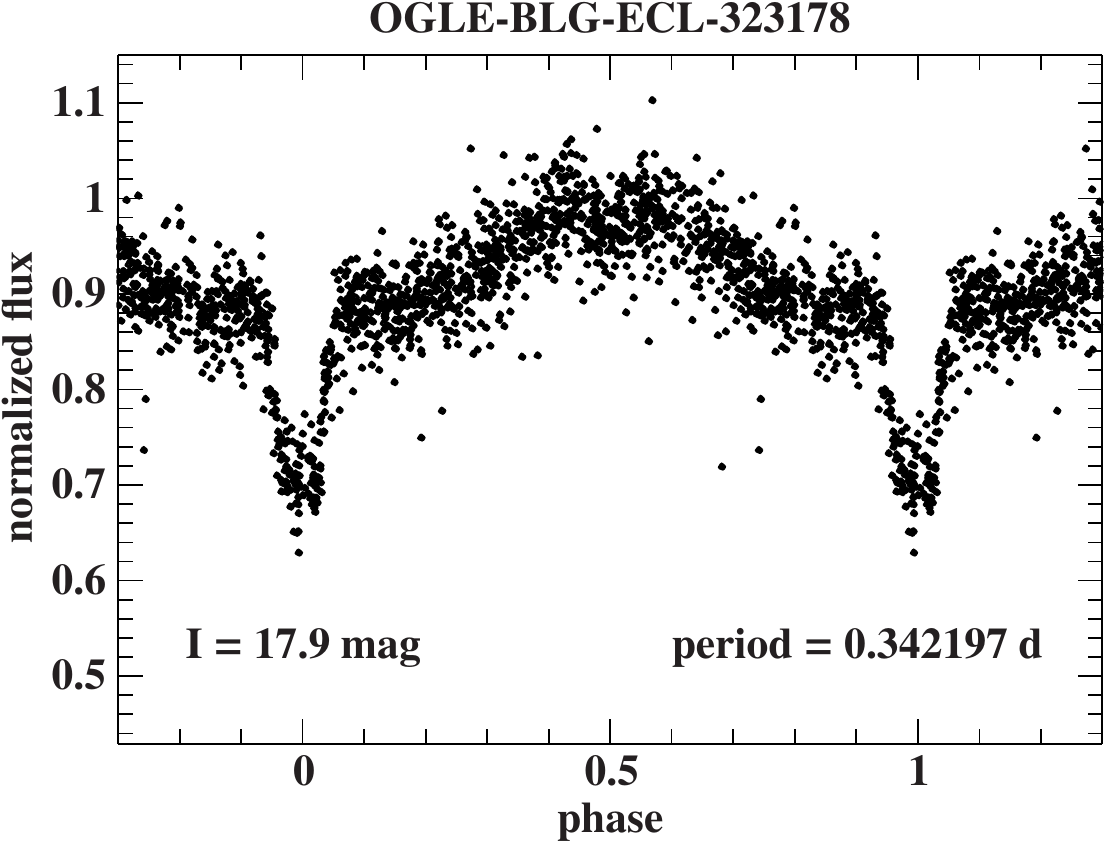}\hfill
		\includegraphics[width=0.25\linewidth]{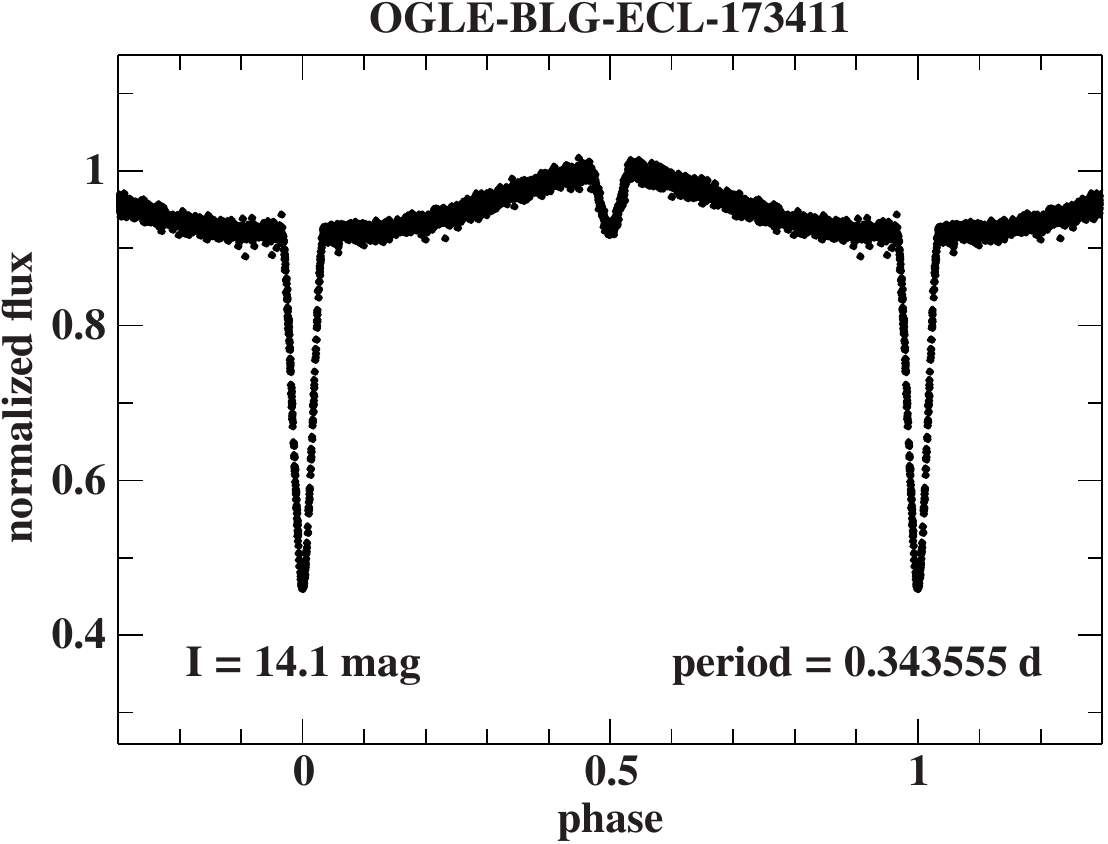}\hfill
		\includegraphics[width=0.25\linewidth]{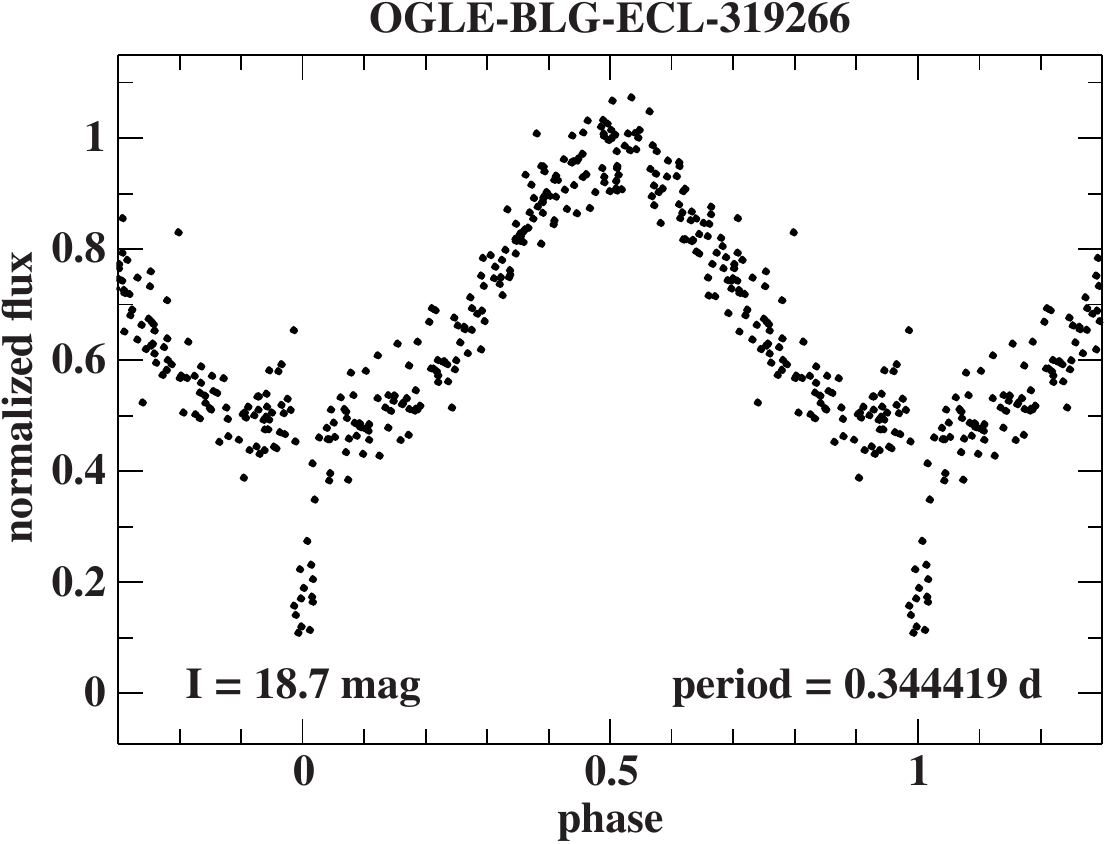}\hfill
		\includegraphics[width=0.25\linewidth]{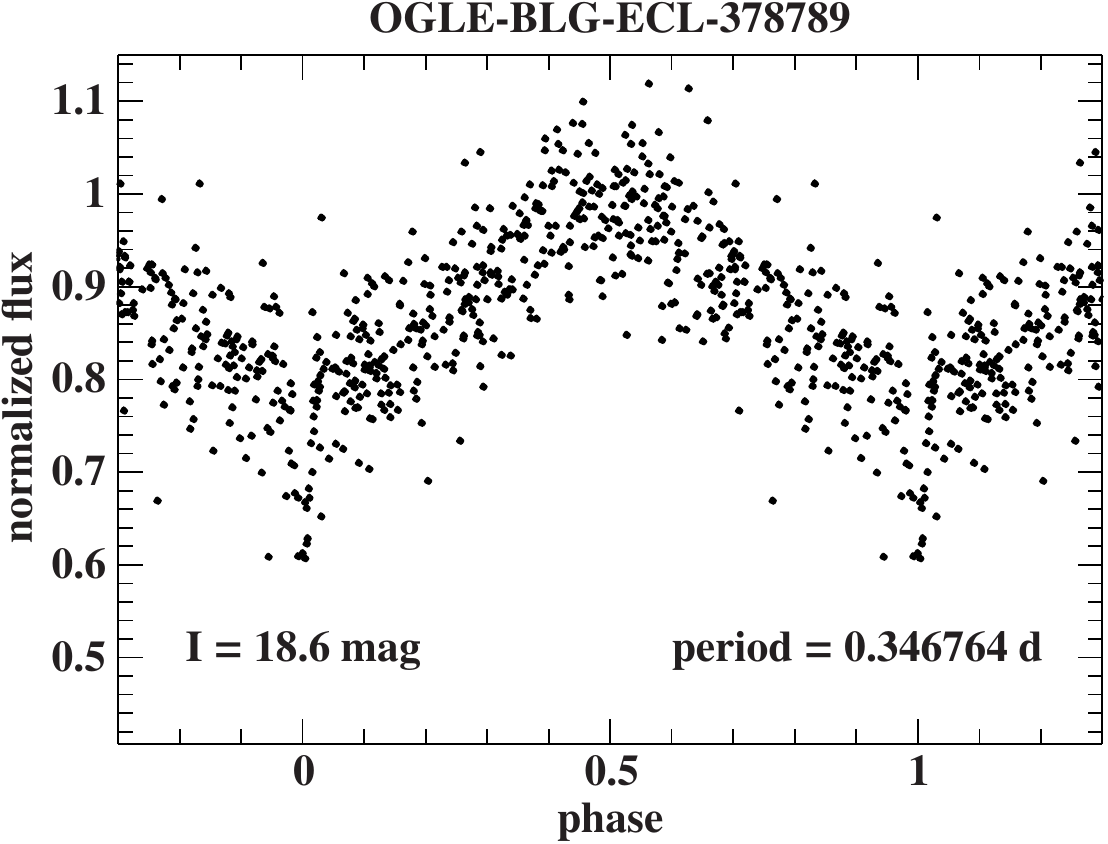}\hfill
		\includegraphics[width=0.25\linewidth]{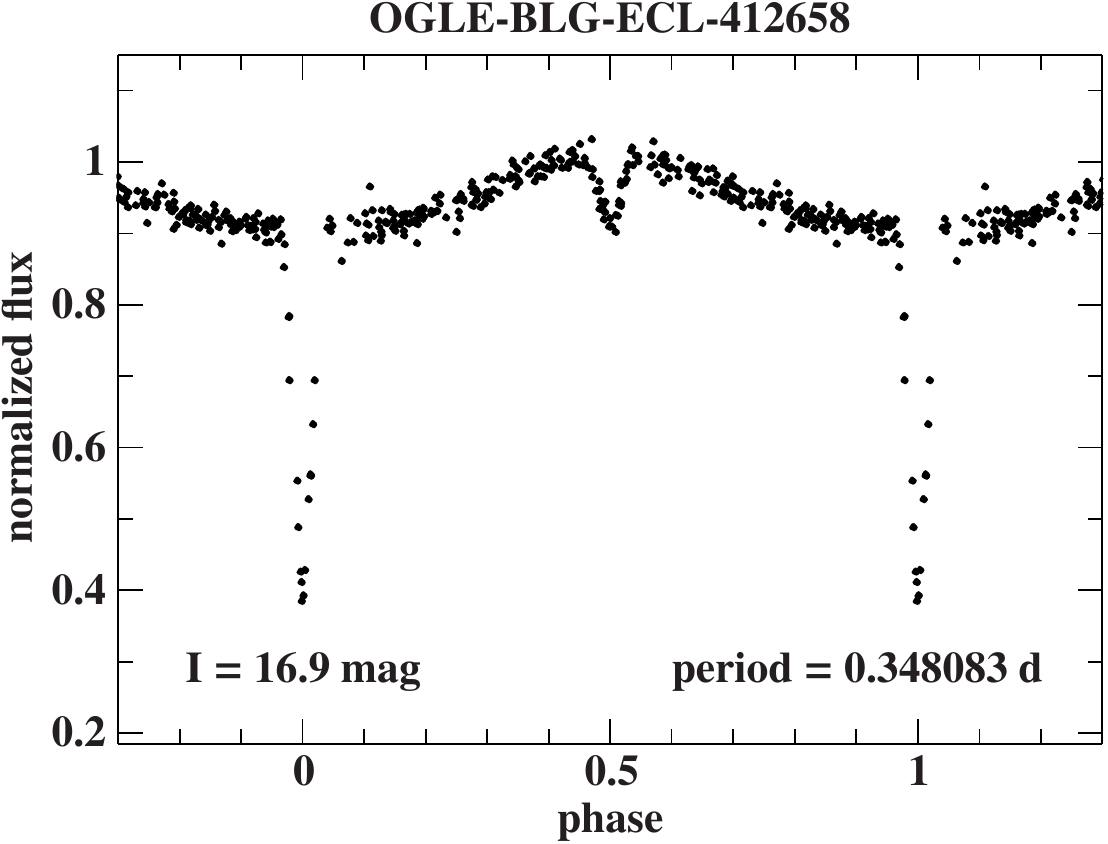}\hfill
		\includegraphics[width=0.25\linewidth]{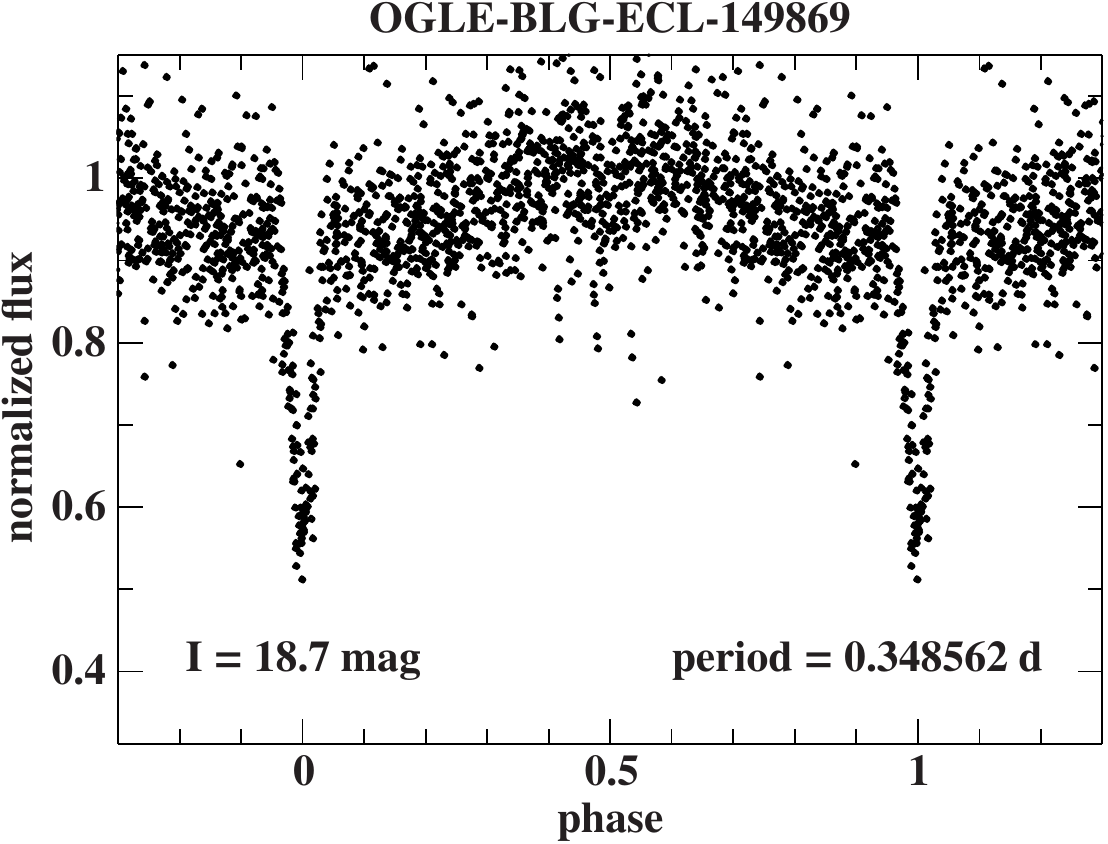}\hfill
		\includegraphics[width=0.25\linewidth]{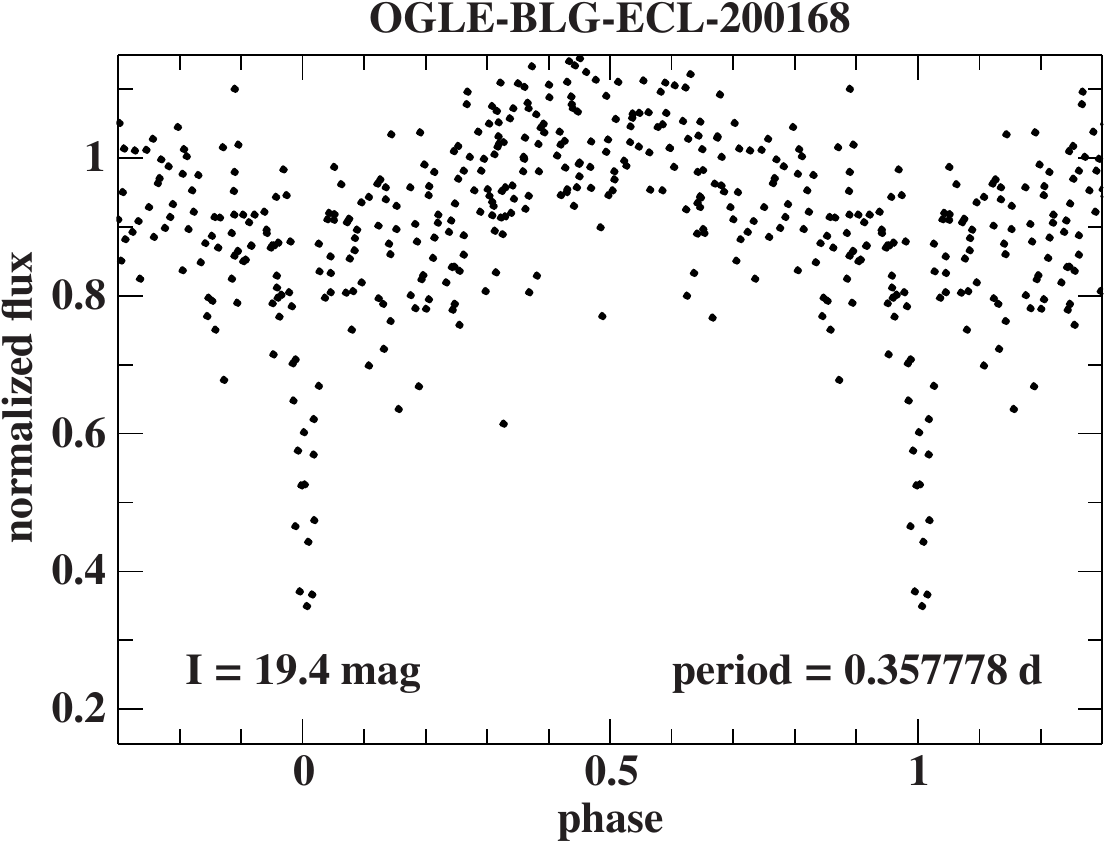}\hfill
		\includegraphics[width=0.25\linewidth]{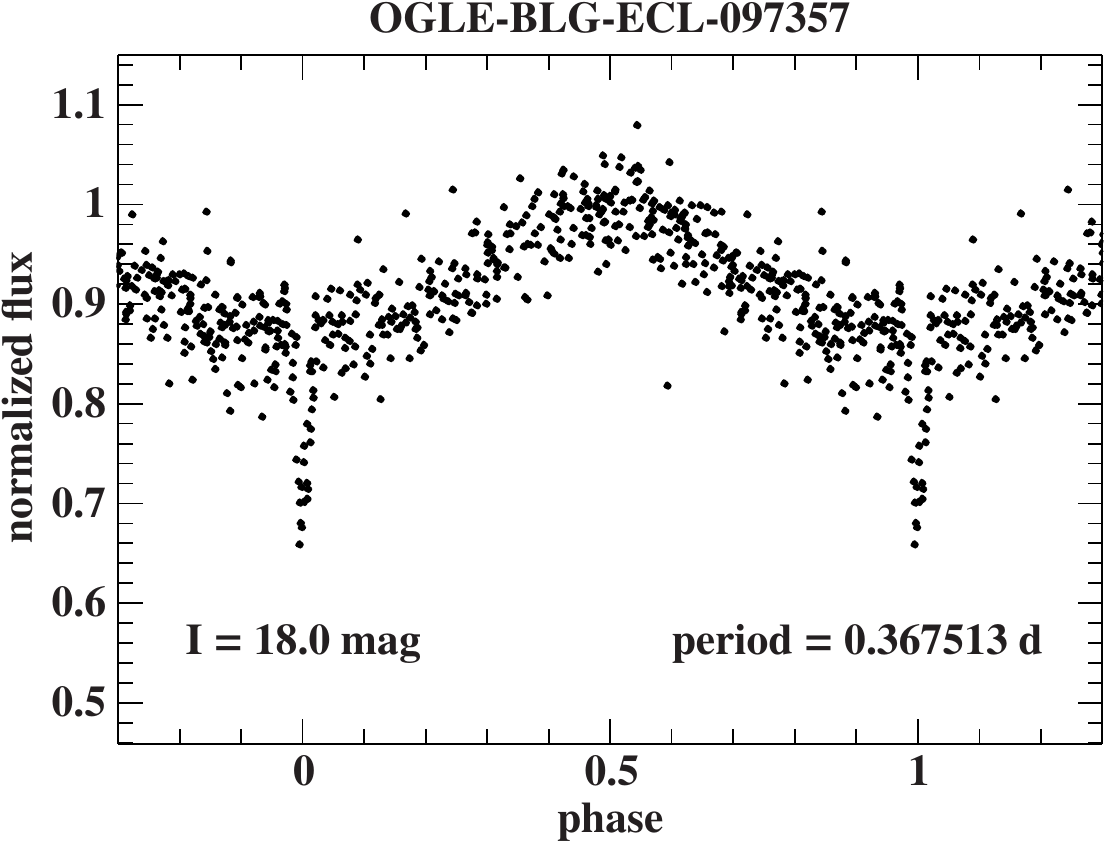}\hfill
		\includegraphics[width=0.25\linewidth]{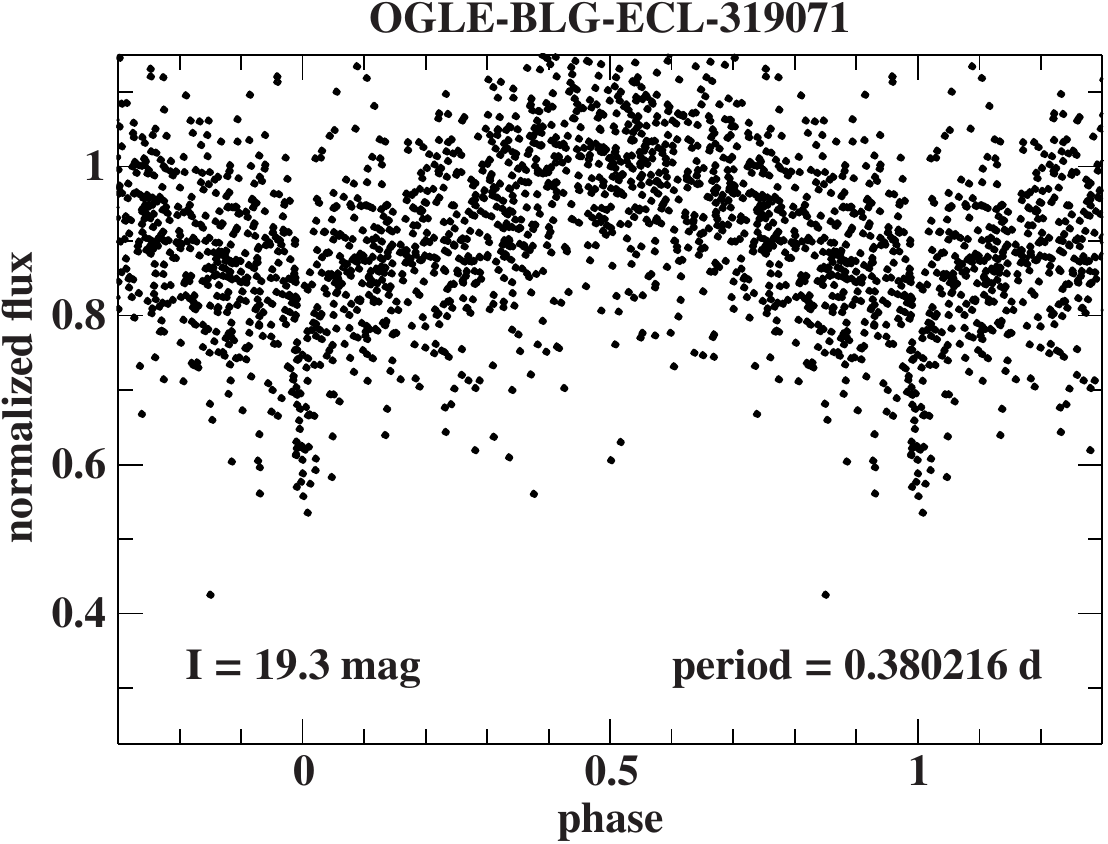}\hfill
		\includegraphics[width=0.25\linewidth]{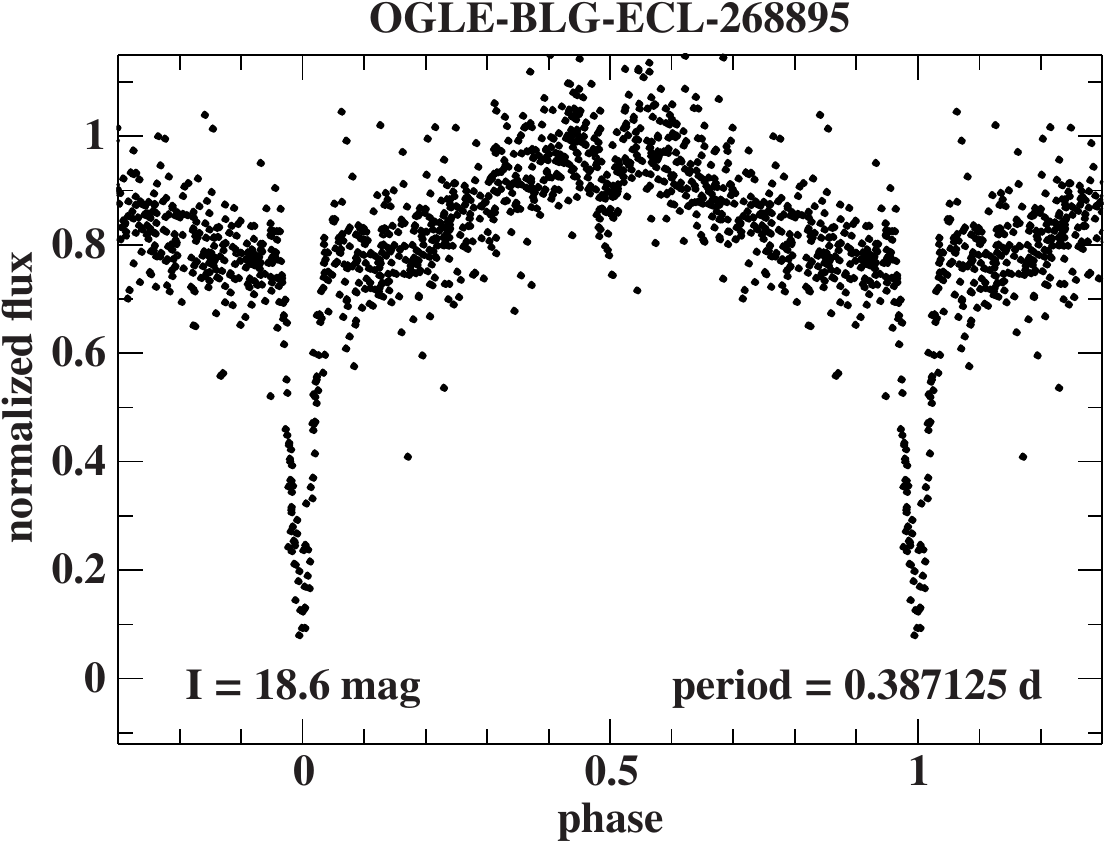}\hfill
		\includegraphics[width=0.25\linewidth]{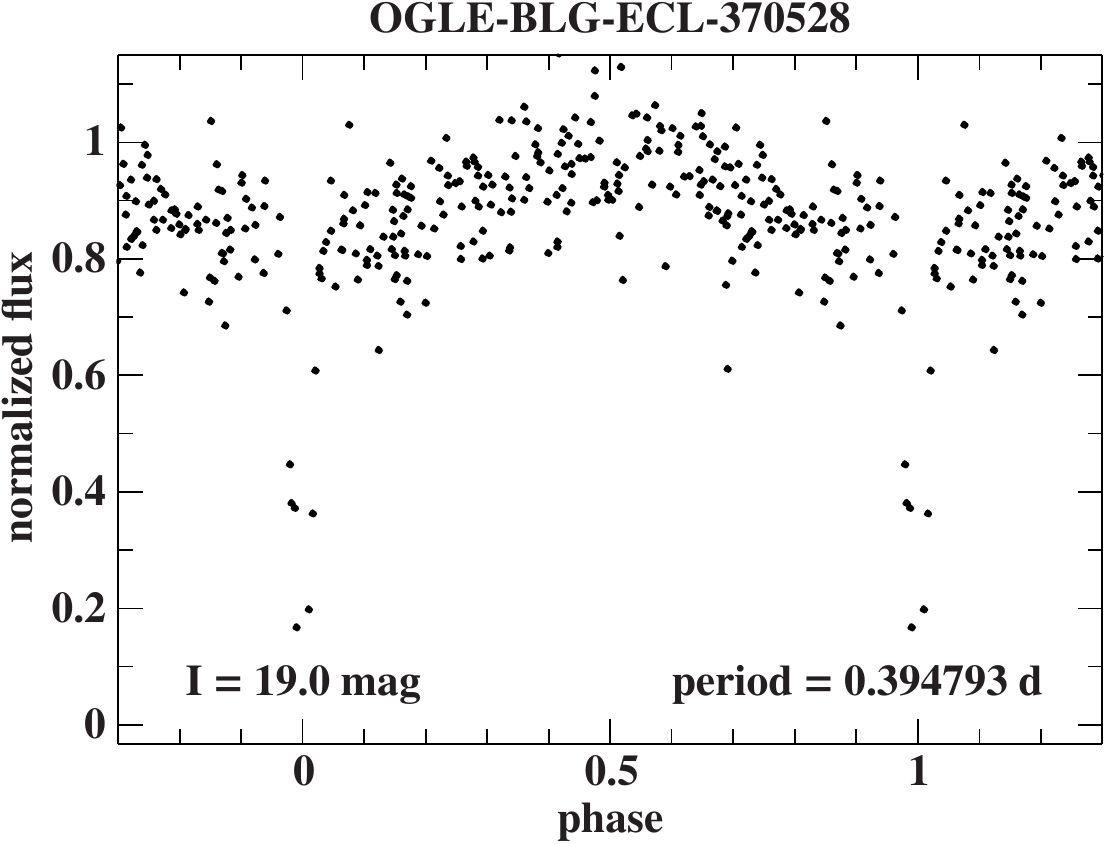}\hfill
		\includegraphics[width=0.25\linewidth]{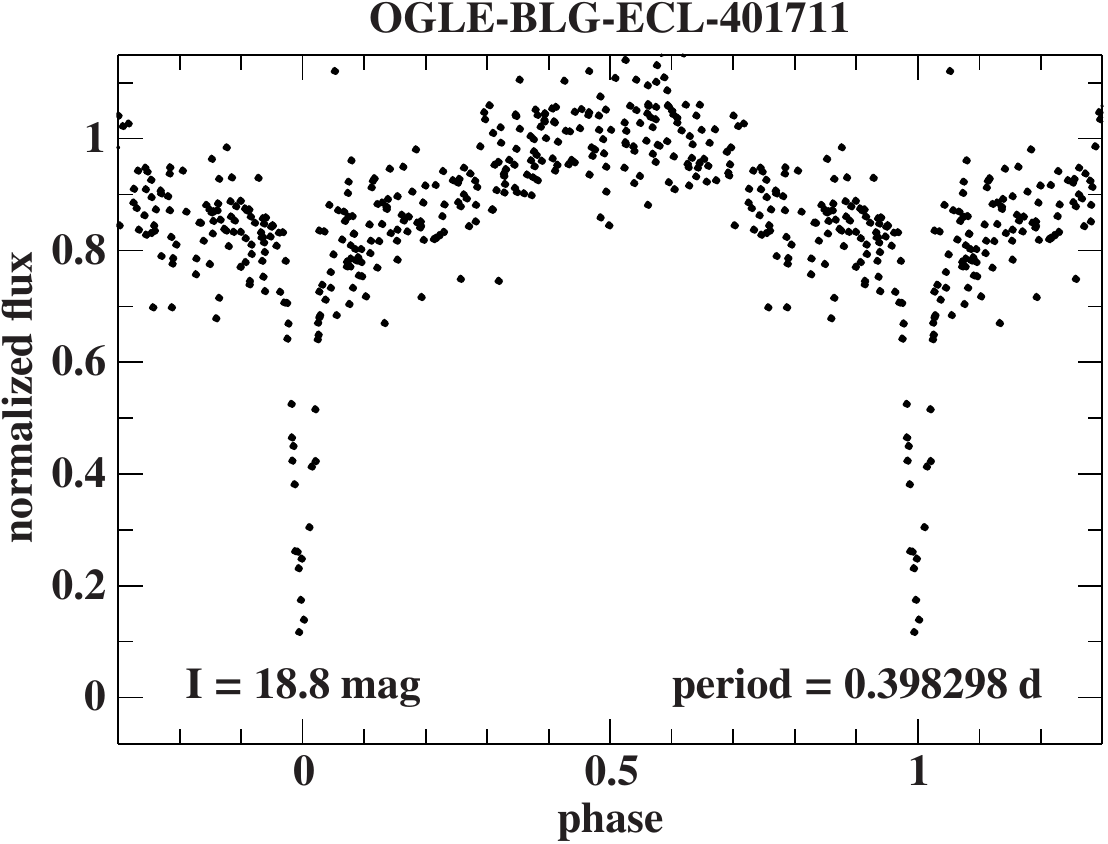}\hfill
		\includegraphics[width=0.25\linewidth]{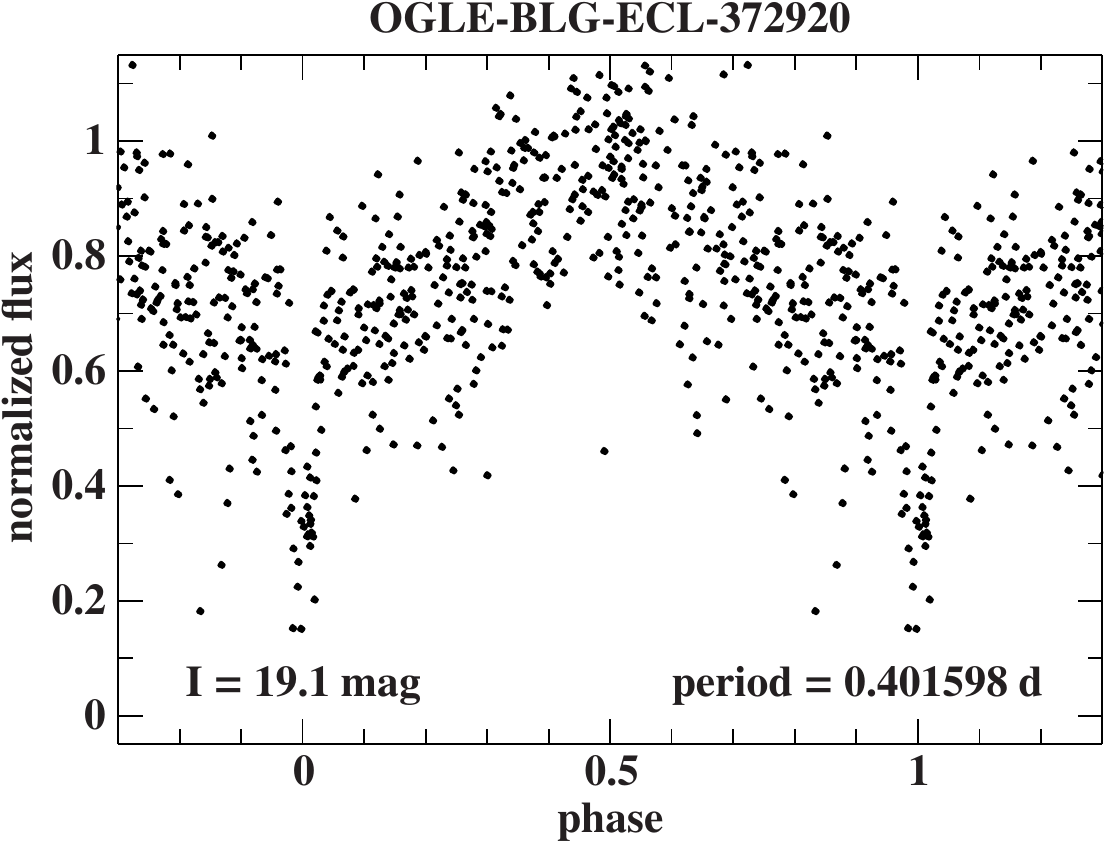}\hfill
		\includegraphics[width=0.25\linewidth]{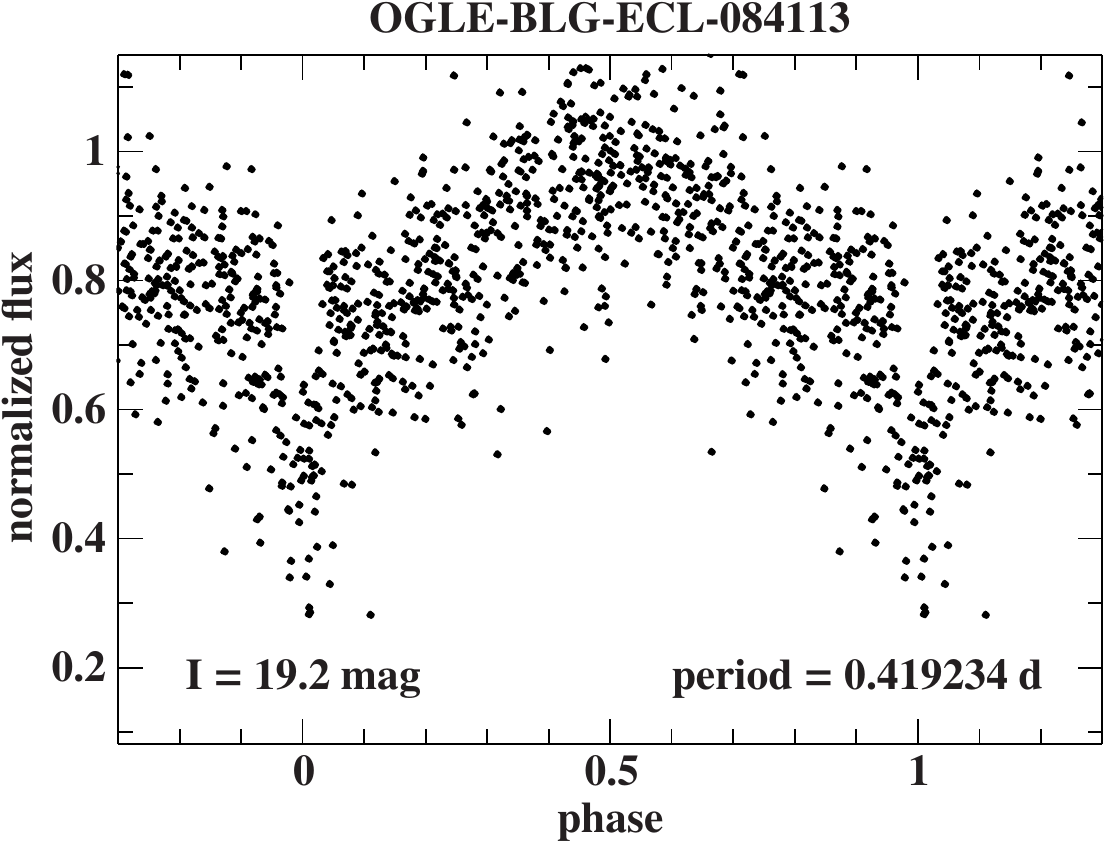}\hfill
	\end{figure}
	\begin{figure}
		\includegraphics[width=0.25\linewidth]{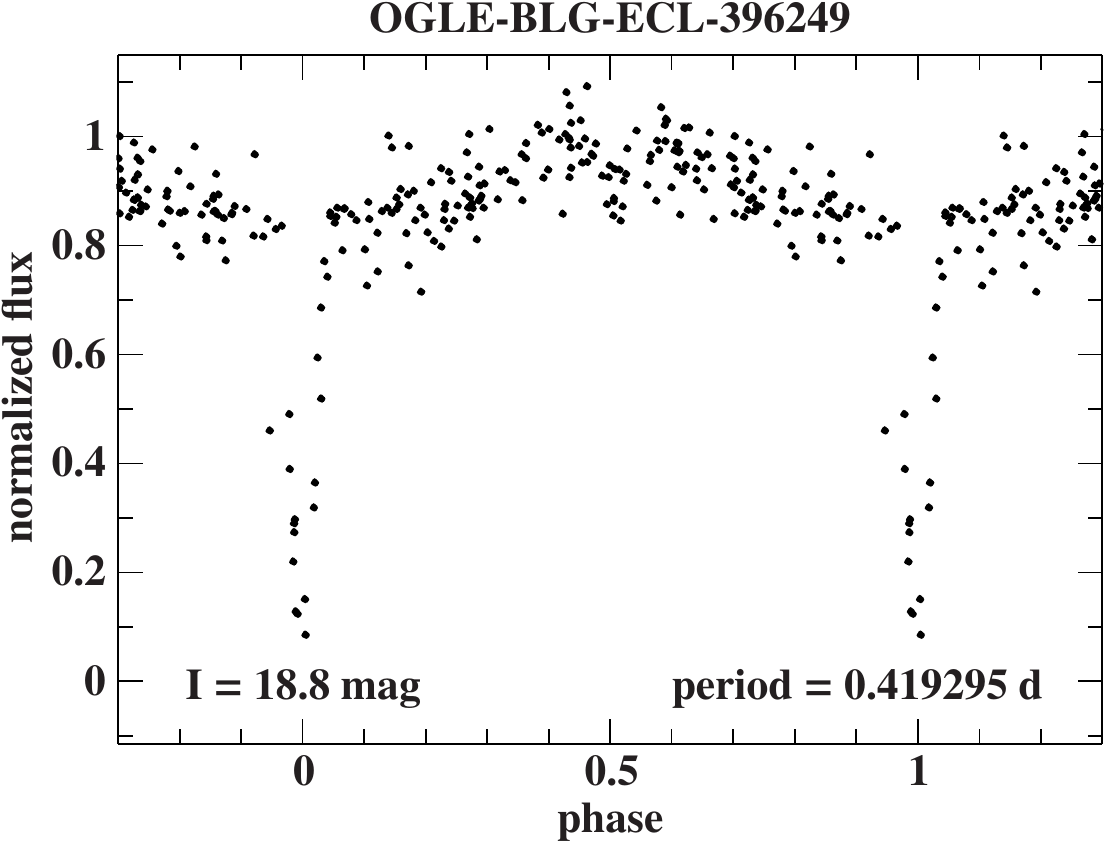}\hfill
		\includegraphics[width=0.25\linewidth]{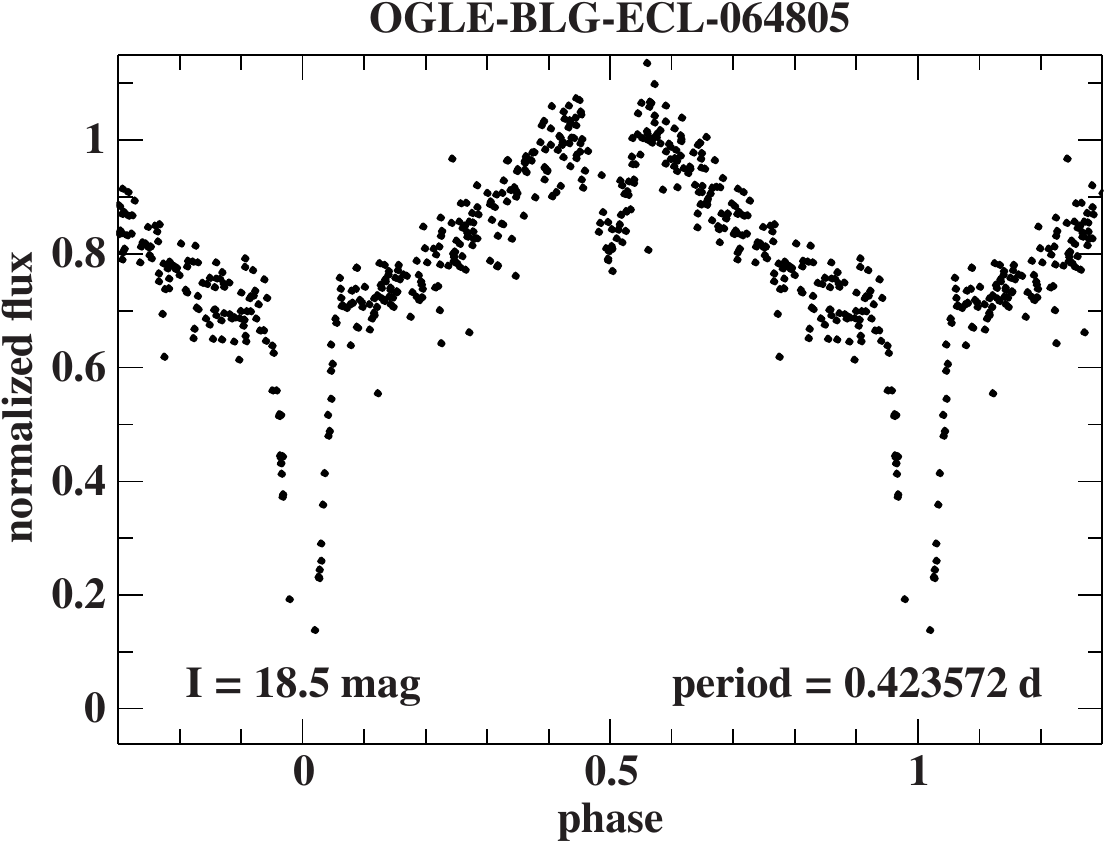}\hfill
		\includegraphics[width=0.25\linewidth]{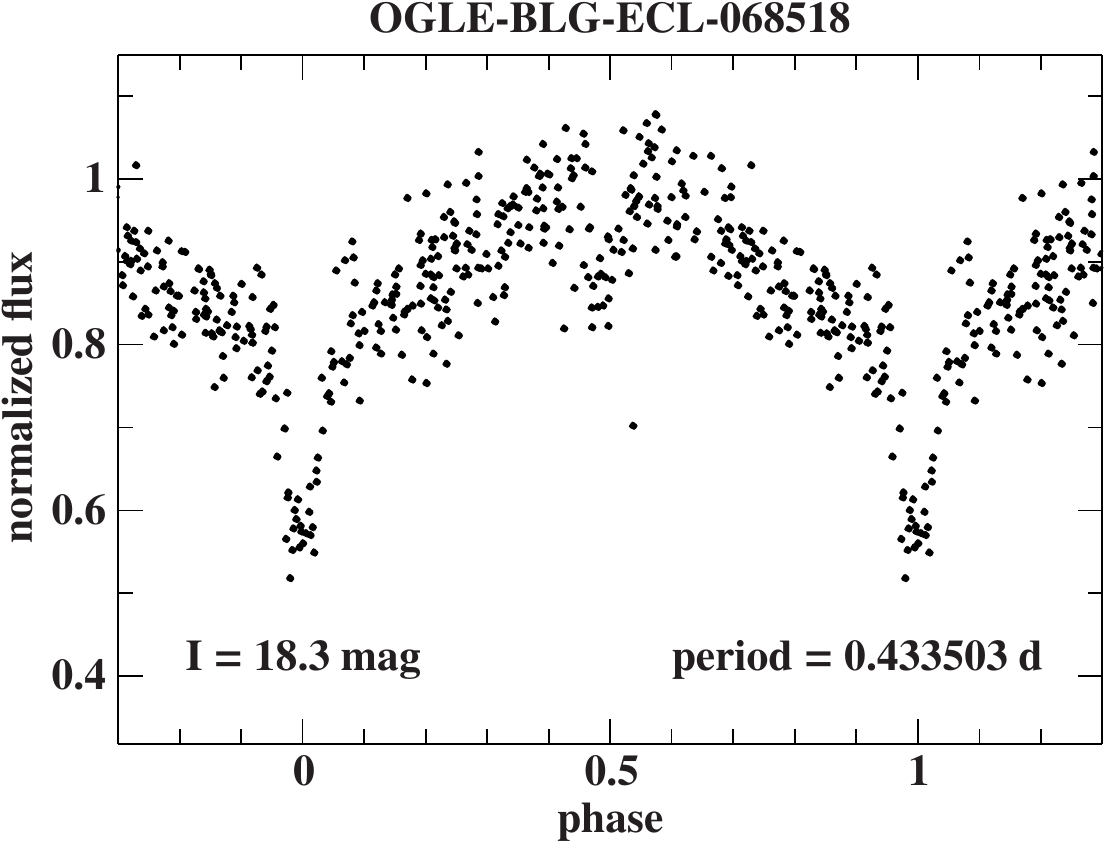}\hfill
		\includegraphics[width=0.25\linewidth]{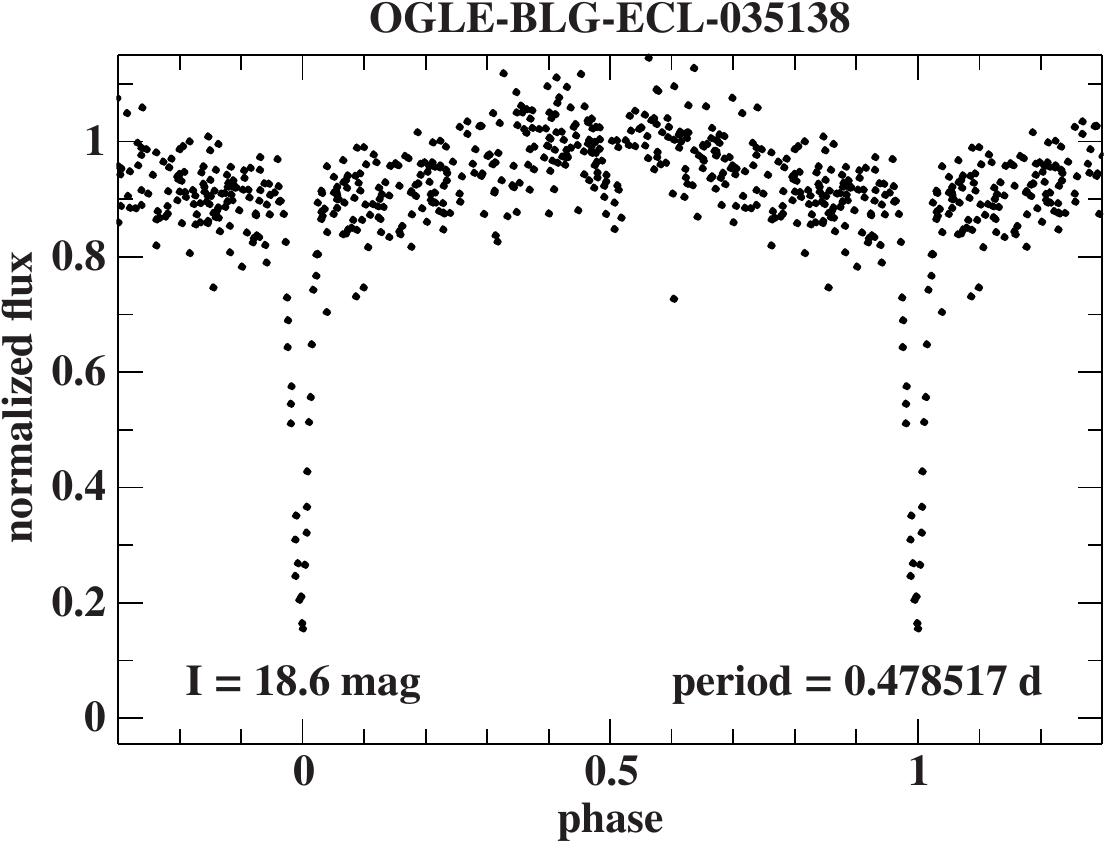}\hfill
		\includegraphics[width=0.25\linewidth]{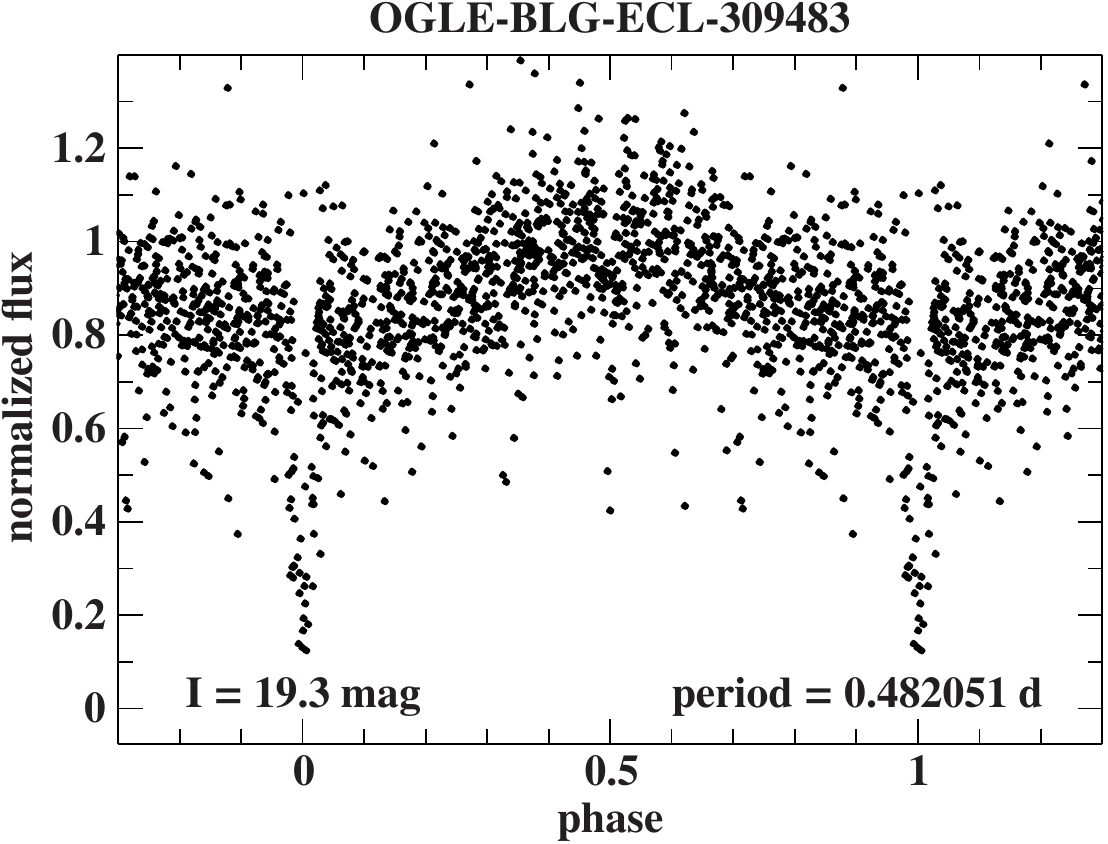}\hfill
		\includegraphics[width=0.25\linewidth]{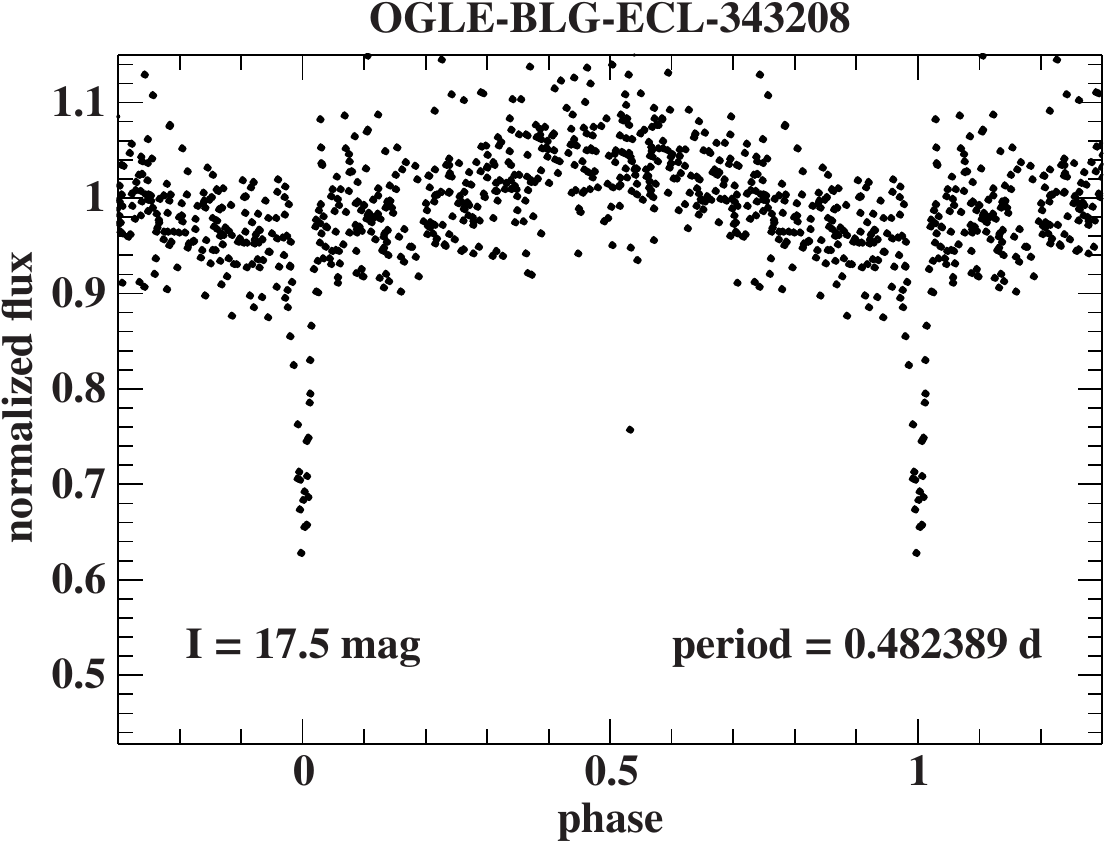}\hfill
		\includegraphics[width=0.25\linewidth]{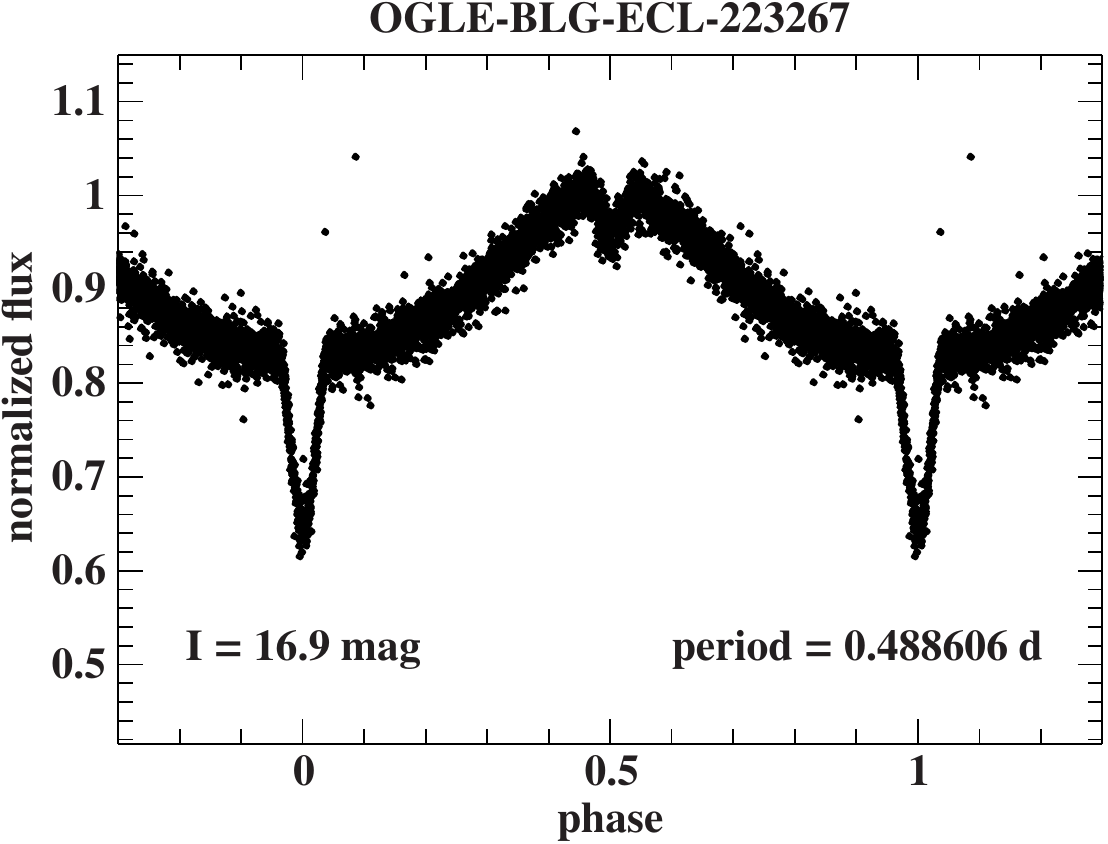}\hfill
		\includegraphics[width=0.25\linewidth]{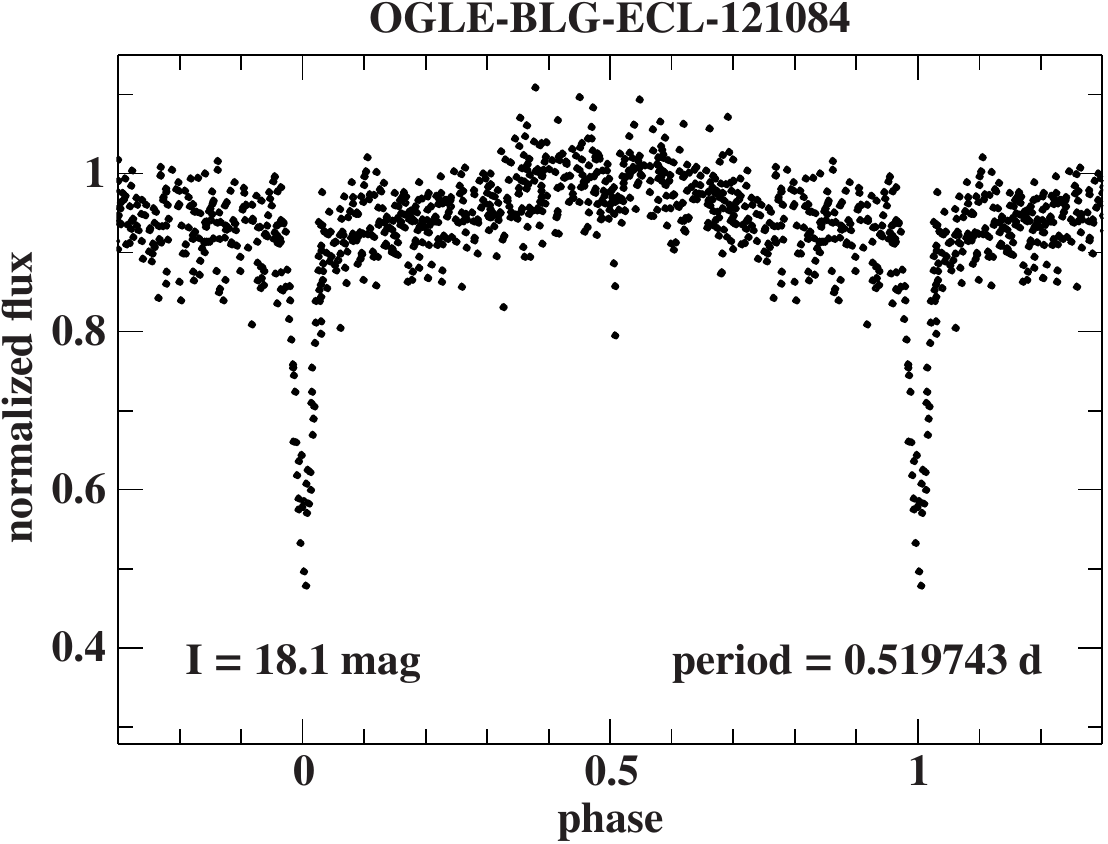}\hfill
		\includegraphics[width=0.25\linewidth]{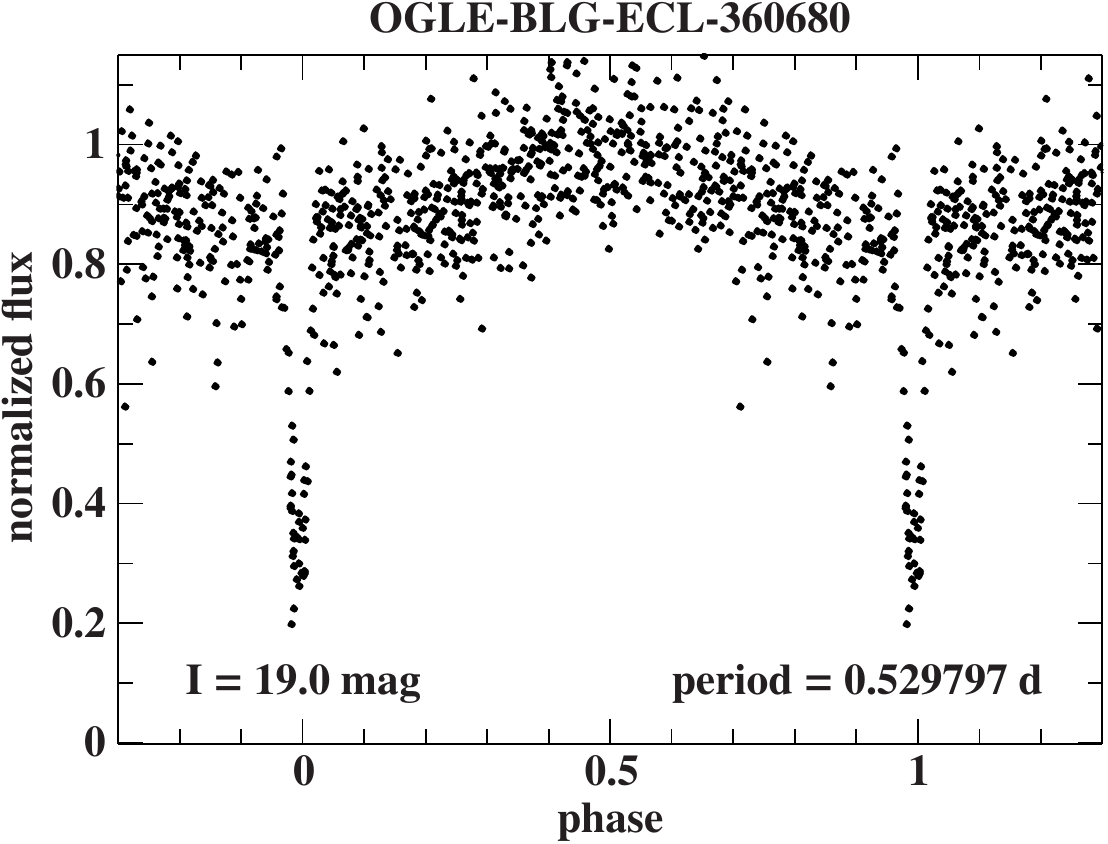}\hfill
		\includegraphics[width=0.25\linewidth]{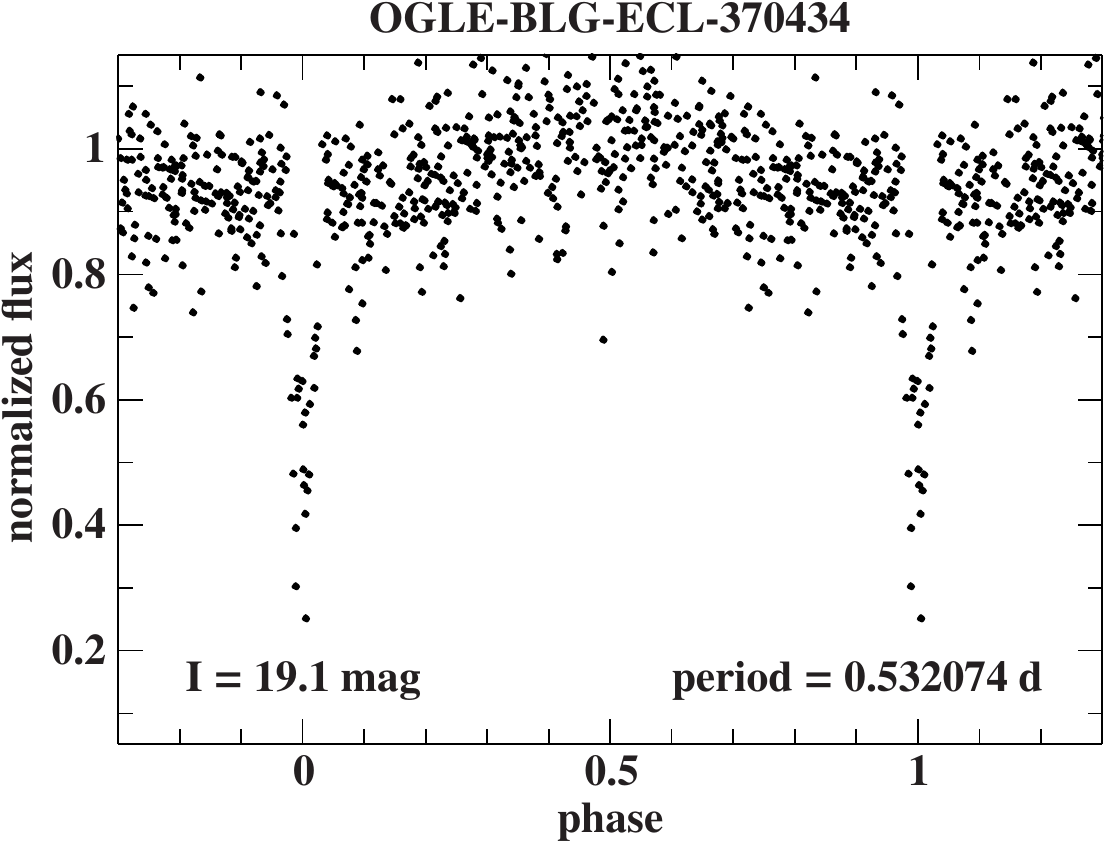}\hfill
		\includegraphics[width=0.25\linewidth]{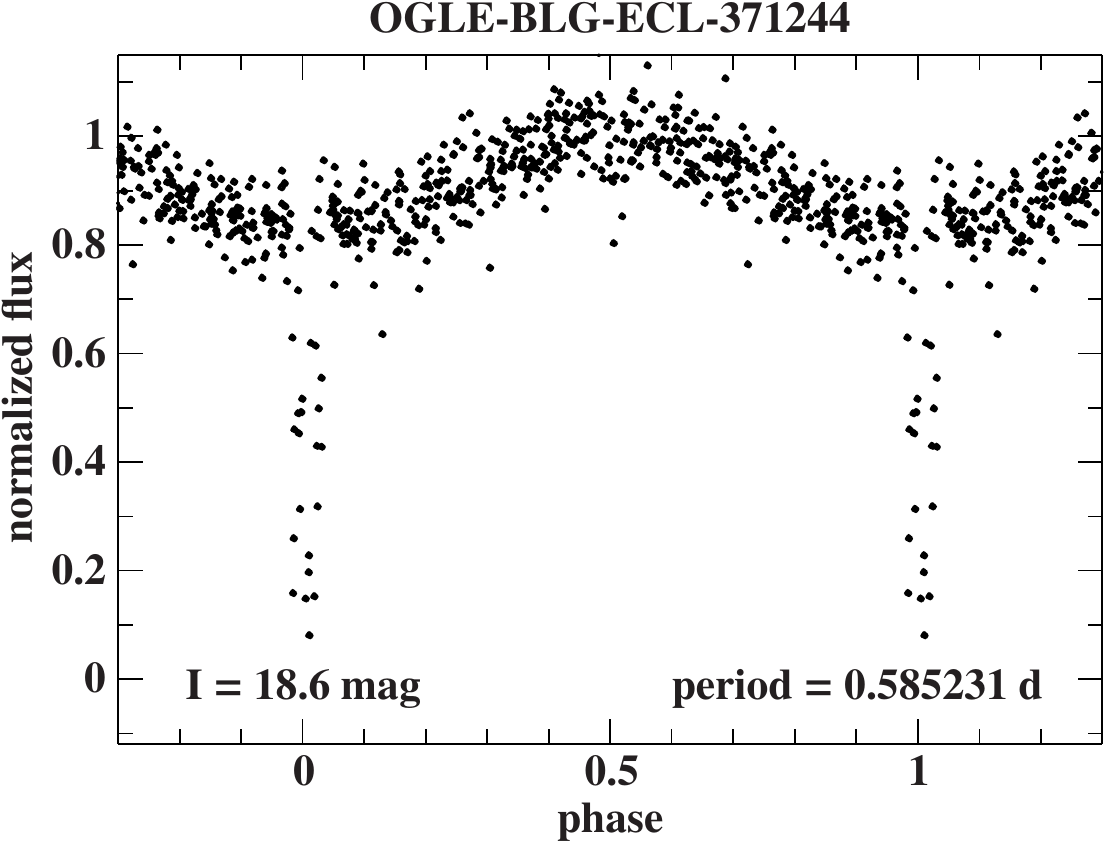}\hfill
		\includegraphics[width=0.25\linewidth]{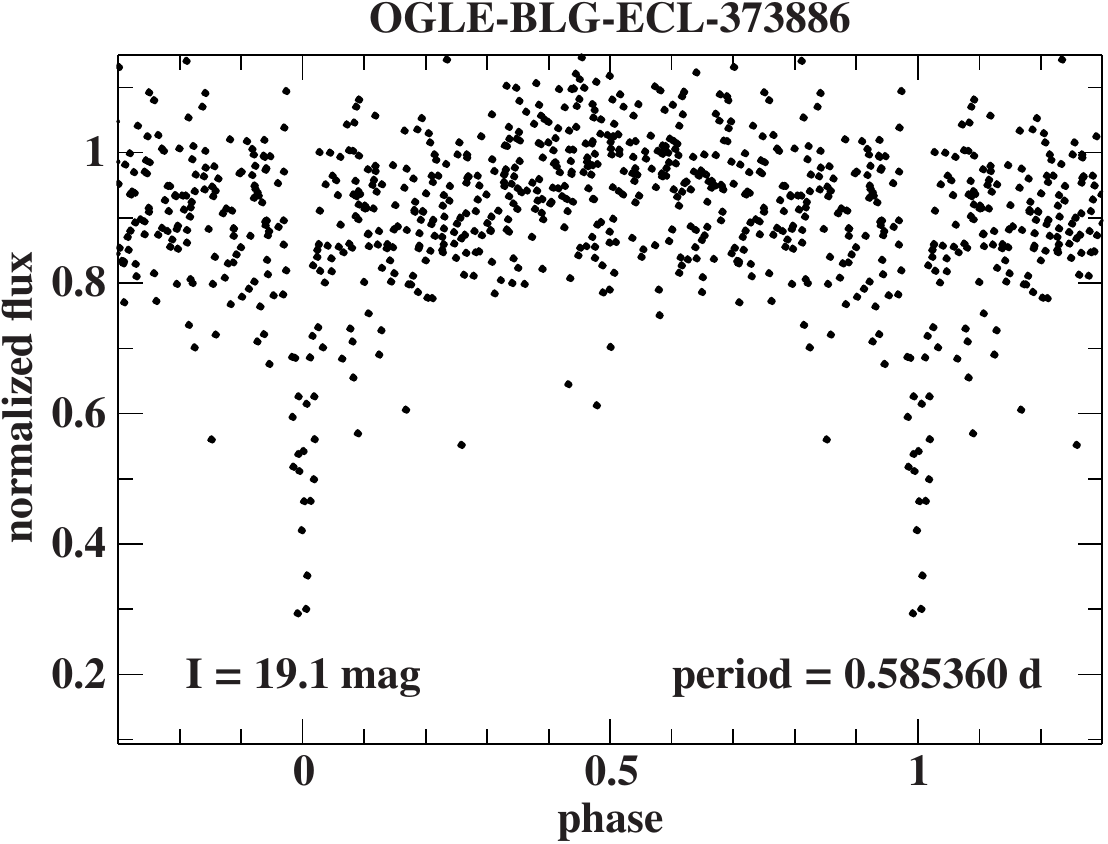}\hfill
		\includegraphics[width=0.25\linewidth]{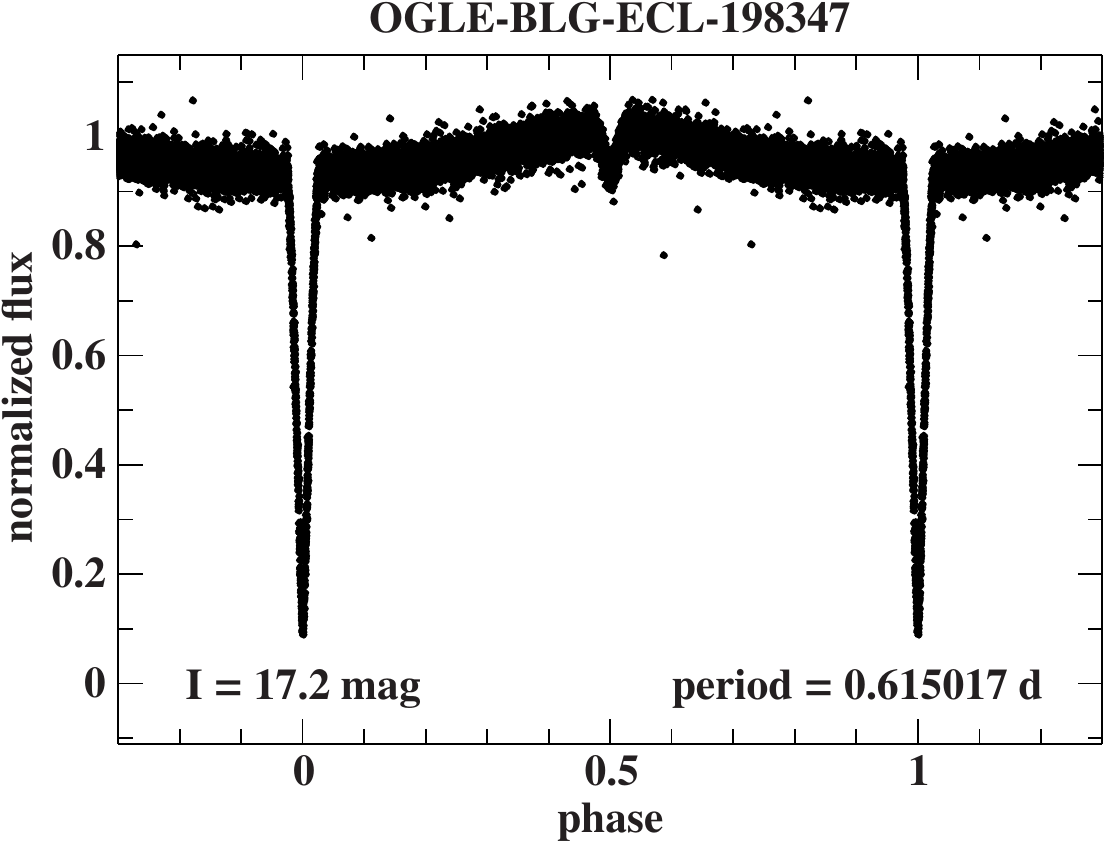}\hfill
		\includegraphics[width=0.25\linewidth]{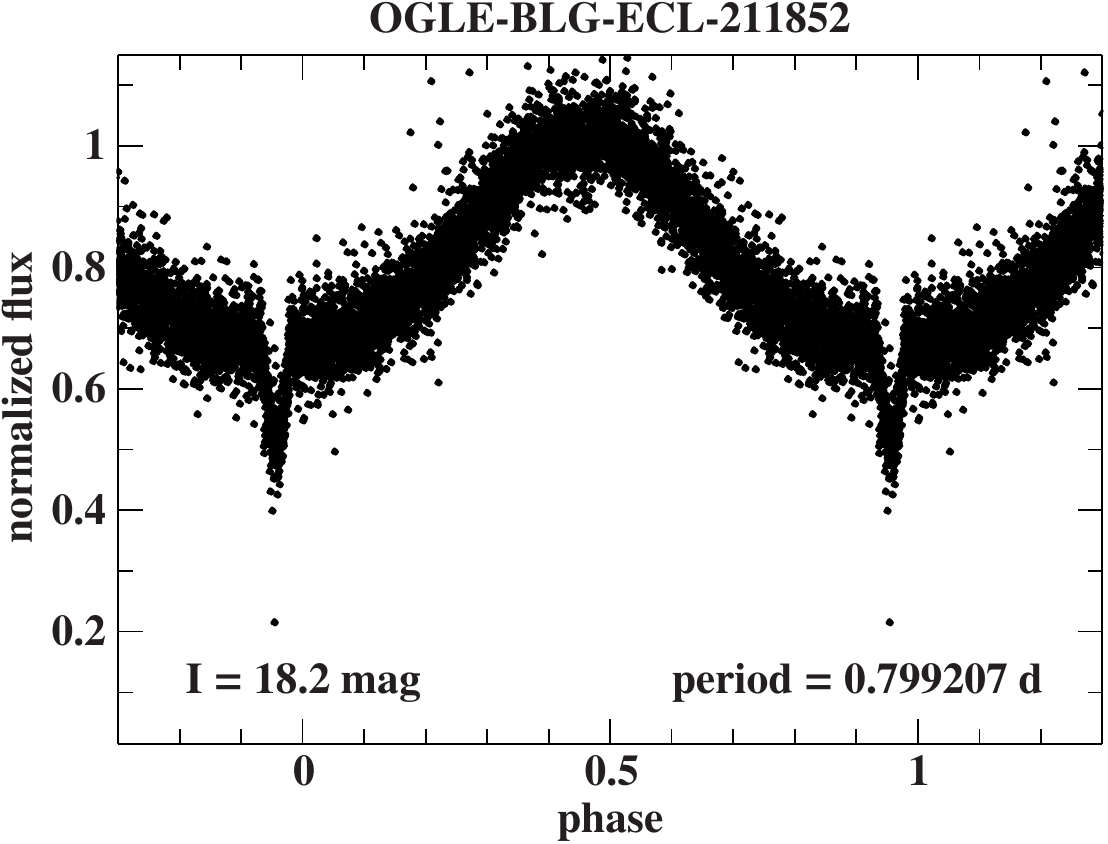}\hfill	
		\includegraphics[width=0.25\linewidth]{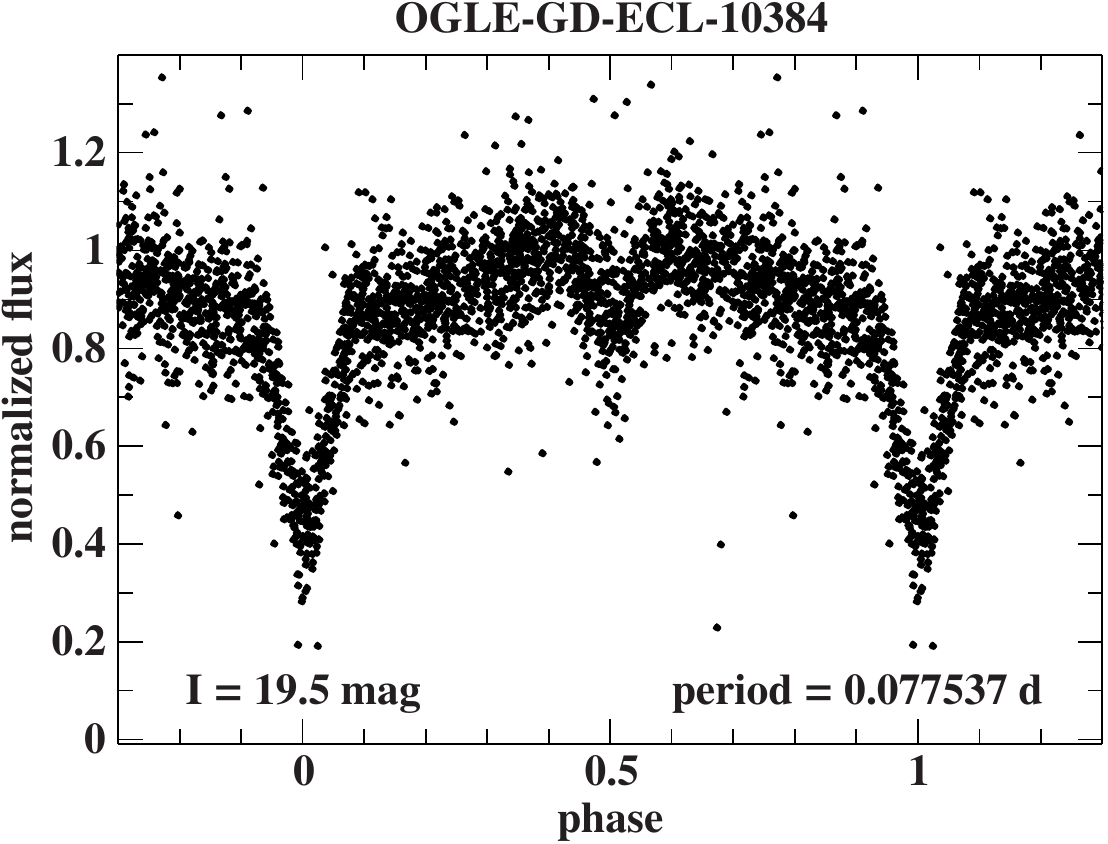}\hfill
		\includegraphics[width=0.25\linewidth]{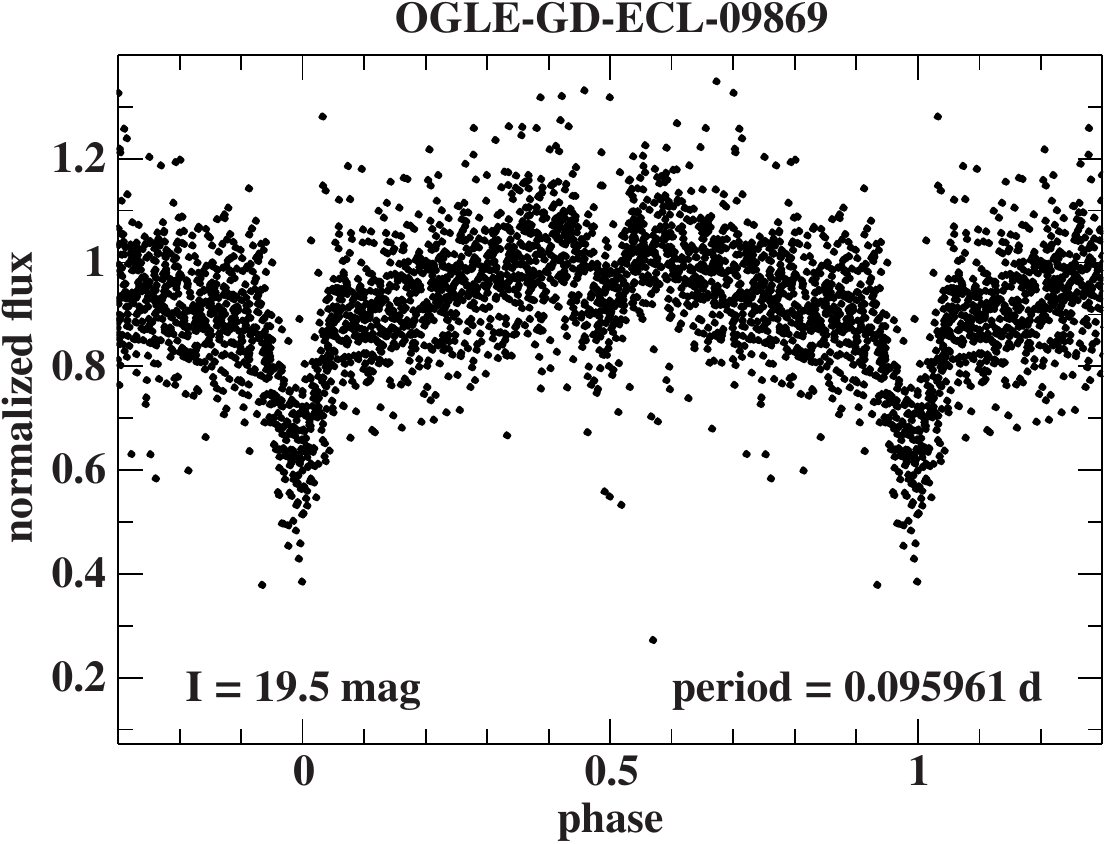}\hfill
		\includegraphics[width=0.25\linewidth]{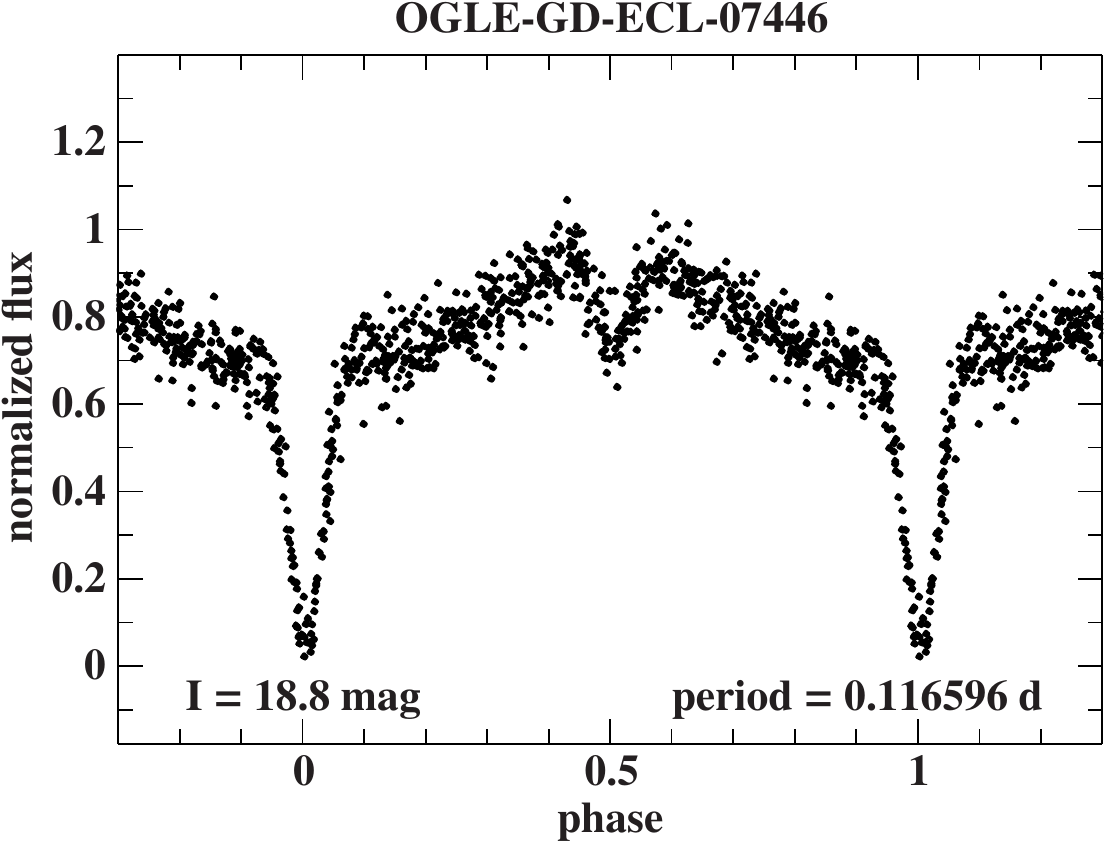}\hfill
		\includegraphics[width=0.25\linewidth]{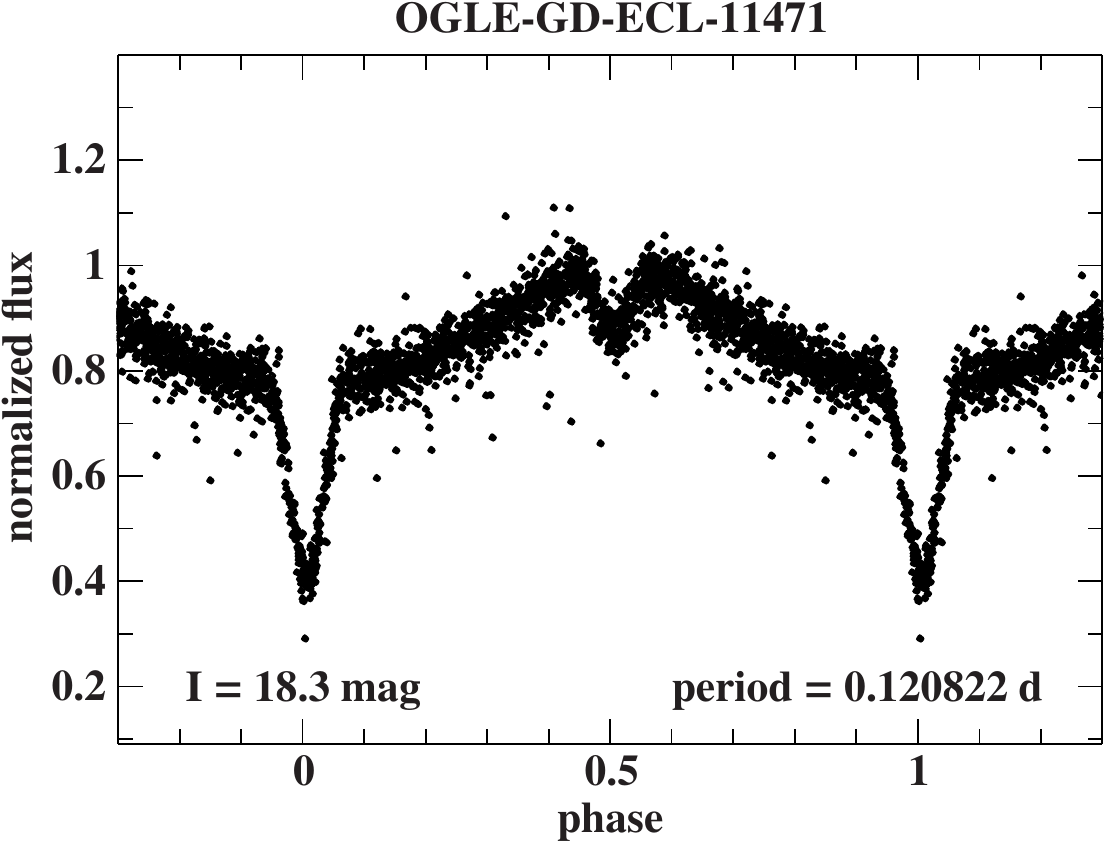}\hfill
		\includegraphics[width=0.25\linewidth]{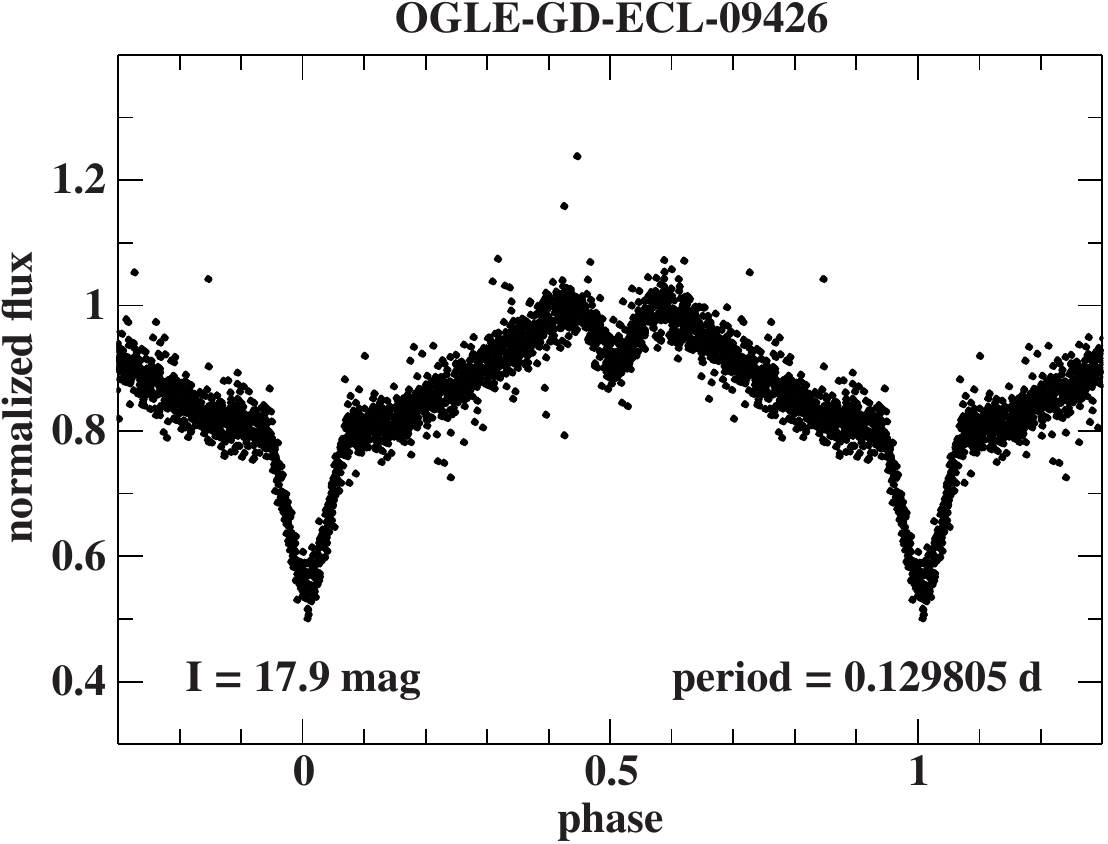}\hfill
		\includegraphics[width=0.25\linewidth]{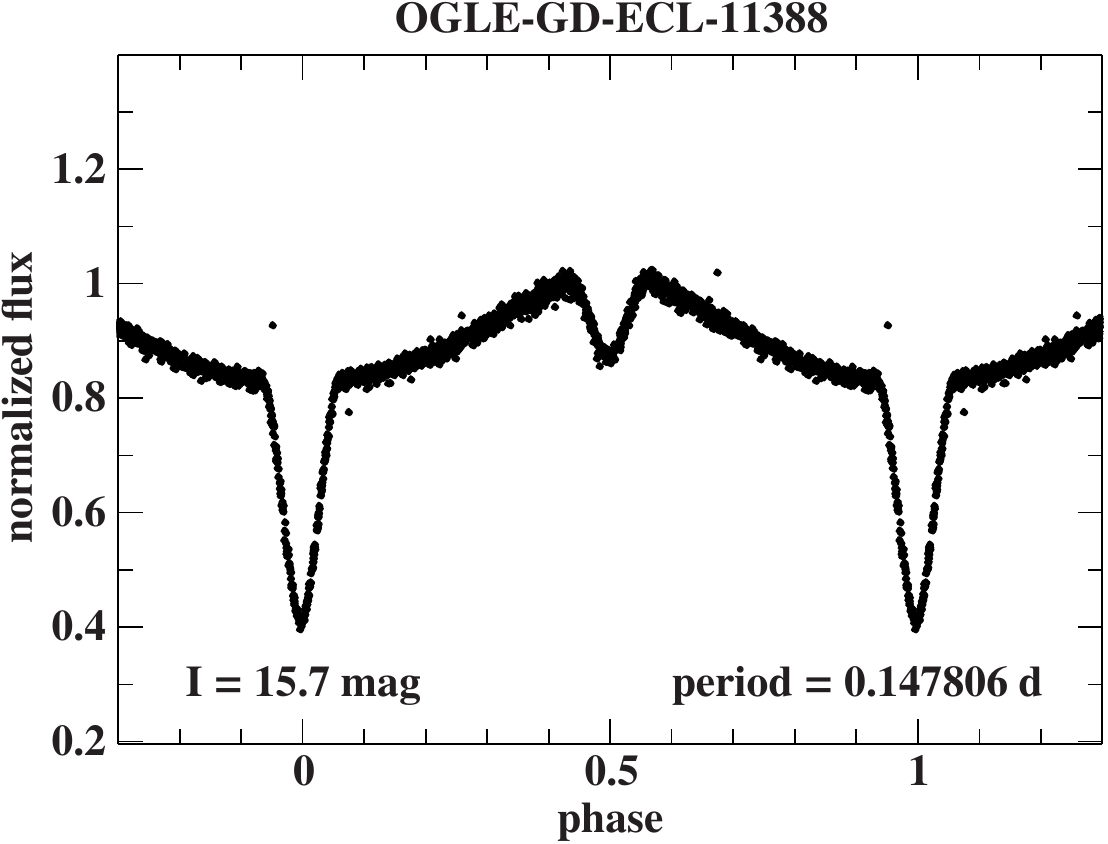}\hfill
		\includegraphics[width=0.25\linewidth]{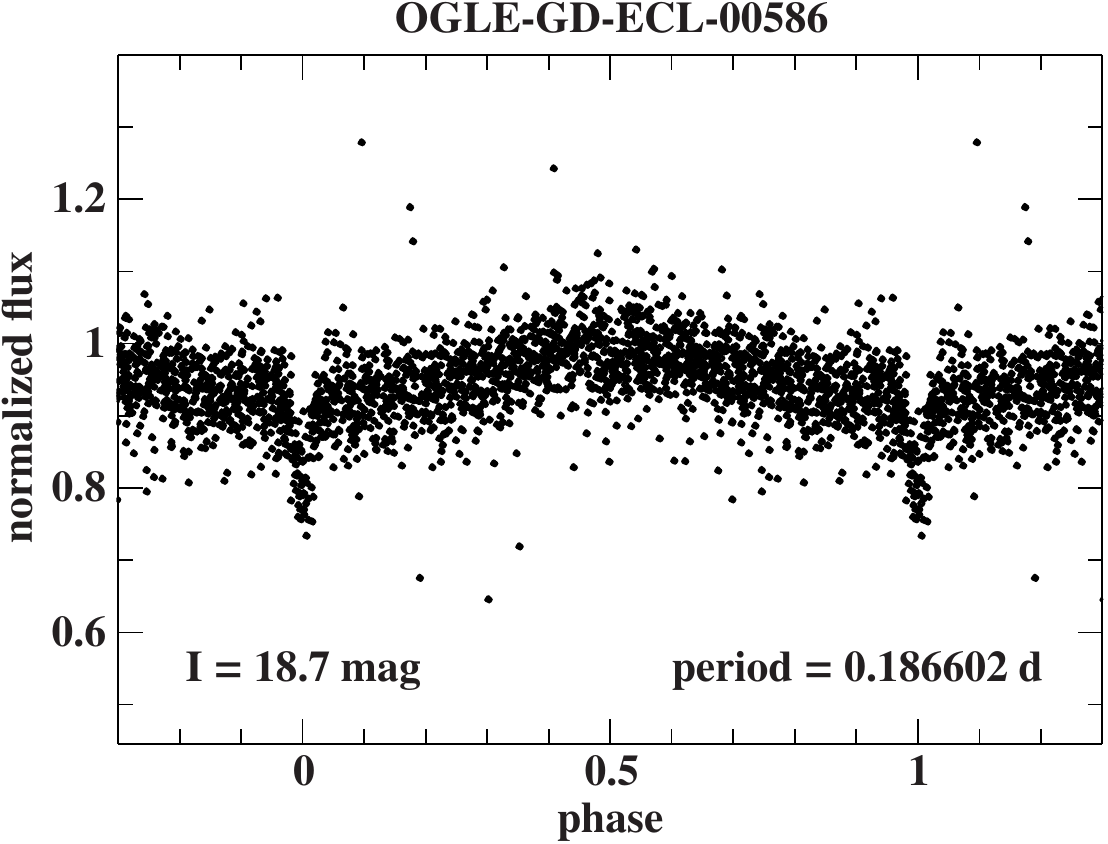}\hfill
		\includegraphics[width=0.25\linewidth]{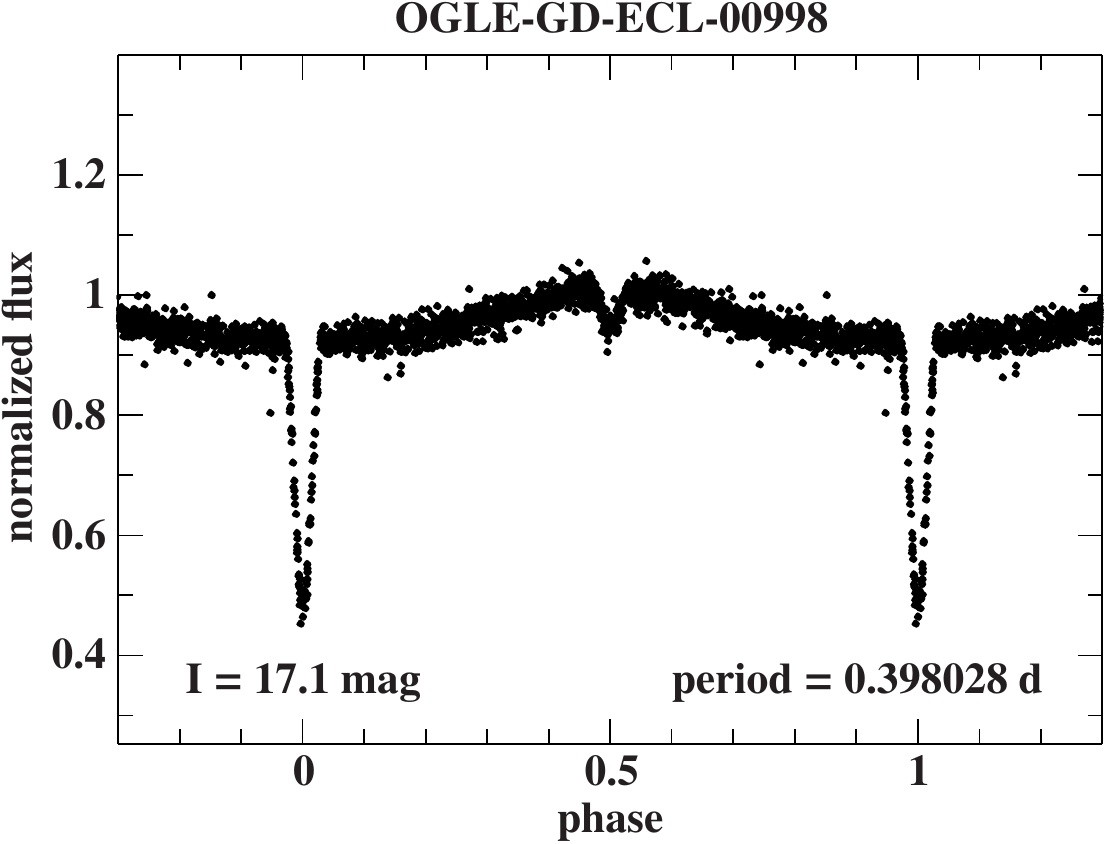}\hfill
		\includegraphics[width=0.25\linewidth]{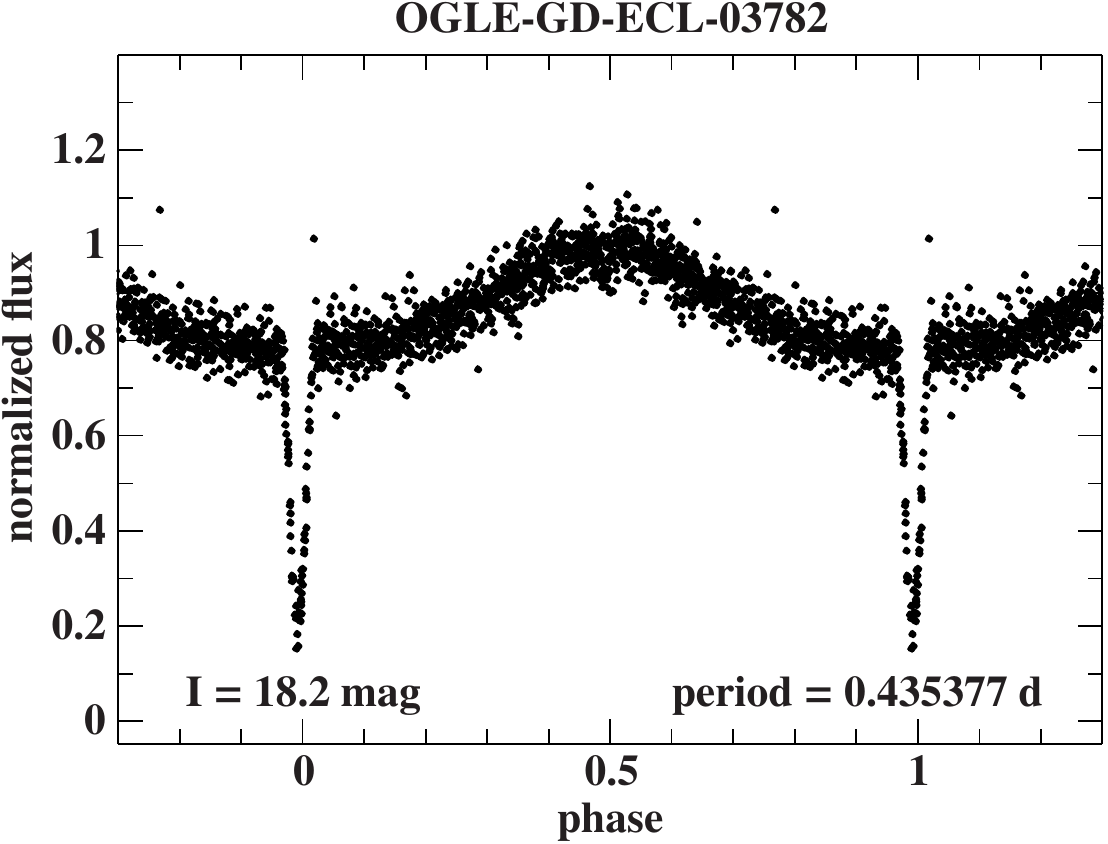}\hfill
		\includegraphics[width=0.25\linewidth]{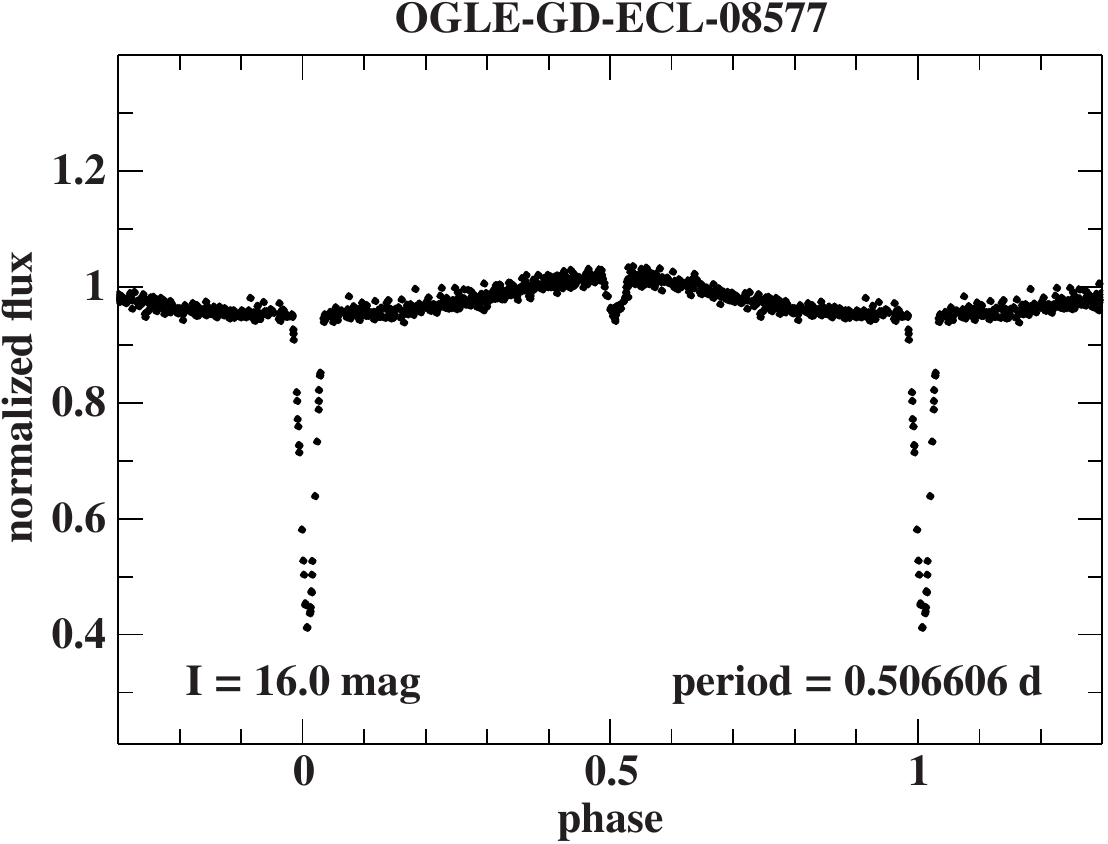}\hfill
	\end{figure}
	\begin{figure}
		\caption{Phased light curves of all our HW Vir candidates from the ATLAS survey.}
		\label{lc2}	
		
		\includegraphics[width=0.25\linewidth]{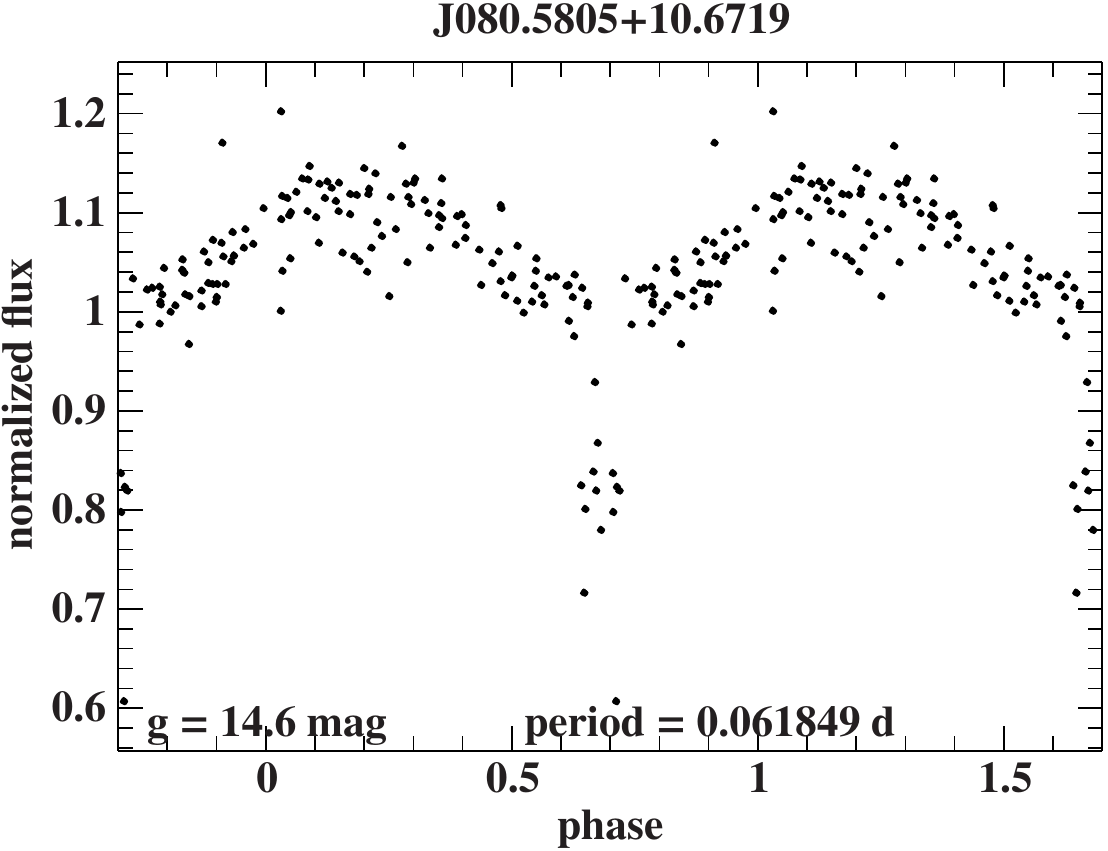}\hfill
		\includegraphics[width=0.25\linewidth]{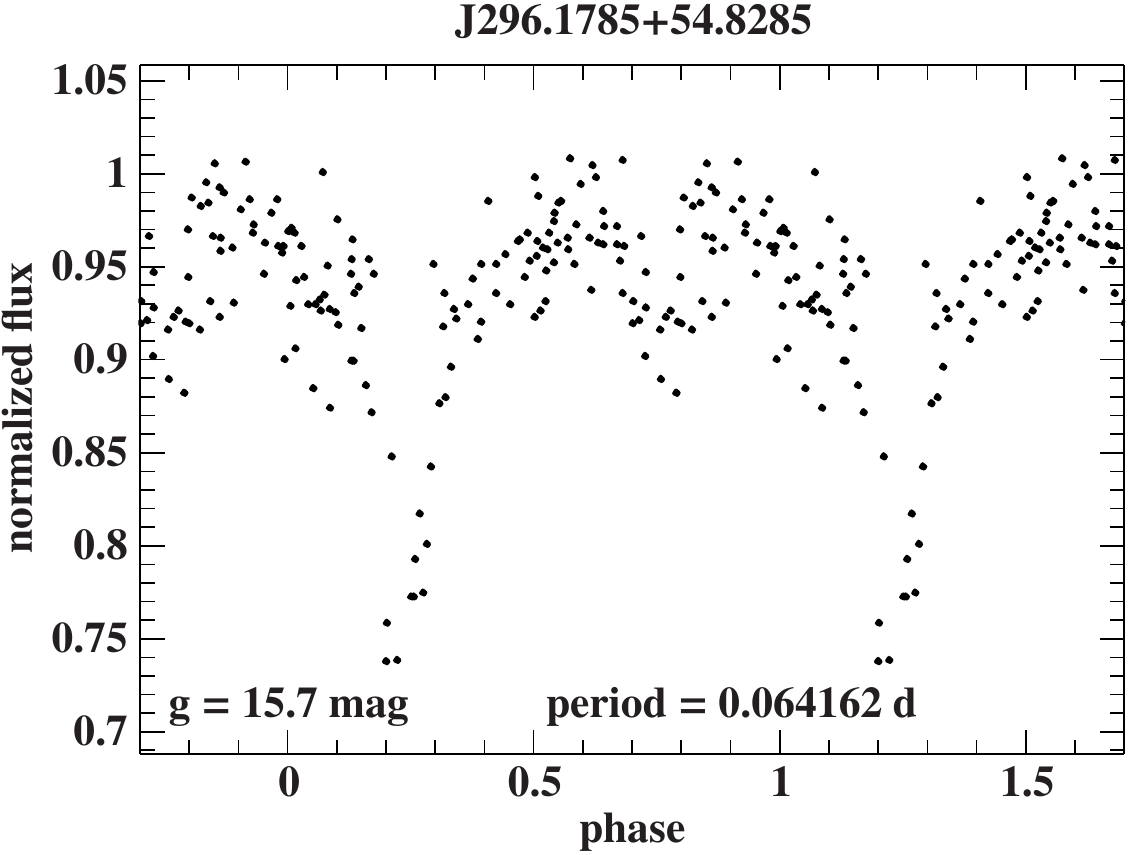}\hfill
		\includegraphics[width=0.25\linewidth]{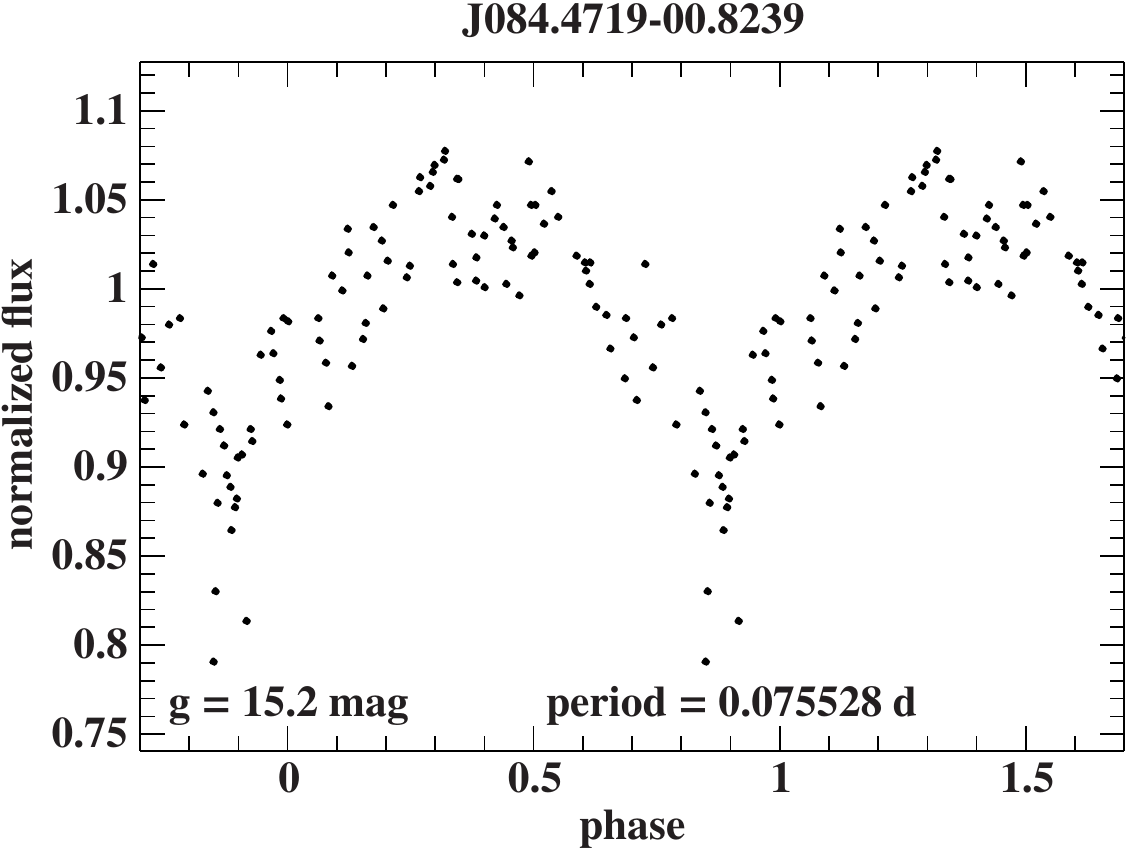}\hfill
		\includegraphics[width=0.25\linewidth]{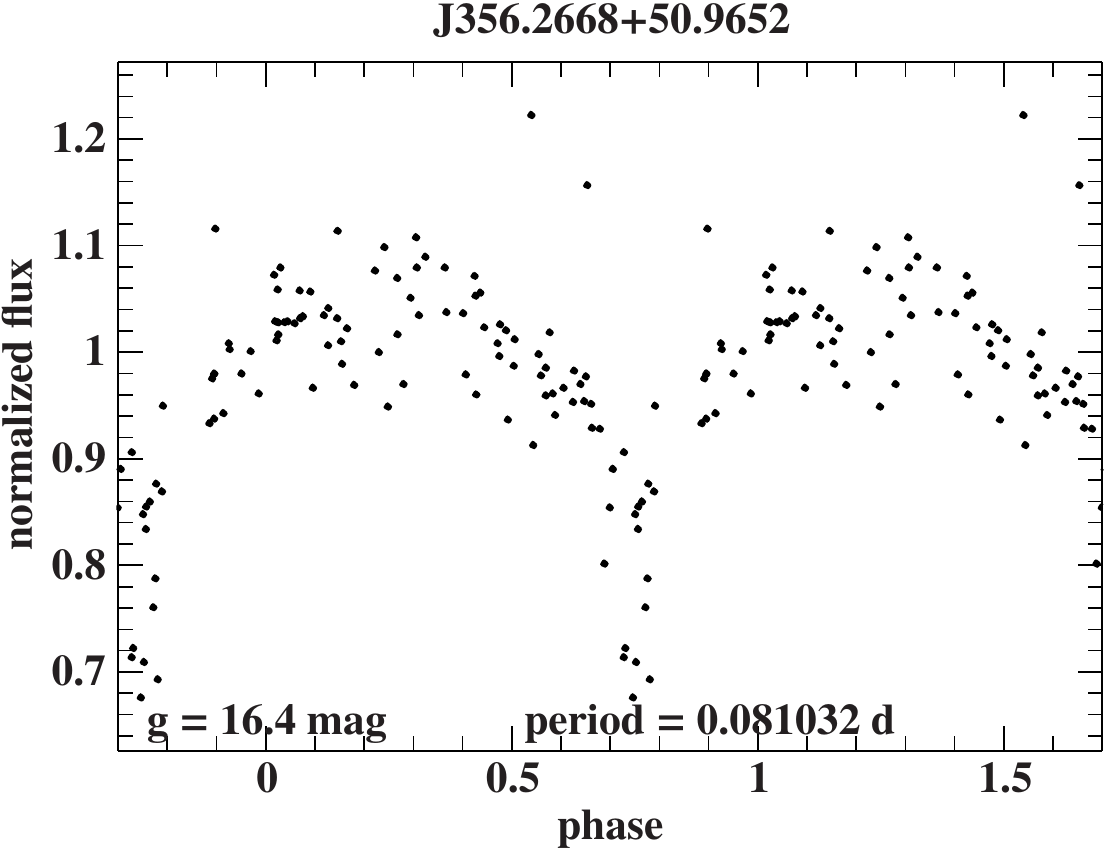}\hfill
		\includegraphics[width=0.25\linewidth]{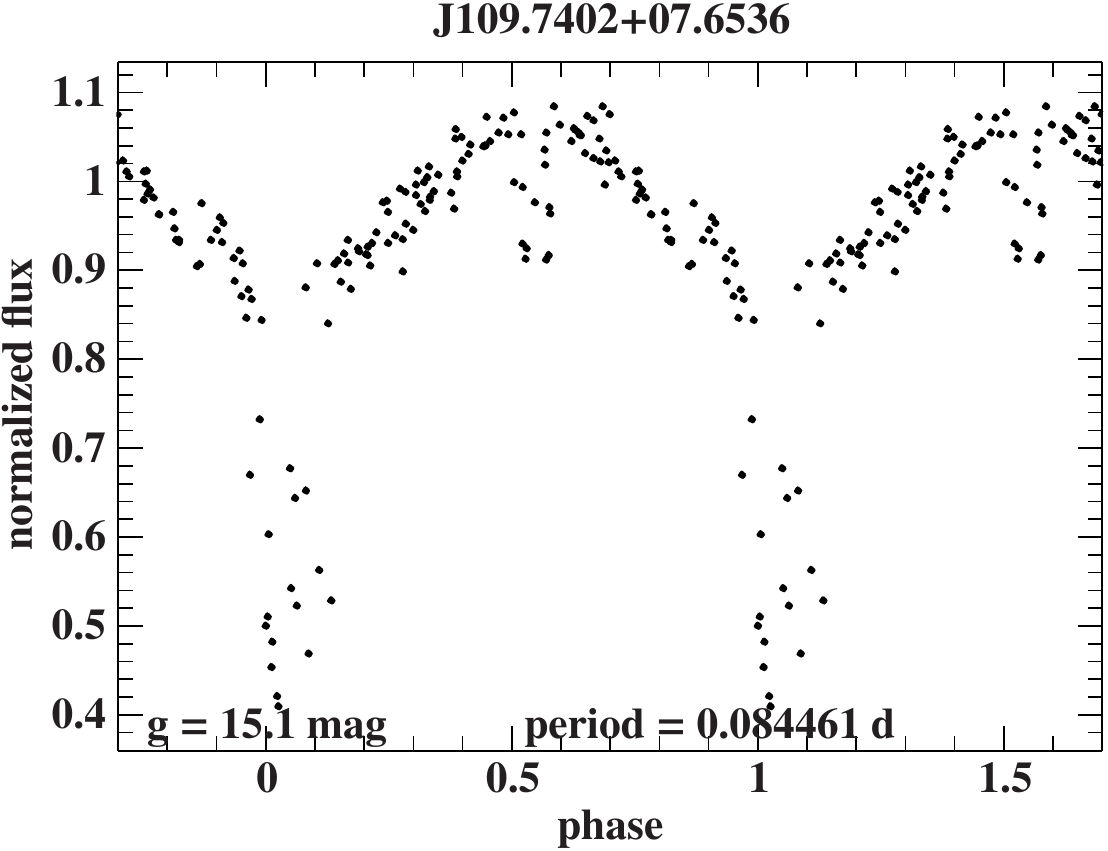}\hfill
		\includegraphics[width=0.25\linewidth]{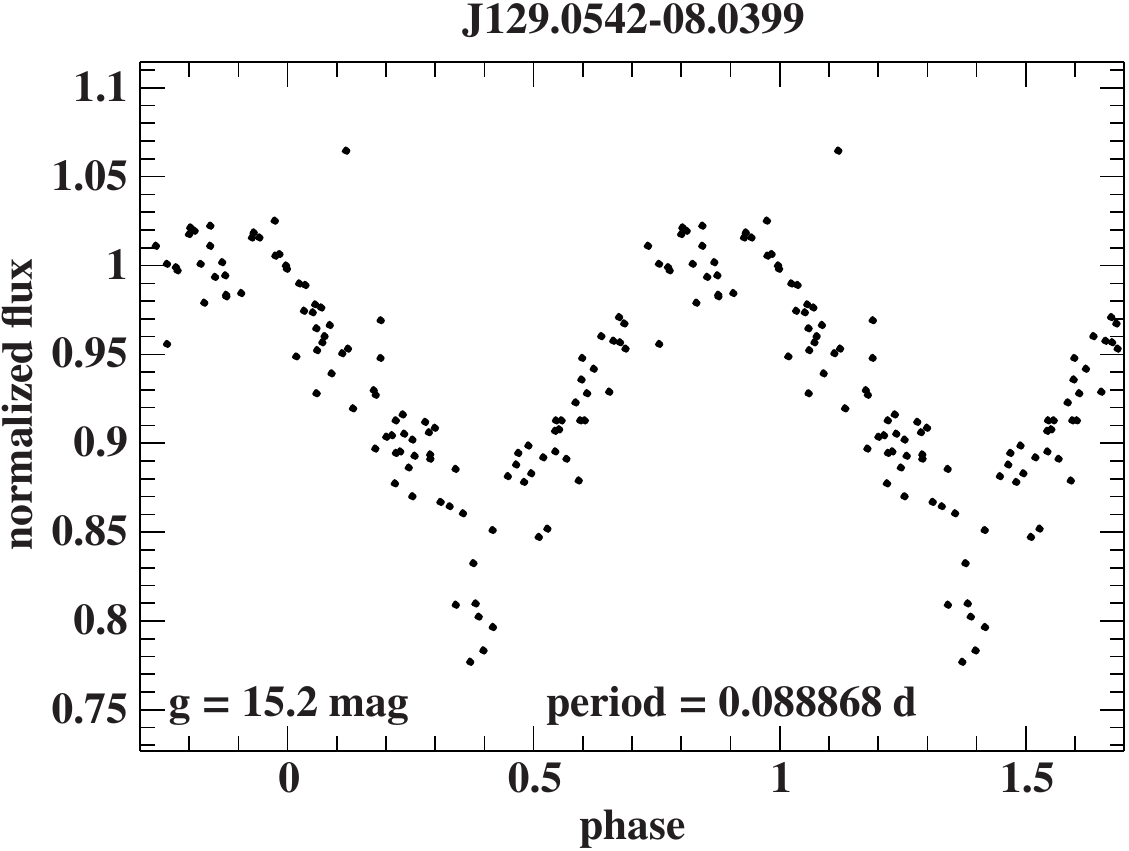}\hfill
		\includegraphics[width=0.25\linewidth]{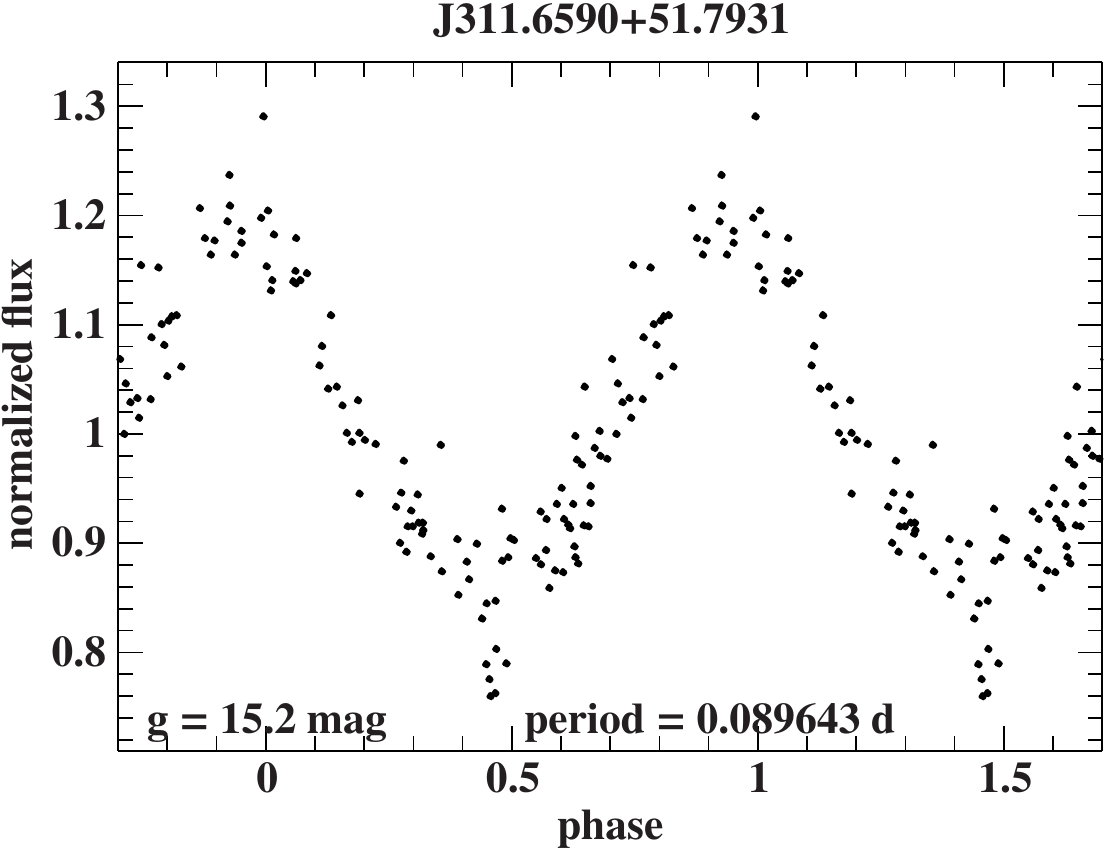}\hfill
		\includegraphics[width=0.25\linewidth]{figures/{J311.6590+51.7931_c_lc}.pdf}\hfill
		\includegraphics[width=0.25\linewidth]{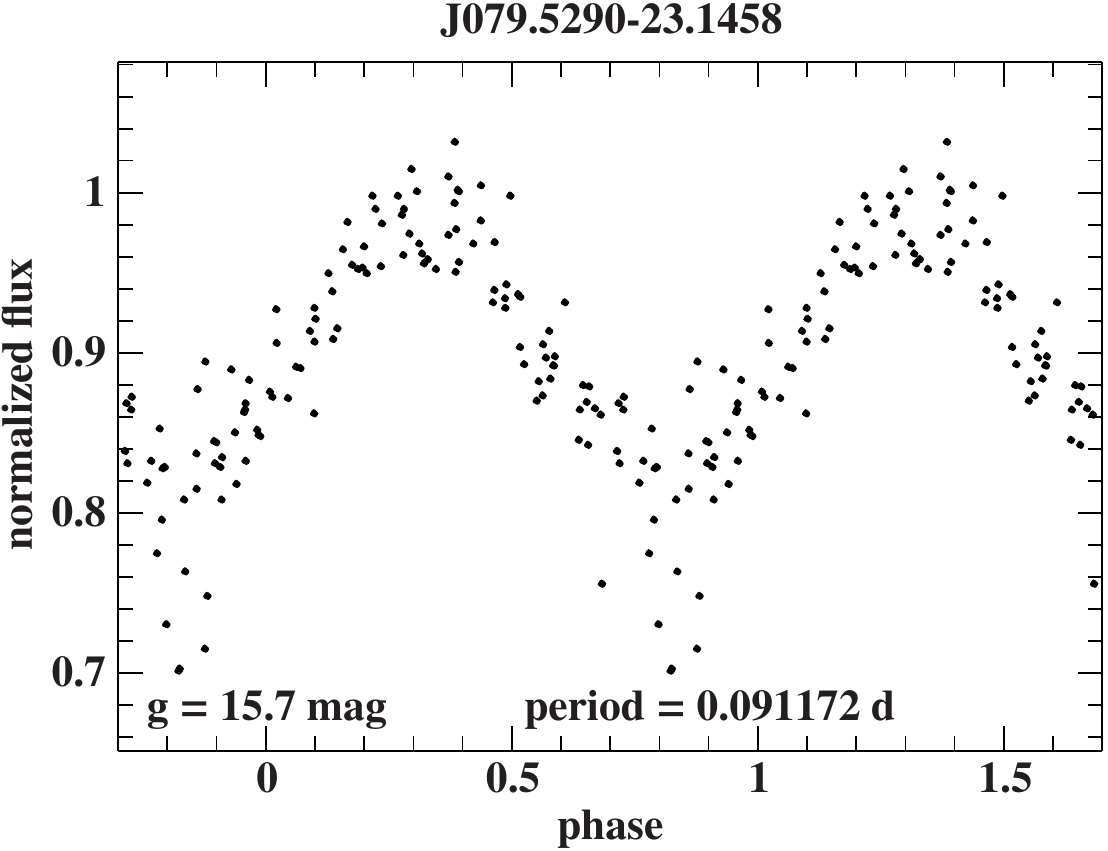}\hfill
		\includegraphics[width=0.25\linewidth]{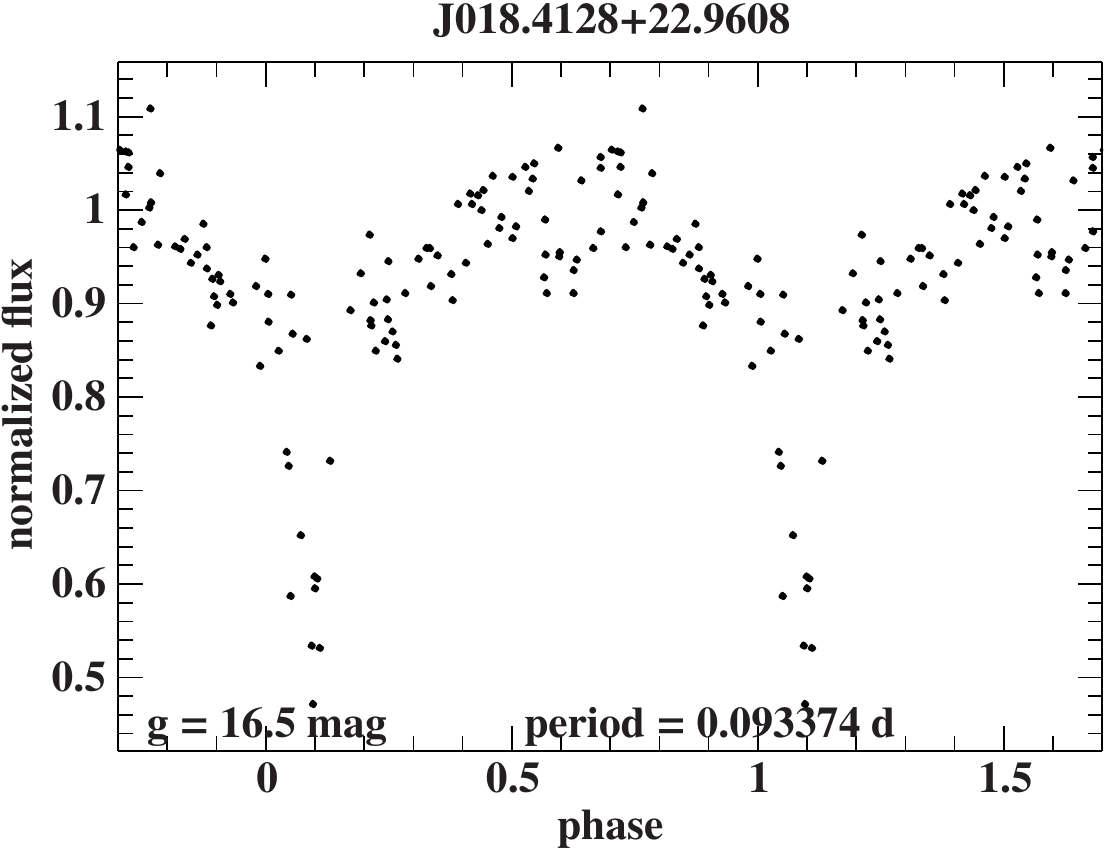}\hfill
		\includegraphics[width=0.25\linewidth]{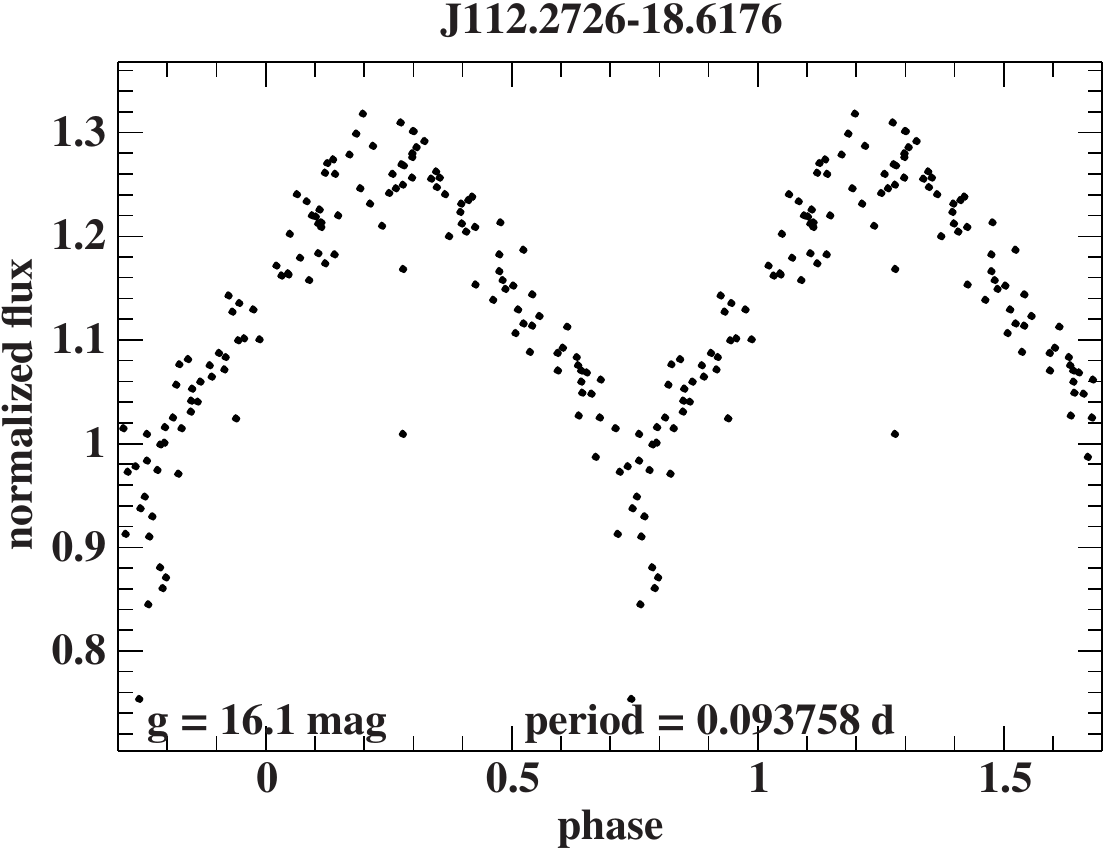}\hfill
		\includegraphics[width=0.25\linewidth]{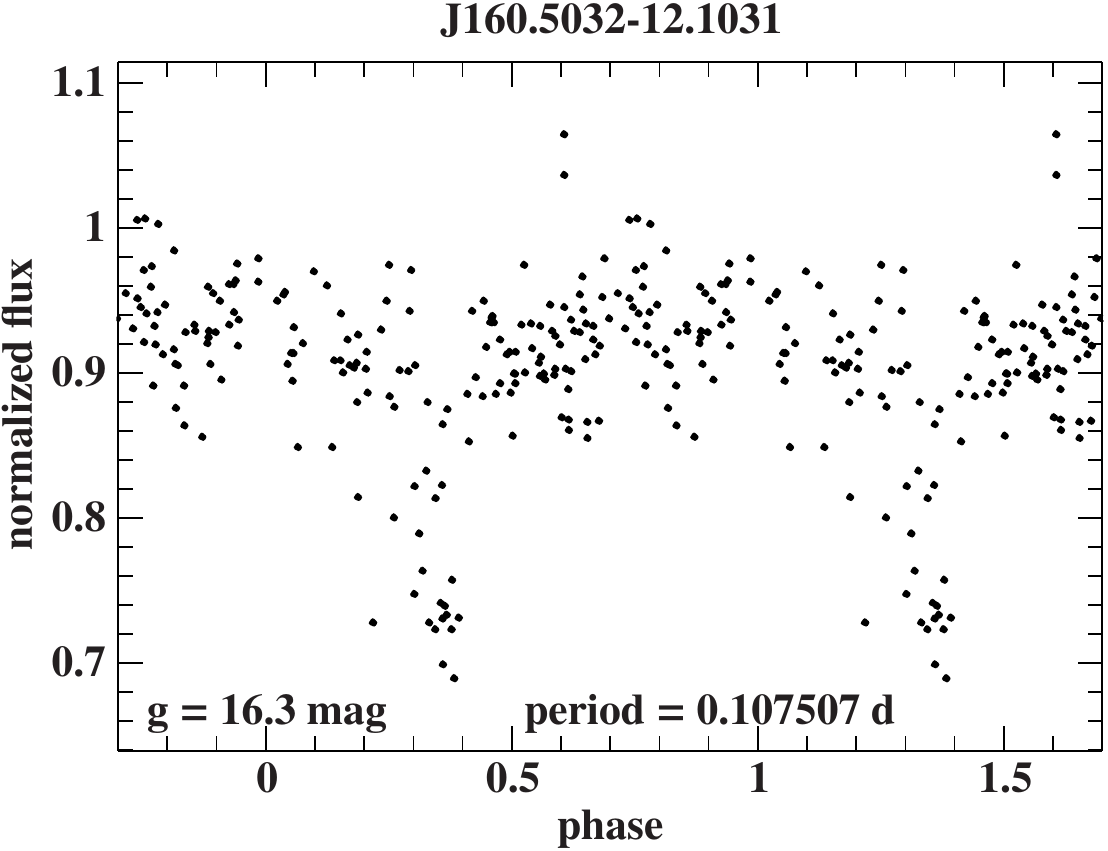}\hfill
		\includegraphics[width=0.25\linewidth]{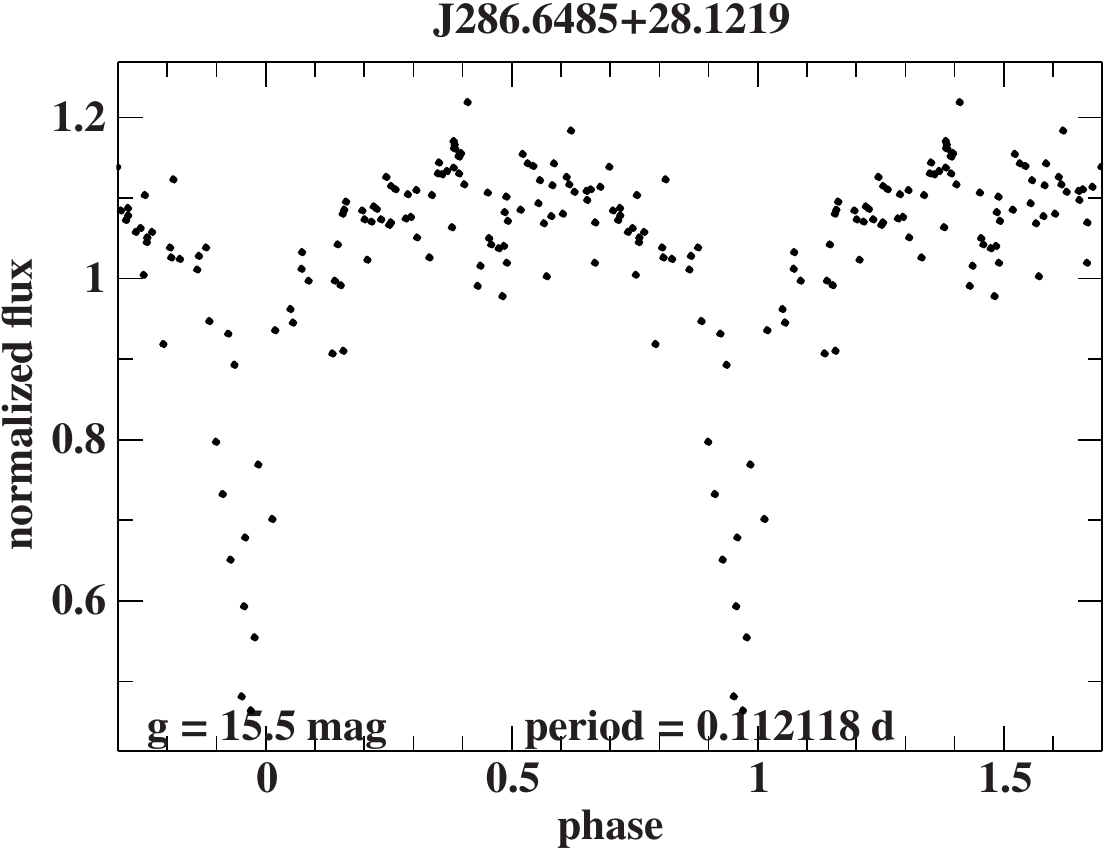}\hfill
		\includegraphics[width=0.25\linewidth]{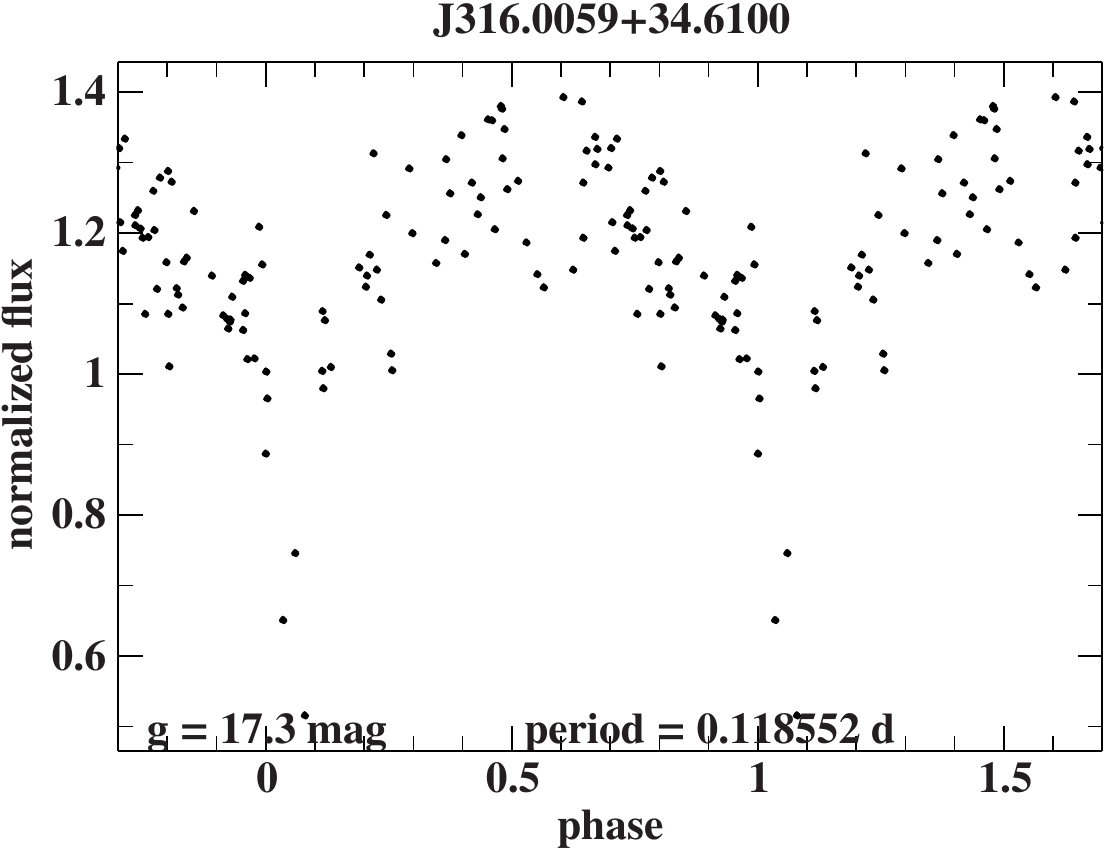}\hfill
		\includegraphics[width=0.25\linewidth]{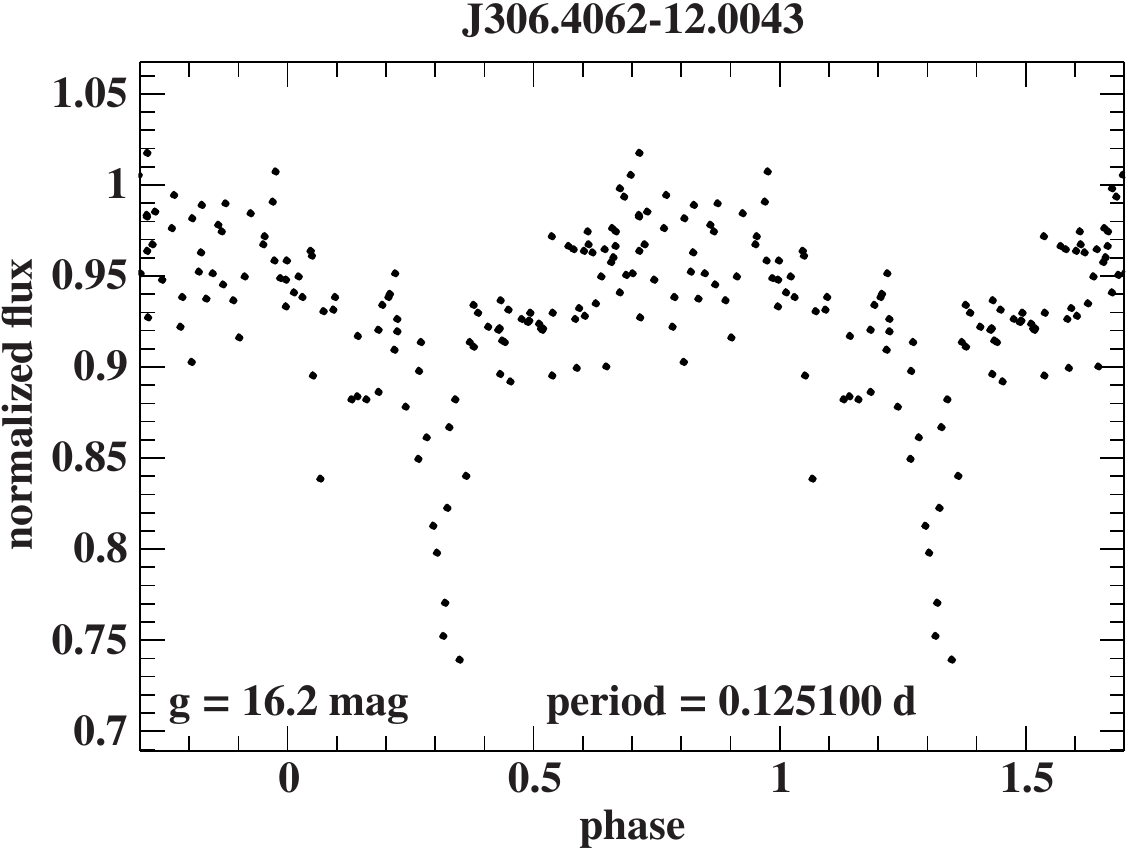}\hfill
		\includegraphics[width=0.25\linewidth]{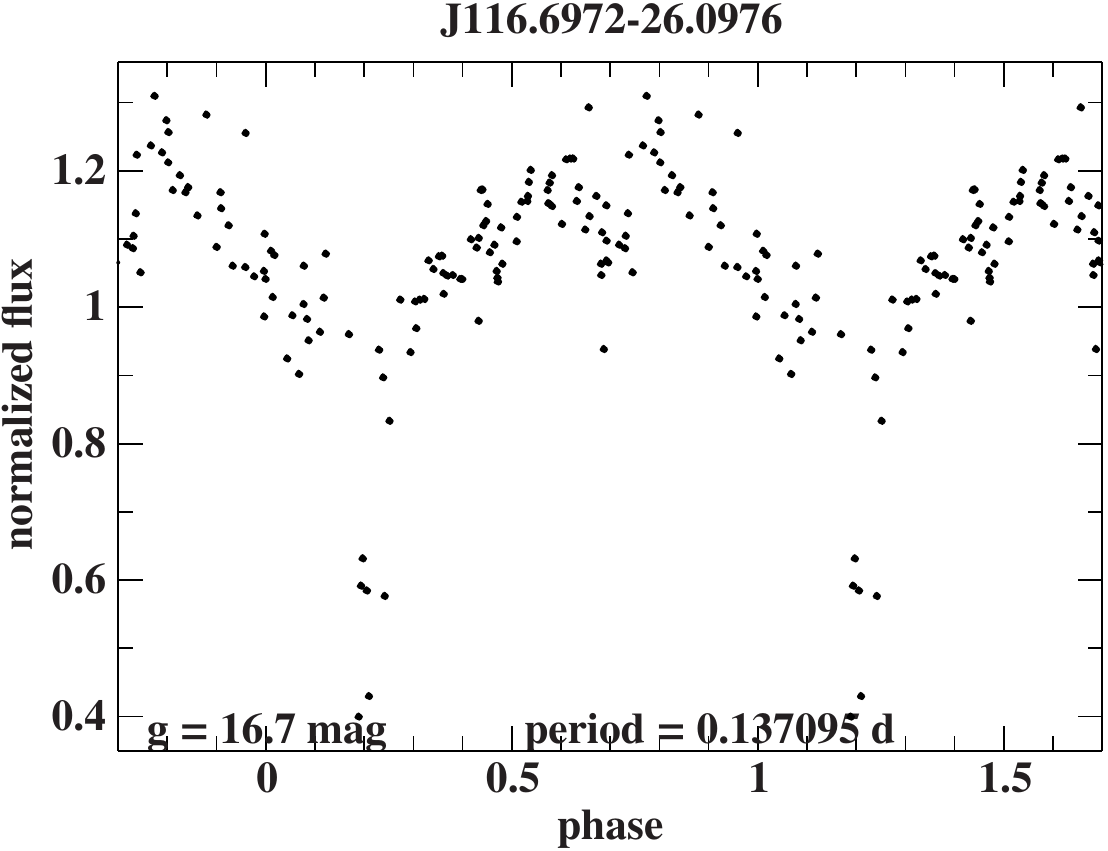}\hfill
		\includegraphics[width=0.25\linewidth]{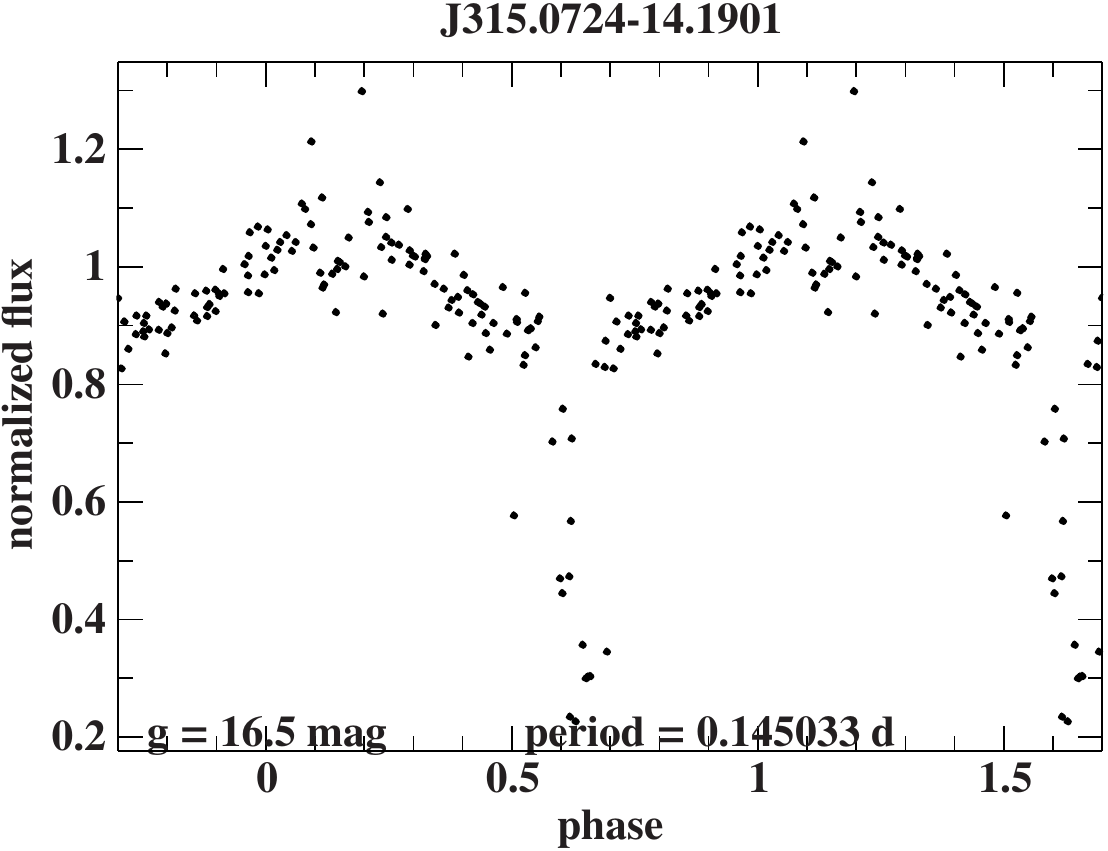}\hfill
		\includegraphics[width=0.25\linewidth]{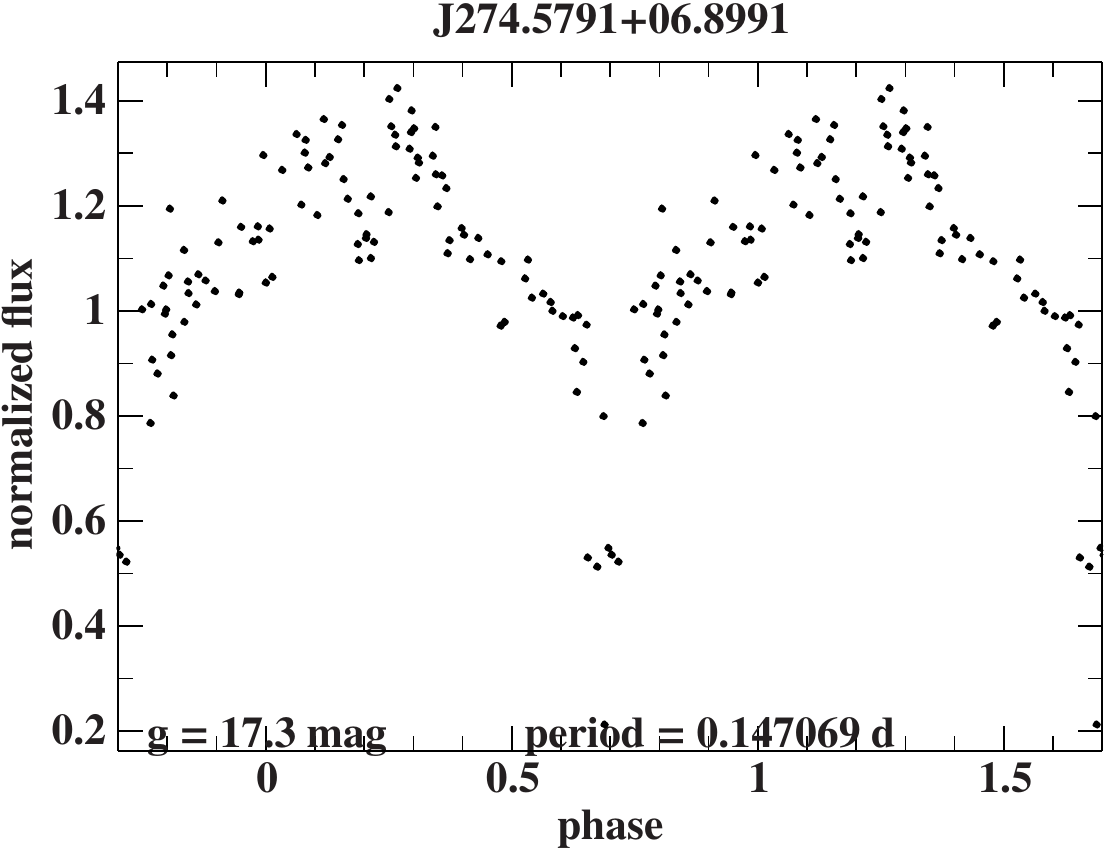}\hfill
		\includegraphics[width=0.25\linewidth]{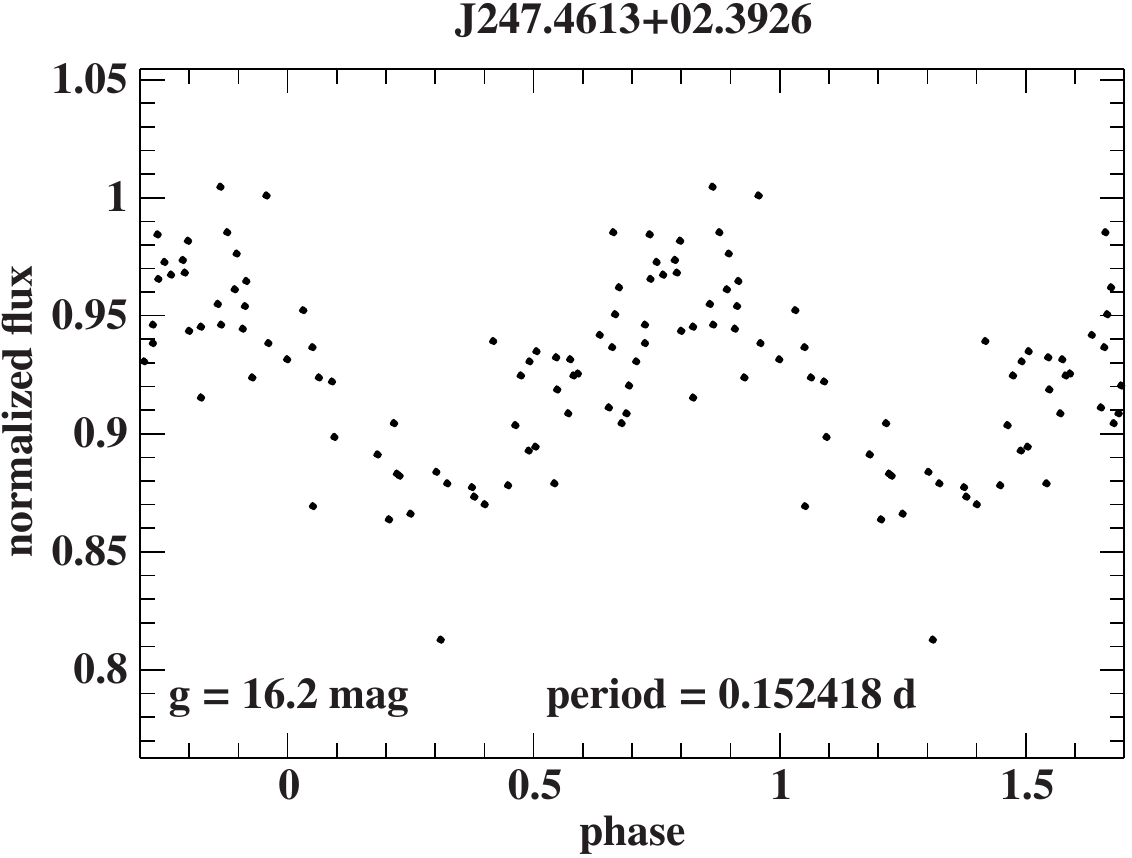}\hfill
		\includegraphics[width=0.25\linewidth]{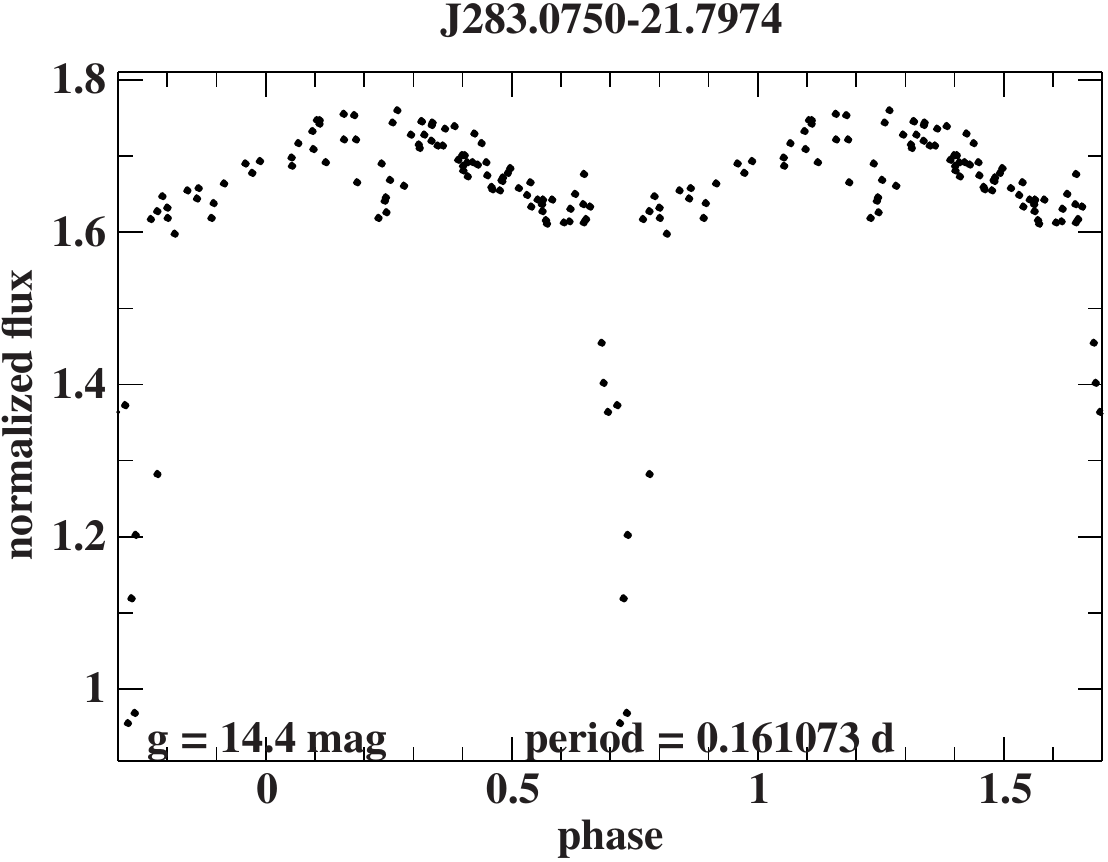}\hfill
		\includegraphics[width=0.25\linewidth]{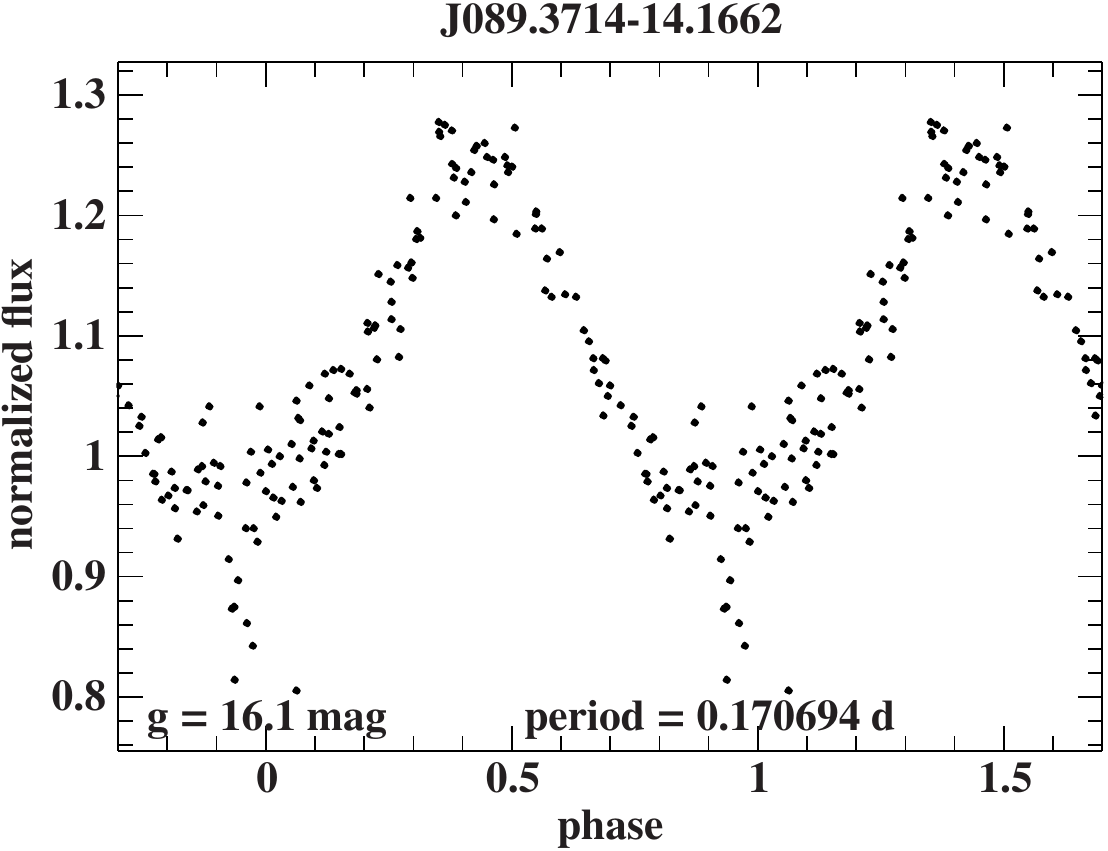}\hfill
		\includegraphics[width=0.25\linewidth]{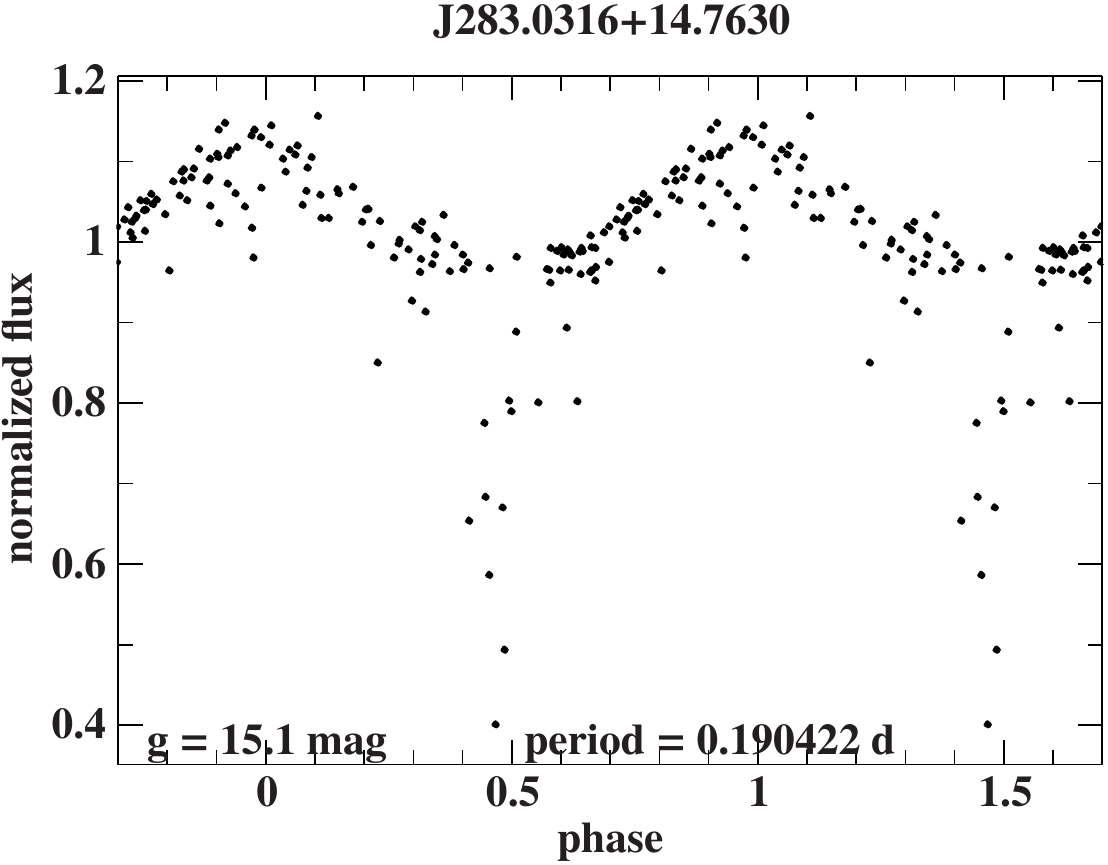}\hfill
		\includegraphics[width=0.25\linewidth]{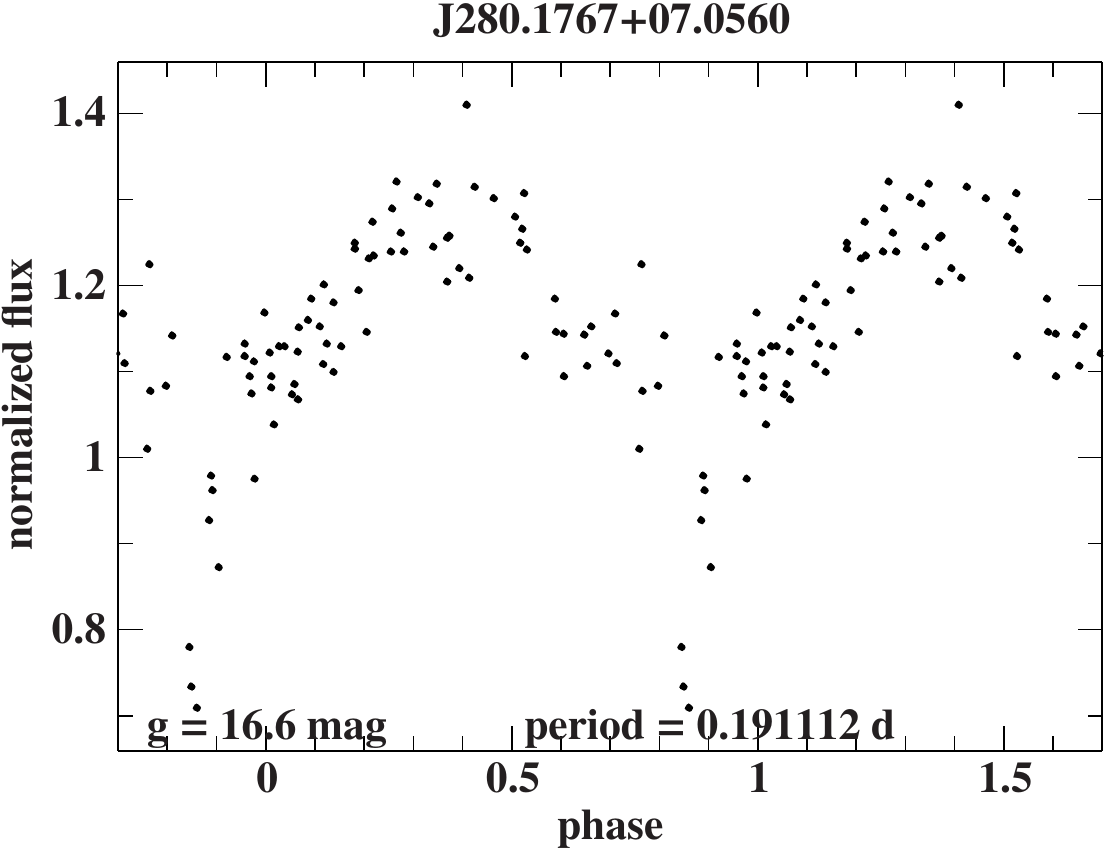}\hfill
		\includegraphics[width=0.25\linewidth]{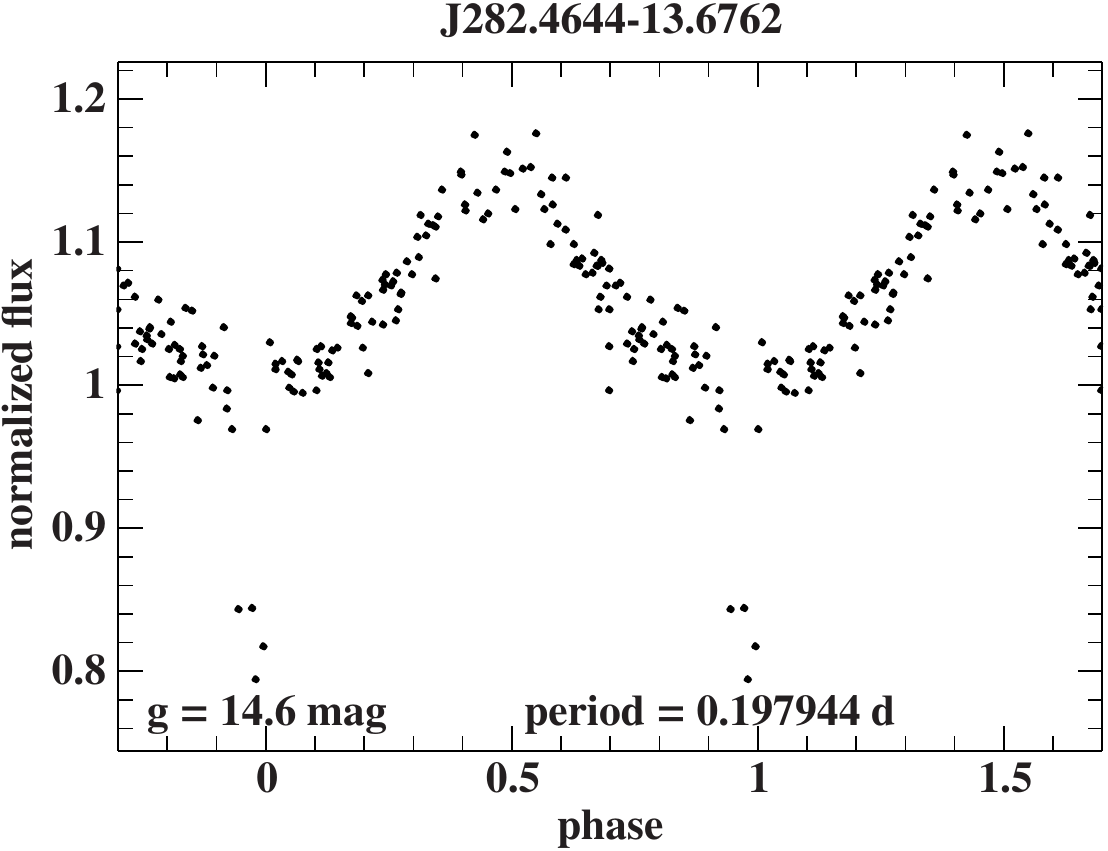}\hfill
		\includegraphics[width=0.25\linewidth]{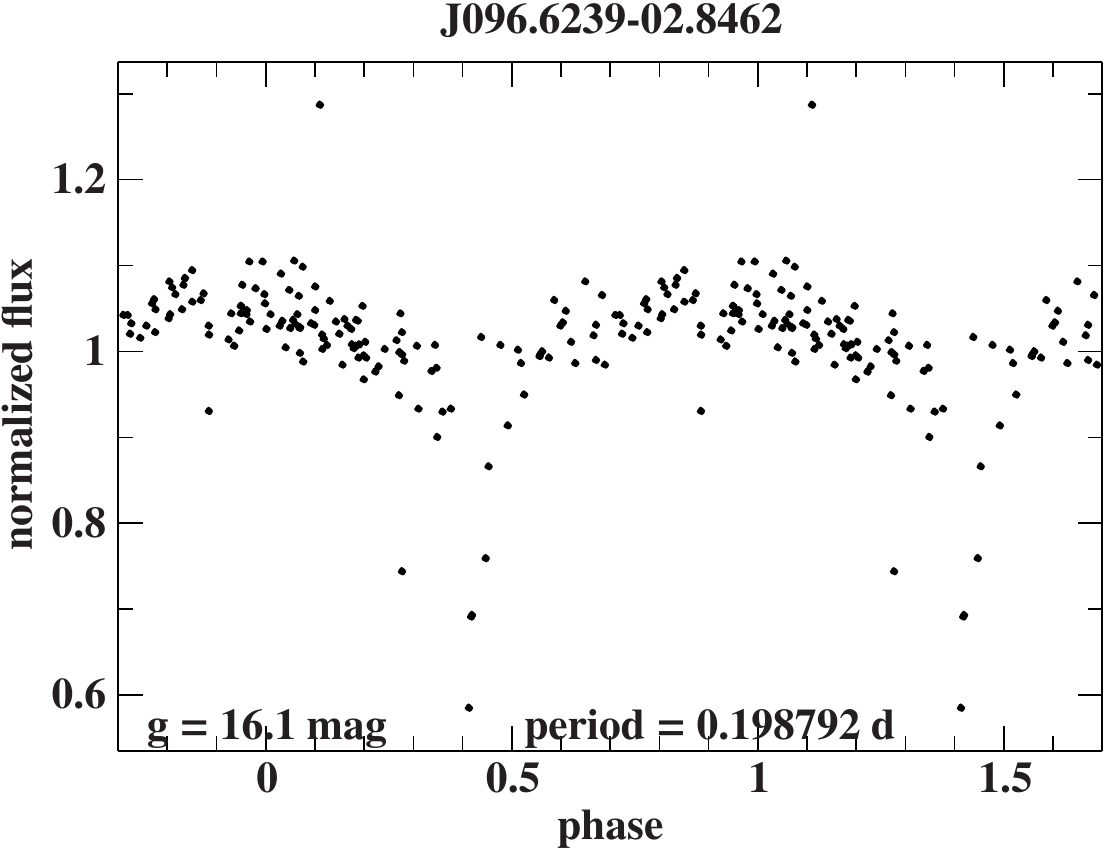}\hfill
		\includegraphics[width=0.25\linewidth]{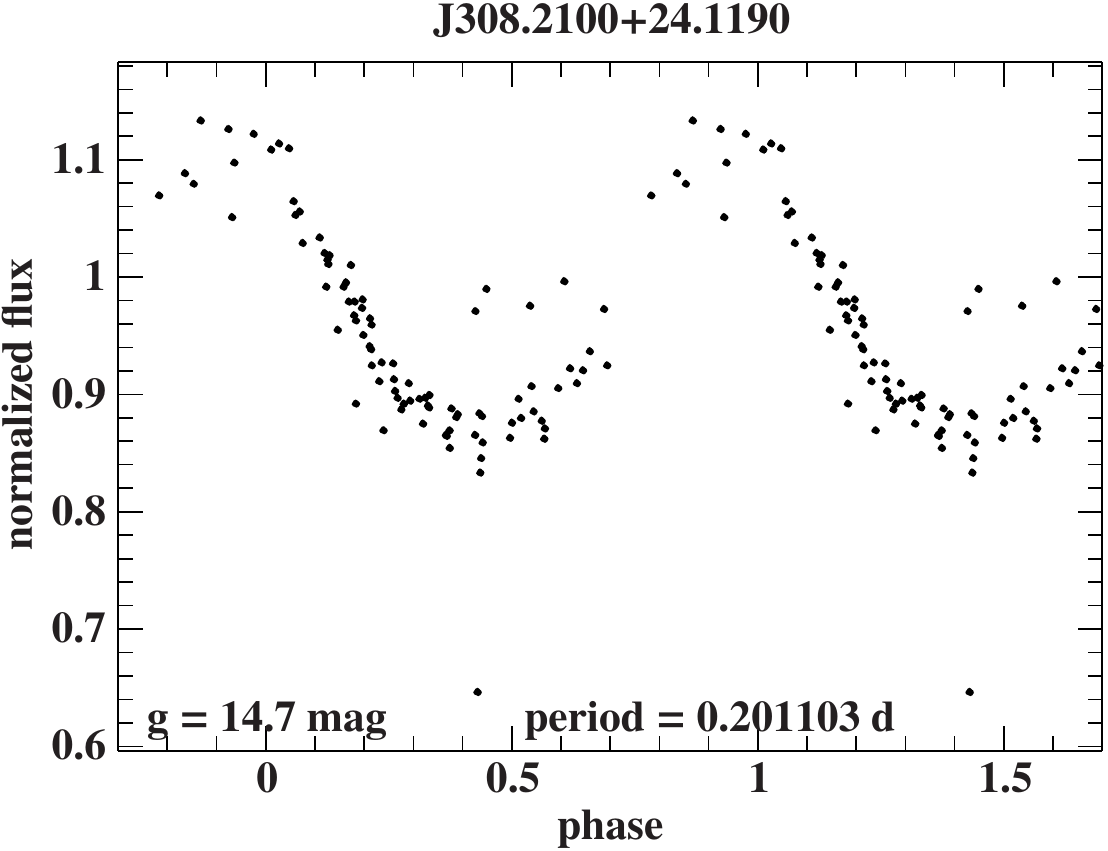}\hfill
		\includegraphics[width=0.25\linewidth]{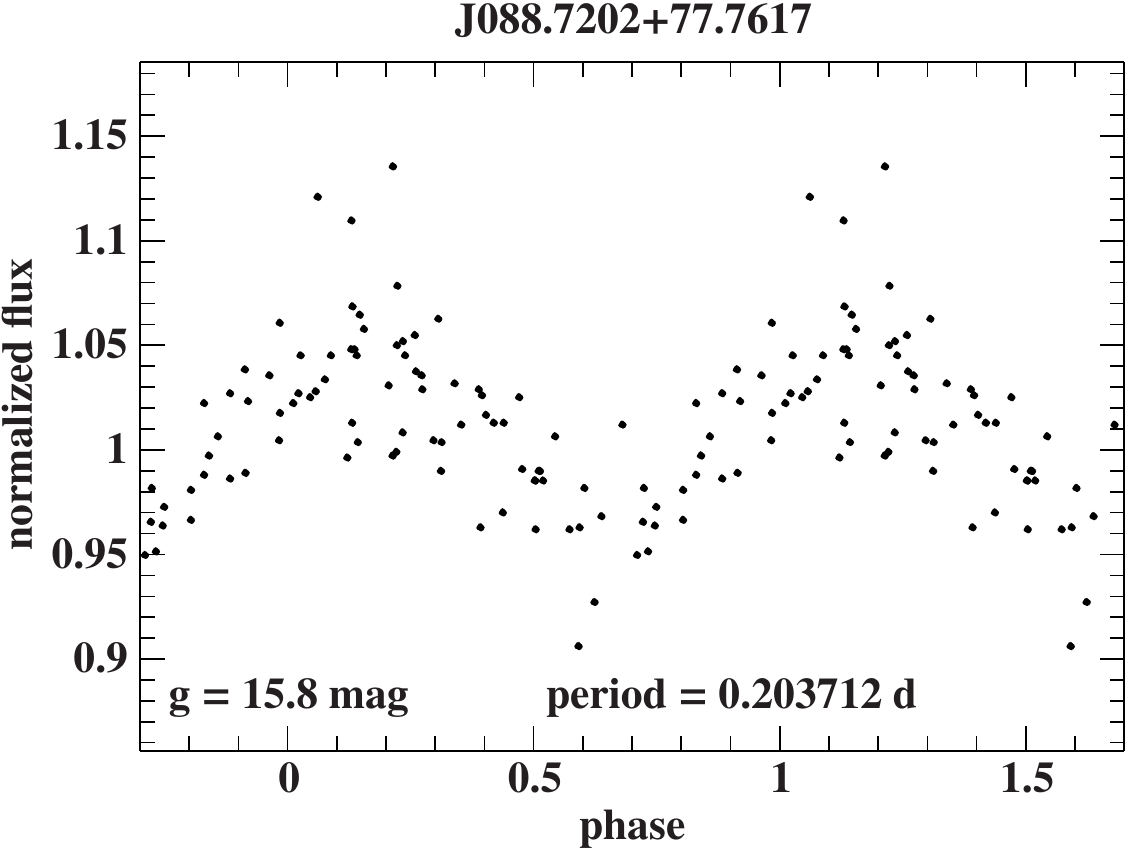}\hfill
		\includegraphics[width=0.25\linewidth]{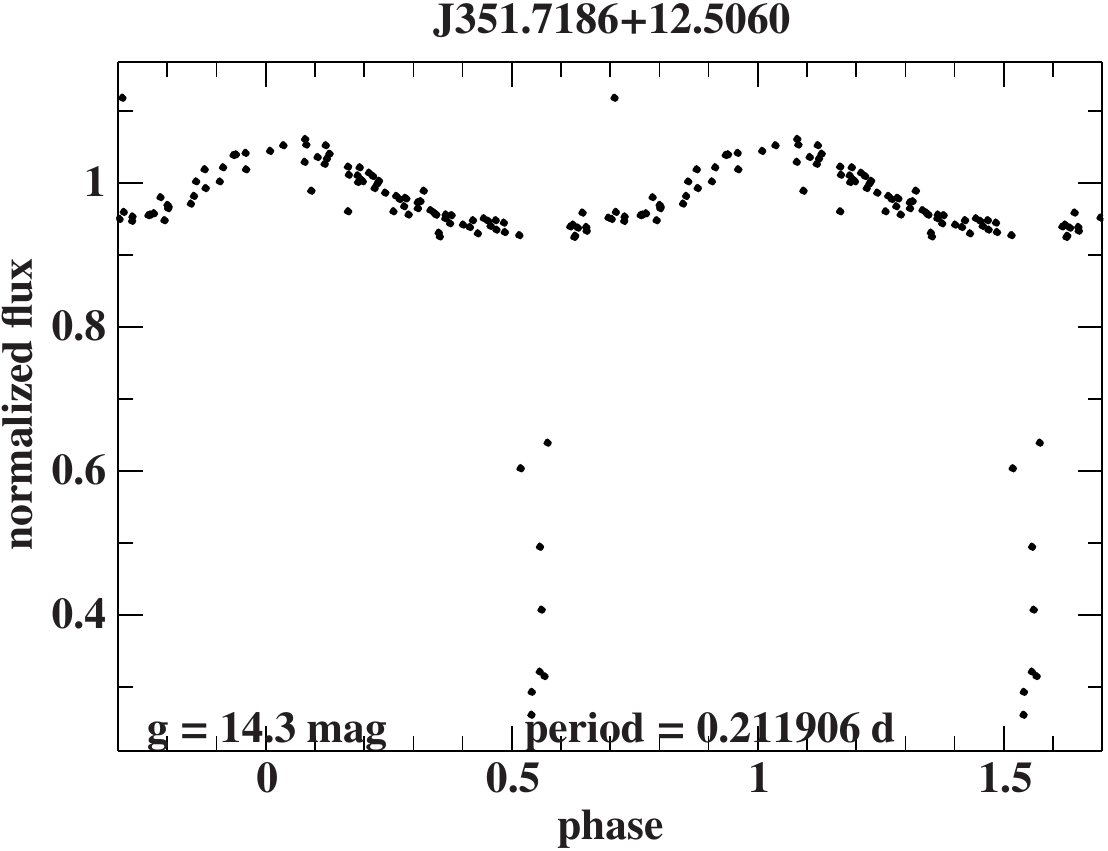}\hfill
	\end{figure}
	\begin{figure}
		\includegraphics[width=0.25\linewidth]{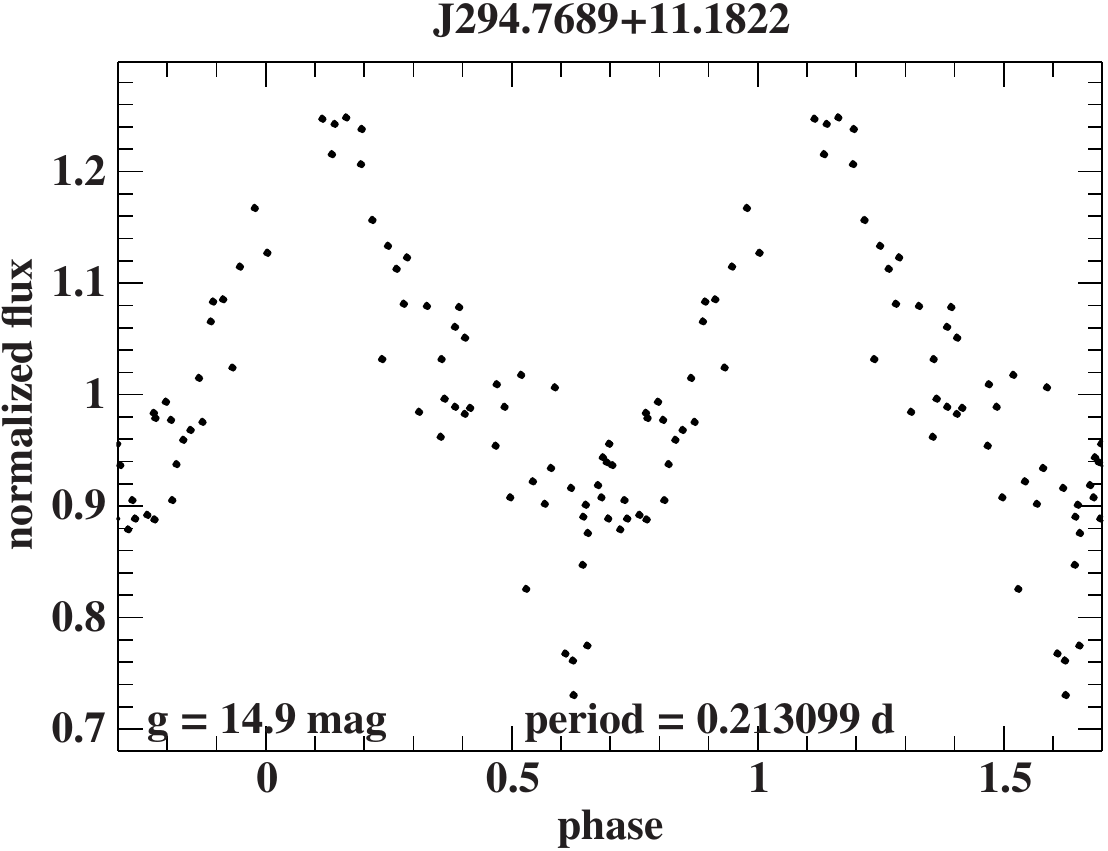}\hfill
		\includegraphics[width=0.25\linewidth]{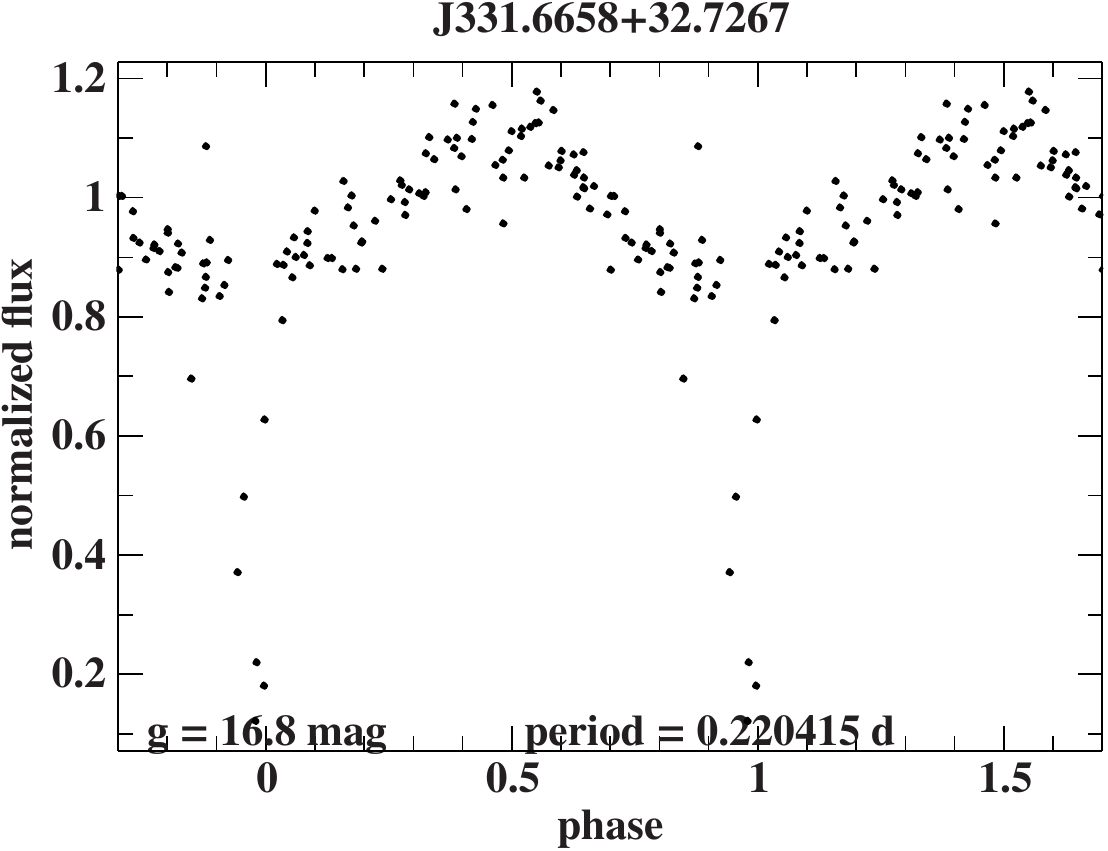}\hfill
		\includegraphics[width=0.25\linewidth]{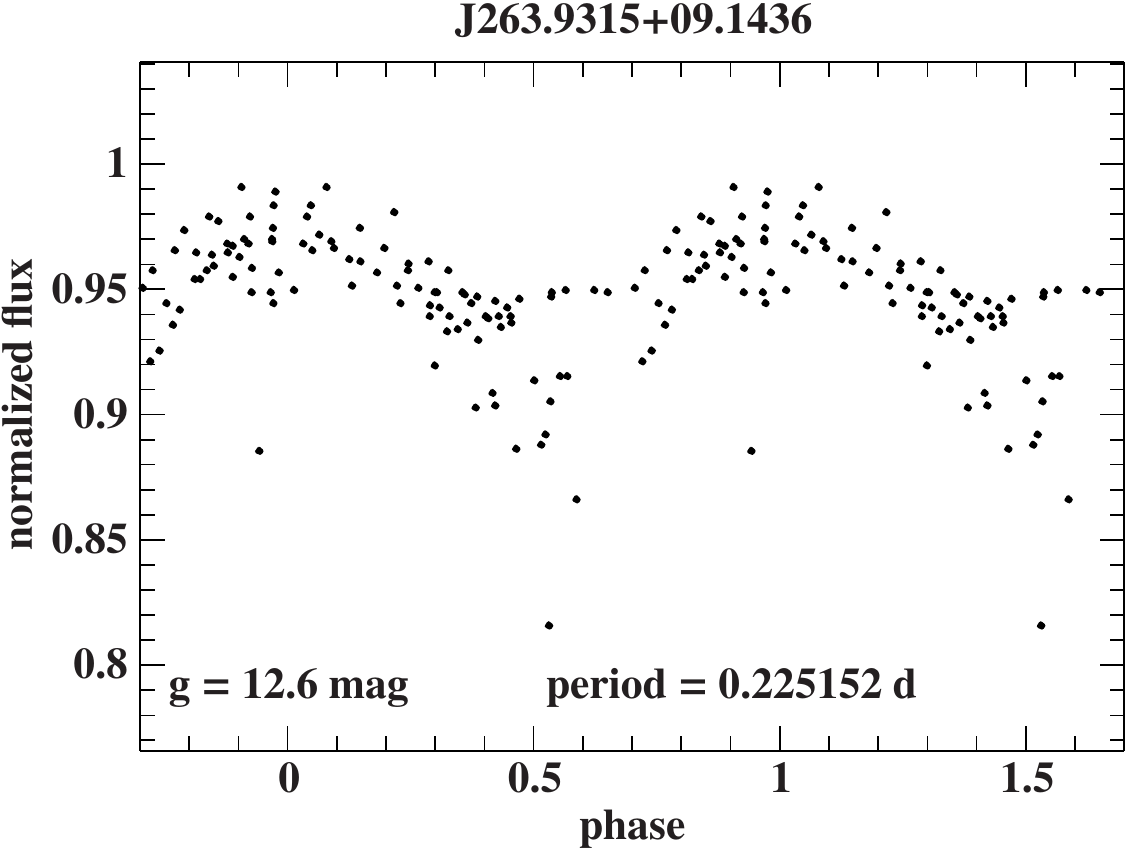}\hfill
		\includegraphics[width=0.25\linewidth]{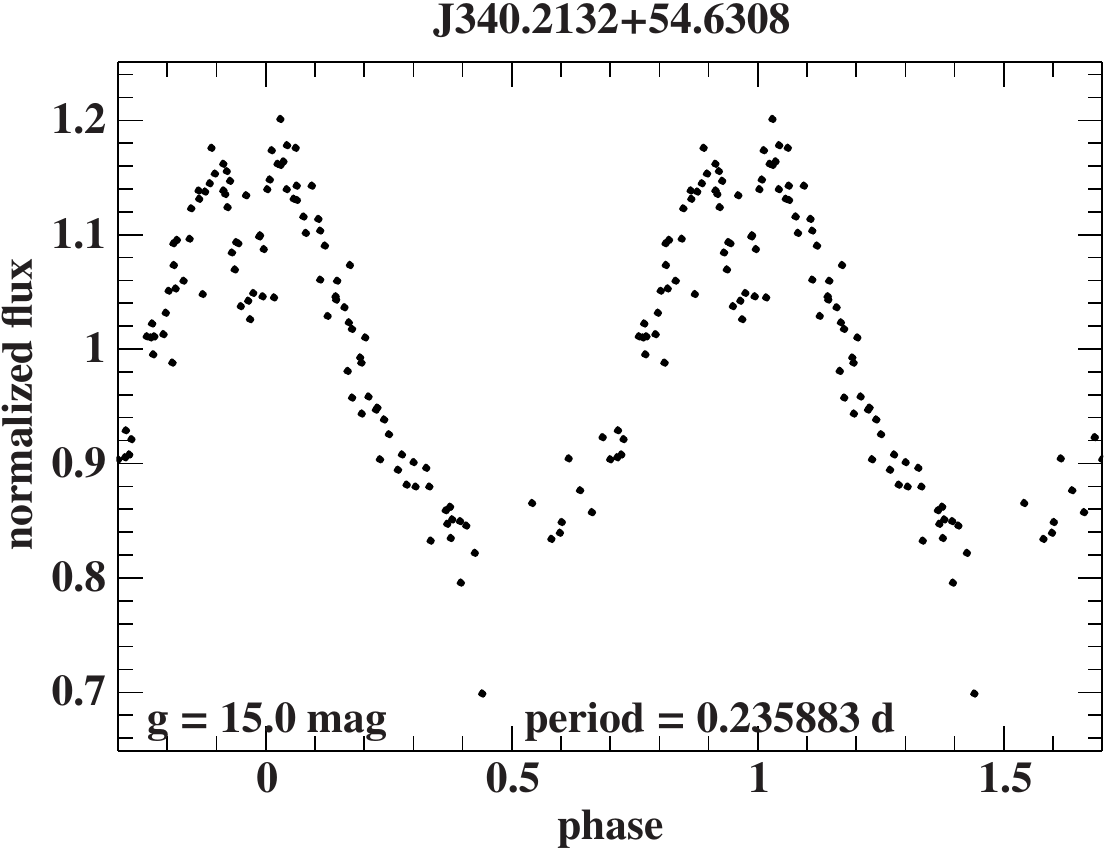}\hfill
		\includegraphics[width=0.25\linewidth]{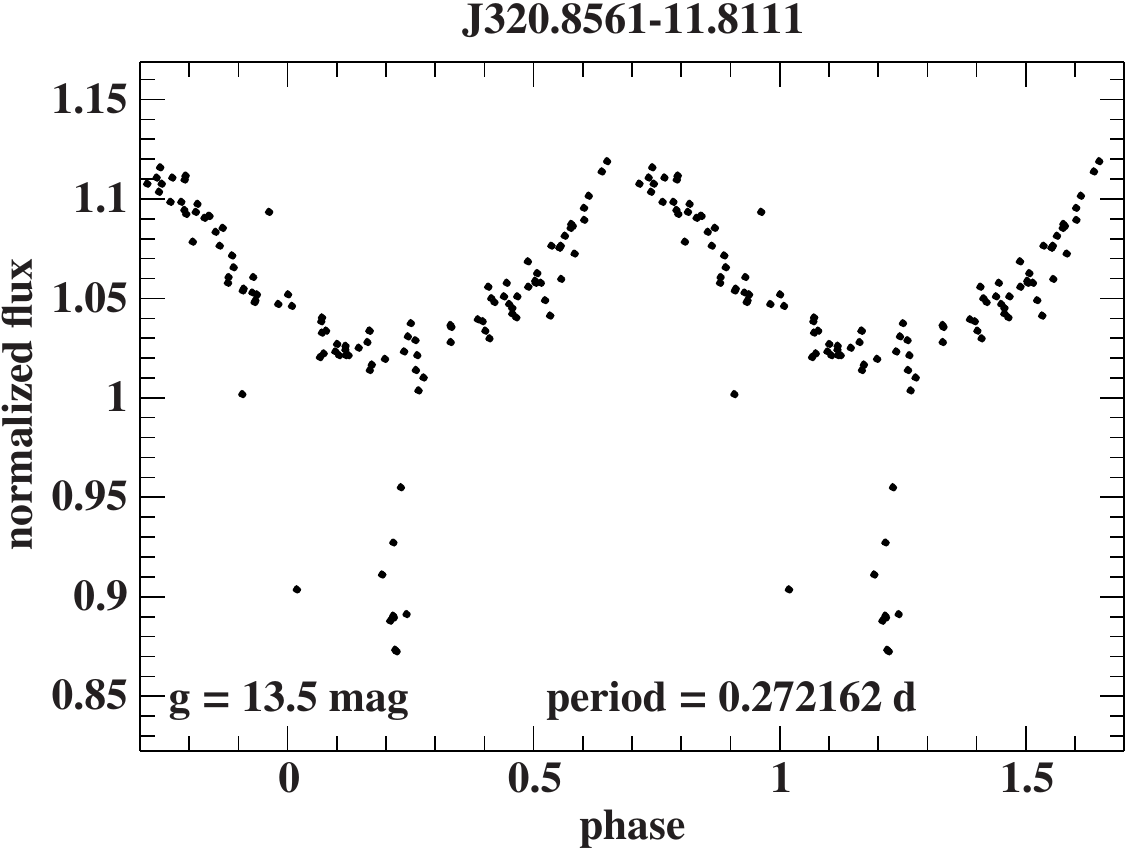}\hfill
		\includegraphics[width=0.25\linewidth]{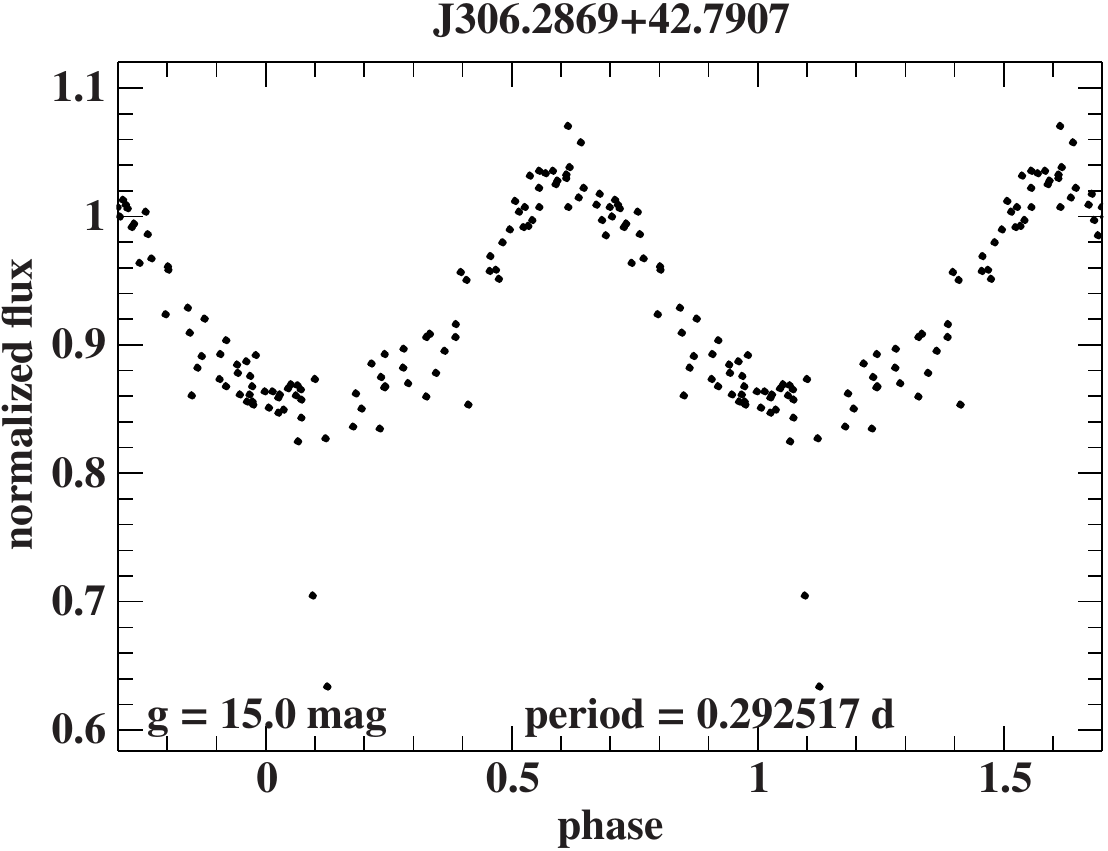}\hfill
		\includegraphics[width=0.25\linewidth]{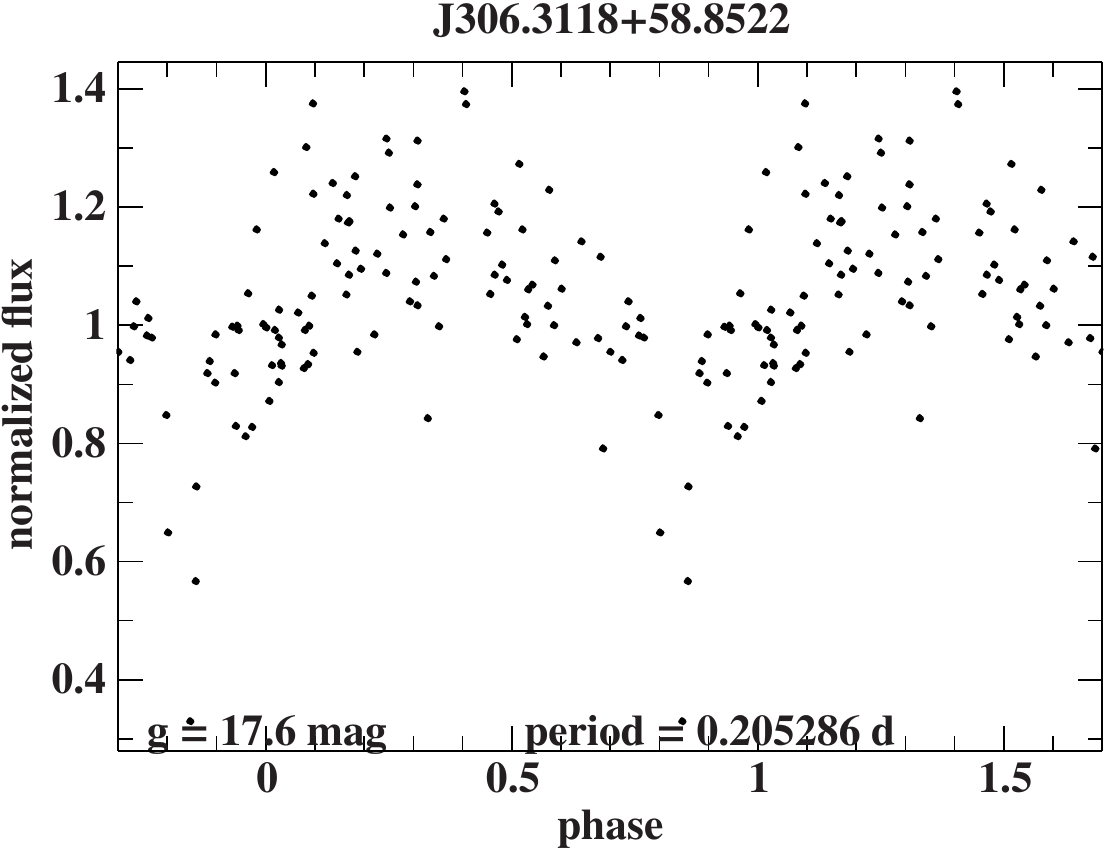}\hfill
		\includegraphics[width=0.25\linewidth]{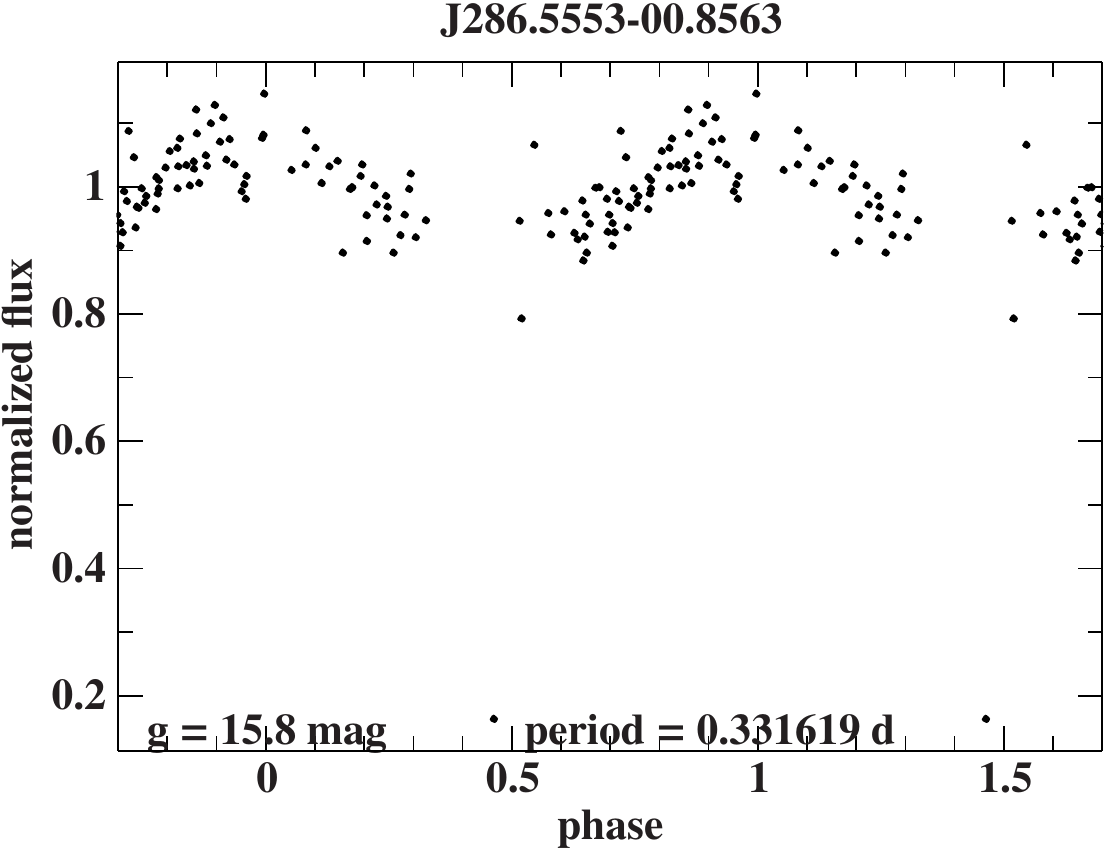}\hfill
		\includegraphics[width=0.25\linewidth]{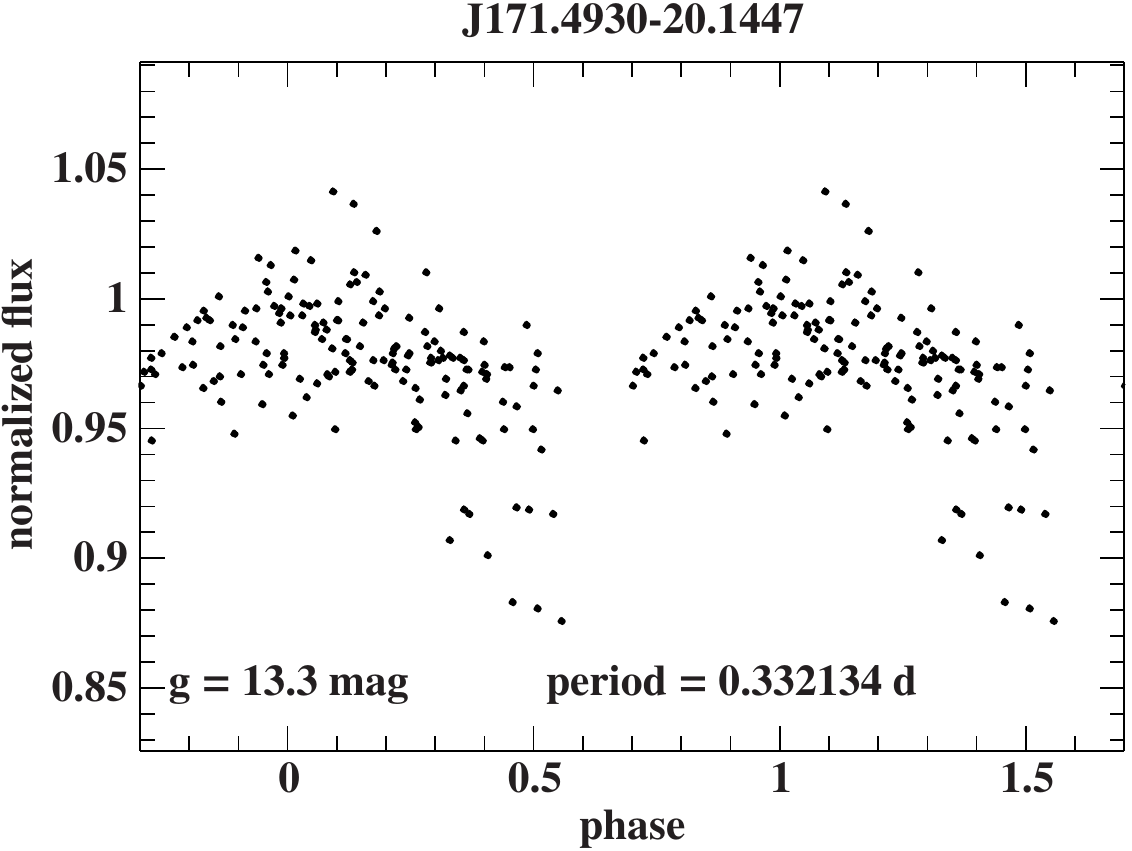}\hfill
		\includegraphics[width=0.25\linewidth]{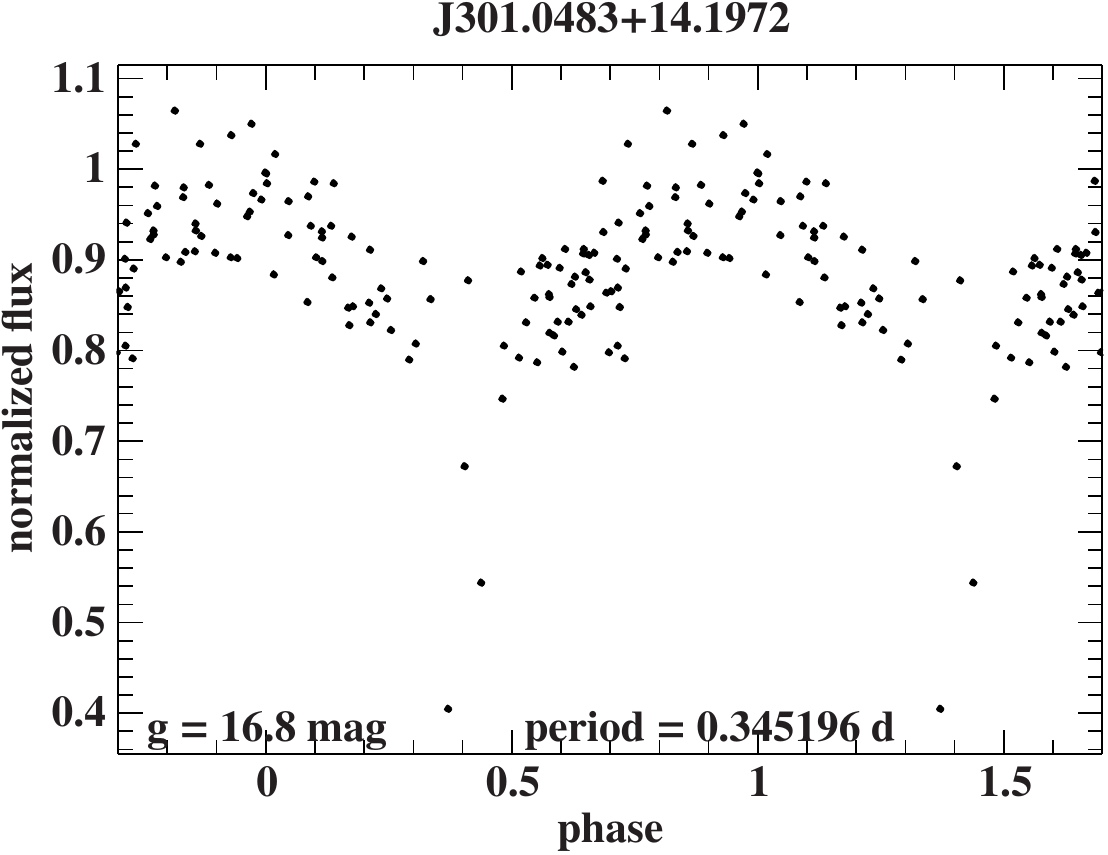}\hfill
		\includegraphics[width=0.25\linewidth]{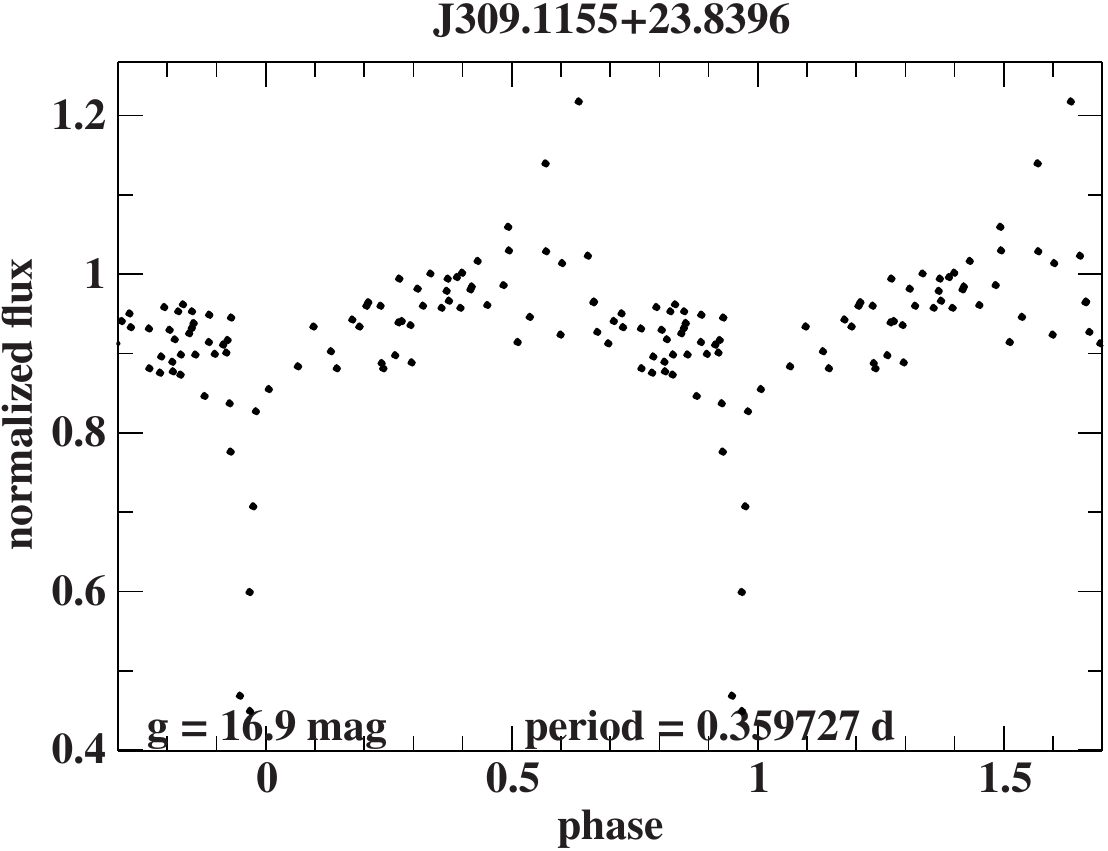}\hfill
		\includegraphics[width=0.25\linewidth]{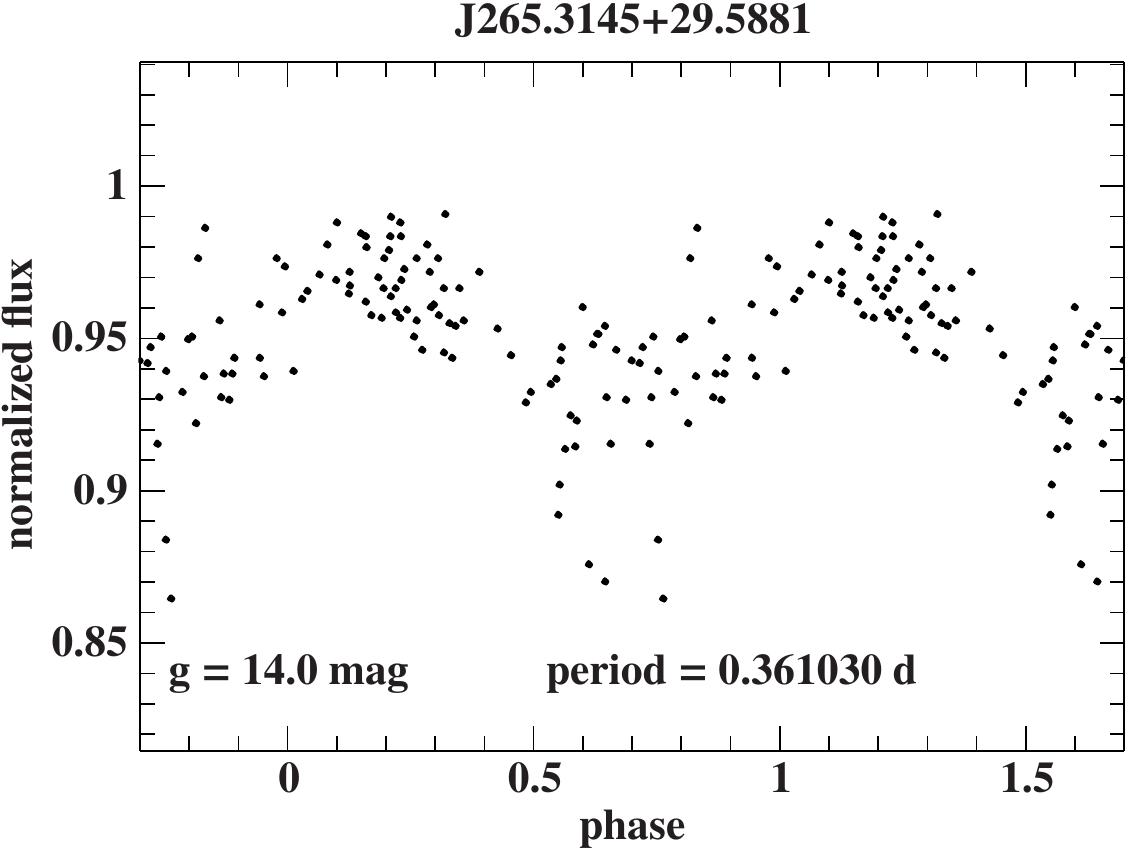}\hfill
		\includegraphics[width=0.25\linewidth]{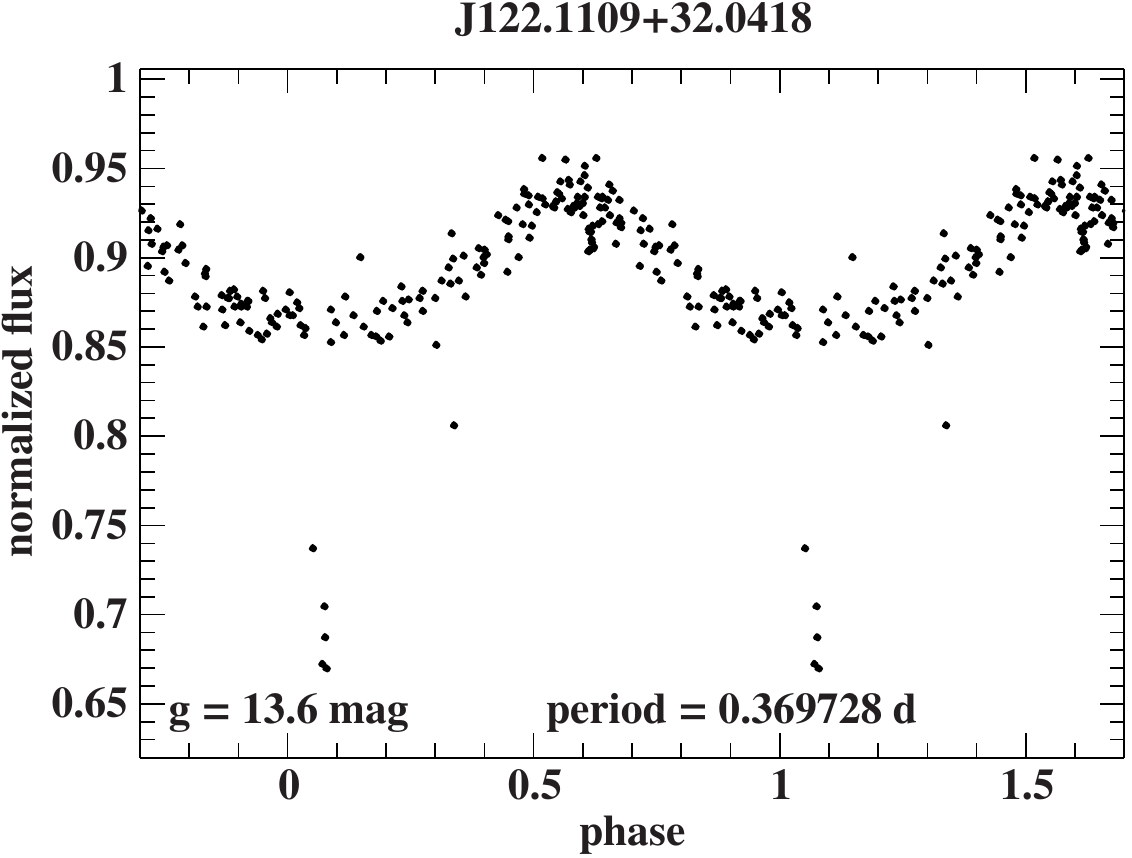}\hfill
		\includegraphics[width=0.25\linewidth]{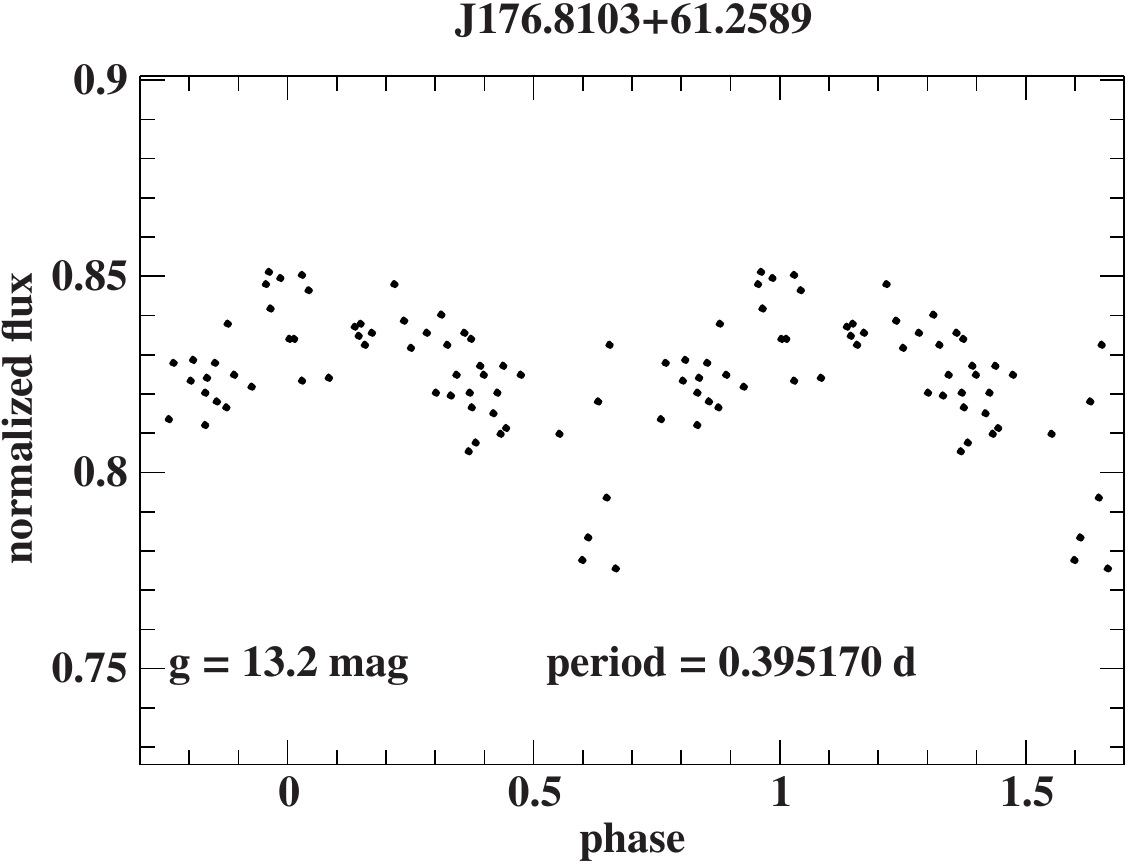}\hfill
		\includegraphics[width=0.25\linewidth]{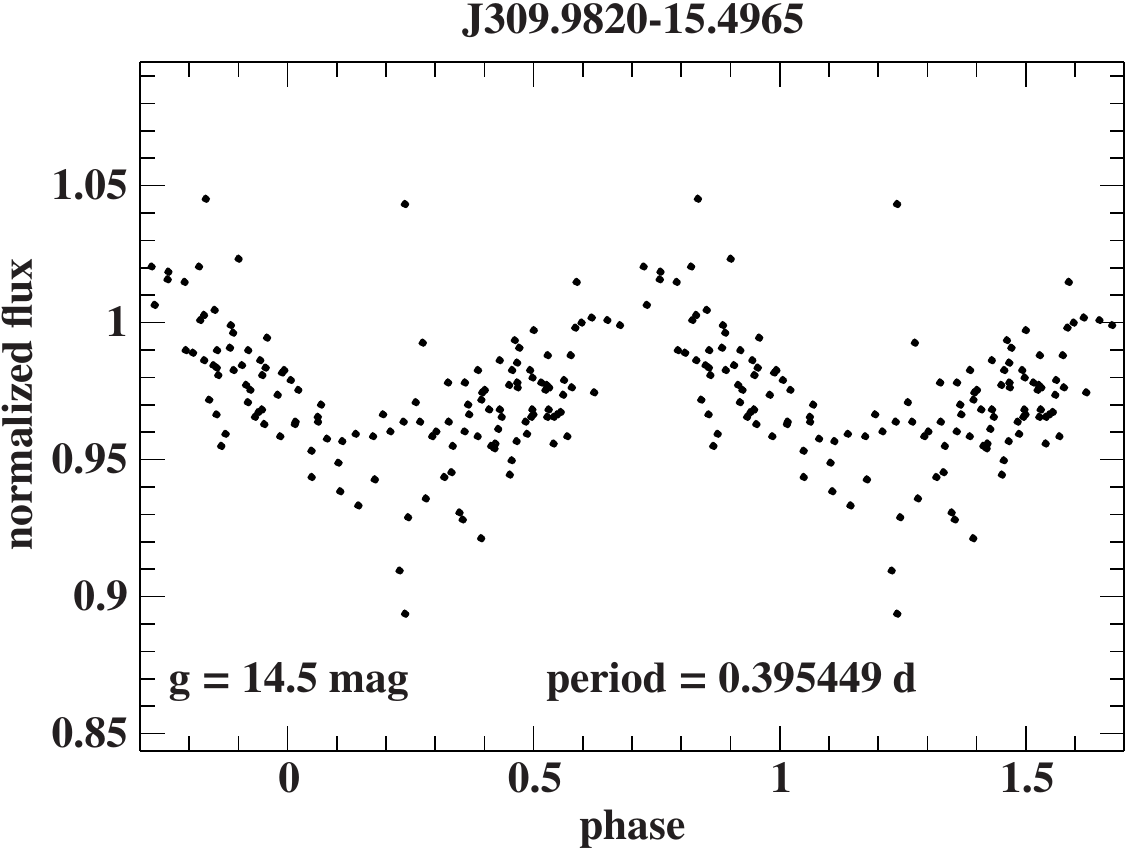}\hfill
		\includegraphics[width=0.25\linewidth]{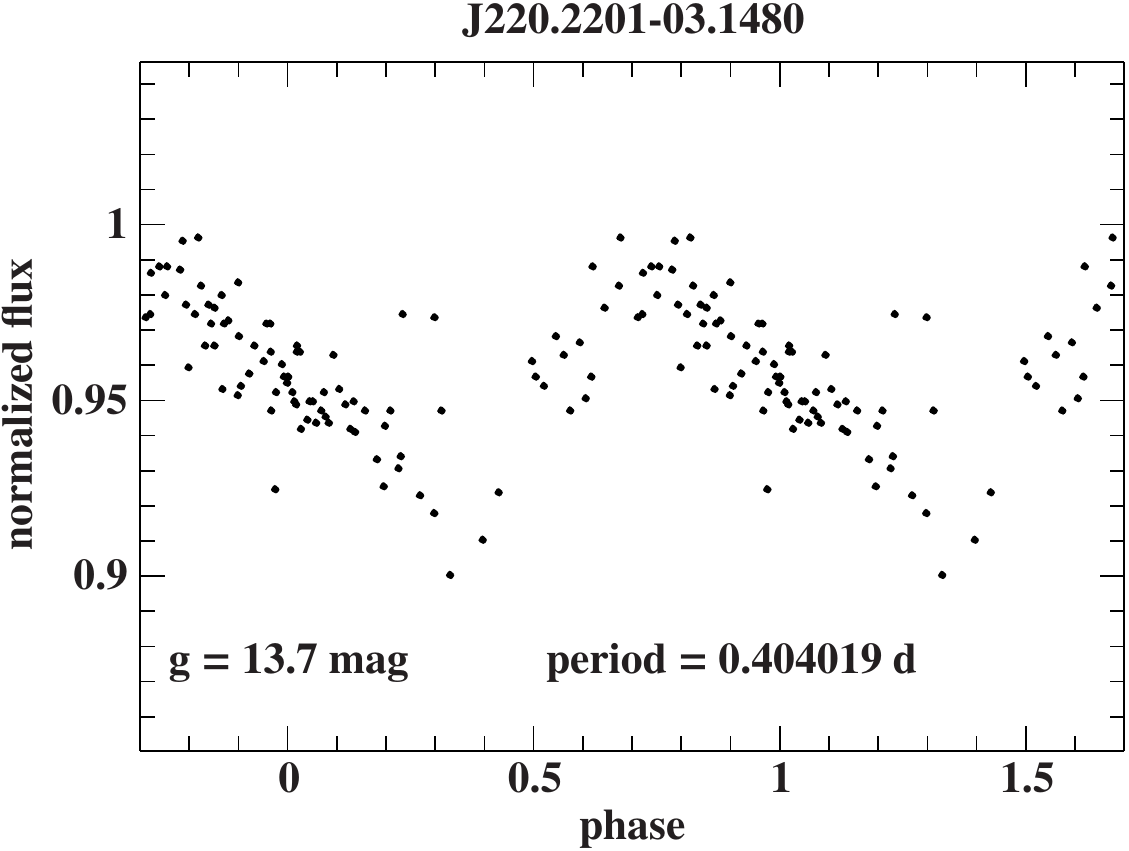}\hfill
		\includegraphics[width=0.25\linewidth]{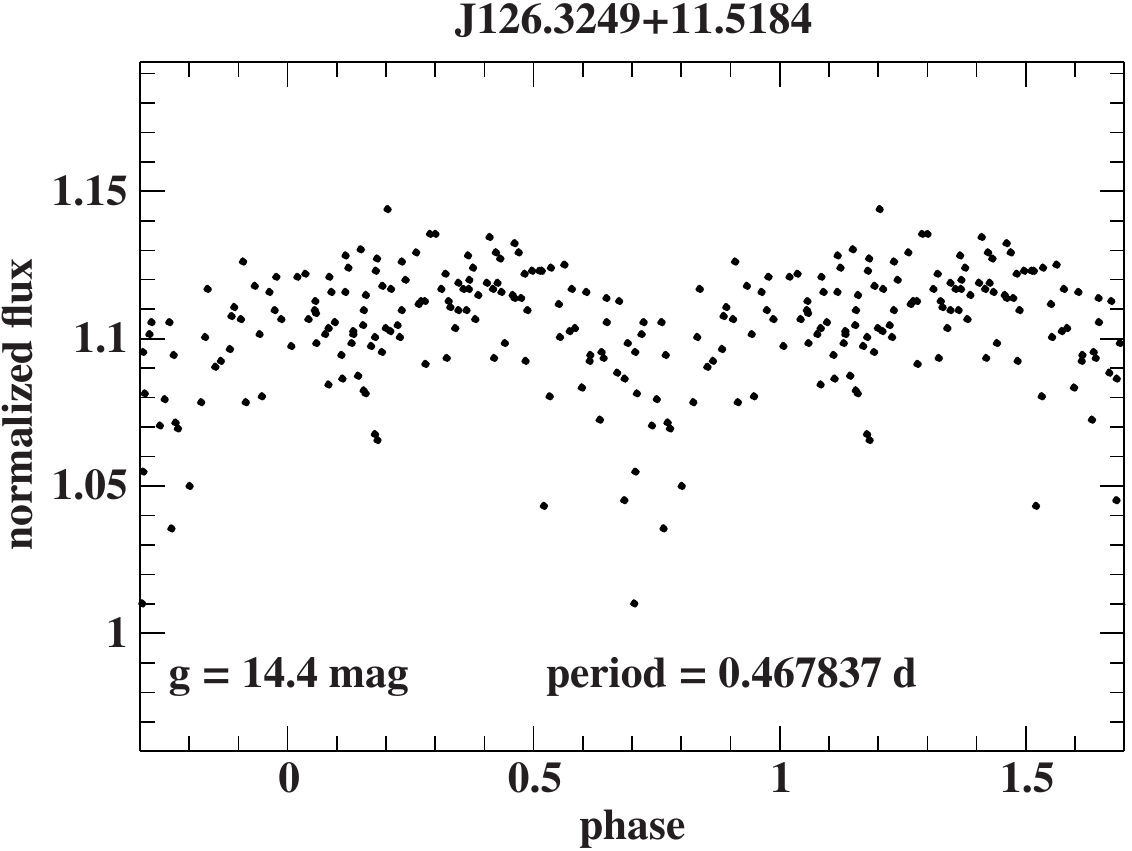}\hfill
		\includegraphics[width=0.25\linewidth]{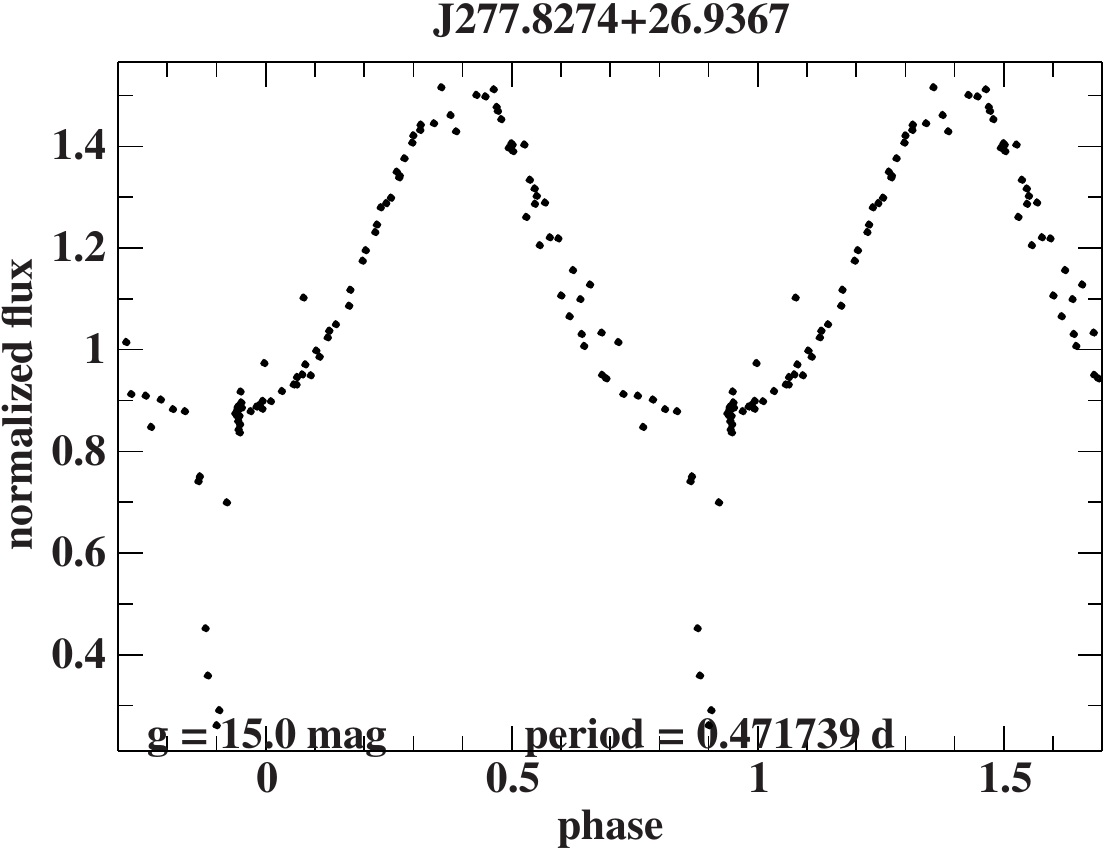}\hfill
		\includegraphics[width=0.25\linewidth]{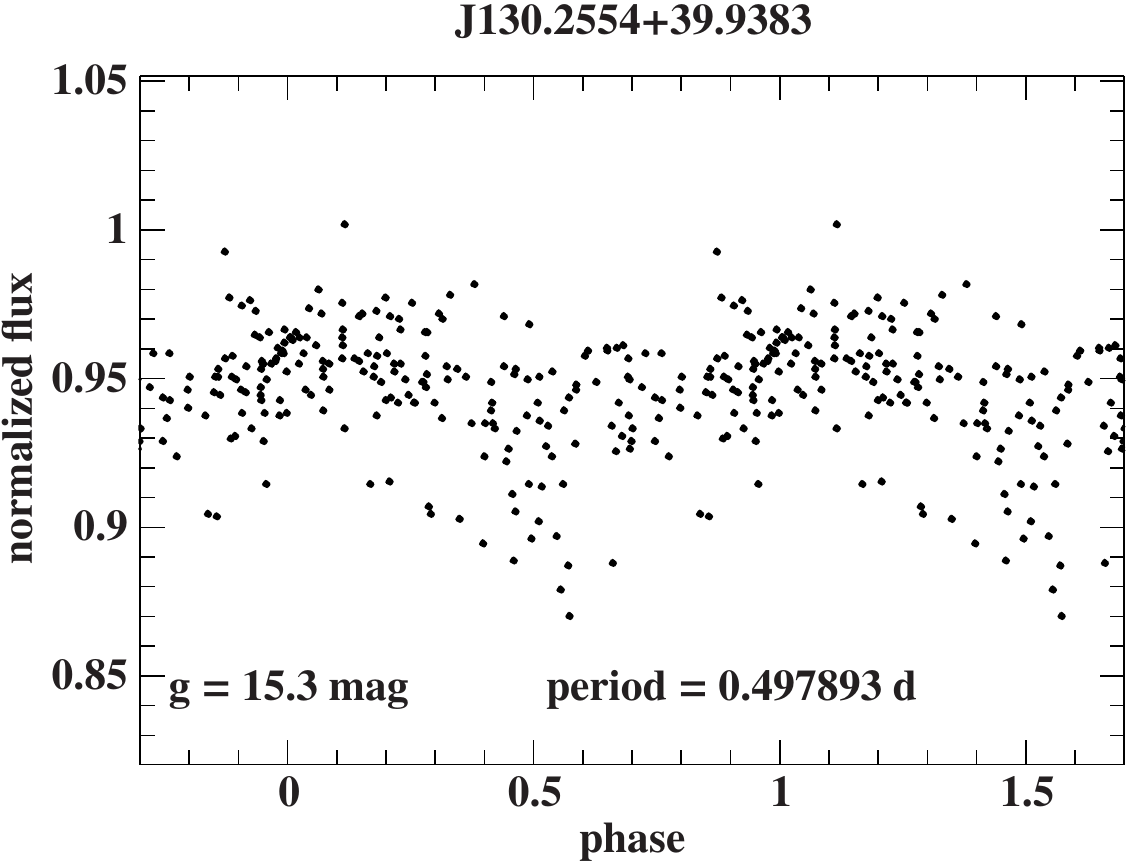}\hfill
		\includegraphics[width=0.25\linewidth]{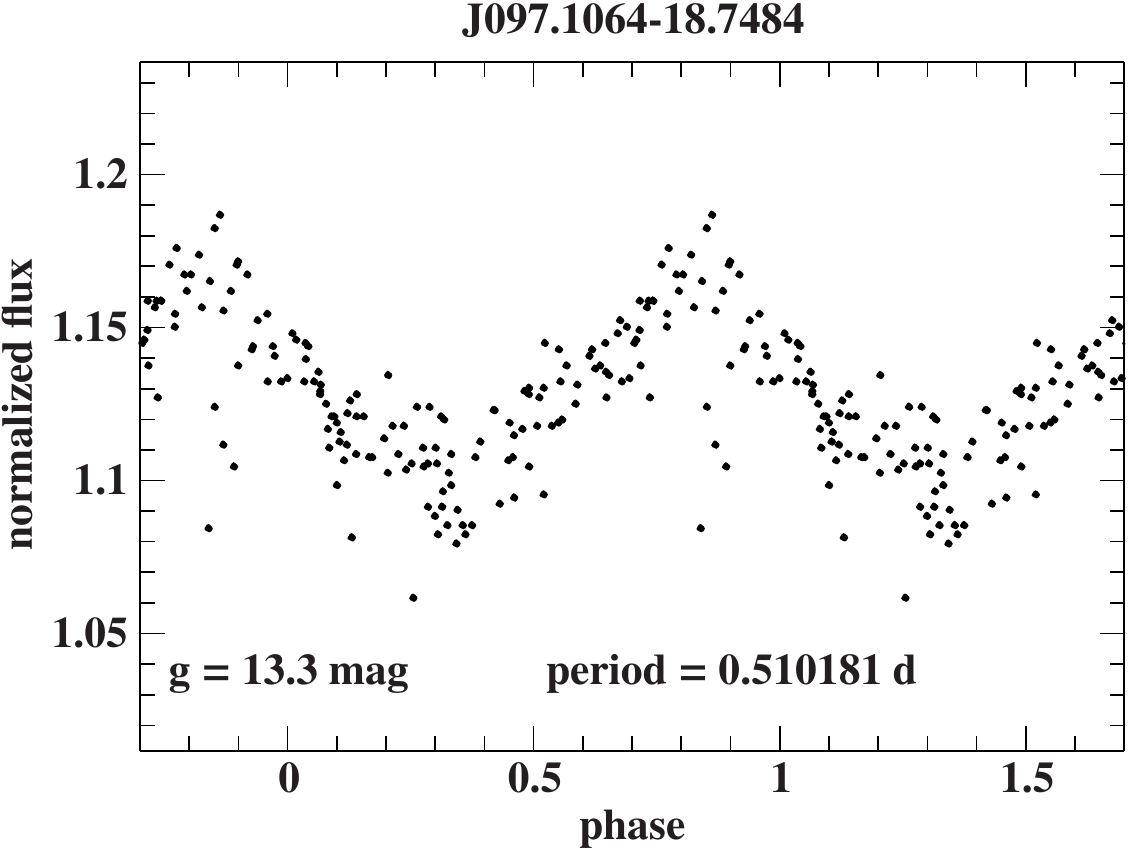}\hfill
		\includegraphics[width=0.25\linewidth]{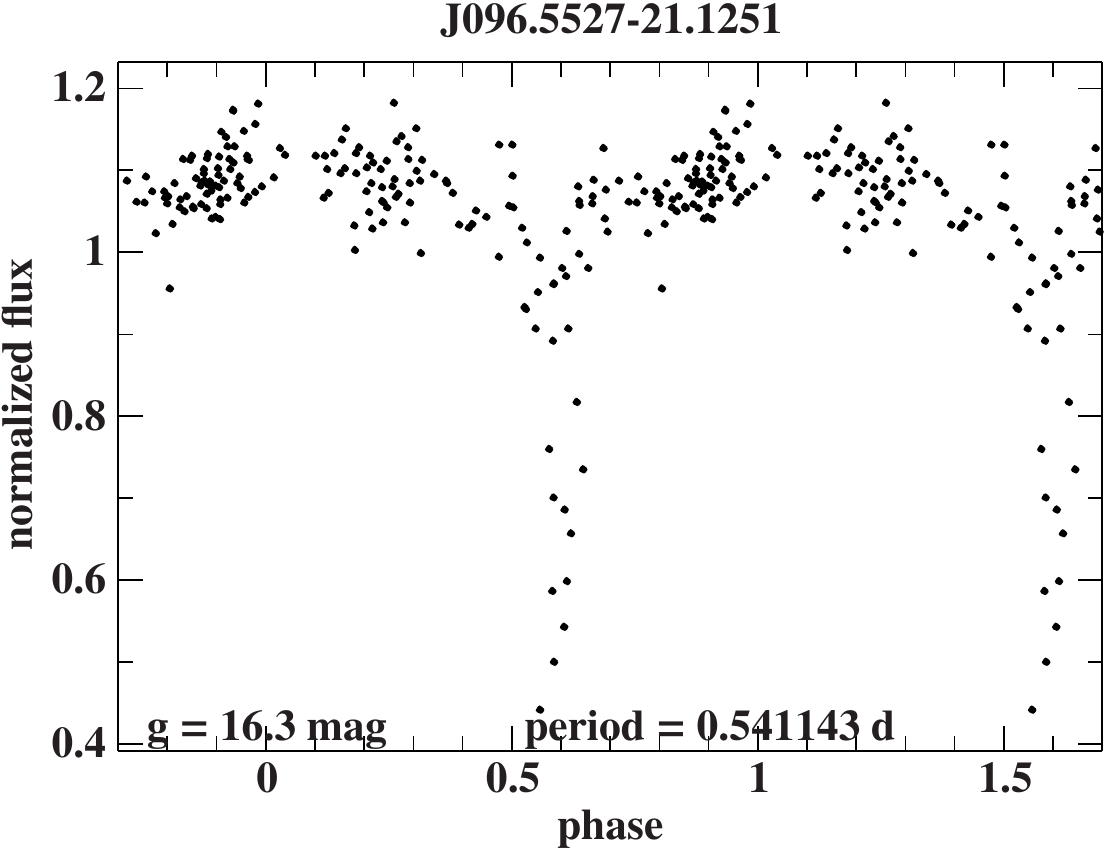}\hfill
		\includegraphics[width=0.25\linewidth]{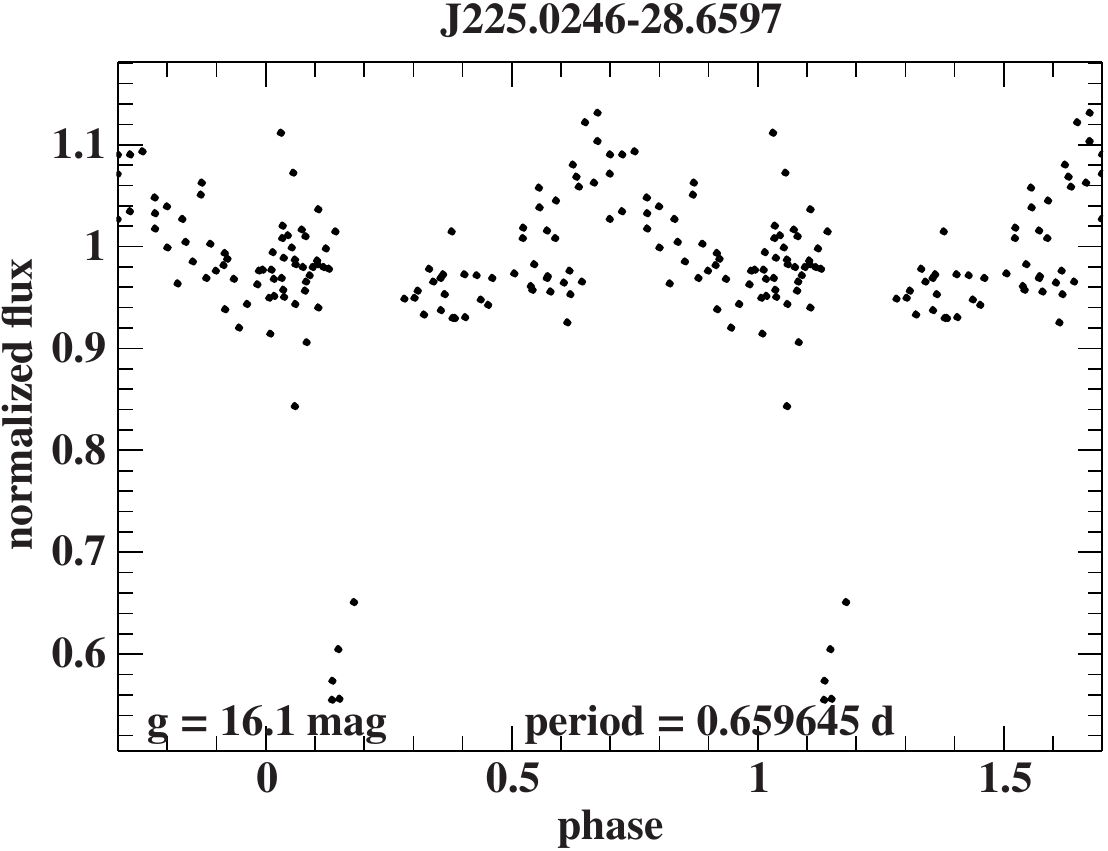}\hfill
		\includegraphics[width=0.25\linewidth]{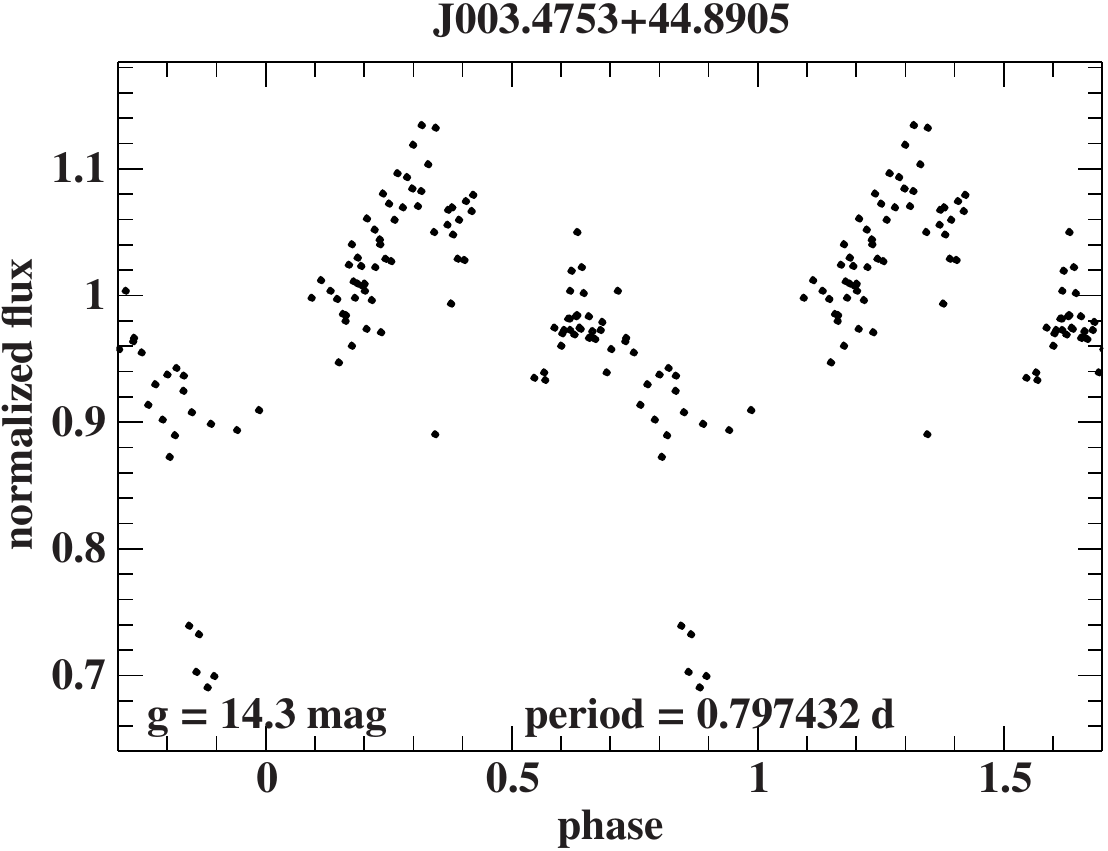}\hfill
		\includegraphics[width=0.25\linewidth]{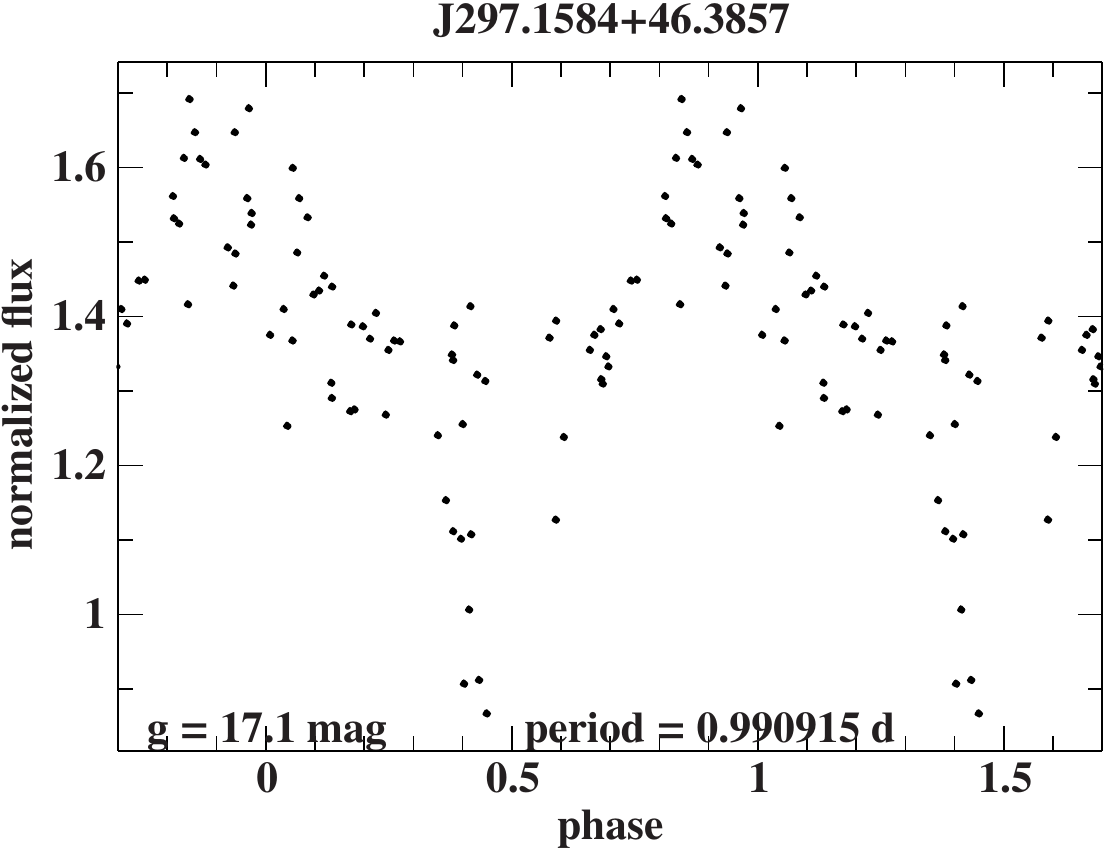}\hfill
		\includegraphics[width=0.25\linewidth]{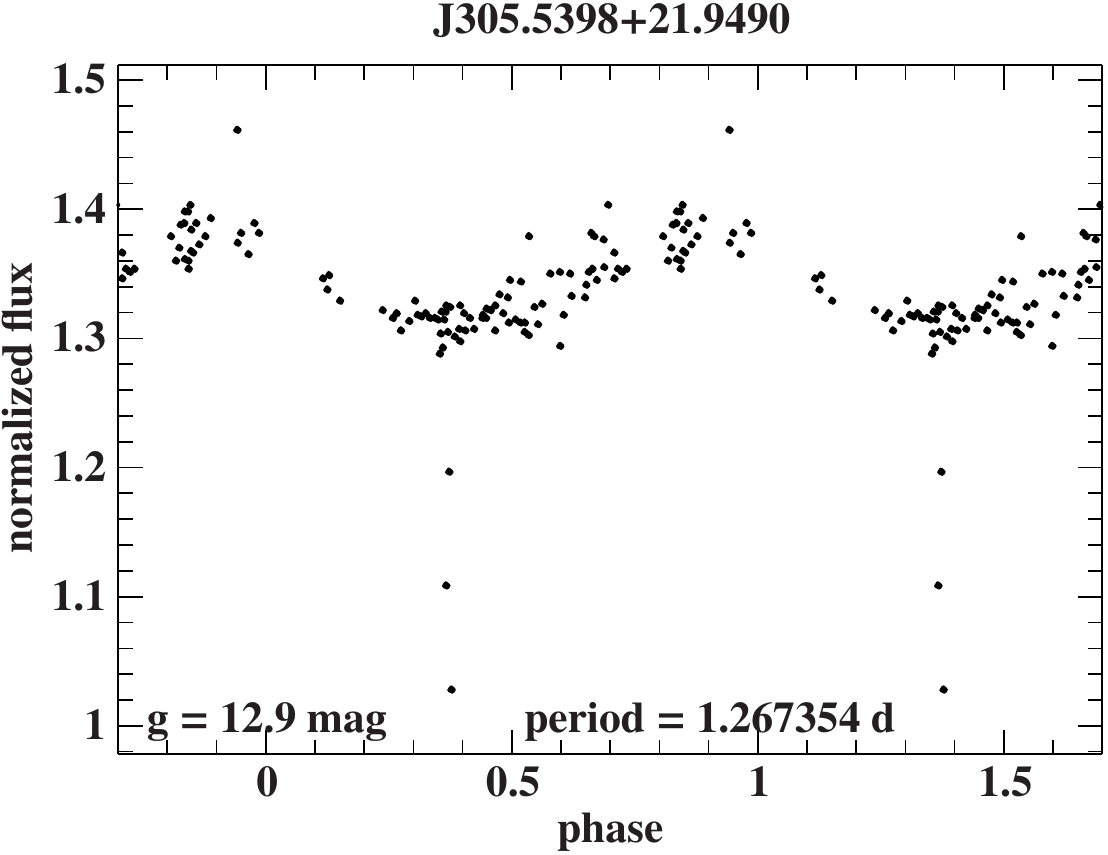}\hfill
		
	\end{figure}

	
	\setlength{\tabcolsep}{15pt}
	\begin{landscape}
		\tiny
		\setlength{\tabcolsep}{0.5mm}
\begin{longtable}{lccccccccccccccc}
			\caption{Orbital parameters, Gaia and OGLE magnitudes and coordinates of the HW Vir candidates and published HW Vir systems (objects in bold face were confirmed by deriving the atmospheric parameters)}\\
			\toprule \hline
			Target & RA (OGLE) & DEC (OGLE) & RA (Gaia) & RA(Gaia)& I & V & G & $\Delta \rm G$& $\rm G_{\rm BP}$ & $\Delta \rm G_{\rm BP}$& $\rm G_{\rm RP}$ & $\Delta \rm G_{\rm RP}$& $\rm G_{\rm BP}$-$\rm G_{\rm RP}$& P & $T_0$\\
			& $[\mathrm{h}]$ & $[^\circ]$ & $[\mathrm{h}]$ & $[^\circ]$ &$[\mathrm{mag}]$&$[\mathrm{mag}]$&$[\mathrm{mag}]$&$[\mathrm{mag}]$&$[\mathrm{mag}]$&$[\mathrm{mag}]$&$[\mathrm{mag}]$&$[\mathrm{mag}]$&$[\mathrm{mag}]$ &d&HJD-2400000\\  \midrule
			\multicolumn{16}{c}{Discovered by \citet{ogle_II}}\\ \midrule
			OGLE-BLG-ECL-000036 & 17:41:20.55 & -34:34:48.5 & 17:41:20.5412 & -34:34:48.546 & 18.945 & 19.623 & 19.5374 & 0.0154 & - & - & - & - & - & 0.05194084 & 7000.0499\\
			OGLE-BLG-ECL-000052 & 17:47:43.31 & -23:31:22.1 & 17:47:43.3091 & -23:31:22.252 & 18.499 & 19.808 & 19.5879 & 0.0091 & 19.5473 & 0.0777 & 17.9754 & 0.0826 & 1.5719 & 0.05258043 & 7000.0167\\
			OGLE-BLG-ECL-000176 & 18:12:13.63 & -33:09:43.7 & 18:12:13.6385 & -33:09:43.583 & 19.730 & 19.594 & 19.7050 & 0.0133 & 19.1867 & 0.1426 & 19.8247 & 0.9303 & -0.6380 & 0.05274168 & 7000.0339\\
			\textbf{OGLE-BLG-ECL-000212} & 18:36:22.73 & -23:15:45.0 & 18:36:22.7387 & -23:15:45.340 & 19.006 & 19.325 & 19.6910 & 0.0120 & - & - & - & - & - & 0.06258789 & 7000.0820\\
\end{longtable}
		\begin{longtable}{lccccccccccccccc} 	
			\caption{Orbital parameters, OGLE and Gaia magnitudes and coordinates of the white dwarf candidates (objects in bold face were confirmed by deriving the atmospheric parameters)}\\ 
			\toprule \hline
			Target & RA (OGLE) & DEC (OGLE) & RA (Gaia) & RA(Gaia)& I & V & G & $\Delta \rm G$& $\rm G_{\rm BP}$ & $\Delta \rm G_{\rm BP}$& $\rm G_{\rm RP}$ & $\Delta \rm G_{\rm RP}$& $\rm G_{\rm BP}$-$\rm G_{\rm RP}$& P & $T_0$\\
			
			& $[\mathrm{h}]$ & $[^\circ]$ & $[\mathrm{h}]$ & $[^\circ]$ &$[\mathrm{mag}]$&$[\mathrm{mag}]$&$[\mathrm{mag}]$&$[\mathrm{mag}]$&$[\mathrm{mag}]$&$[\mathrm{mag}]$&$[\mathrm{mag}]$&$[\mathrm{mag}]$&$[\mathrm{mag}]$ &d&HJD-2400000\\  \midrule
			
			\textbf{OGLE-BLG-ECL-000091$^1$} & 17:56:38.74 & -31:16:28.8 & 17:56:38.8029 & -31:16:28.883 & 18.254 & - & 19.1304 & 0.0109 & 18.5850 & 0.0853 & 17.1033 & 0.1069 & 1.4818 & 0.06742161 & 7000.0016\\
			OGLE-BLG-ECL-207454$^2$ & 17:57:20.62 & -29:33:50.3 & 17:57:20.3861 & -29:33:51.328 & 18.525 & 19.310 & 19.4611 & 0.0130 & - & - & - & - & - & 0.1848801 & 7000.1690\\
			OGLE-BLG-ECL-192029$^2$ & 17:56:01.82 & -27:13:24.6 & 17:56:01.8191 & -27:13:24.458 & 19.022 & 19.663 & 19.5699 & 0.0059 & 19.8330 & 0.0504 & 18.9572 & 0.1026 & 0.8758 & 0.1358285 & 7000.0366\\
			OGLE-GD-ECL-09869$^5$ & 13:24:02.36 & -64:41:21.9 & 13:24:02.3654 & -64:41:21.859 & 19.473 & - & 20.1570 & 0.0080 & 19.9348 & 0.0772 & 19.3598 & 0.1178 & 0.5750 & 0.09596148 & 3836.77432\\
			OGLE-GD-ECL-07446$^5$ & 11:31:05.84 & -61:04:36.4 & 11:31:05.8451 & -61:04:36.393 & 18.797 & - & 19.5351 & 0.0091 & - & - & - & - & - & 0.1165955 & 2735.17078\\
\end{longtable}
		\begin{longtable}{lcccccccccccccccc} 	
			\caption{Orbital parameters, OGLE and Gaia magnitudes and coordinates of central stars of planetary nebula from \citet{cat_pn}}\\ 
			\toprule \hline
			Target & Planetary nebula& RA (OGLE) & DEC (OGLE) & RA (Gaia) & RA(Gaia)& I & V & G & $\Delta \rm G$& $\rm G_{\rm BP}$ & $\Delta \rm G_{\rm BP}$& $\rm G_{\rm RP}$ & $\Delta \rm G_{\rm RP}$& $\rm G_{\rm BP}$-$\rm G_{\rm RP}$& P & $T_0$\\
			
			&& $[\mathrm{h}]$ & $[^\circ]$ & $[\mathrm{h}]$ & $[^\circ]$ &$[\mathrm{mag}]$&$[\mathrm{mag}]$&$[\mathrm{mag}]$&$[\mathrm{mag}]$&$[\mathrm{mag}]$&$[\mathrm{mag}]$&$[\mathrm{mag}]$&$[\mathrm{mag}]$&$[\mathrm{mag}]$ &d&HJD-2400000\\  \midrule
			OGLE-BLG-ECL-000045$^1$ &	Pe 1-9  & 17:45:36.77 & -23:02:26.1 & 17:45:36.7590 & -23:02:26.307 & 17.736 & 19.166 & 18.7573 & 0.0071 & 18.4291 & 0.1355 & 17.3009 & 0.0376 & 1.1283 & 0.13985716 & 7000.0072\\
			OGLE-BLG-ECL-000111$^1$  & BMP 1800-3408 & 18:00:26.16 & -34:08:02.3 & 18:00:26.1596 & -34:08:02.348 & 17.825 & 18.132 & 18.3643 & 0.0228 & 18.0724 & 0.0605 & 17.5176 & 0.0542 & 0.5548 & 0.14477890 & 7000.1238\\
			OGLE-BLG-ECL-159262$^3$  & H 2-29&17:53:16.82 & -32:40:38.5 & 17:53:16.8248 & -32:40:38.464 & 18.308 & 19.313 & 19.0919 & 0.0084 & - & - & - & - & - & 0.2441101 & 7000.2000\\
			OGLE-BLG-ECL-211852$^3$ & PHR 1757-2824& 17:57:42.29 & -28:24:07.8 & 17:57:42.2886 & -28:24:07.924 & 18.194 & 19.688 & 19.4525 & 0.0359 & - & - & - & - & - & 0.7992066 & 7000.7602\\
			J277.8274+26.9367$^4$ & PNA6646/V477Lyrae & - & - & 18 31 18.5911 &	+26 56 12.110 & -& -&14.9397 &	0.0241 &	14.7389 &	0.0589 &	15.0386 &	0.0702 &	-0.2997 &  0.471739 & \\
			\bottomrule
			\label{orbit_pn}
		\end{longtable}
		\footnotetext[1]{Discovered by \citet{ogle_II}}
		\footnotetext[2]{Discovered by visual inspection}
		\footnotetext[3]{Discovered by machine-learning}
		\footnotetext[4]{Discovered by \citet{ogle}}
		\footnotetext[5]{Discovered by cross-match of the sdB candidate catalogue \citep{gaia_catalog} with the ATLAS survey}

		\newpage
		\begin{longtable}{lcccccccccccccccc} 
			\caption{Gaia parallaxes, proper motions and calculated reduced proper motions, distances, absolute G magnitudes and reddening of the HW Vir candidates and published HW Vir systems}\\
			\toprule \hline
			Target & parallax  & proper motion in RA & proper motion in DEC & $\rm M_G$ & reduced proper motion & distance & distance \citep{distances}&$A_G$ & E(B-V) & reddening corrected $M_G$\\
			& mas  &mas/yr &mas/yr&mag&&kpc&kpc\\  \midrule	
			\multicolumn{16}{c}{Discovered by \citet{ogle_II}}\\ \midrule
			
			OGLE-BLG-ECL-000036 & -1.6483 $\pm$ 1.0676 & 2.780 $\pm$ 1.490 &0.424 $\pm$ 1.406 &- &11.78$\pm$1.24 &- & $4.814^{+3.884}_{-2.395}$&2.15&0.7&$3.97_{-1.28}^{+1.49}$\\
			OGLE-BLG-ECL-000176 & 0.5533 $\pm$ 0.8016 & -3.889 $\pm$ 1.854 &-6.848 $\pm$ 1.823 &8.41$\pm$3.14 &14.18$\pm$0.30 &1.807$\pm$2.618 & $4.391^{+4.276}_{-2.541}$&0.87&0.25&$5.62_{-1.47}^{+1.87}$\\
			OGLE-BLG-ECL-000095 & -1.0688 $\pm$ 0.8665 & -3.316 $\pm$ 1.539 &-5.131 $\pm$ 1.273 &- &13.34$\pm$0.37 &-& $5.388^{+4.396}_{-2.699}$&0.85&0.30&$4.90_{-1.29}^{+1.50}$\\
			\textbf{OGLE-BLG-ECL-000124} & 0.6039 $\pm$ 0.1953 & 0.801 $\pm$ 0.269 &-4.044 $\pm$ 0.226 &6.66$\pm$0.70 &10.83$\pm$0.48 &1.655$\pm$0.535 & $1.829^{+1.958}_{-0.643}$&0.85&0.297&$5.59_{-1.58}^{+0.94}$\\
\end{longtable}
		
		\newpage
		\begin{longtable}{lcccccccccccccccc}
			\caption{Gaia parallaxes, proper motions and calculated reduced proper motions, distances, absolute G magnitudes and reddening of the white dwarf candidates}\\	
			\toprule \hline
			Target & parallax  & proper motion in RA & proper motion in DEC & $\rm M_G$ & reduced proper motion & distance & distance \citep{distances}&$A_G$ & E(B-V) & reddening corrected $M_G$\\
			& mas  &mas/yr &mas/yr&mag&&kpc&kpc\\  \midrule
			\textbf{OGLE-BLG-ECL-000091$^1$} & 2.4371 $\pm$ 0.4993 & -8.415 $\pm$ 0.815 &-11.326 $\pm$ 0.709 &11.0$\pm$0.44 &14.87$\pm$0.13 &0.410$\pm$0.084 & $0.441^{+0.191}_{-0.101}$&0.30&0.128&$10.6_{-0.78}^{+0.57}$\\	
			OGLE-BLG-ECL-207454$^2$ & 6.3403 $\pm$ 1.4567 & -7.147 $\pm$ 1.491 &-7.039 $\pm$ 1.391 &13.4$\pm$0.49 &14.46$\pm$0.26 &0.157$\pm$0.036 & $0.177^{+5.471}_{-0.177}$&0.18&0.065&$13.0_{-7.51}^{+inf}$\\
			OGLE-BLG-ECL-192029$^2$ & 2.0833 $\pm$ 0.4635 & -2.160 $\pm$ 1.046 &-14.402 $\pm$ 0.883 &11.1$\pm$0.48 &15.38$\pm$0.14 &0.480$\pm$0.106 & $0.522^{+0.304}_{-0.139}$&0.415&0.157&$10.5_{-0.99}^{+0.67}$\\
			OGLE-GD-ECL-09869$^5$ & 0.6784 $\pm$ 0.5749 & -10.260 $\pm$ 1.476 &-2.394 $\pm$ 1.111 &9.31$\pm$1.84 &15.27$\pm$0.29 &1.474$\pm$1.249 & $2.753^{+3.088}_{-1.548}$&1.32&0.45&$6.63_{-1.63}^{+1.79}$\\
			OGLE-GD-ECL-07446$^5$ & 0.5187 $\pm$ 0.3764 & -8.549 $\pm$ 0.886 &1.546 $\pm$ 0.637 &8.10$\pm$1.57 &14.22$\pm$0.22 &1.927$\pm$1.398 & $2.384^{+2.505}_{-1.110}$&0.80&0.275&$6.84_{-1.55}^{+1.36}$\\
\end{longtable}
		\begin{longtable}{lcccccccccccccccc}
			\caption{Gaia parallaxes, proper motions and calculated reduced proper motions, distances, absolute G magnitudes and reddening of central stars of planetary nebula from \citet{cat_pn}}\\	
			\toprule \hline
			Target & parallax  & proper motion in RA & proper motion in DEC & $\rm M_G$ & reduced proper motion & distance & distance \citep{distances}&$A_G$ & E(B-V) & reddening corrected $M_G$\\
			& mas  &mas/yr &mas/yr&mag&&kpc&kpc\\  \midrule
			
			OGLE-BLG-ECL-000111$^1$ & -0.6078 $\pm$ 0.3030 & -1.016 $\pm$ 0.451 &-4.317 $\pm$ 0.403 &- &11.59$\pm$0.43 &- & $7.238^{+4.519}_{-2.906}$&1.48&0.5&$2.58_{-1.05}^{+1.11}$\\
			OGLE-BLG-ECL-159262$^3$ & 0.8984 $\pm$ 0.5485 & -6.094 $\pm$ 0.961 &-8.566 $\pm$ 0.739 &8.85$\pm$1.32 &14.20$\pm$0.19 &1.113$\pm$0.679 & $3.196^{+4.243}_{-2.045}$&1.49&0.5&$5.07_{-1.83}^{+2.21}$\\
			J277.8274+26.9367$^4$ & 0.3286 $\pm$	0.0440 &	-1.936 $\pm$	0.052 &	0.692 $\pm$	0.070 & $2.52 \pm 0.29$ & 6.5 $\pm$ 0.39 & $3.099 \pm 0.415$ & $2.768^{+0.394}_{-0.310}$\\
			
			
			\bottomrule
			\label{gaia_pn}
		\end{longtable}
		
		\footnotetext[1]{Discovered by \citet{ogle_II}}
		\footnotetext[2]{Discovered by visual inspection}
		\footnotetext[3]{Discovered by machine-learning}
		\footnotetext[4]{Discovered by \citet{ogle}}
		\footnotetext[5]{Discovered by cross-match of the sdB candidate catalogue \citep{gaia_catalog} with the ATLAS survey}
		
	\end{landscape}

\end{document}